\documentclass{aa}  

\usepackage{graphicx}
\usepackage{txfonts}

\usepackage{hyperref}
\hypersetup{
     colorlinks = true,
     linkcolor = blue,
     anchorcolor = blue,
     citecolor = blue,
     filecolor = blue,
     urlcolor = blue
}
\usepackage{bm}
\usepackage{caption}
\usepackage{subcaption}
\usepackage{bigstrut}
\usepackage{xspace}

\mathchardef\mhyphen="2D  

\newcommand{\revisionA}[1]{#1}

\newcommand{\revisionC}[1]{#1}
\newcommand{\hh}{$\mathrm{H_2}$\xspace}
\newcommand{\shtg}{HUV-LH-$\mathrm{T_{g}}$\xspace}
\newcommand{\shteff}{HUV-LH-$\mathrm{T_{a}}$\xspace}
\newcommand{\shb}{HUV-B14-$\mathrm{T_{a}}$\xspace}
\newcommand{\sltg}{LUV-LH-$\mathrm{T_{g}}$\xspace}
\newcommand{\slteff}{LUV-LH-$\mathrm{T_{a}}$\xspace}
\newcommand{\slb}{LUV-B14-$\mathrm{T_{a}}$\xspace}

\newcommand{\mhlh}{M-HUV-LH\xspace}
\newcommand{\mhb}{M-HUV-B14\xspace}
\newcommand{\mllh}{M-LUV-LH\xspace}
\newcommand{\mlb}{M-LUV-B14\xspace}

\begin{document}

   \title{Impact of Size-dependent Grain Temperature on Gas-Grain Chemistry in Protoplanetary Disks: the case of low mass star disks }

\titlerunning{Impact of Size-dependent Grain Temperature on Chemistry in Protoplanetary Disks}

 \author {
       S. Gavino \inst{1}
        \and A. Dutrey \inst{1}
        \and V.Wakelam \inst{1}        
        \and S. Guilloteau \inst{1}
        \and J.Kobus \inst{2}   
        \and S.Wolf \inst{2}   
        \and W. Iqbal \inst{3}
        \and E. Di Folco  \inst{1} 
        \and E. Chapillon \inst{1,4}             
	\and V. Pi\'etu \inst{4}
	 }      
 \institute{
		Laboratoire d'Astrophysique de Bordeaux, Universit\'e de Bordeaux, CNRS, B18N, 
			  All\'ee Geoffroy Saint-Hilaire, F-33615 Pessac
		 \and University of Kiel, Institute of Theoretical Physics and Astrophysics, Leibnizstrasse 15, 24118 Kiel, Germany 
		 \and South-Western Institute for Astronomy Research (SWIFAR), Yunnan University (YNU), Kunming 650500, People's Republic of China
		 \and IRAM, 300 rue de la piscine, F-38406 Saint Martin d'H\`eres Cedex, France
}

\offprints{ Sacha Gavino\\
\email{sacha.gavino@nbi.ku.dk}}

   \date{Received June 30, 2020; accepted May 31, 2021}

 
  \abstract
{Grain surface chemistry is key to the composition of protoplanetary disks around young stars.} 
{The temperature of grains depends on their size. We evaluate the impact of this temperature dependence on the disk chemistry.} 
{We model a moderately massive disk with 16 different grain sizes. We 
use the 3D Monte-Carlo POLARIS code to calculate the dust grain 
temperatures and the local UV flux.  We model the chemistry using the 
3-phase astrochemical code NAUTILUS. Photo processes are handled using 
frequency-dependent cross-sections, and a new method to account for 
self and mutual shielding. The multi-grain model outputs are compared 
to those of single-grain size models (0.1 $\mu$m), with two different 
assumptions for their equivalent temperature.}
{We find that the Langmuir-Hinshelwood (LH) mechanism at equilibrium temperature is not efficient to form H$_2$ at 3-4 scale heights ($H$), and adopt a 
parametric fit to a stochastic method to model H$_2$ formation instead. 
We find the molecular layer composition (1-3\,$H$) to depend on the 
amount of remaining H atoms. Differences in molecular surface densities 
between single and multi-grain models are mostly due to what occurs 
above 1.5\,$H$. At 100 au, models with colder grains produce H$_2$O and 
CH$_4$ ices in the midplane, and warmer ones produce more CO$_2$ ices, both
allowing efficient depletion of C and O as soon as CO sticks on grain surfaces.
Complex organic molecules (COMs) production is enhanced by the presence of warmer grains in
the multi-grain models. \revisionC{Using a single grain model mimicking grain growth and dust settling fails to reproduce the complexity of gas-grain chemistry.}}
{\revisionA{Chemical models with a single grain size are sensitive to the adopted
grain temperature, and  cannot account for all 
expected effects.
A spatial spread of the snowlines is expected to result from
the ranges in grain temperature. The amplitude
of the effects will depend on the dust disk mass.}}



   \maketitle
\section{Introduction}
\label{sec:intro}

As planets form in protoplanetary disks, studying the physical and 
chemical evolution of their gas and dust content along the whole disk 
lifetime is important in order to investigate how planetary formation 
proceeds. In the early stages of disks (the so-called protostellar 
phase) small dust grains are essentially dynamically well coupled to 
the gas. \revisionA{When} grain growth occurs  (up to at least cm-sized particles) 
larger particles, which dynamically decouple from the gas-phase, 
settle toward the disk midplane. This vertically changes the 
gas-to-dust ratio in the disk. Moreover, grain growth also allows UV to 
penetrate deeper in the disk, changing its chemistry.

As a consequence of the complex interplay of physical and chemical 
processes, it is now demonstrated that the disk structure can be 
separated vertically in three distinct regions: from top to bottom, the 
irradiated tenuous disk atmosphere (above $3-4\,H$ or scale heights) where 
the gas can be hot and is ionized or in atomic form, the warm denser 
molecular layer ($1-3\,H$) where most molecules reside and below $1\,H$, the 
midplane. Beyond the CO snowline, which is typically located at a 
radius about 20 au in a TTauri disk, the midplane is very dense and 
cold ($\leq 20$ $\mathrm{K}$). This area is essentially shielded from the
stellar radiation, with both low ionization and turbulence levels. In 
this cold region, most molecules are stuck onto dust grains which 
exhibit icy mantles that can still be processed by cosmic rays.

Thanks to ALMA, several high angular resolution and sensitive  
observations of disks orbiting either low or intermediate mass young 
stars (TTauri and Herbig Ae stars) have quantitatively confirmed this 
general scheme. CO observations of HD 163296 have clearly shown the CO 
gas-phase depletion in the dark, cold disk midplane \citep{ 
De-Gregorio-Mionsalvo+etal_2013}, revealing also that the CO gas is 
located in the intermediate warm molecular layer.  More recently, the 
ALMA observations of the edge-on TTauri disk called the Flying Saucer 
\citep{Dutrey+etal_2017} directly show the molecular layer at 
intermediate scales ($1-2 H$) with CS, the denser gas tracer being 
located at about $1 H$, below the CO emission ($1-3 H$) which better traces 
the whole envelope of the molecular layer. 

In the last 10 years, several astrochemical models have emerged that 
attempt to incorporate the observed complexity of disk physics \revisionA{in order} to 
provide more accurate molecular abundance and surface density 
predictions. While the very first models included only the gas-phase 
chemistry, gas-grain coupling was added to take into account firstly 
adsorption and desorption for a few molecules 
\citep[e.g.][]{Woitke+etal_2009} and then reactions on the grain 
surface \citep[e.g.][]{Semenov+Wiebe_2011, Walsh+etal_2014}.  Another 
step was finally achieved with models which include 3 material phases: 
the gas-phase and 2 phases for the gas-grain chemistry with grains 
modeled by an icy surface layer and an active mantle 
\citep{Ruaud+Gorti_2019, Wakelam+etal_2019}. Thermochemical models  
which calculate the density and thermal structures in a self-consistent 
manner have been also developed \citep[e.g.][]{Hollenbach+Gorti_2009}. 
Recently, \citet{Ruaud+Gorti_2019} have also coupled such a 
thermochemical model \citep{Gorti+etal_2011} to a 3-phase chemical 
model.      

An important improvement in chemical models was to introduce more 
realistic dust disk structures by taking into account grain growth and 
dust settling. Grain growth reduces the dust cross-section and 
therefore increases the UV penetration while dust settling locally 
changes the gas-to-dust ratio. Grain growth was first introduced in 
disk chemistry by  \citet{Aikawa+Nomura_2006}. Dust settling is often 
mimicked by adding larger grains in the midplane 
\citep[e.g.][]{Wakelam+etal_2019}. 

\revisionA{Several authors have already studied the effect of multiple 
grain sizes onto chemistry. As an example,  \citet{Acharyya+etal_2011}  
used five different grain sizes in their models but the grain 
temperature was the same for all grains. Following this work, 
\citet{Pauly+Garrod_2016} tested the impact of a simple temperature
distribution, with $T(a) \propto a^{-(1/6)}$, $a$ being the grain size,
while \citet{Ge+etal_2016} explored a small range of temperatures
(14.9--17.9) for the Cold Neutral Medium. More recently, \citet{Iqbal+Wakelam_2018} 
also developed a model handling different grain sizes.}

\revisionA{On another hand, \citet{Akimkin+etal_2013} developed 
a disk model where they couple the dust grain time evolution (grain 
growth and settling) to chemistry. Another improvement was achieved 
with the development of grids of thermochemical models allowing 
multiwavelength fitting of dust and molecular lines such as DIANA 
\citep{Woitke+etal_2019}. 
However, these elaborate approaches use a single grain temperature 
which is either fixed or derived from a dust radiative transfer 
simulation while the temperature of a grain depends on its size 
\citep[e.g.][]{Wolf_2003}, larger grains being colder. 
\citet{Chapillon+etal_2008} found in the disk of CQ Tau, whose 
temperature is well above the CO snowline temperature, an important CO 
depletion. They suggested that CO may remain trapped onto larger grains 
that are cold enough to prevent thermal CO desorption. 
\citet{Heese+etal_2017} have investigated the dust temperature 
variations with grain size and position (radius and altitude above the 
midplane) in a typical TTauri disk using the 3D dust radiative 
transfer code Mol3D \citep{Ober+etal_2015}.  The dust temperature 
variations in the molecular layer appear to be significant 
enough to affect the disk chemistry. \revisionA{This dependency questions
the applicability of elaborate, self-consistent, thermochemical
models that rely on a single grain size and temperature.} }

\revisionA{We explore here the impact of the variations of the grain temperature 
with their size onto the chemistry of a representative TTauri disk. Our goal is not to 
build a complete chemical model that incorporates the best of all 
previous studies, neither to make grid models that could provide relevant comparisons 
with the observations. Instead, we only intend to check the limits of
an assumption that is used in most previous models, and explore the 
main consequences of dust grain temperature dependence with size.}

For this purpose, we have coupled the 3D Monte-Carlo continuum 
radiative transfer code \textsc{POLARIS} 
\citep{Reissl+Wolf+Brauer_2016} to the NAUTILUS Multi Grain Code 
\citep{Hersant+etal_2009, Ruaud+etal_2016} using the 3-phase version 
which also takes into account a grain size distribution 
\citep{Iqbal+Wakelam_2018}. We present the two codes and how we 
parametrize the gas and dust disk in Section \ref{sec:mod}. Section 
\ref{sec:model_param} deals with the disk model description and global 
results. We discuss in more detail the results in Section 
\ref{sec:discuss} and we state our conclusions in Section 
\ref{sec:concl}.

\section{Model description}
\label{sec:mod}

\begin{figure*}
  \centering
  \includegraphics[width=0.7\linewidth]{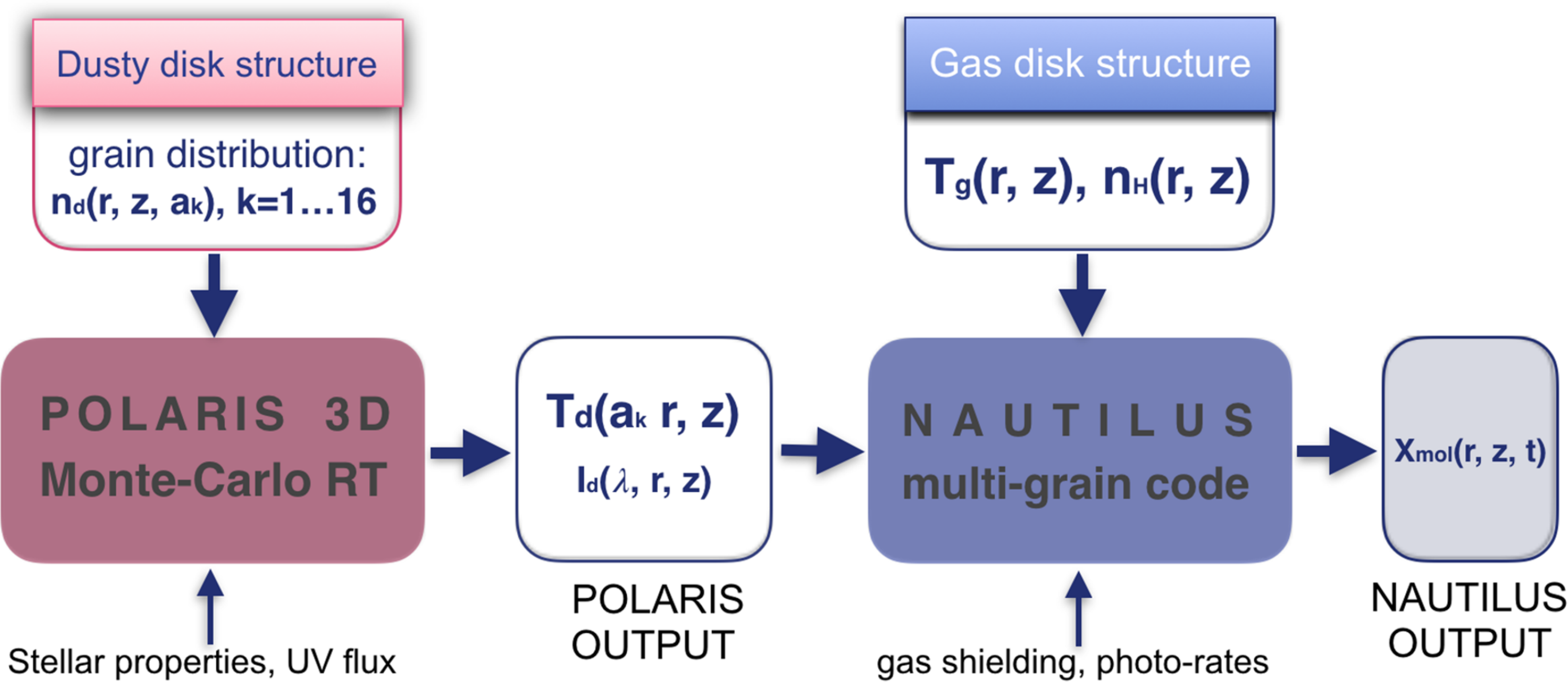}   
  \caption{Multi-grain model: Coupling scheme between NAUTILUS and POLARIS. }
\label{fig:coupling}
\end{figure*}

\revisionA{To estimate the impact of the grain temperature onto the 
chemistry, we couple the radiative 3D Monte-Carlo code POLARIS with the 
astrochemistry code NAUTILUS. This requires to independently assume a 
gas and dust disk structure, a gas temperature, a dust distribution and 
settling. Moreover, the coupling between the two codes requests for 
NAUTILUS a new method to calculate the dust extinction, the self and 
mutual shielding using the UV flux calculated by POLARIS in each disk 
cell. Figure \ref{fig:coupling} is a scheme summarizing how we proceed.  
}

\subsection{Disk Model}
\revisionA{Our disk structure is derived from a simple disk model
where truncated power  laws describe most of the quantities: surface density, midplane 
temperature, velocity field, with sharp inner and outer radii. A
detailed model description is given in Appendix \ref{app:model}} 

\revisionA{We adopt a power-law grain size distribution $dn(a) = C a^{-d} da$ with $dn(a)$, the number of grain 
of size $a$ and $C$ a normalization constant that can be derived from the dust mass 
(see Appendix \ref{app:dust}). The resulting distribution is 
discretized into 16 logarithmically distributed intervals, with grain 
sizes ranging from $a_{min} = 5\,$nm to $a_{max}= 1\,$mm. \revisionA{The fraction of mass and area per bin
are given in Table \ref{tab:grains}.}}

\revisionA{Our goal being to evaluate the impact of dust grain temperatures,
we assume a common gas temperature structure for all our simulations (including single and multi-grain models),
in order to avoid chemical effects that could be due to gas only.
We selected values derived from the observations of the Flying Saucer TTauri disk \citep[see][]{Dutrey+etal_2017}, 
whose edge-on orientation allow a direct measure of the 2D gas
temperature distribution from ALMA observations of the optically thick CO J=2-1 line. 
Values are given in Table \ref{tab:param}.}

\revisionA{The vertical distribution of gas is self-consistently
derived from the imposed temperature law.  However, to account for
dust settling, we assume grains of a given size follow a simple
Gaussian vertical profile, whose scale height is simply related
to the gas hydrostatic scale at the disk midplane temperature
according to a simple prescription (see Appendix \ref{app:dust}).}
\revisionA{Our model is therefore not fully consistent, since very small grains should follow the vertical profile of the gas,
but should be adequate for the larger grains observed in disks.
To compute the settling, we assume an $\alpha$ viscosity of 0.01 and a Schmidt number of $S_c = 1$. 
Although often assumed in disk modeling, these values are somewhat arbitrary,
\revisionA{and observations with ALMA are not yet accurate enough to allow for a quantitative
description of dust settling.} }

\begin{figure*} 
  \centering
  \includegraphics[width=1\linewidth]{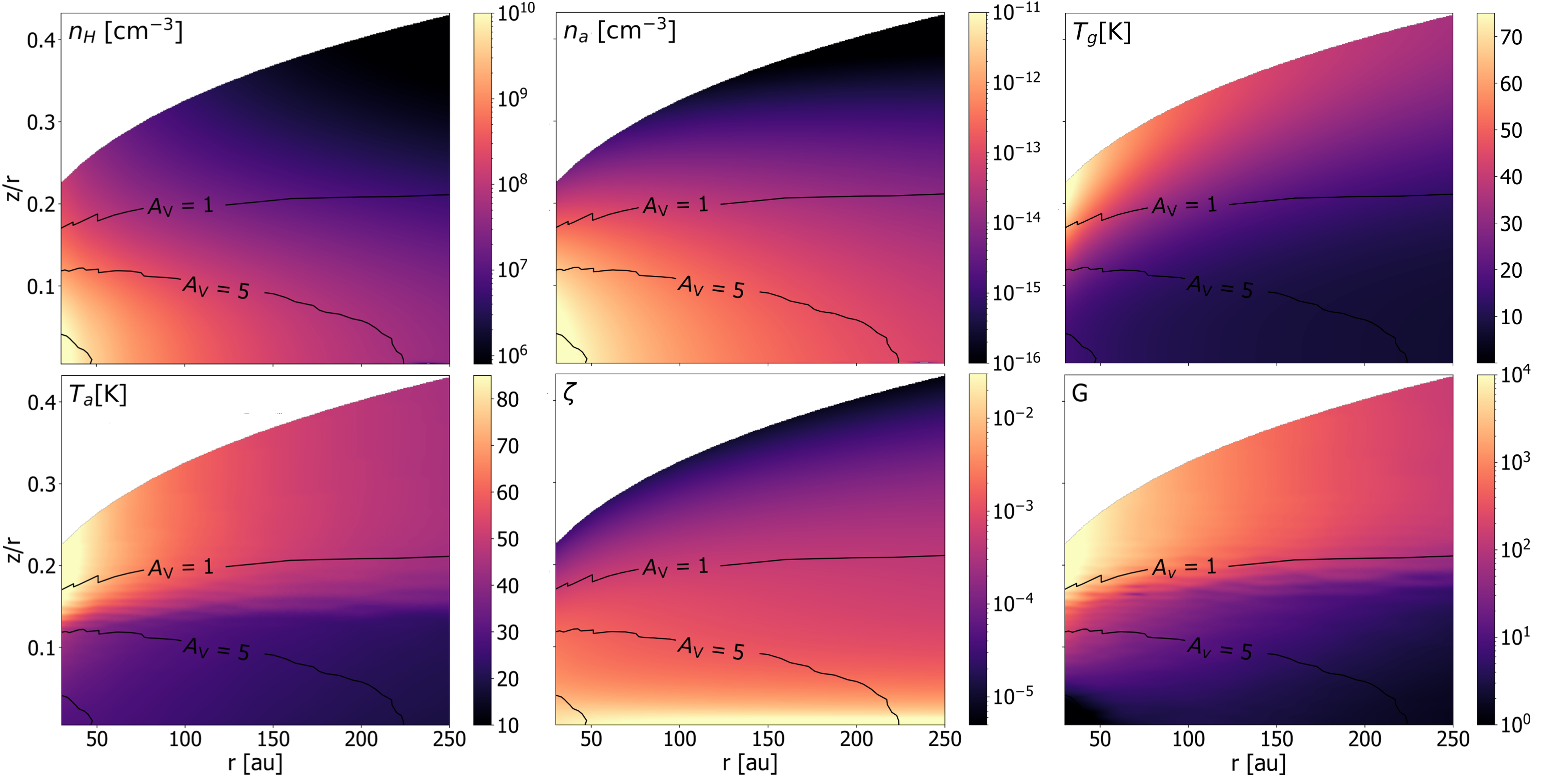}   
\caption{Physical structure of the multi-grain models incorporating a 
full dust distribution. $T_g$ is the gas temperature, $n_H$ is the gas 
number density, $G$ is the local stellar and interstellar field in the 
case of the HUV models (see Appendix \ref{app:model}), $\zeta$ is the 
dust to gas mass ratio, \revisionA{$T_a$ (Eq.\ref{eq:ta}) is the area-weighted
dust temperature, and $n_a$ is the area-weighted grain number density.
A$_V$ is the visual extinction counted from disk surface toward the disk midplane.}
\label{fig:struct}}
\end{figure*}

\revisionA{Figure \ref{fig:struct} is a 2D representation of the disk 
structure in density and in temperature for the gas and for the dust 
(multi-grain model with settling).  }

\begin{table}
\centering
\caption{Overview of the disk model parameters \label{tab:param}}
\begin{tabular}{p{0.68\linewidth}r}
\hline
\noalign{\smallskip}
\multicolumn{2}{c}{\textbf{Disk Parameters}}\\
\noalign{\smallskip}
\hline
 \noalign{\smallskip}
\bm{$T_\mathrm{\star,eff}$} (star temperature)            	& 3900 K \\
\bm{$L_\star$} (star luminosity)       & 0.75 $\mathrm{L_{\odot}}$  \\
\bm{$M_\star$} (star mass)        & 0.58 $\mathrm{M_{\odot}}$ \\
\textbf{ISRF}								  & DRAINE \\

\bm{$\Sigma_{g,0}$} (gas (H + H$_2$) surface density at $R_0$) & 0.335 $\mathrm{g.cm^{-2}}$ \\
\bm{$p$} (surface density exponent) & 1.5 \\
\bm{$R_\mathrm{in}$} (innermost radius)  & 1 $\mathrm{au}$ \\
\bm{$R_\mathrm{out}$} (outermost radius) & 250 $\mathrm{au}$ \\
\bm{$R_0$} (reference radius)& 100 $\mathrm{au}$  \\
\bm{$\zeta$} (dust to gas mass ratio) & 0.01 \\

\bm{$T_\mathrm{mid,0}$} (observed T$_\mathrm{mid}$ at $R_0$)& 10 $\mathrm{K}$\\
\bm{$T_\mathrm{atm,0}$} (observed T$_\mathrm{atm}$ at $R_0$)& 50 $\mathrm{K}$\\
\bm{$\sigma$} (stiffness of the vertical T profile) & 2 \\
\bm{$q$} (radial T profile exponent) & 0.4 \\
\bm{$S_{c}$} (Schmidt number) & 1 \\
\bm{$\alpha$} (viscosity coefficient) & $0.01$ \\

\bm{$d$} (grain size distribution exponent) & $3.5$ \\

\bm{$a_\mathrm{min}$} (smallest grain radius)  & 5 $\mathrm{nm}$ \\
\bm{$a_\mathrm{max}$} (biggest grain radius) & 1 $\mathrm{mm}$ \\
\textbf{number of bins} & 16 \\
\bm{$\rho_\mathrm{grain}$} (material density of dust)  & 2.5 $\mathrm{g.cm^{-3}}$ \\
\bm{$\mathrm{M_{disk}}$} (total disk mass) & 7.5.10$^{-3}$ $\mathrm{M_{\odot}}$\\
\end{tabular}
\end{table}

\subsection{Radiation fields and dust temperature}
\label{sec:rad}
In this new approach, we use the dust radiative transfer code to 
calculate  throughout the disk the UV field  and the dust temperature 
for each of the 16 grain sizes, in each cell of the disk (in radius and 
altitude). The dust temperatures are then used to compute 
the  chemistry with NAUTILUS. 

\revisionA{However, the UV field derived above only accounts for dust extinction and
scattering. A proper account for self and mutual shielding of molecules
and atoms is required to handle the photo-chemical processes.}

\subsubsection{Radiation sources}  
\label{sec:sources}
As the central heating source we assume a low mass pre-main sequence 
star with a mass of $M_\star = 0.58$ $\mathrm{M}_\odot$, which radiates as a black 
body with an effective temperature of $T_\mathrm{{\star, eff}}$ = 3900 
$\mathrm{K}$ and a luminosity $L_\star$ = 0.75 $\mathrm{L_\odot}$. 
Additionally, we consider the interstellar radiation field (ISRF) with 
a spectral energy distribution (SED) from \citet{Draine_1978} between 
91.2 $\mathrm{nm}$ and 200 $\mathrm{nm}$ and the extension of 
\citet{Dishoeck+Black_1982} for longer wavelengths. Besides the stellar 
radiation field, TTauri stars exhibit a significant UV excess coming 
from the inner disk boundary and accretion shocks on the stellar 
surface. This contribution is highly wavelength-dependent, with 
spectral lines like Lyman $\mathrm{\alpha}$ having substantial 
intensities \citep{France+etal_2014}. As concerns the thermal balance of dust, the frequency 
dependence is largely irrelevant, and can be absorbed by ensuring that 
the luminosity used in the model contains the UV excess contribution. 
However, photoprocesses being wavelength dependent, the details of the UV spectrum shape 
strongly affect the chemistry. For this reason, we take as stellar 
input spectrum for our radiative transfer and chemical model the
sum of the stellar black body spectrum and the UV excess typically 
found in the in TTauri stars. Further details about the 
radiation fields are given in Section \ref{sec:set}.

\subsubsection{Radiative transfer simulations}\label{sec:RT_sim} 

The radiation field is computed using the 3D Monte-Carlo continuum and 
line radiative transfer code \textsc{POLARIS} 
\citep{Reissl+Wolf+Brauer_2016}.  
As in \citet{Heese+etal_2017}, 
we calculate  the dust temperature distribution of 16 grain size 
intervals (see Fig.\,\ref{fig:RTsimulation}). 
\revisionA{We assume spherical grains with a size independent composition}
consisting of a 
homogeneous mixture of 62.5 \% silicate and 37.5 \% graphite. 
Optical properties from \citet{Draine_1984} and 
\citet{Laor+Draine_1993} are used to calculate the scattering 
and absorption coefficients based on Mie scattering \citep{Mie_1908} 
using the \textsf{MIEX} code \citep{Wolf+Voshchinnikov_2004}. 
\revisionA{The radiative transfer is solved using 100 wavelengths
logarithmically spaced between 50 nm and 2 mm.
The spatial grid involves 300 radii, logarithmically spaced by a factor \revisionA{1.05}
between 1 and 300 au, and 181 angles with the same factor between 0 and \revisionA{$\pi$}.
}

The temperature in the upper optically thin disk layers strongly 
depends on the grain size (Fig.\,\ref{fig:RTsimulation}, middle panel). 
The dust temperature increases with grain size for sizes in the range 
0.007 $\mu$m to 0.070 $\mu$m, then decreases for sizes in range 0.32 
$\mu$m to 15 $\mu$m, and increases again, although only slightly, with 
larger grain sizes. As explained in \citet{Heese+etal_2017},
this can be understood by looking at the ratios of 
the absorption cross-section for wavelengths at which the star radiates 
($\approx 1$ $\mu$m) and the absorption cross-section for wavelengths at 
which the dust emits ($\approx 20$ $\mu$m, Fig.\,\ref{fig:CabsRatio}). The 
ratio increases up to grain sizes of 0.07 $\mu$m, meaning that the 
ratio of absorbed to emitted radiation increases, leading to higher 
grain temperatures.  For larger grain sizes, the behavior is reversed 
until grains become larger than about 15 $\mu$m. Using these radiative 
transfer simulations we can obtain a distribution of dust temperatures 
$T_d(a_i, r, z)$ that fully depend on the size and location of 
the grains. 
To compare our set of models with those consisting of a 
single grain size (see Section \ref{sec:single}), it is also convenient 
to introduce an area-weighted temperature defined as:

\begin{equation}
\label{eq:ta}
	T_{a}(r, z) =  \frac{ \sum_i a_i^2 T_d(a_i, r, z) n_d(a_i, r, z) }{ \sum_i a_i^2 n_d(a_i, r, z) }.
\end{equation}

\noindent We note that the area-weighted dust density $n_\mathrm{a}(r, 
z)$ (Fig.\,\ref{fig:struct}) is defined by an equation of the same 
form. Figure \ref{fig:weighted_100} shows the dust temperature profiles 
at 100 $\mathrm{au}$ used by all our different models.

Our disk model is less massive than that used by \citet{Heese+etal_2017},
leading to smaller opacities, and thus the temperature differences between
grains of different sizes is much larger in Figs.\ref{fig:RTsimulation} and \ref{fig:weighted_100}
than in that previous work.

\begin{figure} 
\centering

\includegraphics[width=1.05\linewidth]{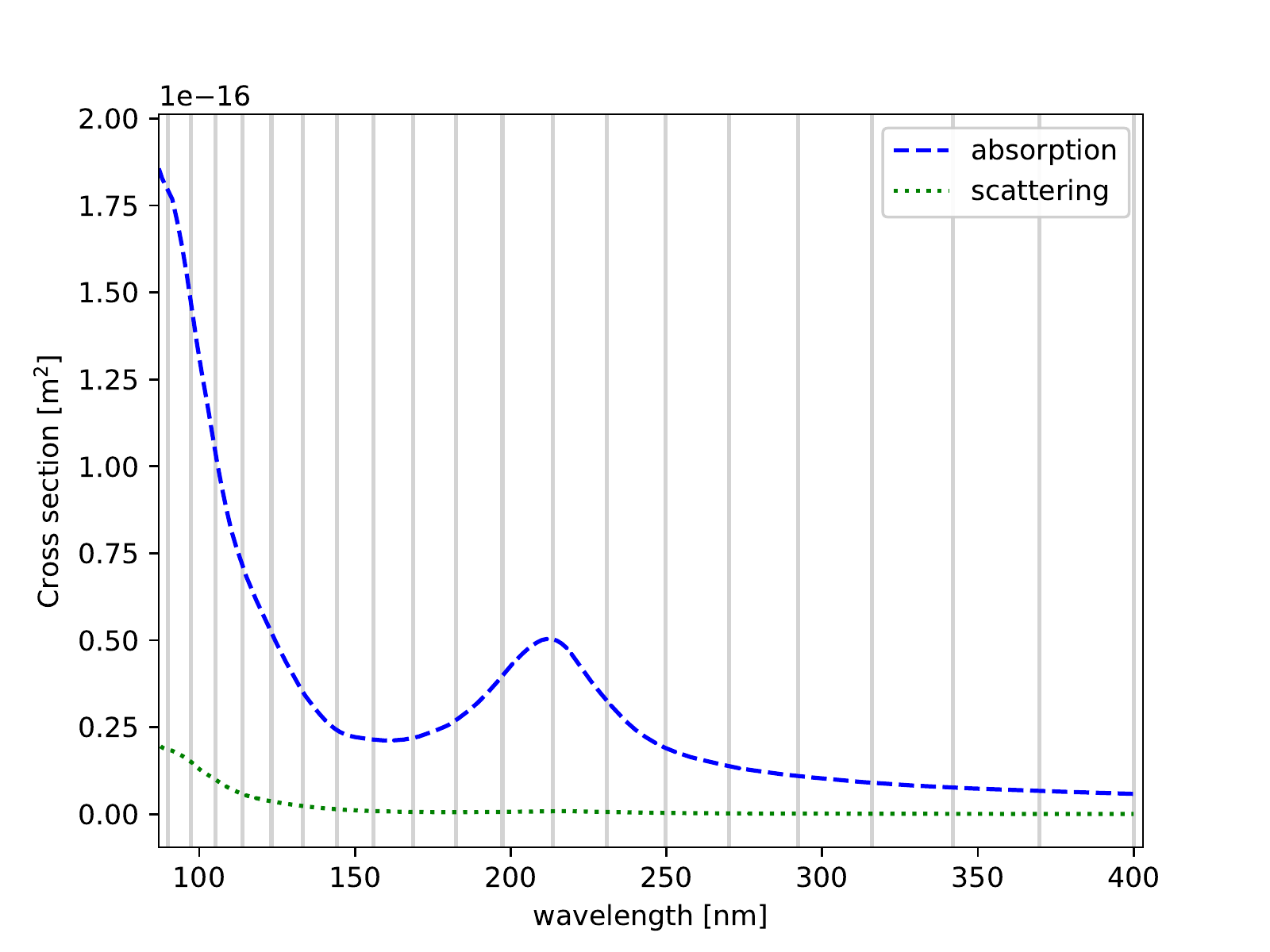}
\caption{Absorption (\textit{blue dashed line}) and scattering 
cross-sections (\textit{green dotted line}) for the first grain size 
bin (7 $\mathrm{nm}$). The vertical gray lines mark the wavelengths for 
which the radiation fields are simulated. 
\label{fig:opticalProperties}}
\end{figure}

\begin{figure} 
\centering

\includegraphics[width=0.95\linewidth]{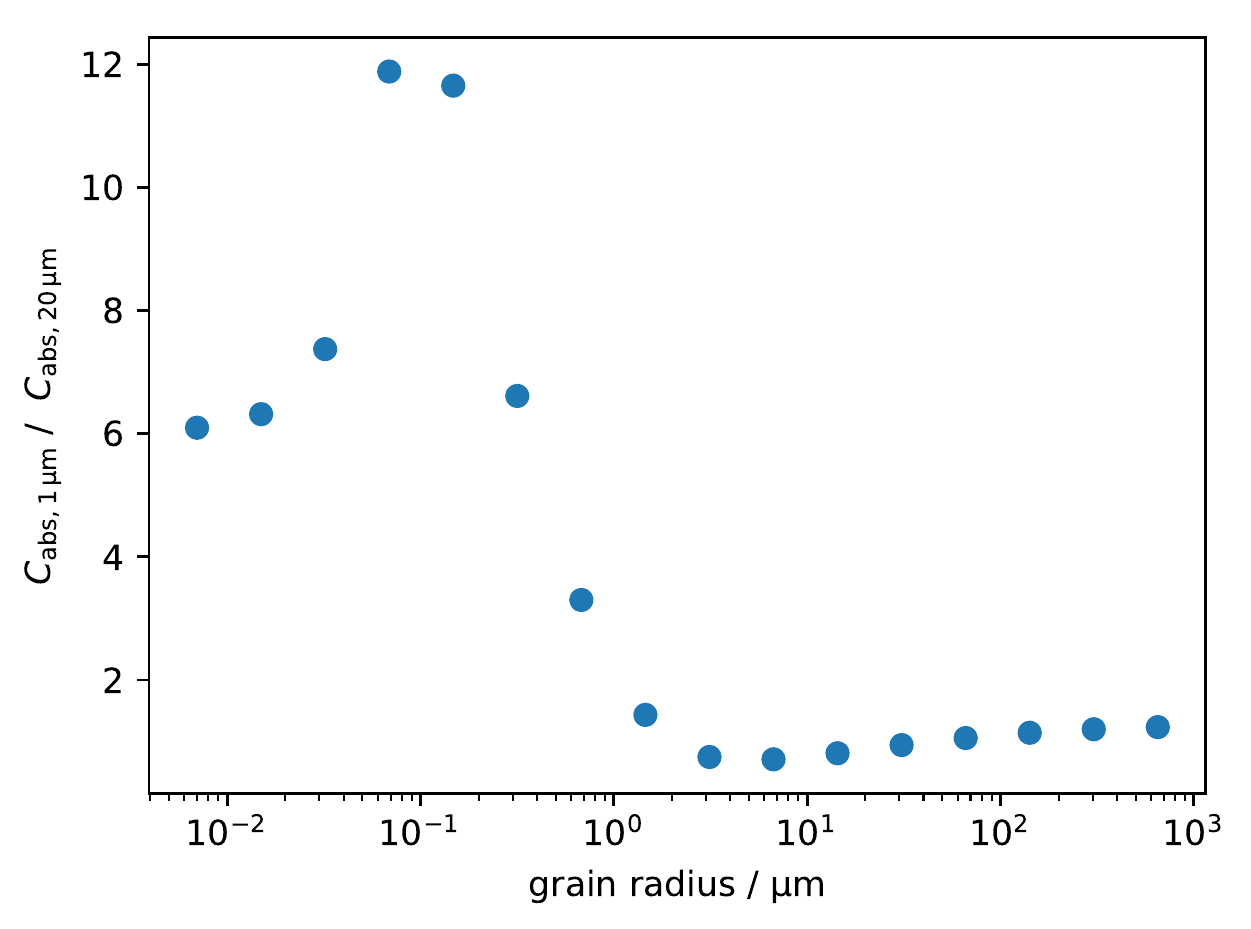}
\caption{Ratio of the absorption cross-sections for wavelengths characteristic for stellar radiation ($\mathrm{C_{abs, 1 \mu m}}$) and dust emission ($\mathrm{C_{abs, 20 \mu m}}$).     \label{fig:CabsRatio}}
\end{figure}

\begin{figure*} 
\includegraphics[width=1.0\textwidth]{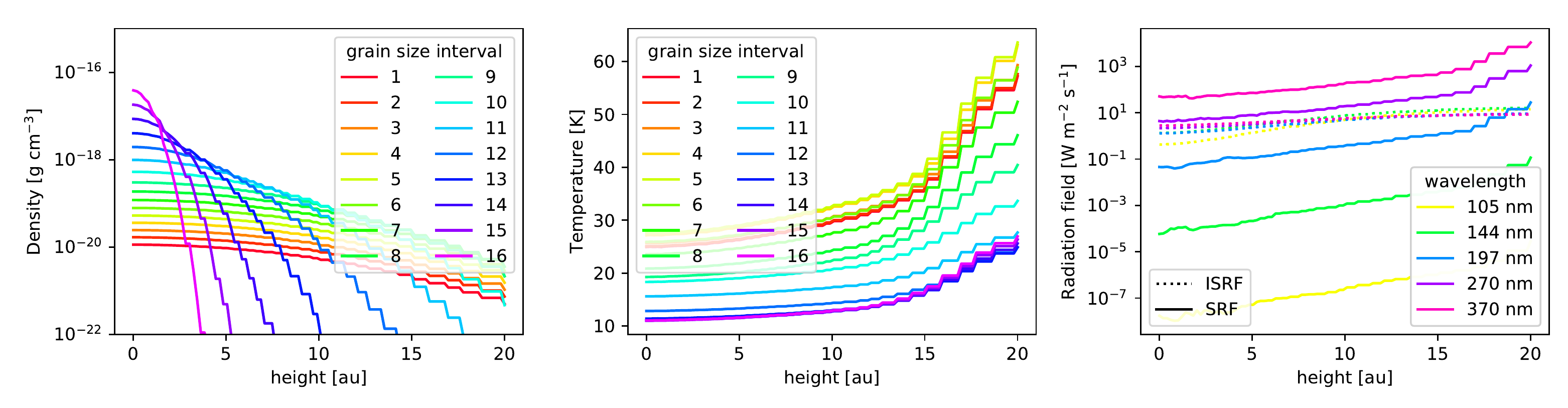}
\caption{Results of the radiative transfer simulation in vertical direction at a disk radius of $r$ = 100 $\mathrm{au}$. \textit{Left:} Dust density distribution for the 16 grain size intervals as described in Appendix \ref{app:distrib} and \ref{app:settle}. \textit{Middle:} Dust temperature of the 16 grain size intervals. \textit{Right:} Stellar (solid lines) and interstellar radiation field (dotted lines).
\label{fig:RTsimulation}}
\end{figure*}

\begin{figure} 
\includegraphics[width=\linewidth]{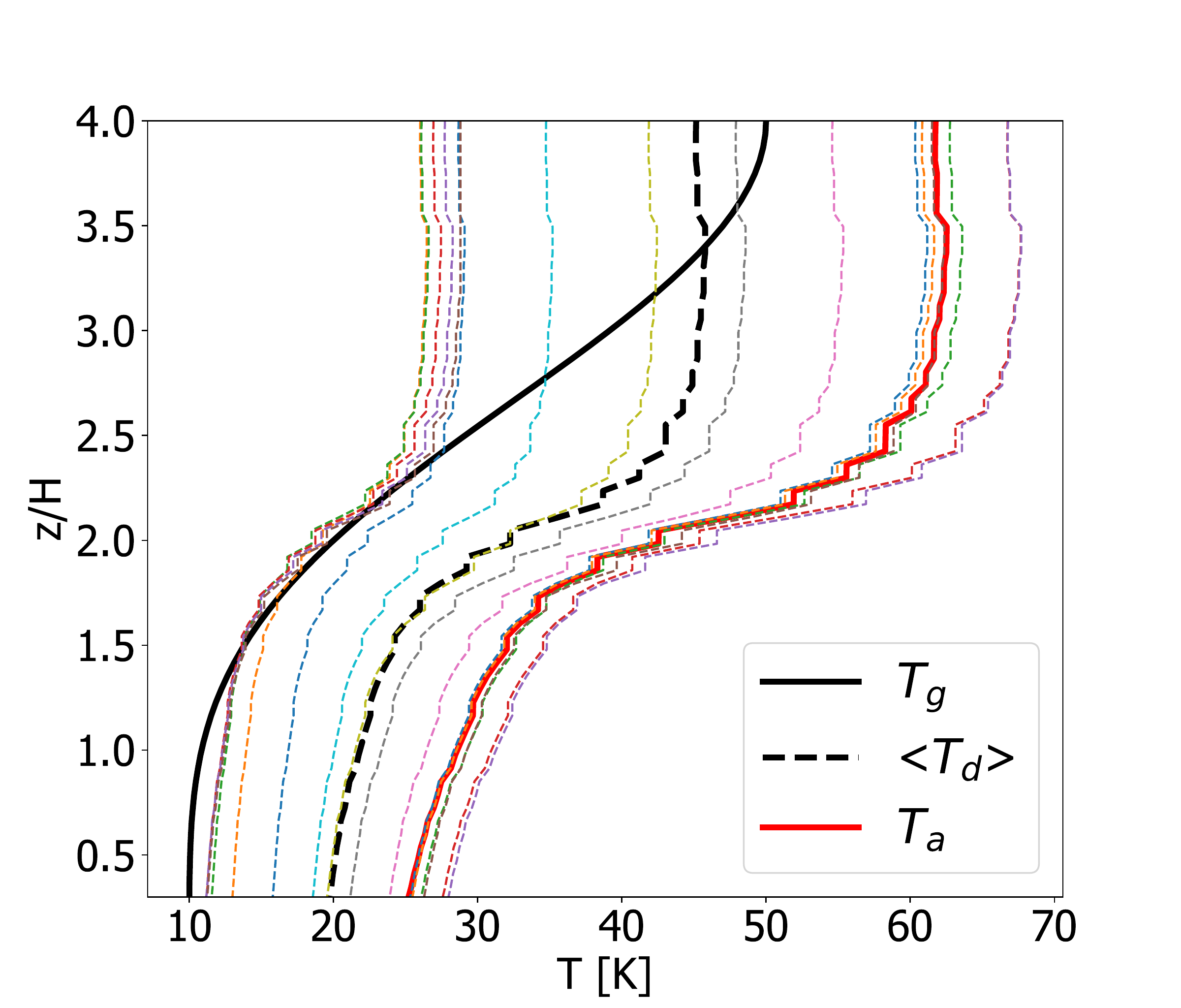}
\caption{\label{fig:weighted_100} Vertical profiles of temperatures at 100 $\mathrm{au}$. The thick black line is the gas vertical temperature profile in Kelvin. The dotted colored lines are temperatures of each grain population and the thick red line is the area-weighted temperature $T_{a}$. We see that temperatures are roughly constant for $z/H$ < 1.5 and for $z/H$ > 2.5. Between these two altitudes the temperatures exhibit a strong transition.}
\end{figure}

\subsubsection{Dust extinction, self and mutual shielding}  
\label{sec:UVfield} 
The radiation field is a contribution of both the interstellar 
radiation field (ISRF) and the stellar radiation field (SRF). Radiation 
from these two sources encounters molecular or atomic gas and dust 
grains that affects its propagation through extinction and scattering. 
The effect of the dust is computed by the 3D Monte-Carlo continuum and 
line radiative transfer code POLARIS as described in 
Sec.\,\ref{sec:RT_sim} \revisionA{(Fig.\ref{fig:opticalProperties} shows the sampling
in the 90 to 400 $\mathrm{nm}$ wavelengths range relevant for photoprocesses)}.
Figure \ref{fig:RTsimulation} shows results at a 
disk radius of $r =100$ $\mathrm{au}$.

In chemical codes, the effect of extinction due to gas, known as self 
and mutual shielding, is often estimated using empirical attenuation 
factors as a function of line-of-sight visual extinction 
\cite[e.g.][for a discussion]{Lee+etal_1996}. \revisionA{The situation for 
disks is more complex, due to the two  sources of UV radiation (stellar 
neighborhood and ISRF) and varying dust properties. Empirical 
attenuation factors have been derived for disks \citep[see for 
example][]{Visser+etal_2009,Heays+etal_2017}. They are in general used 
in a 1+1D  approach, where the two sources of radiation are treated 
independently and scattering is neglected. Furthermore, they  depend 
implicitely on the dust settling that was assumed during their 
derivation, as well as on the shape of the incident UV field.}

\revisionA{For better consistency}, we use here a procedure that 
builds on our knowledge of the dust-attenuated UV field. The extinction 
produced by the gas requires to know the opacity generated by the 
molecules through which the radiation travels from outside the disk to the 
local cell. 


\revisionA{We show in Appendix \ref{app:self} that a 
first order approximation for self and mutual shielding is
obtained  by applying the attenuation of the vertical opacity due to 
molecules to the radiation field computed with dust only by the POLARIS 
code:}
\begin{equation}
\label{eq:uvfield}
	I_L(\lambda, r, z) =  e^{-\tau^V_m} I_d(\lambda, r, z)
\end{equation}
where $I_d(\lambda, r, z)$ is the radiation field given by POLARIS, which 
explicitly includes the impact of dust scattering, \revisionA{and $\tau^V_m$ is the
total opacity due to molecules from the $(r,z)$ point to $(r,+\infty$).}

\subsection{Chemistry simulations} \label{sec:chemistry}

The NAUTILUS gas-grain code \citep{Ruaud+etal_2016} is used to perform
chemistry simulations. NAUTILUS is dedicated to computing chemical 
evolution in different interstellar environments and uses the rate 
equation approach \citep{Hasegawa+etal_1992, Hasegawa+Herbst_1993b} to calculate 
the abundance as a function of time. The NAUTILUS gas-grain code is a 
3-phase model \citep{Ruaud+etal_2016} that includes gas-phase chemistry and chemically active grain surfaces and mantles. 
Thus, aside from the gas-phase chemistry, NAUTILUS covers the physisorption of neutral 
species on the surface, the diffusion of these species and their 
reactions, desorption of species from the surface and repopulation of 
the surface by species from the mantle as the species on the surface 
evaporate. The latest version of NAUTILUS, the Nautilus Multi Grain Code (NMGC)\footnote{Code available here: \url{https://github.com/sgvn/NMGC}} 0D model 
\citep{Iqbal+Wakelam_2018}, can perform simulations using a full 
grain distribution in size. 
\revisionA{Each grain size bin is treated independently of the others, and only interact with gas.}
\revisionA{Compared to the single-grain model (model using a single grain size), a multi-grain model changes 
chemical rates in two ways. First, accretion rates of species depend on 
grain population. In a multi-grain model, accretion rates vary on 
different grain sizes according to their total surface area (Table 
\ref{tab:grains} shows the relative surface areas as function of grain sizes).
Second, dust temperature depends on its 
size (see Fig.\,\ref{fig:RTsimulation}). Species have higher reaction 
rates on grains which are hotter due to higher hopping and desorption 
rates. In general, grains which are hotter have lower abundances of 
lighter species such as CN, CH$_2$, CO, etc. and more of heavier 
species such as H$_2$O, CH$_3$OH, etc\ldots \citep[see][for details]{Iqbal+Wakelam_2018}. 
In our current work,} we generalize this multi-grain 
capability to a 1D situation to represent a protoplanetary disk as 
described in previous Sections. 
NMGC computes chemistry in all cells given in \revisionA{cylindrical} coordinates $(r,z)$ where  
we provide as input the gas temperature (Eq. \ref{eq:verticalT}), 
the local radiation field $I_d(\lambda,r,z)$ and dust temperatures 
$T_d(a_i, r, z)$ given by the radiative transfer simulations, 
and the number density of each dust population $n_d(a_i, r, z)$ in order to compute the local 
grain abundances relative to the number density of Hydrogen nuclei 
$n_H(r, z)$:
\begin{equation}
\label{eq:rapport}
	AB_d(a_i, r, z) = \frac{n_d(a_i, r, z)}{n_H(r, z)}.
\end{equation}
\revisionA{The $z$ direction is treated as in \citet{Hersant+etal_2009}, but using 64 points.} 

\subsection{Photorates} 

Radiations in the UV range have a critical impact on the disk chemistry 
since UV photons have the necessary energy to photoionize or 
photodissociate molecules in the disk. For this reason, characterizing 
the photorates [s$^{-1}$] at which molecules are dissociated or ionized 
is of high importance in chemistry models and necessary for a full 
interpretation of observations.

We evaluate the photoprocess rate as in \citet{Heays+etal_2017}:
\begin{equation}
\label{eq:rate}
	k_{p}(X, r, z) =  \int_\lambda  \sigma_{p}(X, \lambda) I_L(\lambda, r,z) d\lambda
\end{equation}
where $\sigma_{p}(X, \lambda)$ is the photoprocess cross-section of 
species $X$ at wavelength $\lambda$ and $I_L(\lambda,r,z)$ is the local 
UV radiation. The index $p$ equals $i$ when it stands for ionization 
and equal $d$ when it stands for dissociation. We perform our 
integration over the 91.2 to 400 $\mathrm{nm}$ wavelength range. To 
evaluate the rate at each point in the disk, we thus need the local UV 
radiation field and the (space-independent) cross-sections. Following
Section \ref{sec:UVfield}, the UV field is derived using Eq.
\ref{eq:uvfield}.

The molecular cross-sections used in this study are 
extracted from the Leiden University 
website\footnote{\url{https://home.strw.leidenuniv.nl/~ewine/photo/index.php?file=cross_sections.php}} 
\citep{Heays+etal_2017}. They are collected from experimental and 
theoretical studies. \revisionA{Eighty molecules are treated in this
way.} 
\revisionA{The photorates are recomputed at each explicit time step of the chemical evolution.} 

\revisionA{For all other gas constituents \citep[atoms or molecules unavailable in][]{Heays+etal_2017}, we
use approximate shielding factors from \cite{Visser+etal_2009}}.

\subsection{Molecular Hydrogen formation at the disk surface} 
\label{sec:h2_formation}

\subsubsection{The atomic H problem}
The NAUTILUS code initially implemented the \hh formation through 
the Langmuir-Hinshelwood (LH) mechanism that considers physisorption on 
grain surfaces. The LH mechanism is efficient only over a relatively 
narrow temperature range 
\citep[5-15\,K on flat surfaces][]{Katz+etal_1999,Vidali+etal_2004,Vidali+etal_2005}.

\revisionC{In the case of multiple grain sizes, dust settling implies that only the small grains remain in the PDR regions of the disk. These small grains illuminated by
the UV field from the star can get significantly warmer at the
disk surface, and the \hh formation via the LH mechanism as treated in most astrochemistry codes} becomes inefficient, leading
to much smaller \hh formation rate than in the case of a single,
larger, equivalent-area, grain size with lower surface temperature. The low formation rate of \hh leaves
a significant amount of atomic Hydrogen that can severely affect the
chemical balance in the disk upper layer (see Section \ref{sec:discuss} for a
more detailed discussion).

Observations of \hh~in unshielded regions \citep{Habart+etal_2004, 
Habart+etal_2011} where dust can reach high 
temperatures ($\geq20$ K), show that the column density can be 
significantly higher than what is expected by simple equilibrium rate 
equation and PDR model predictions. This implies that the formation mechanisms 
of \hh can be efficient in a wider domain than predicted by the LH 
mechanism at equilibrium temperature.

\revisionC{To evaluate the \hh formation rate, many studies adopt a canonical value for interstellar clouds derived by \citet{Jura_1975}, which is defined by taking half the rate at which hydrogen atoms stick to the grain surface  \citep[e.g.][]{Walsh+etal_2014, Agundez+etal_2018}}.

Several \revisionC{other} studies proposed more sophisticated mechanisms that theoretically produce comparable 
\hh column densities to those derived by observations 
\citep[e.g.][]{Duley_1996,Cazaux+Tielens_2004,Cuppen+Herbst_2005,Iqbal+etal_2012, 
Iqbal+etal_2014,Thi+etal_2020}. We follow here the approach
of \citet{Bron+etal_2014}, hereafter B14, which \revisionC{considers the stochastic fluctuations of both the grain temperature and the H-atom surface population.}

\subsubsection{Effects of temperature fluctuations} 

Small grains ($\leq$ 0.01 $\mu$m), given their 
very small heat capacity, have their temperature fluctuating widely 
even when a relatively small quantity of energy is absorbed. In typical 
unshielded regions, a small grain can absorb UV photons originated from 
the interstellar field with a rather high probability, causing a sudden 
spike in temperature. If the interval of time between UV photon 
absorptions is sufficiently long, the grain which has undergone 
transient heating has the time to cool down to temperatures smaller than 
the equilibrium temperature, giving the opportunity for \hh to form on 
the surface. Then, the rapid heating of the grain by the absorption of 
a UV photon will thermally desorb most species on the surface, 
releasing \hh molecules to the gas phase.  

\citet{Bron+etal_2014} have studied the effect of these temperature fluctuations on the formation of \hh in PDRs and 
high UV radiation regions using a statistical approach. By numerically 
solving the master equation \citep{LeBourlot+etal_2012}, they show 
\citep[][see their Fig.6]{Bron+etal_2014} that the LH mechanism alone is 
sufficient to reach the typical \hh formation rates observed  in PDRs 
when the temperature fluctuations of small grains are accounted for.

\subsubsection{Implementing the \hh formation rates} 
The investigation of B14 lies in a 
domain of gas temperature (100 K) and gas density (10$^0$ to 10$^6$ 
$\mathrm{cm^{-3}}$) distinct from our study 
but this difference can easily be 
corrected as both temperature and gas density only affect the results 
through the collision rate of atomic Hydrogen with the grains, which 
scales as $n_H\sqrt{T_{g}}$. Thus, we can estimate the rate for a 
given density $n_H$ and a given temperature $T_g$ by extracting or 
interpolating from B14's data the value corresponding to an equivalent 
gas density $n_{eq}$ equal to:
\begin{equation}
\label{eq:neq}
	n_{eq} =  n_H\sqrt{\frac{T_g}{100\,\mathrm{K}}}\frac{s(T_g)}{s(100\,\mathrm{K})}
\end{equation}
where s($T_g$) is the sticking coefficient at temperature $T_g$ as defined by \citet{LeBourlot+etal_2012}: 

\begin{equation}
\label{eq:sticking}
	s(T) = \left(1+\frac{T}{T_2}^{\beta}\right)^{-1}
\end{equation}
with $T_2= 464$\,K and $\beta=1.5$, which approximately gives the same result as in \citet{Sternberg+Dalgarno_1995}. 

\begin{figure}
  \centering
  \includegraphics[width=\linewidth]{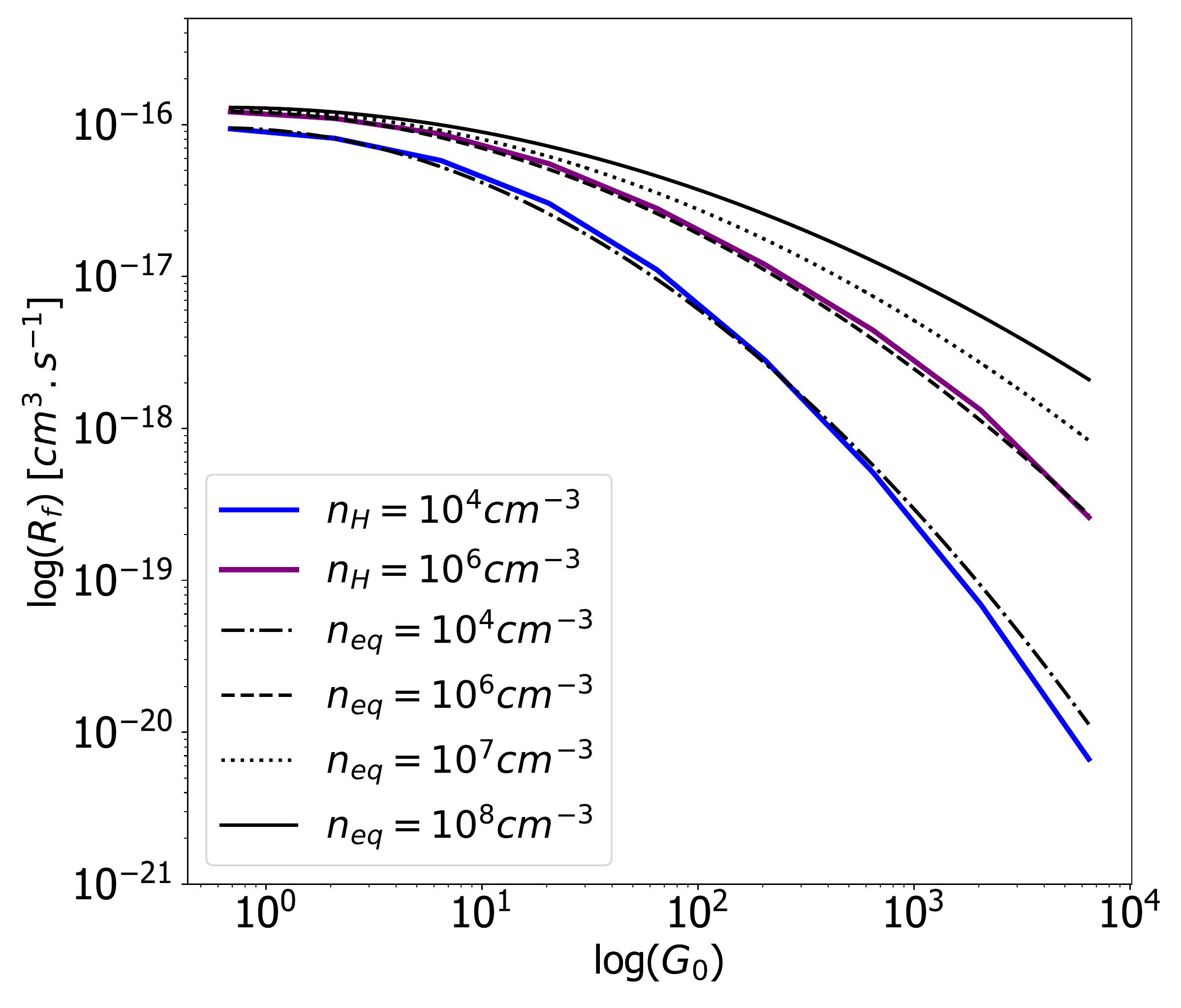}   
\caption{\label{fig:b14} 
\hh~ formation rate as a function of the UV field intensity
(in units of the Habing interstellar field) for various proton number density. Data adapted from Fig.\,6 of 
\citet{Bron+etal_2014} are the blue (10$^{4}$ cm$^{-3}$) and 
purple (10$^{6}$ cm$^{-3}$) solid lines while black lines correspond 
to our analytic fit 
for four different number densities (10$^{4}$, 10$^{6}$, 10$^{7}$, 10$^{8}$ cm$^{-3}$).}
\end{figure}

We interpolated B14 data as a function of proton density and UV field
intensity using an analytic polylogarithmic function. The interpolation
results are displayed in Fig.\ref{fig:b14}. The PDR layers of 
protoplanetary disks are dense, typically above 
$10^6 -- 10^8$\,cm$^{-3}$ so that essentially the upper curves in 
Fig.\ref{fig:b14} apply to our case.

It should be noted that the B14 calculations assume a power-law 
size distribution from 1 nm to 300 nm composed of only carbonaceous 
grains. 
Changing the composition and/or size distribution
would affect the precise values of the rates and of the fitted 
coefficients.
However, the asymptotic behaviors of the rates 
should remain similar. The rate is proportional to the density at low 
densities, and progressively saturates for larger densities as the 
fraction of atomic H decreases because the UV flux is insufficient
to dissociate H$_2$. As a function of the UV field intensity the (unsaturated) rate is expected to decrease as $1/G$ since the 
time spent by any grain in the relevant temperature range where the 
LH formation is efficient is inversely proportional to the UV photon flux.

\section{Model parameters and results}
\label{sec:model_param}


\subsection{Detailed description of models}\label{sec:set}

We implement a set of \revisionC{twelve} different models (see Table \ref{tab:full3}). 
All models \revisionA{share a common set of disk structure and physical parameters (see 
Fig.\,\ref{fig:struct}) as described in Table \ref{tab:param}, but
differ for a limited number of assumptions on grain sizes and
temperatures,  UV field or H$_2$ formation mechanism.}

The set is divided into \revisionC{three groups, i.e. the single-grain models,  
the multi-grain models and an intermediate category of models with single grains locally varying
to mimic the grain distribution used in the multi-grain model.}  
 We describe their specific characteristics in 
detail in Section \ref{sec:single}, \ref{sec:multi} and  \ref{sec:interm}. As mentioned in 
Section \ref{sec:sources}, the excess of UV radiation exhibited by 
typical TTauri stars, given the strong wavelength-dependence of 
photoprocesses, inevitably impacts the chemistry of the disk. In order 
to have a broad view on how our chemical models react to different 
intensities of UV flux, we adopt two regimes of radiation. One regime 
uses the spectrum of a relatively high-UV emitting TTauri star and 
another uses the spectrum of a relatively low UV emitting one, 
hereafter called HUV and LUV, respectively. 
For the HUV regime we use 
the spectrum of TW Hya \citep{Heays+etal_2017} while for the LUV we use 
the one of DM Tau. Both FUV spectra are provided in the CTTS FUV 
spectra 
database\footnote{\url{https://cos.colorado.edu/~kevinf/ctts_fuvfield.html}} 
\citep{France+etal_2014}. 
The blue solid line in Fig.\,\ref{fig:flux_approx} exhibits the spectrum 
of TW Hya as provided by \citet{France+etal_2014} before it is 
attenuated by the disk.

The chemical abundances in the disk are calculated using the 
time-dependent equation embedded in NAUTILUS. We use atomic initial 
abundances described in Table \ref{tab:init_ab} and all the chemical 
output are extracted after an \revisionC{evolution time of $5.10^6$ yrs. We note that using molecular initial abundances would give almost the same final results for disks of this age  \citep[see][]{Wakelam+etal_2019}}. All 
models incorporate the same chemical network and chemical processes 
described in Section \ref{sec:chemistry}. \revisionC{ We ignore X-rays because of the large cumulative column densities found around the midplane, the main interest of this paper. X-rays would mostly affect the disk surface layers. We run all models using a standard cosmic-ray ionization rate of 1.3$\times 10^{-17}$ s$^{-1}$ in the entire disk \citep[see][for a detailed discussion on the ionization rate]{Cleeves+etal_2015}}. 

\subsubsection{Single-grain models}\label{sec:single} 

We use single-grain models as a comparison to our more sophisticated 
multi-grain models described in Section \ref{sec:multi}. 
\revisionA{These models are treated as in \citet{Wakelam+etal_2016}, in
particular with respect to the UV field penetration and photoprocesses \citep[see][]{Hersant+etal_2009}.
We use the same gas density distribution and same overall gas-to-dust ratio (and hence the same dust mass) as in the multi-grain models.}
All single-grain models have a single dust size of 0.1 $\mathrm{\mu m}$. 
The grains are settled toward the midplane assuming a dust scale 
height as defined in \autoref{eq:rhod}. However, 0.1 $\mathrm{\mu m}$ 
grains exhibit little settling, as it can be seen from Fig.\ref{fig:hd} 
and  such models are virtually equivalent to non-settled dust models 
\citep{Wakelam+etal_2016}.  

\paragraph{Model 1: xUV-LH-$\mathrm{T_{g}}$}
The dust temperature ($T_d$) profile follows that of the gas ($T_g$) 
which is given by the solid black line in Fig. \ref{fig:weighted_100}), 
$T_d$ is equal to $T_g$ everywhere in the disk. $\mathrm{x}$ means either $\mathrm{H}$ for high or $\mathrm{L}$ for low, hence Model 1 is either 
\shtg or \sltg, depending on whether the stellar UV field is high (HUV) 
or low (LUV). The differences are detailed in Section \ref{sec:set}. 

\paragraph{Model 2: xUV-LH-$\mathrm{T_{a}}$}
Model 2 is similar to Model 1, except for the dust temperature. In 
Model 2, the grain temperature is given by Eq. \ref{eq:ta} and shown in 
Fig.\,\ref{fig:weighted_100} (red solid line). The temperature $T_d$ is 
hence the area-weighted temperature from the grain size distribution of 
the multi-grain models and it is called $T_a$. As $T_a$ is 
significantly higher than $T_g$, we expect depletion to occur at lower 
elevations than in Model 1. 

\paragraph{Model 3: xUV-B14-$\mathrm{T_{a}}$}
In Model 3, the prescription of \citet{Bron+etal_2014} as described in 
Section \ref{sec:h2_formation} is used to form \hh. The expected effect 
is a higher production of \hh in the PDRs region of the disk than in 
the other models. 

\subsubsection{Multi-grain models}\label{sec:multi} 
Multi-grain models, as opposed to single-grain ones, include the full 
grain distribution detailed in Table \ref{tab:grains} where each grain 
population has a specific temperature (Fig. \ref{fig:weighted_100}), a 
specific settling factor and is chemically active. For all multi-grain 
models, we use a size range of [5 $\mathrm{mm}$, 1 $\mathrm{mm}$] with 
a size exponent $d=-3.5$ (Eq. \ref{eq:da}) and divide the grain size 
interval into 16 logarithmically distributed subintervals using Eq. 
\ref{eq:interval}. An overview of the grain size interval with their 
respective relative dust mass and surface area is given in 
\autoref{tab:grains} while Fig.\,\ref{fig:hd} shows the scale height of 
each 16 grain size population as a function of the radius. Multi-grain 
models are composed of two categories. 

\paragraph{Model 4: M-xUV-LH}
Either called \mhlh or \mllh depending on the flux, they use the classical LH mechanisms for \hh formation without considering small grains temperature fluctuation. 

\paragraph{Model 5: M-xUV-B14}
We add the \hh formation mechanisms prescribed in B14 (see Section \ref{sec:h2_formation}). 
These models, with grain growth, dust settling, one temperature per grain size and H$_2$ formation derived from \citet{Bron+etal_2014} are the most complex models we compute.

\subsubsection{Intermediate models}\label{sec:interm} 

 These models use locally a single grain size mimicking the grain growth and dust settling of the multi-grain models (see Appendix.\,\ref{app:interm} for details).  At each point, the single grain size, which is chosen in order to obtain
the same grain mass and area, are summed up over the grain size distribution of the multi-grain models. 
The purpose of these models is to better illustrate the impact of a dust temperature spread onto the 
chemistry, therefore we only run these models at radii 50, 100 and 200 au.

 \paragraph{Model 6: Set-HUV-y-$\mathrm{T_{a}}$}
There are two models. The dust temperature is T$_a$ while the H$_2$ formation (y) either follows the LH mechanism or the B14 approximation.
\\\\
\revisionA{Table \ref{tab:full3} summarizes the various models.}
\revisionA{Table \ref{tab:fig} lists the different figures in the text in relation with their respective models. 
A few figures relevant of the HUV cases are presented in Appendix \ref{app:huv},
while all figures for the LUV cases appear in Appendix \ref{app:luv}.}

\begin{table}
\caption{List of figures. \label{tab:fig}}
\centering
\begin{tabular}{ l c }

   \hline
   \noalign{\smallskip}
  \textbf{Models} &  \textbf{Figures} \\ 
  \noalign{\smallskip}
 \hline
 \noalign{\smallskip}
   \textbf{single, HUV} & \ref{fig:compare_high}, \ref{fig:100cumul_huv},  \ref{fig:reservoirs}, \ref{fig:s-s-maps-huv},  \ref{fig:ab_surface},  \ref{fig:s-water_maps}, \ref{fig:s-100profile_high}    \\
   \noalign{\smallskip}
   \hline
   \noalign{\smallskip}
    \textbf{single, LUV} & \ref{fig:100cumul}, \ref{fig:s-maps-luv}, \ref{fig:s-100profile_low}  \\
    \noalign{\smallskip}
   \hline
   \noalign{\smallskip}
   \hline
   \noalign{\smallskip}
   \textbf{multi, HUV} &  \ref{fig:compare_high}, \ref{fig:100cumul_huv}, \ref{fig:m-maps},   \ref{fig:reservoirs}, \ref{fig:m-s-maps-huv}, \ref{fig:ab_surface}, \ref{fig:m-water_maps}, \ref{fig:methyl_compare}, \ref{fig:CH3OH_compare}, \ref{fig:surfdensmap}, \ref{fig:m-100profile_high}    \\
   \noalign{\smallskip}
      \hline
      \noalign{\smallskip}
   \textbf{multi, LUV} & \ref{fig:100cumul}, \ref{fig:m-100profile_low}    \\
      \hline
      \noalign{\smallskip}
   \textbf{Set, HUV} &\ref{fig:compare_high}, \ref{fig:reservoirs}, \ref{fig:ab_surface} , \ref{fig:interm-100profile_LH}, \ref{fig:interm-100profile_B14}   \\
\end{tabular}
\end{table}


\subsection{Results} 

\revisionA{We present here some general issues
about the UV field, the dust temperature and the H$_2$ formation. A more thorough discussion of} 
the chemistry is done in the next section.

\begin{table*}  
\centering
 \caption{Description of models}
 \begin{tabular}{l  c  c  c  c  c  c}
   \hline
    & \textbf{bins} & \textbf{sizes} (\bm{$\mathrm{\mu m}$}) & \bm{$\mathrm{T_{d}}$} & \textbf{HUV} & \textbf{LUV} & \textbf{H$_2$ formation} \bigstrut \\
   \hline

   \textbf{HUV-LH-$\bm{\mathrm{T_{g}}}$} & 1 & 0.1 & $T_{g}$ & \checkmark &  & LH \bigstrut \\
   \hline
   \textbf{HUV-LH-$\bm{\mathrm{T_{a}}}$}  & 1 & 0.1 & $T_{a}$ & \checkmark &  & LH  \bigstrut \\
   \hline
    \textbf{HUV-B14-$\bm{\mathrm{T_{a}}}$} & 1 & 0.1 & $T_{a}$ & \checkmark &  & B14  \bigstrut \\
   \hline
   
    \textbf{LUV-LH-$\bm{\mathrm{T_{g}}}$} & 1 & 0.1 & $T_{g}$ &  & \checkmark & LH  \bigstrut \\
   \hline
    \textbf{LUV-LH-$\bm{\mathrm{T_{a}}}$} & 1 & 0.1 & $T_{a}$ &  & \checkmark & LH  \bigstrut  \\
   \hline
     \textbf{LUV-B14-$\bm{\mathrm{T_{a}}}$} & 1 & 0.1 & $T_{a}$ &  & \checkmark & B14 \bigstrut \\
   \hline
   \hline
   \textbf{M-HUV-LH} & 16 & [$5.10^{-3} - 10^{3}$] & $T_{i}$ $ _{\{i = 1 ... 16\}}$ & \checkmark &  & LH  \bigstrut \\
      \hline
   \textbf{M-HUV-B14} & 16 & [$5.10^{-3} - 10^{3}$] & $T_{i}$ $ _{\{i = 1 ... 16\}}$ & \checkmark &  & B14  \bigstrut \\
      \hline
   \textbf{M-LUV-LH} & 16 & [$5.10^{-3} - 10^{3}$] & $T_{i}$ $ _{\{i = 1 ... 16\}}$ &  & \checkmark & LH  \bigstrut \\
      \hline
   \textbf{M-LUV-B14} & 16 & [$5.10^{-3} - 10^{3}$] & $T_{i}$ $ _{\{i = 1 ... 16\}}$ &  & \checkmark & B14  \bigstrut \\
      \hline
     \textbf{Set-HUV-LH-$\bm{\mathrm{T_{a}}}$} & 1 & See appendix C  &  $T_{a}$ &\checkmark & & LH  \bigstrut \\
      \hline
     \textbf{Set-HUV-B14-$\bm{\mathrm{T_{a}}}$} & 1 & See Appendix C & $T_{a}$  & \checkmark &  & B14  \bigstrut \\
     \hline
    \end{tabular}
 \tablefoot{HUV stands for high UV 
 field and LUV for low UV field. 
 LH is for the classical traitement of the Langmuir-Hinshelwood 
 mechanism and B14 when the prescription of \citet{Bron+etal_2014} is 
 used instead.  $T_g$ means that the dust temperature equals to that of 
 the gas, $T_a$ is the weighted-area dust temperature as defined by 
 Eq.\,\ref{eq:ta}. $T_i$ is the temperature of i$^\mathrm{th}$ 
 grain population. M stands for multigrain models and Set for dust models mimicking grain growth and dust settling similar to 
 multi-grain models. As an example, the 
 model \mhb is a multi-grain model computed using the B14 method with a 
 high UV field.  \label{tab:full3}}

\end{table*}
\begin{figure*} 
\begin{subfigure}{.48\linewidth}
  \centering
  \includegraphics[width=1\linewidth]{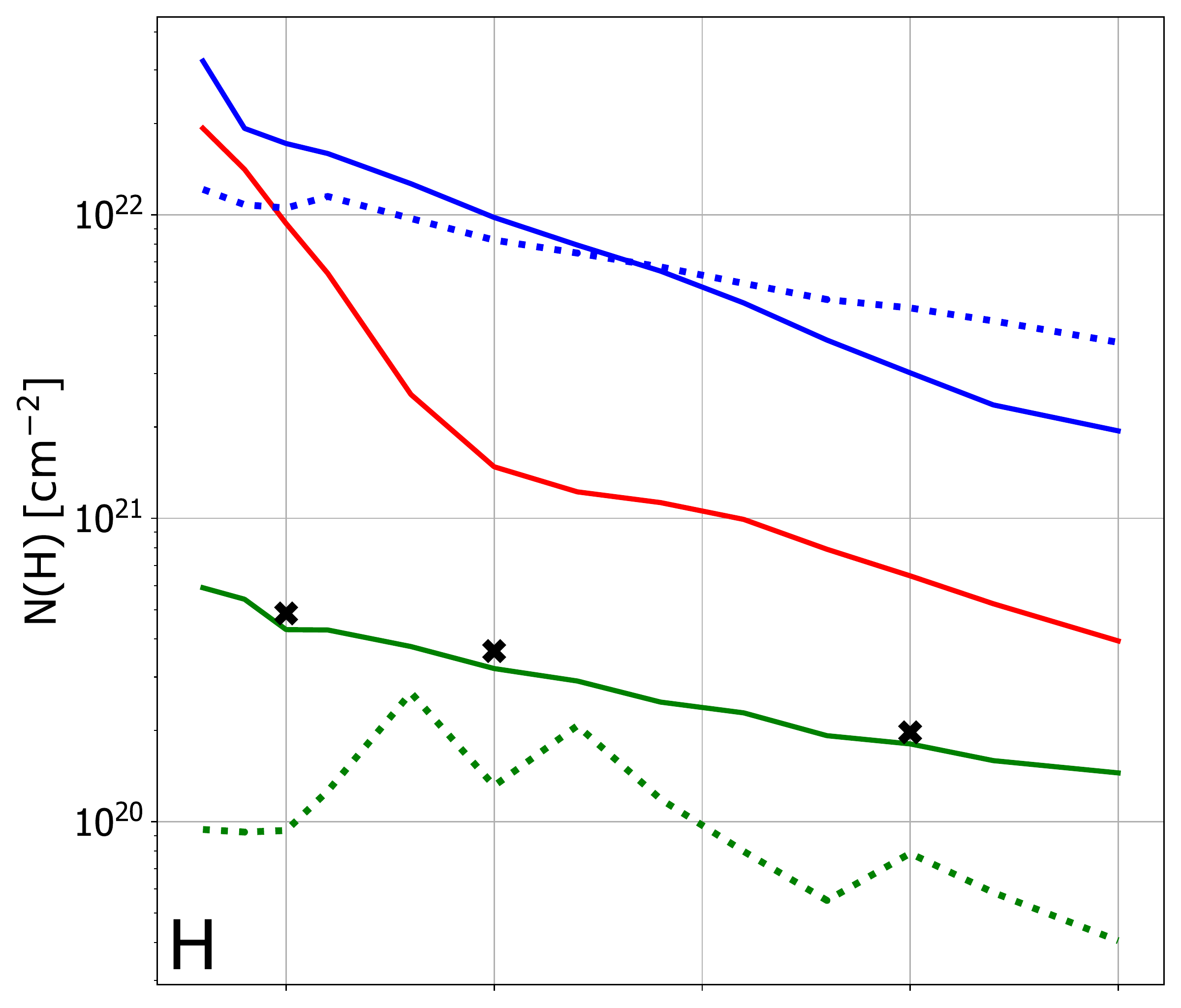}
\end{subfigure}
\begin{subfigure}{.48\linewidth}
  \centering
  \includegraphics[width=1\linewidth]{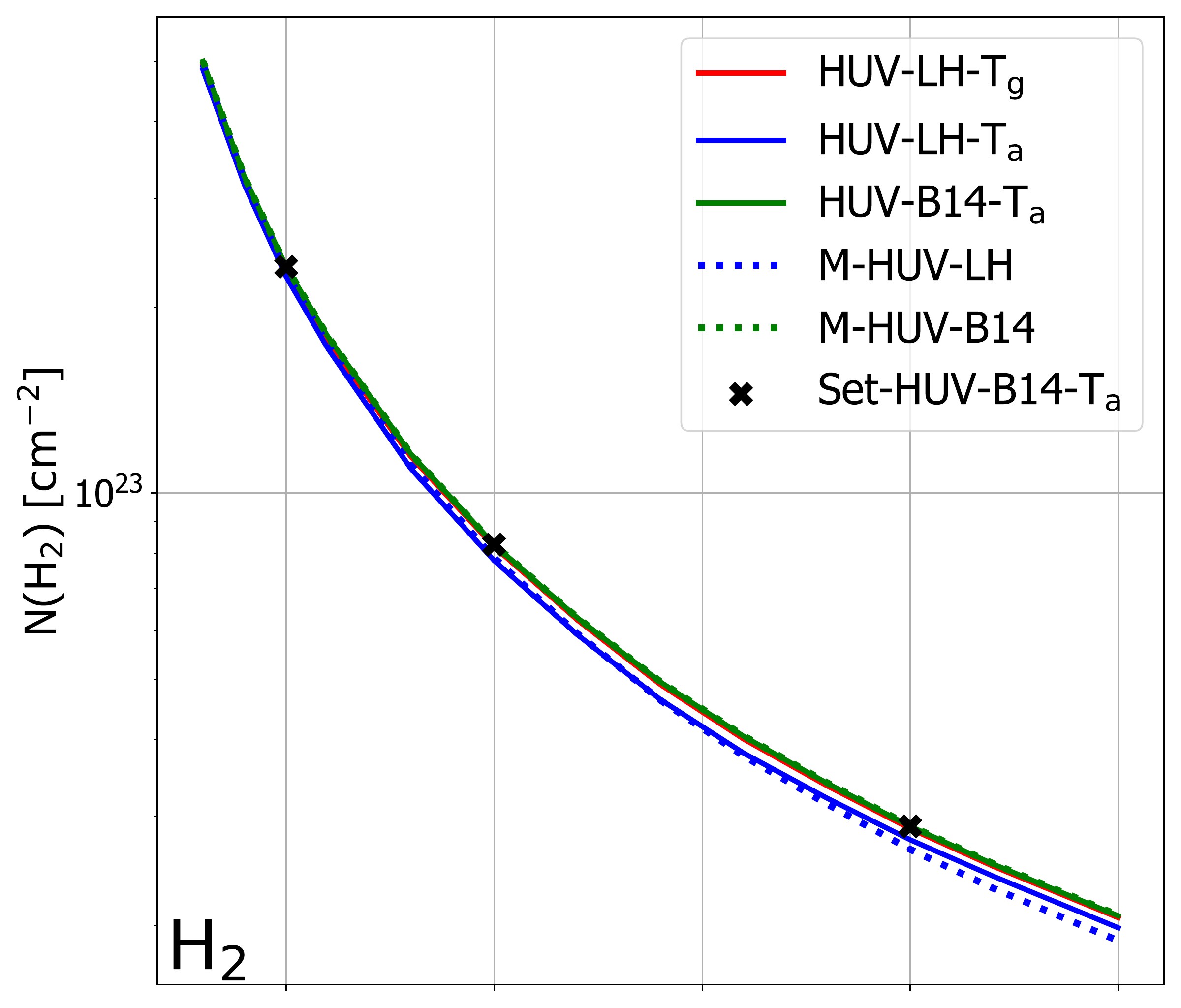}
\end{subfigure}

\begin{subfigure}{.48\linewidth}
  \centering
  \includegraphics[width=1\linewidth]{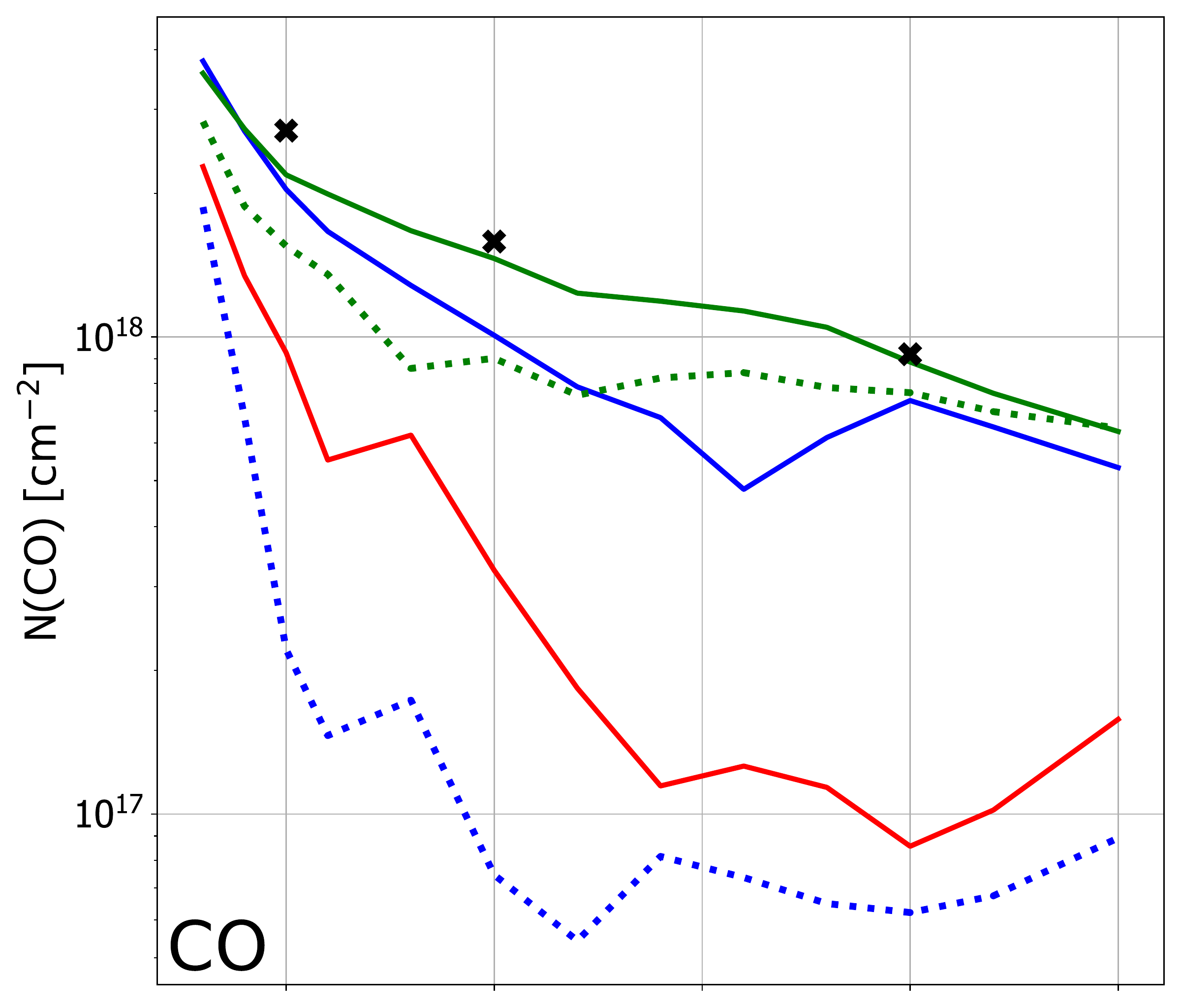}
\end{subfigure}
\begin{subfigure}{.48\linewidth}
  \centering
  \includegraphics[width=1\linewidth]{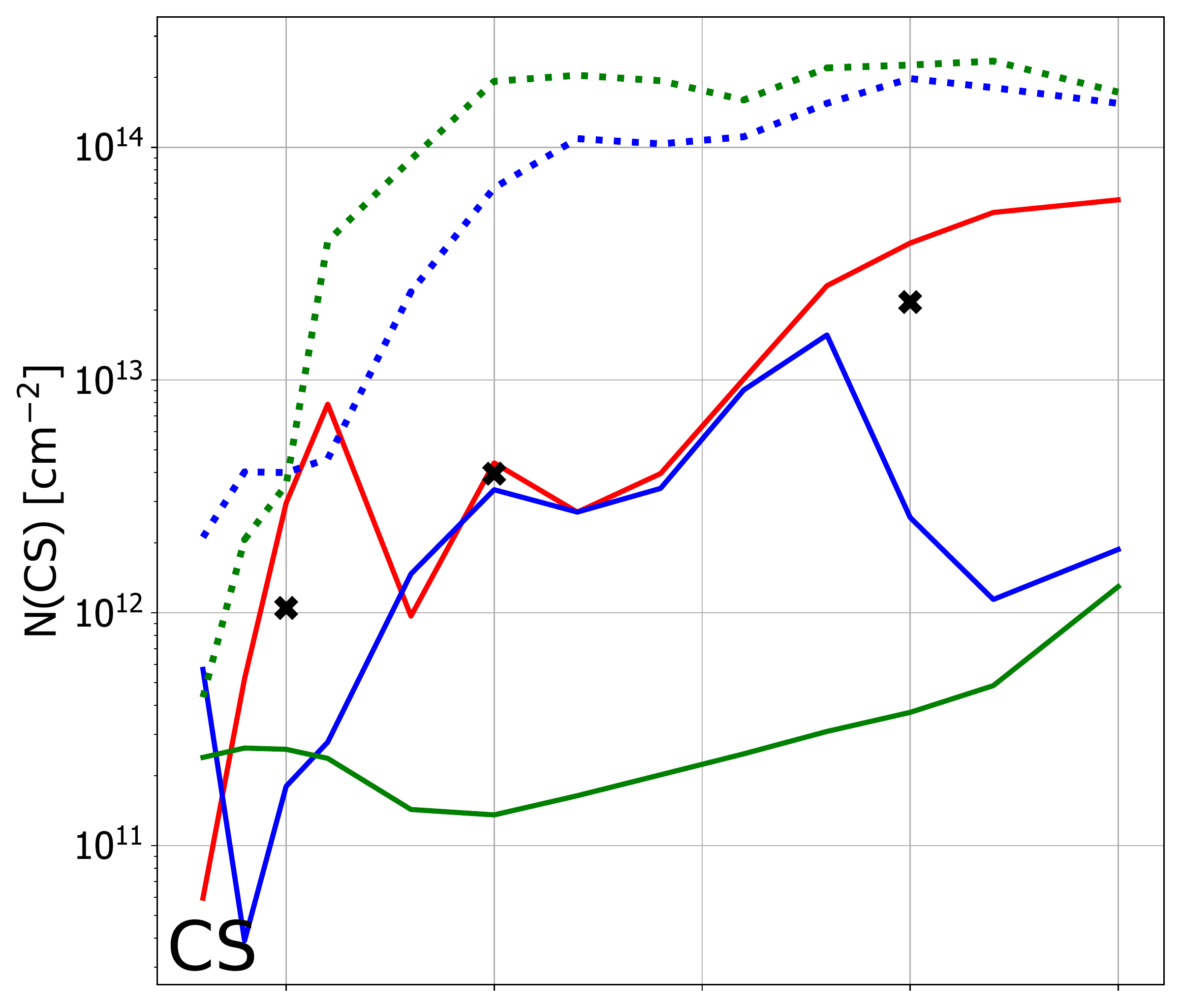}
\end{subfigure}

\begin{subfigure}{.48\linewidth}
  \centering
  \includegraphics[width=1\linewidth]{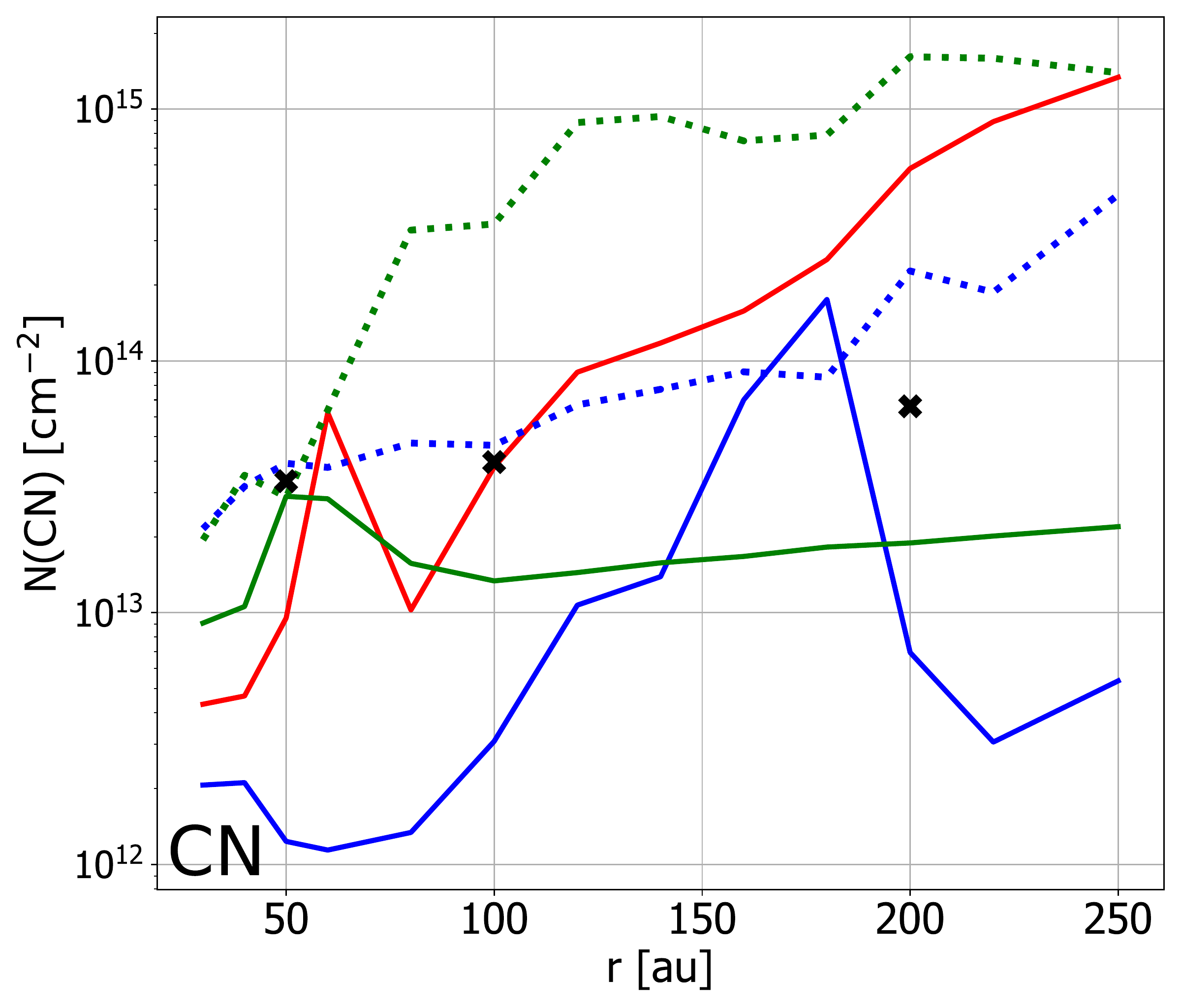}
\end{subfigure}
\begin{subfigure}{.48\linewidth}
  \centering
  \includegraphics[width=1\linewidth]{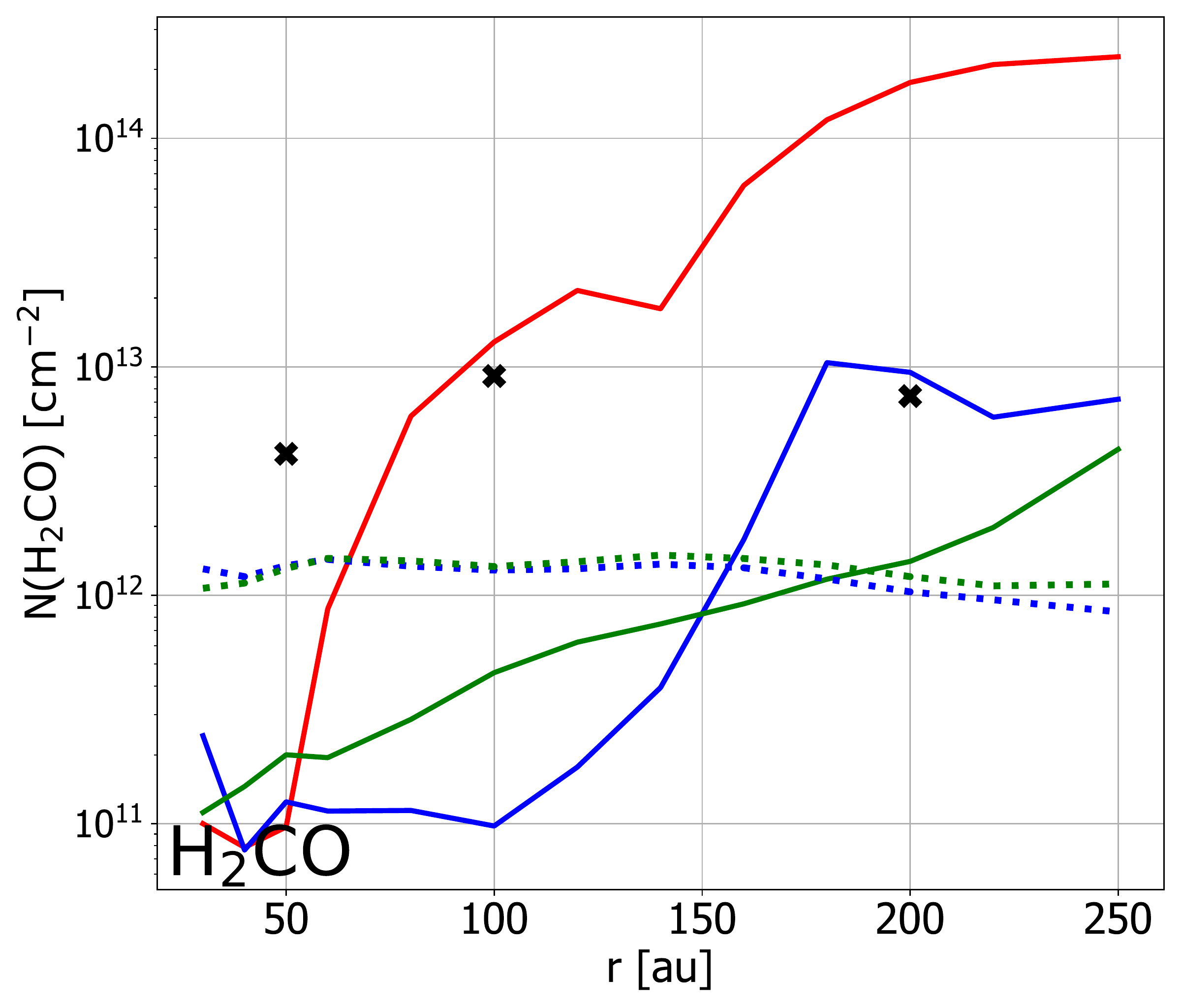}
\end{subfigure}
\caption{Surface densities of the single-grain models (solid lines) and multi-grain models (dotted lines) as a function of the radius in high UV flux regime. \revisionC{Crosses are the intermediate model for a selection of radii}.}
\label{fig:compare_high}
\end{figure*}


 \subsubsection{Impact of UV field} 

\revisionA{A comparison between Tables 
\ref{tab:huv_sigma} and \ref{tab:luv_sigma} shows that, for a given 
model, the integrated column densities (or molecular surface densities) 
have about the same order of magnitude whether the flux is high (HUV) 
or low (LUV). Unlike in the PDR layer, the midplane chemistry is not 
dependent on the UV flux regime. The dust opacity is such that the UV 
flux is sufficiently attenuated in the HUV and LUV models for the 
chemistry to be similar in both cases.}

\revisionA{Figures \ref{fig:s-100profile_high}, 
\ref{fig:m-100profile_high}, \ref{fig:s-100profile_low}, and 
\ref{fig:m-100profile_low}  show that the chemistry is nearly the same 
for all models from the midplane to roughly $1.5\,H$.  We observe more 
\hh, CO, CS and CN above 3 scale heights in the LUV models than in the 
HUV ones because a high UV flux implies more efficient 
photodissociation. Near the midplane, the UV field is similarly 
attenuated and the \revisionA{radiation field} intensities are about the same in all models. 
As a consequence, the only parameter that impacts the chemistry is the 
dust temperature $T_d$ and this applies both to the single-grain models 
and the multi-grain models.}

\subsubsection{Impact of  dust temperature on H$_2$ formation and chemistry} 
 
\revisionA{A first effect of the adopted grain temperature is related to 
the formation of \hh and the remaining amount of H, which is very 
important for the gas-phase chemistry. Adopting the formalism of 
\citet{Bron+etal_2014} to form \hh produces, as expected, more \hh. In 
these cases, Figures \ref{fig:s-100profile_high}  and 
\ref{fig:m-100profile_high} show a shift in the H/\hh transition toward 
the disk surface which implies less gas-phase H and more \hh in the 
upper layers.}

\revisionA{For single-grain models, the gas temperature we take is 
relatively low. Table \ref{tab:huv_sigma} shows that when $T_{d} = 
T_g$, the remaining amount of H is still reasonably low (intermediate), 
contrary to the case where $T_{d} = T_{a}$ (\shteff) which exhibits the 
highest amount of H, together with the multi-grain model \mhlh. 
Using only the LH mechanism to form H$_2$ on grains with high temperatures reduces
significantly the amount of \hh above two scale heights.
This is 
particularly true for the \mhlh Model, where the grain temperature 
depends on the grain size: the reservoir of  ``cold'' grains is not large 
enough to allow for an efficient \hh formation on grains in the upper 
disk layer. } 

\revisionA{On the contrary, near} the midplane, in most models 
the grain temperature is always low enough \revisionA{for the LH mechanism
to efficiently form \hh  and regulate the atomic H abundance.}

\section{Discussion}
\label{sec:discuss}
  
\revisionA{
We first discuss (Sec.\ref{subsec:gas-phase}) 
the vertical variations of abundances for CO, CN and 
 CS. We then investigate the C,N and O reservoirs  and surface 
 chemistry (Sec.\ref{subsec:reservoirs}-\ref{subsec:surface}) before discussing the formation of water and complex organic 
 molecules (Sec.\ref{subsec:water}-\ref{subsec:coms}). We also discuss some possible implications of our results 
 on planet formation and embryos compositions.  We 
 focus the discussion that follows (except for Section\,\ref{subsec:gas-phase}) 
 on the HUV case (corresponding to the UV flux 
 received by the TW Hya disk) because we have seen that most of the 
 analysis of HUV models can be considered valid for LUV models. Figures 
 relevant to the LUV case (similar to what is expected for the DM Tau 
 disk) are  presented in Appendix \ref{app:luv}. All results are presented at the final stage of 
time evolution i.e. 5.10$^6$ yrs.}

\subsection{Vertical \revisionA{distribution} of abundant (easily observed) molecules} \label{subsec:gas-phase}

\subsubsection{Column densities}
\begin{figure*} 
\begin{subfigure}{.48\linewidth}
  \centering
  \includegraphics[width=0.90\linewidth]{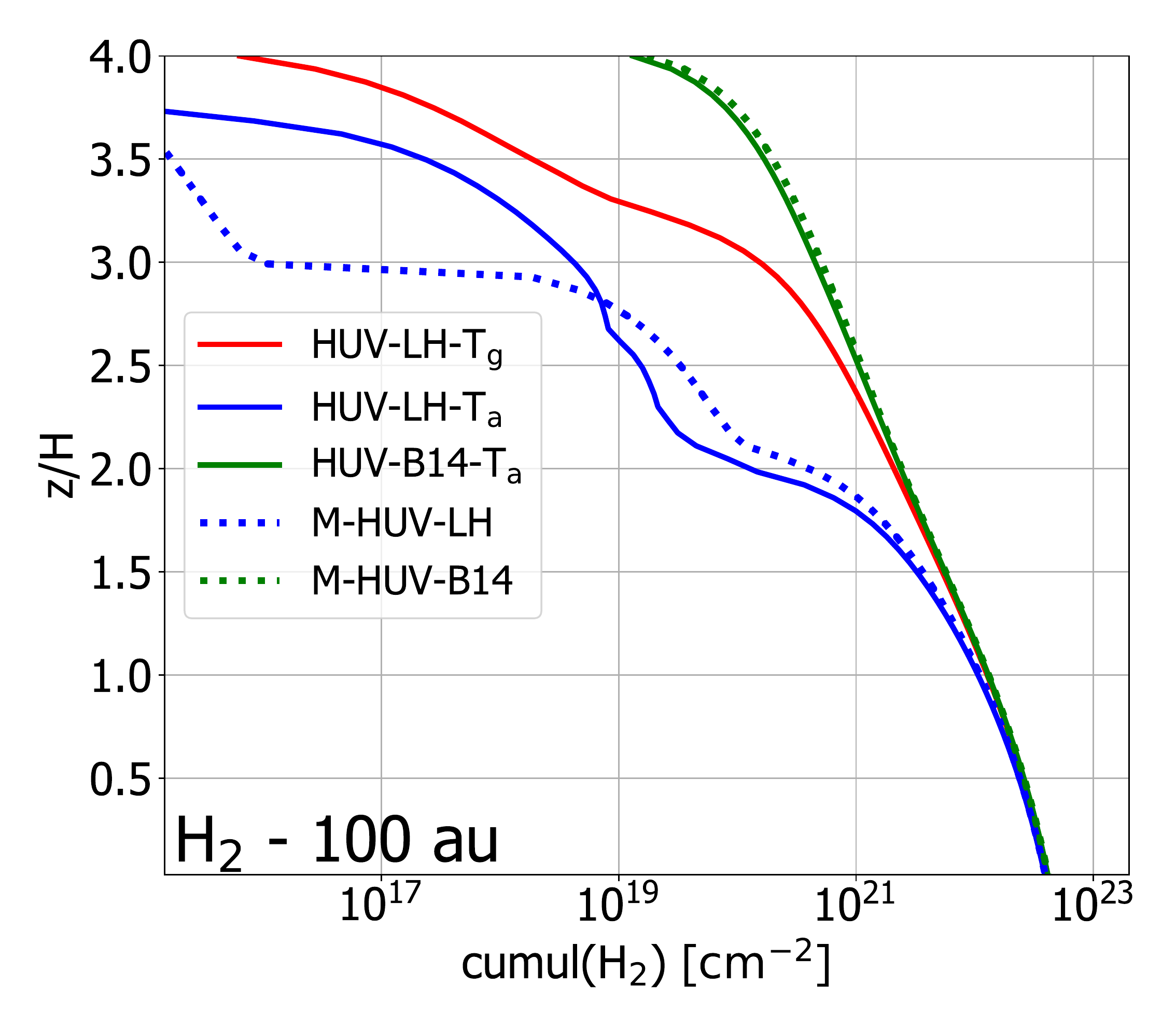}
\end{subfigure}
\begin{subfigure}{.48\linewidth}
  \centering
  \includegraphics[width=0.90\linewidth]{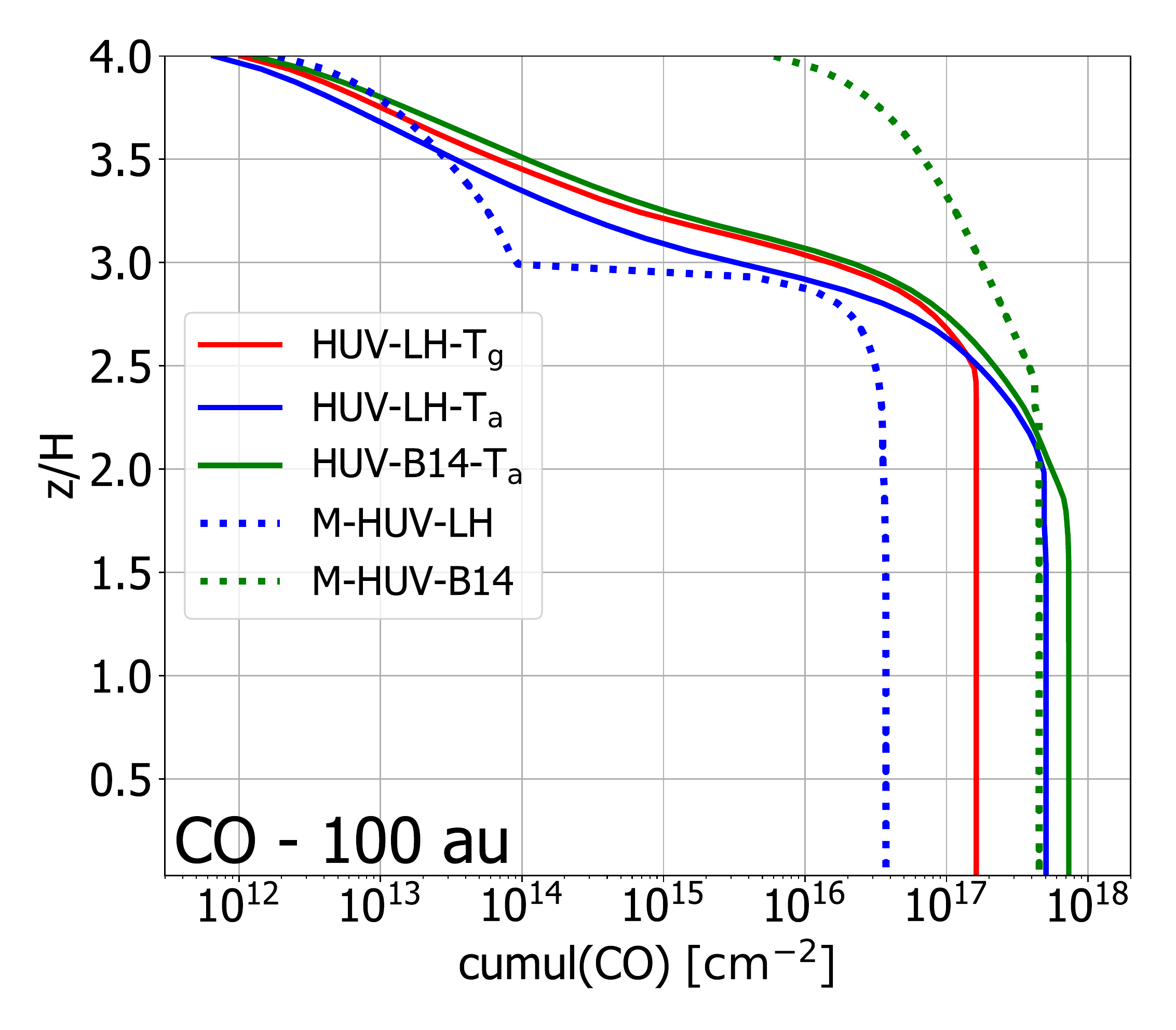}   
\end{subfigure}\\
\begin{subfigure}{.48\linewidth}
  \centering
  \includegraphics[width=0.90\linewidth]{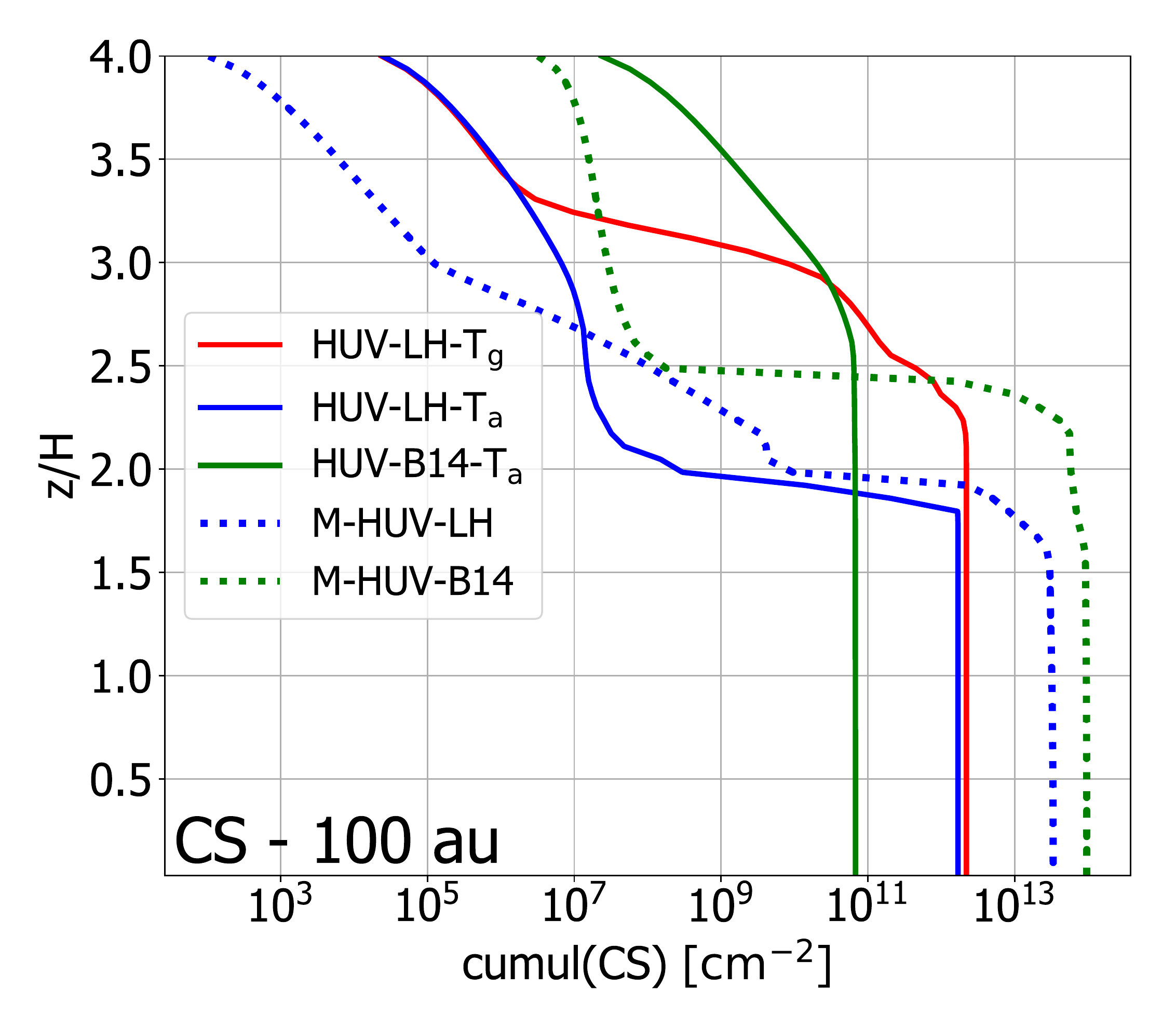}
\end{subfigure}
\begin{subfigure}{.48\linewidth}
  \centering
  \includegraphics[width=0.90\linewidth]{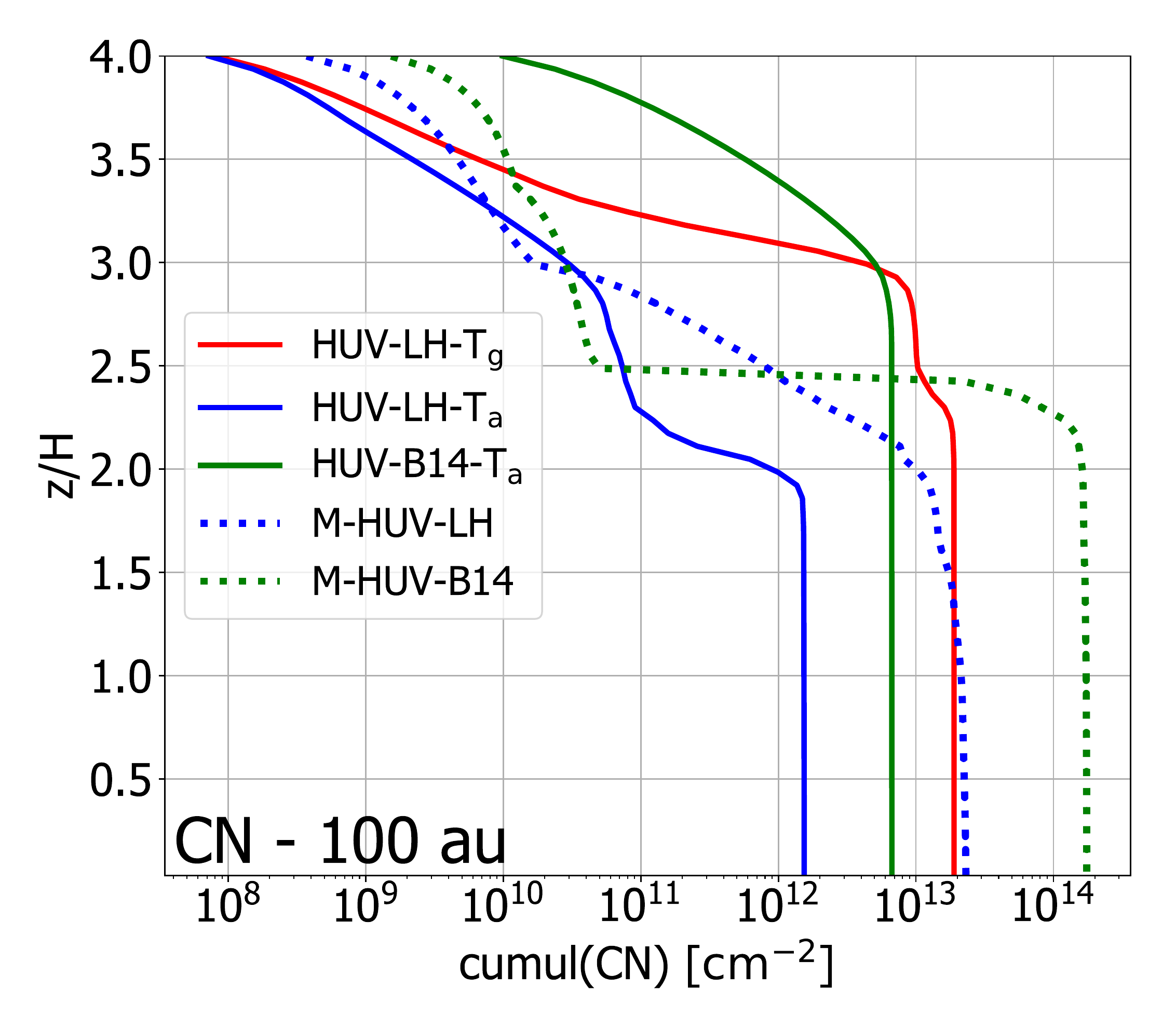}
\end{subfigure}

\caption{Vertical cumulative surface density [$\mathrm{cm^{-2}}$] of \hh, CO, CS and CN  at 100 au from the star in the HUV models.}
\label{fig:100cumul_huv}
\end{figure*}

\revisionA{Figure \ref{fig:compare_high} shows the molecular surface densities of 
a few popular species detected in TTauri disks.}
\revisionA{Table \ref{tab:huv_sigma} and \ref{tab:luv_sigma} present the 
column densities at a radius of 100 au for the most abundant species such as 
CO, CN and CS. These tables reveal that there is no specific trend for a 
given model, \revisionA{although there can be variations of up to 2.5 orders
of magnitude in the column densities.}}
\revisionA{An analysis of Figures 
\ref{fig:100cumul_huv}, \ref{fig:s-100profile_high} and 
\ref{fig:m-100profile_high}, which provide the vertical profile of
molecular densities at 100 au, is required to understand the origin
of these variations.}

\revisionA{For the column density of CO it shall be noted that 1) CO is 
formed in the gas phase and 2) is sensitive to photodissociation. The 
model that produces the most CO is \revisionA{thus \slb because this} model combines both a low 
UV flux, a large column density of H$_2$ that shields CO from the UV 
and a high grain temperature that prevents CO from being adsorbed. The 
second model \revisionA{producing} the most CO is \mlb:  the CO 
column density is slightly lower than in \slb \revisionA{because} in multi-grain       
models a fraction of the grains is significantly settled, involving a 
more effective UV penetration. In that sense, it is not surprising that 
\mhlh is the model that produces the smallest column density:  this 
model combines a high UV flux, more penetration due to settling and a 
small production of H$_2$ which decreases the shielding.}

\revisionA{For the column densities of CS and CN, Table 
\ref{tab:huv_sigma} shows that using predicted molecular surface 
densities is not enough to derive general trends. 
\revisionA{Nevertheless,} we can note that \mhb is the model 
that produces the most CS and CN. On the other hand, \shb produces the 
smallest column density of CS while \shteff is the model that produces 
the smallest column density of CN.  This suggests routes of formation 
which do not simply depend on the grain temperature. }

\revisionA{In the case of the single-grain models (see 
Figs.\,\ref{fig:s-100profile_high} and \ref{fig:s-100profile_low}), all 
models using the same grain temperature ($T_d = T_g$ or $T_a$)
have 
the same vertical profile at altitudes $z < 1.5 H$  regardless of the flux regime. 
Similarly, since all multi-grain models use exactly the same grain 
temperature and size distribution, they all exhibit the same vertical 
density profile at altitudes $z < 1.5 H$ (for a given 
species) regardless of the incident flux and \revisionA{of the assumed \hh formation
mechanism} 
\revisionA{(see for example, the CO abundance profiles for the two multi-grain models
in Figure \ref{fig:m-100profile_high}, middle panels of columns b-c).
Accordingly, in this relatively high mass disk, differences in column 
densities (Tab.\,\ref{tab:huv_sigma}) result from what happens above $z = 1.5 H$.}}

\vspace{0.2cm}

\revisionC{Finally, concerning the intermediate model, Set-HUV-B14-T$\bm{\mathrm{_a}}$, Tab.\,\ref{tab:huv_sigma} shows that
the column densities are rather similar to those found for  model \shteff, with the exception of CS where the column density is closer to
that found for \shtg.  However, these values are integrated column densities and Figs.\,\ref{fig:interm-100profile_LH} and \ref{fig:interm-100profile_B14} 
show that the vertical profiles are more complex than the comparison based on
surface densities may suggest.}

  \begin{table*} 
\centering
\caption{Column densities [cm$^{-2}$] of main molecules at 100 au for HUV models. Last three columns summarize the main properties of the model with respect to the \hh formation and grains (\% of grains, in sites, with a temperature above 20 K between 0 and $1.5 H$ and $1.5$ and $2.5 H$). \label{tab:huv_sigma}}
\begin{tabular}{l c c c c c c c c}
\hline
\noalign{\smallskip}
& \textbf{H} & \textbf{H$_2$} & \textbf{CO} & \textbf{CS} & \textbf{CN} & \hh formation &  0 to $1.5 H$ &  1.5 to $2.5 H$\\
\noalign{\smallskip}
\hline	\hline
\noalign{\smallskip}
\textbf{HUV-LH-T$\bm{\mathrm{_g}}$}  &	 $1.48.10^{21}$ & $8.19.10^{22}$ & $3.24.10^{17}$ & $4.40.10^{12}$ & $3.79.10^{13}$ & "cold" grains &	0\% & 25\% \\
\noalign{\smallskip}
\hline
\noalign{\smallskip}
\textbf{HUV-LH-T$\bm{\mathrm{_a}}$}&        $9.80.10^{21}$ & $7.78.10^{22}$ & $1.01.10^{18}$ & $3.38.10^{12}$ & $3.08.10^{12}$ & more H &100\% & 100\% \\
\noalign{\smallskip}
\hline
\noalign{\smallskip}
\textbf{HUV-B14-T$\bm{\mathrm{_a}}$} &     $3.20.10^{20}$ & $8.25.10^{22}$ & $1.46.10^{18}$ & $1.35.10^{11}$ & $1.34.10^{13}$ & less H & 100\% & 100\%	\\
\noalign{\smallskip}
\hline
\noalign{\smallskip}
\textbf{M-HUV-LH}  &	$8.26.10^{21}$ & $7.86.10^{22}$ & $7.46.10^{16}$ & $6.69.10^{13}$ & $4.62.10^{13}$ & more H &  96\% & 99\%\\
\noalign{\smallskip}
\hline
\noalign{\smallskip}
\textbf{M-HUV-B14}&        $1.31.10^{20}$ & $8.26.10^{22}$ & $9.02.10^{17}$ & $1.92.10^{14}$ & $3.49.10^{14}$ & less H & 96\% & 99\% \\
\noalign{\smallskip}
\hline
\textbf{Set-HUV-B14-T$\bm{\mathrm{_a}}$}&        {\bf $3.65.10^{20}$} & {\bf $8.25.10^{22}$} & {\bf $1.58.10^{18}$} & {\bf $3.96.10^{12}$} &{\bf $3.96.10^{13}$} &  less H & 100\%  & 100\%  \\
\noalign{\smallskip}
\hline
\noalign{\smallskip}
\end{tabular}
\end{table*}


\begin{figure*}
\begin{subfigure}{.33\linewidth}
  \centering
  \includegraphics[width=1.05\linewidth]{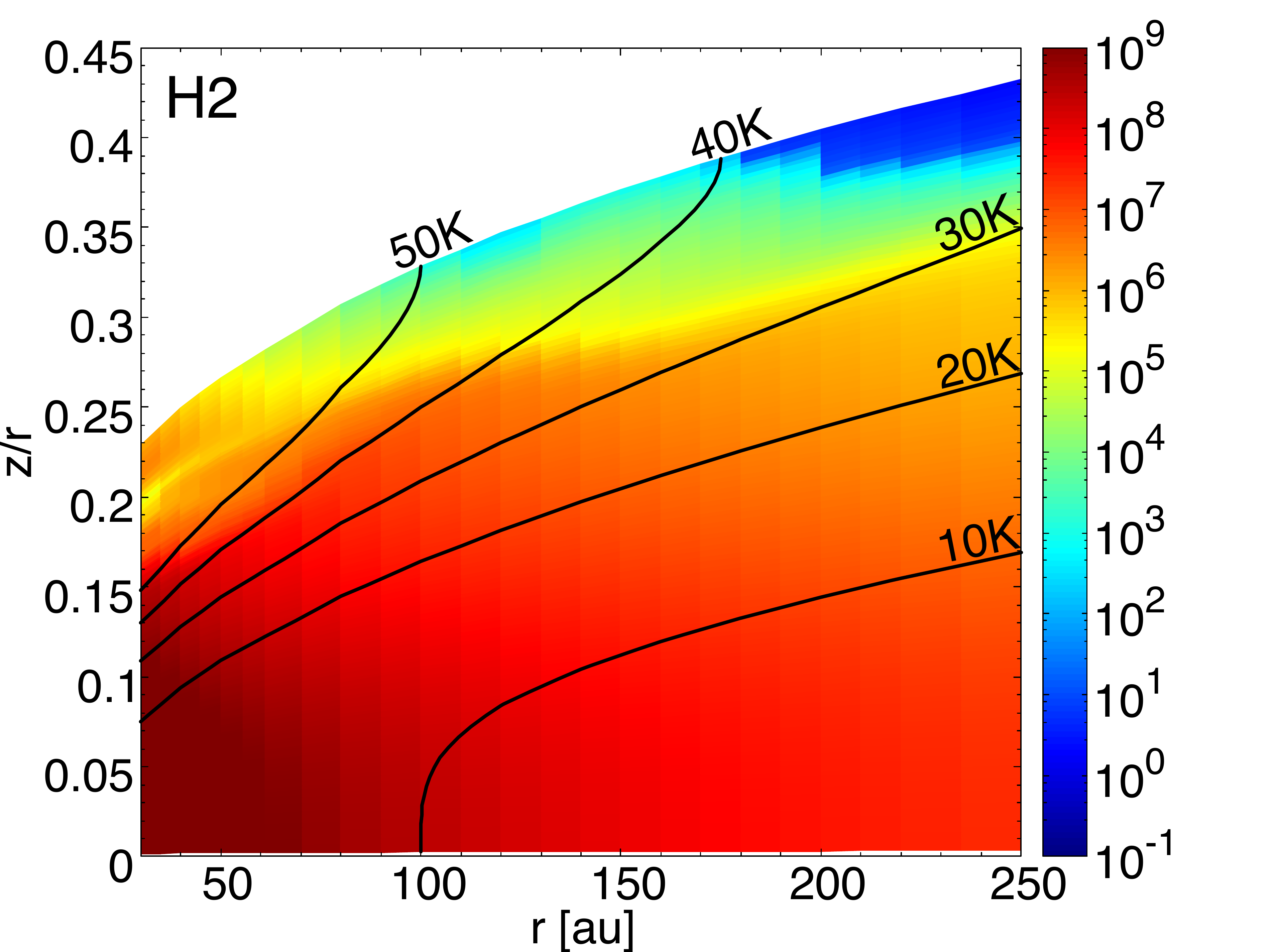}
  
\end{subfigure}
\begin{subfigure}{.33\linewidth}
  \centering
  \includegraphics[width=1.05\linewidth]{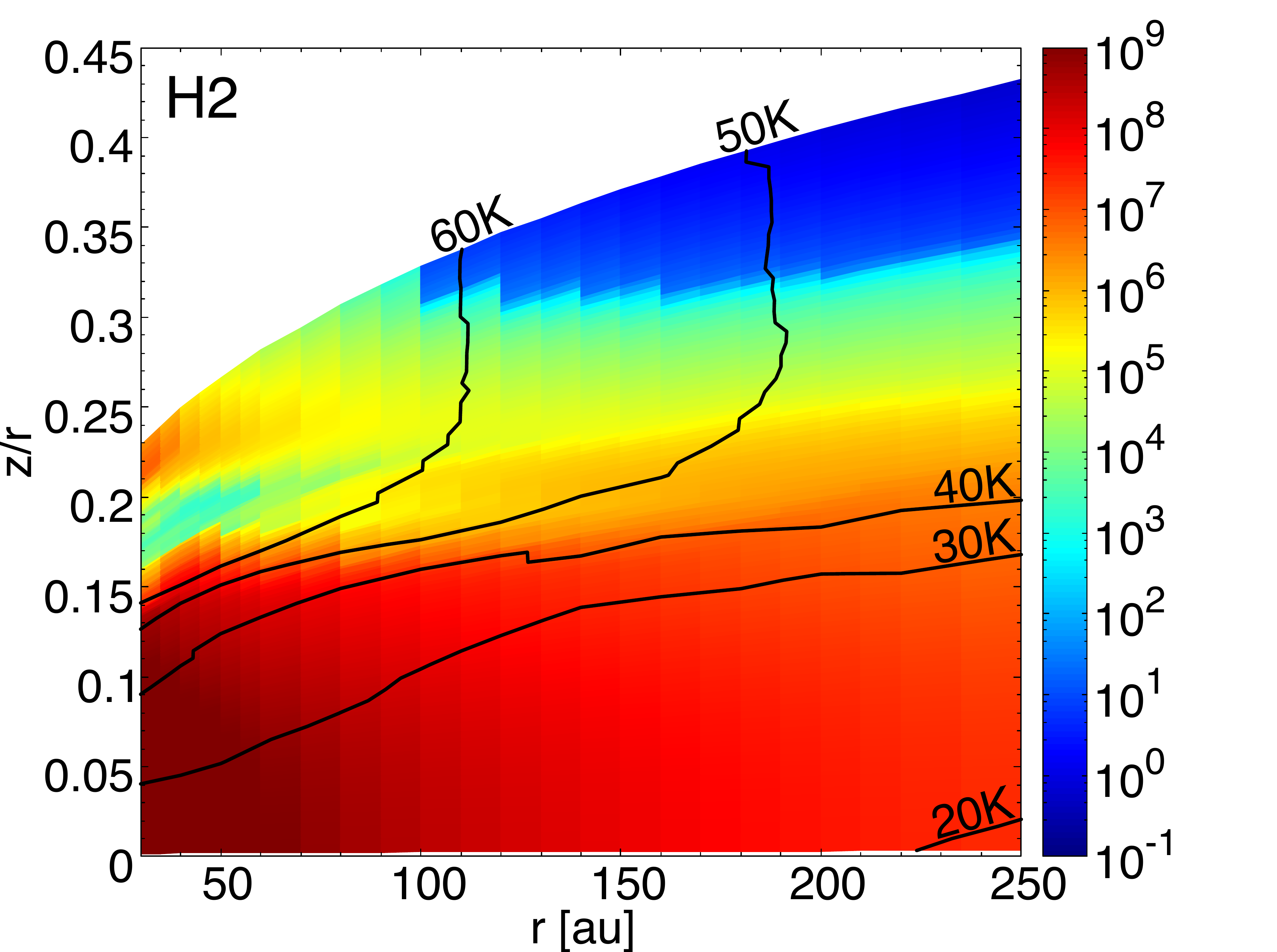}
 
\end{subfigure}
\begin{subfigure}{.33\linewidth}
  \centering
  \includegraphics[width=1.05\linewidth]{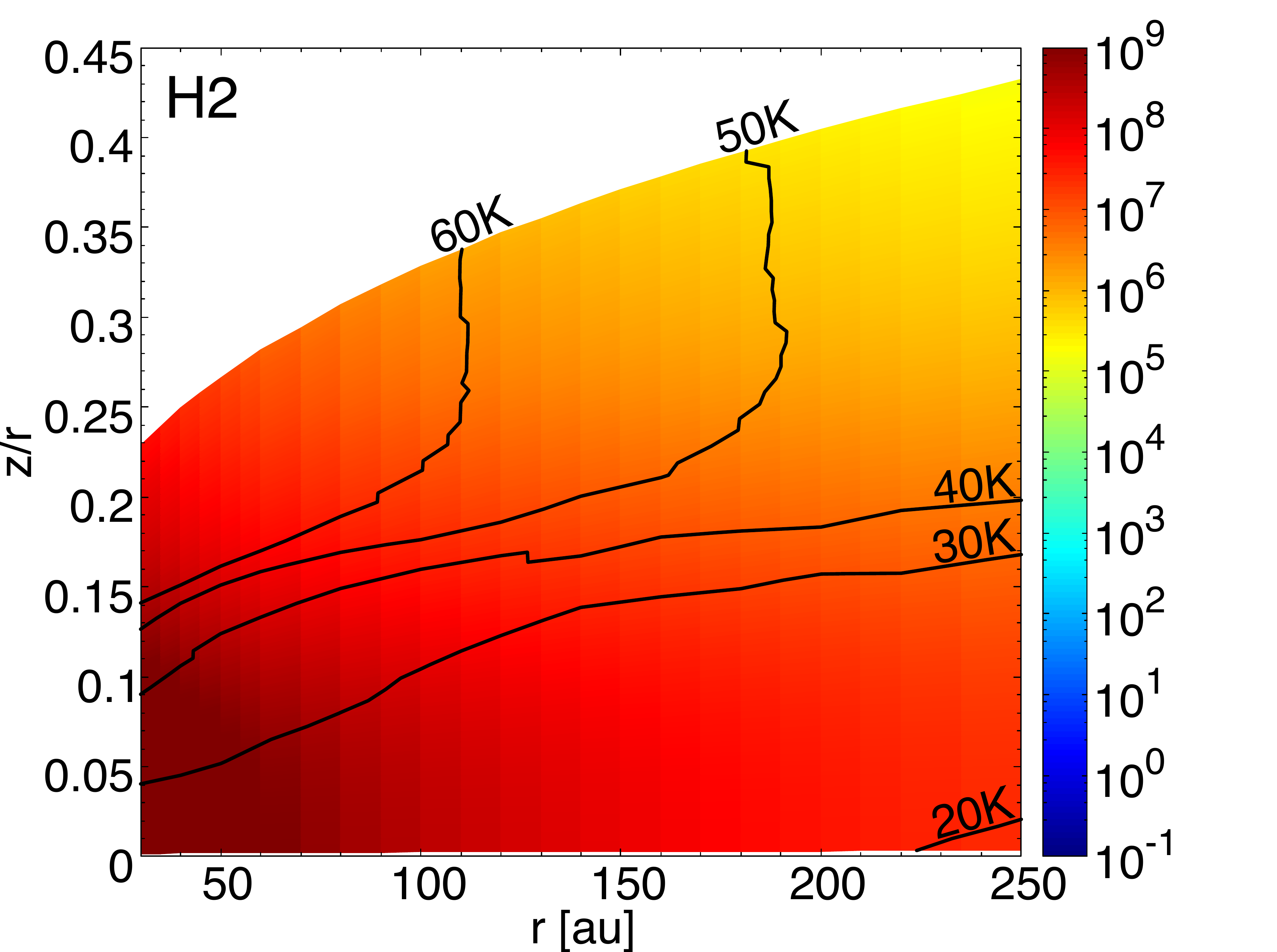}

\end{subfigure}

\begin{subfigure}{.33\linewidth}
  \centering
  \includegraphics[width=1.05\linewidth]{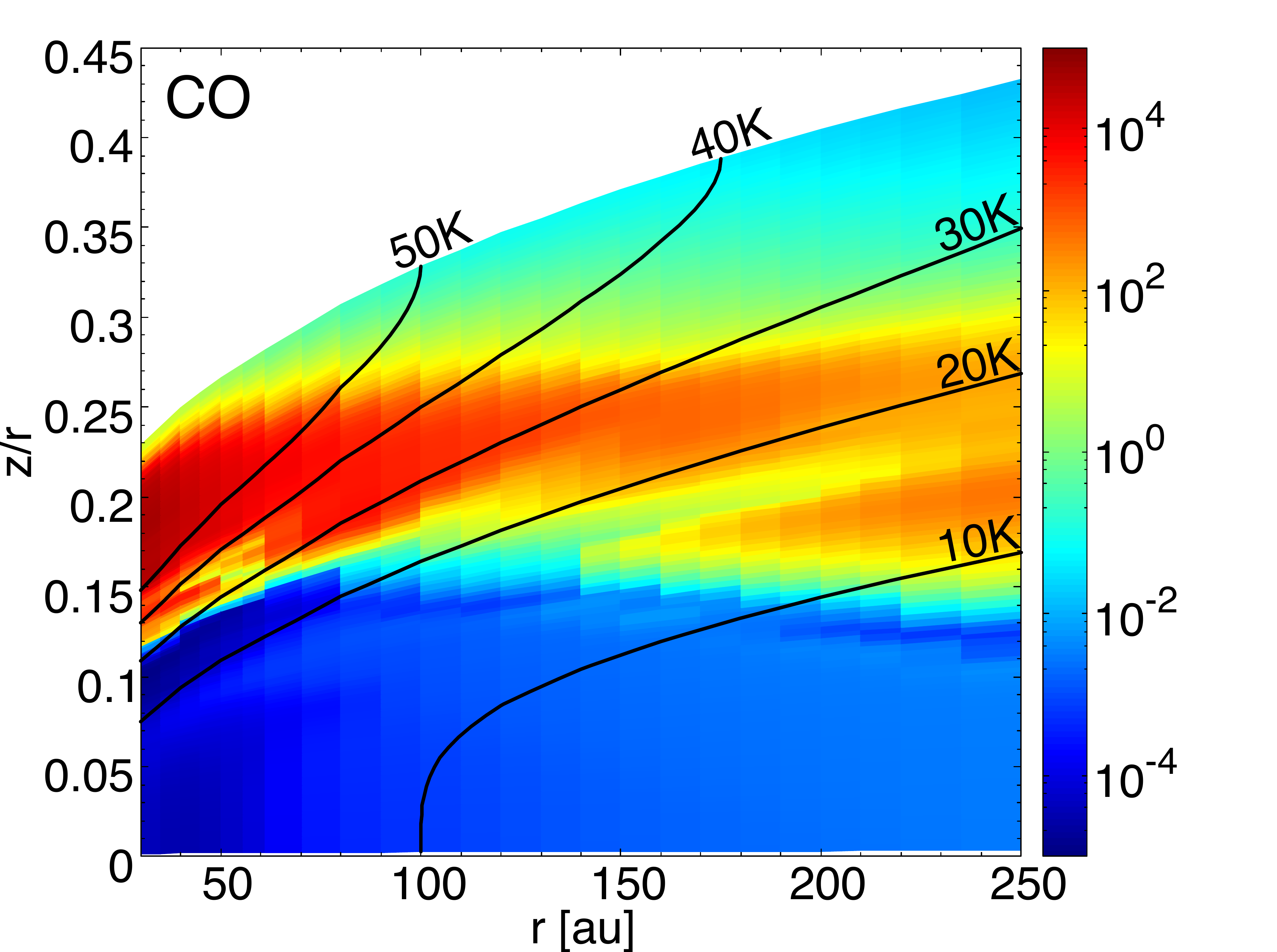}
  
\end{subfigure}
\begin{subfigure}{.33\linewidth}
  \centering
  \includegraphics[width=1.05\linewidth]{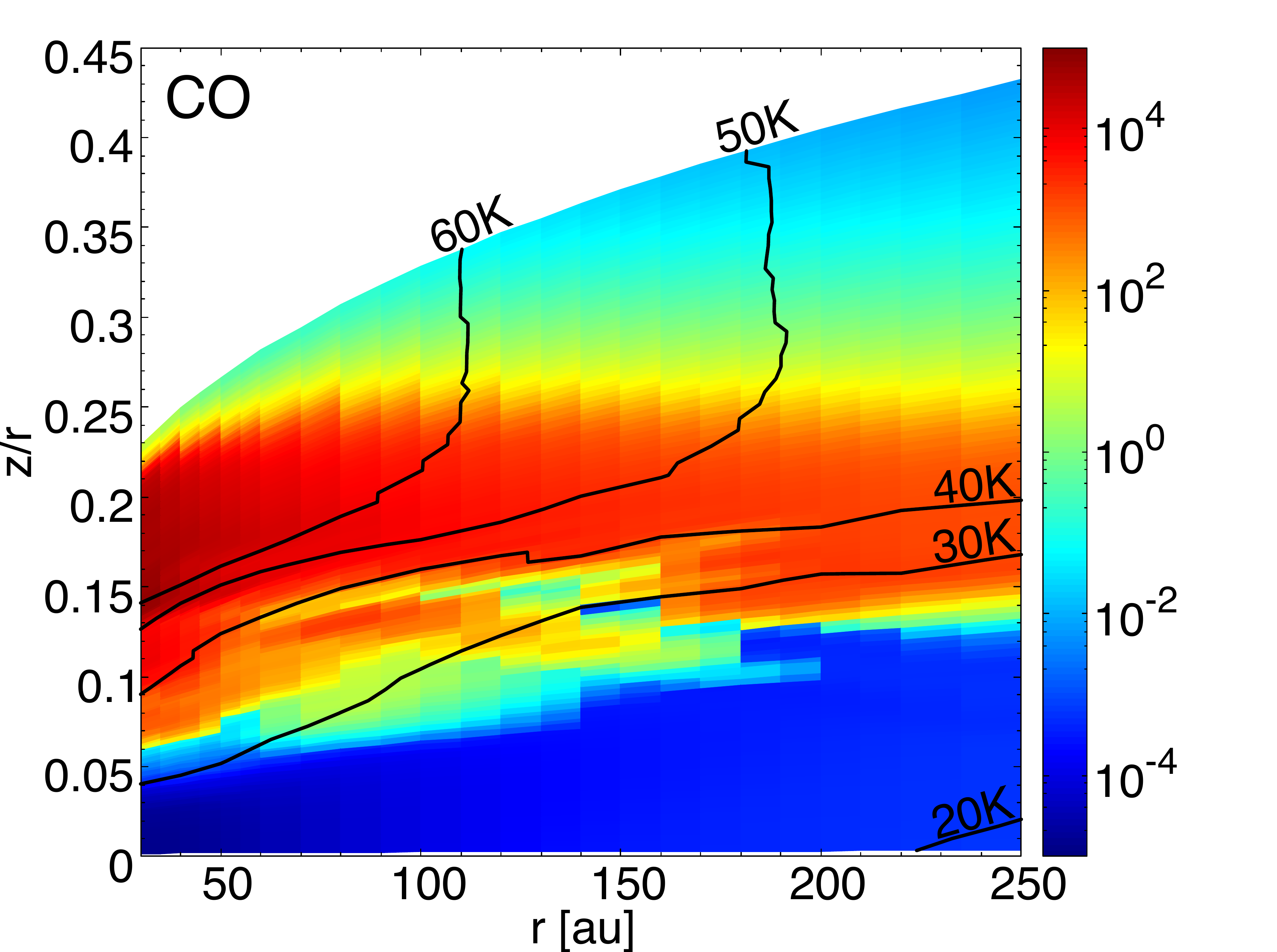}

\end{subfigure}
\begin{subfigure}{.33\linewidth}
  \centering
  \includegraphics[width=1.05\linewidth]{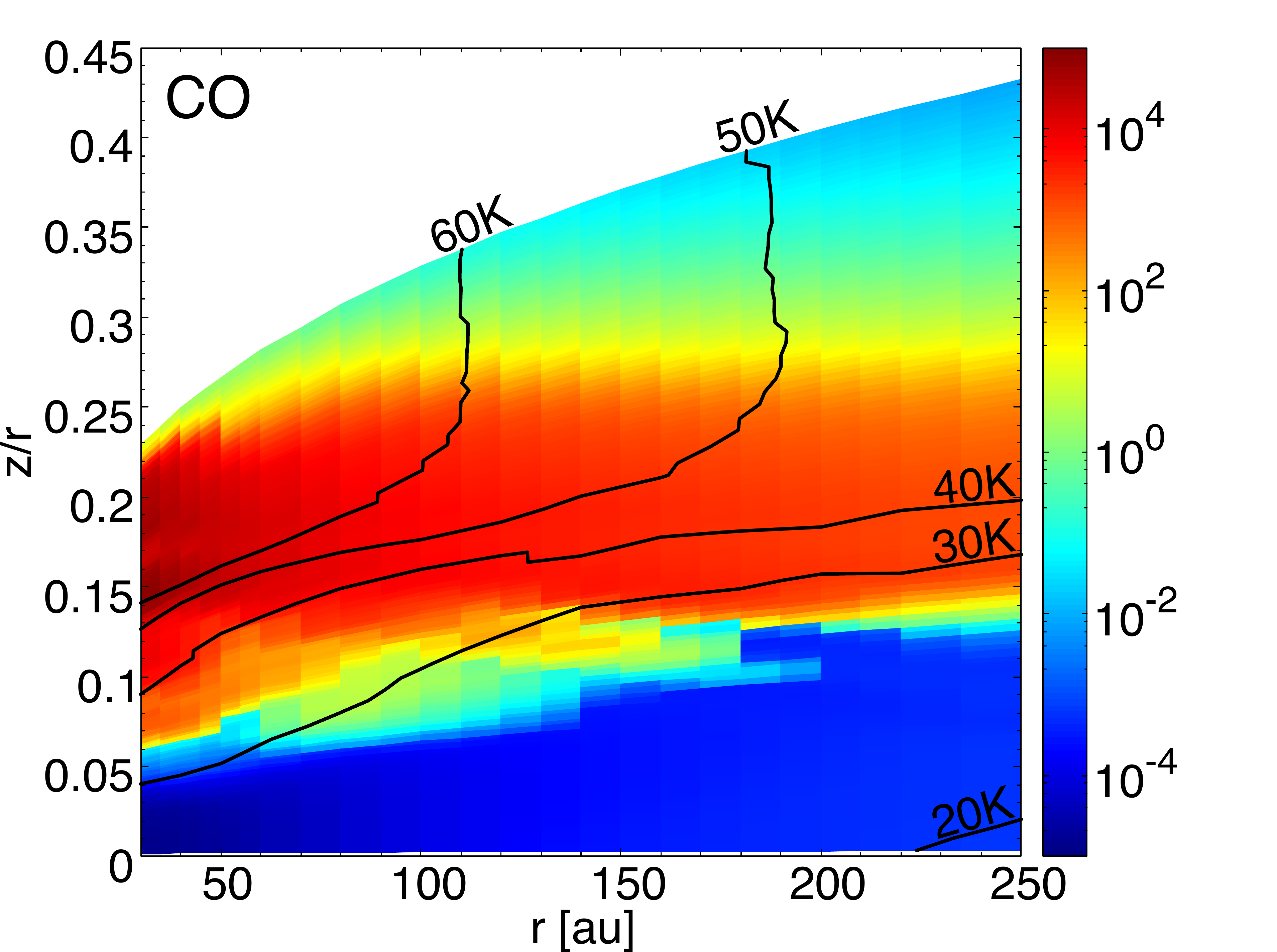}

\end{subfigure}

\begin{subfigure}{.33\linewidth}
  \centering
  \includegraphics[width=1.05\linewidth]{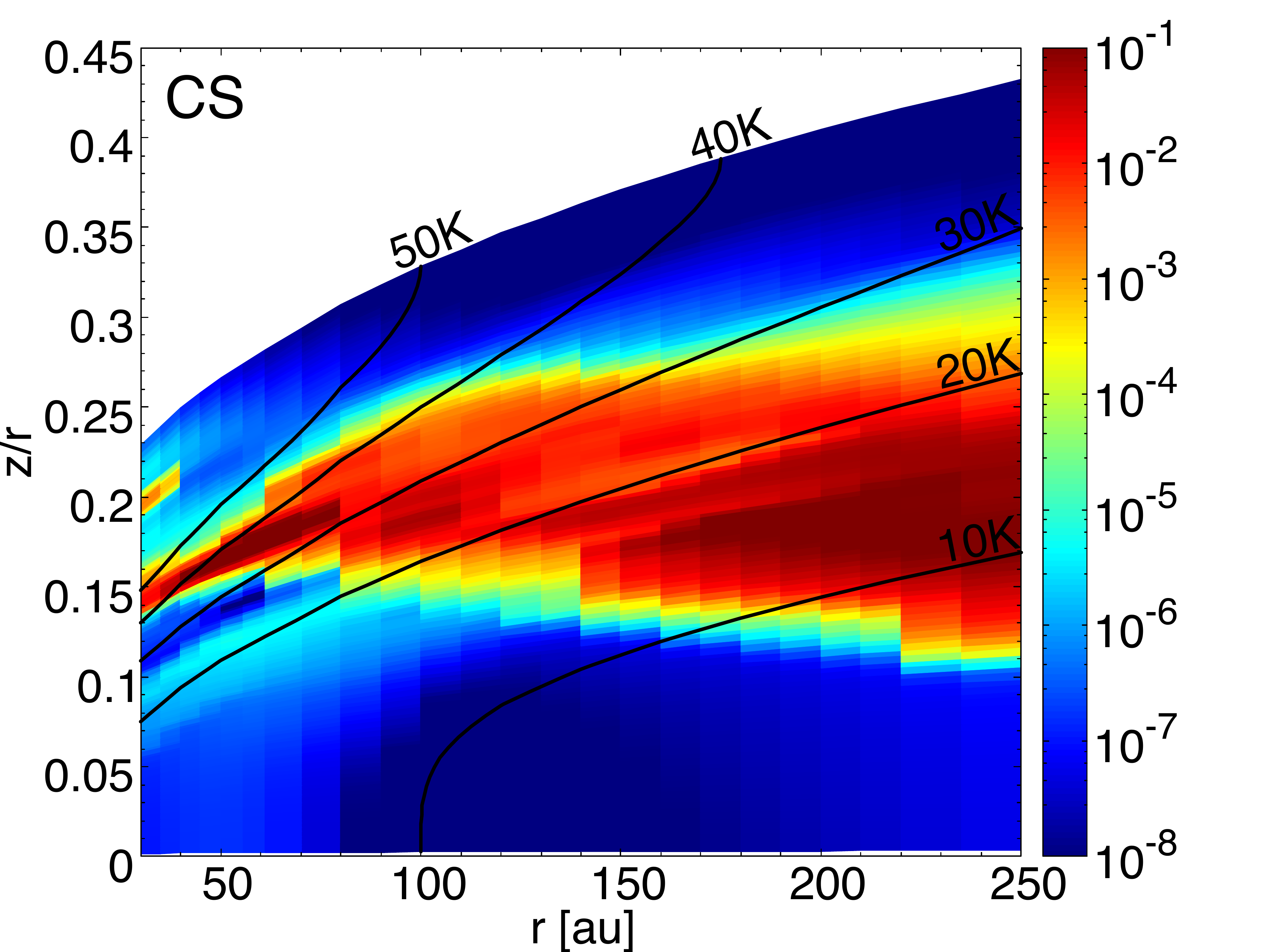}
\end{subfigure}
\begin{subfigure}{.33\linewidth}
  \centering
  \includegraphics[width=1.05\linewidth]{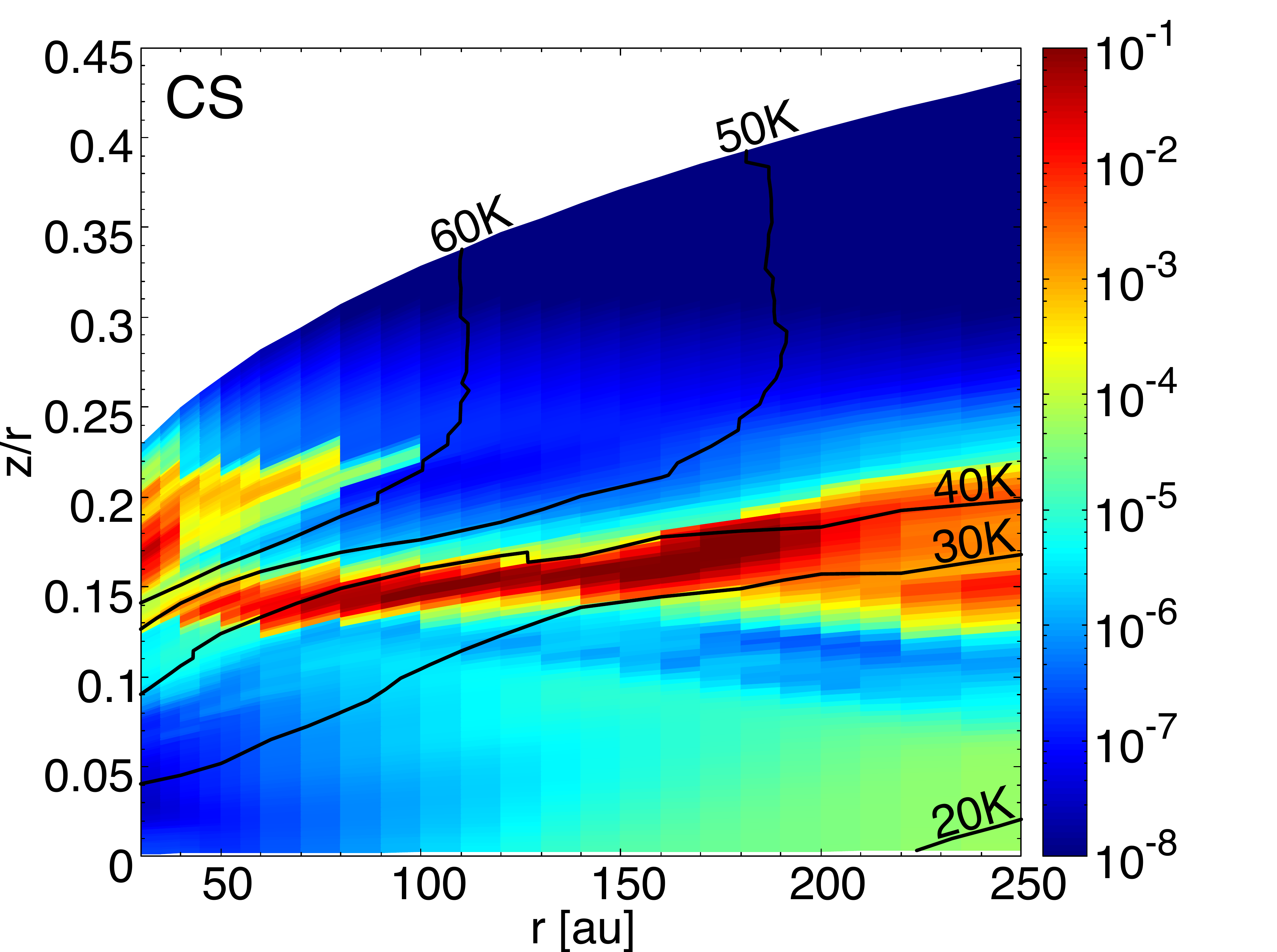}
\end{subfigure}
\begin{subfigure}{.33\linewidth}
  \centering
  \includegraphics[width=1.05\linewidth]{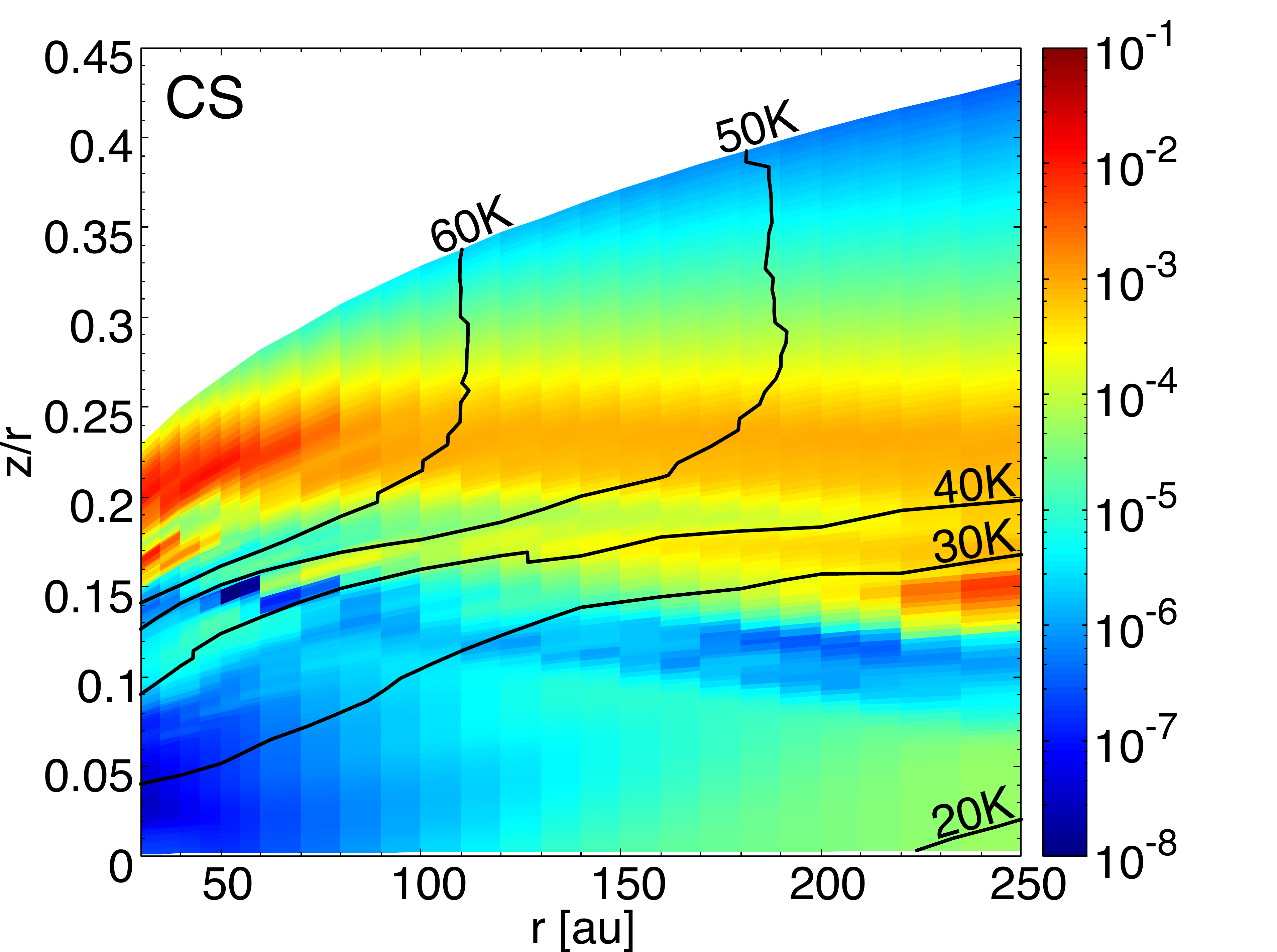}

\end{subfigure}

\begin{subfigure}{.33\linewidth}
  \centering
  \includegraphics[width=1.05\linewidth]{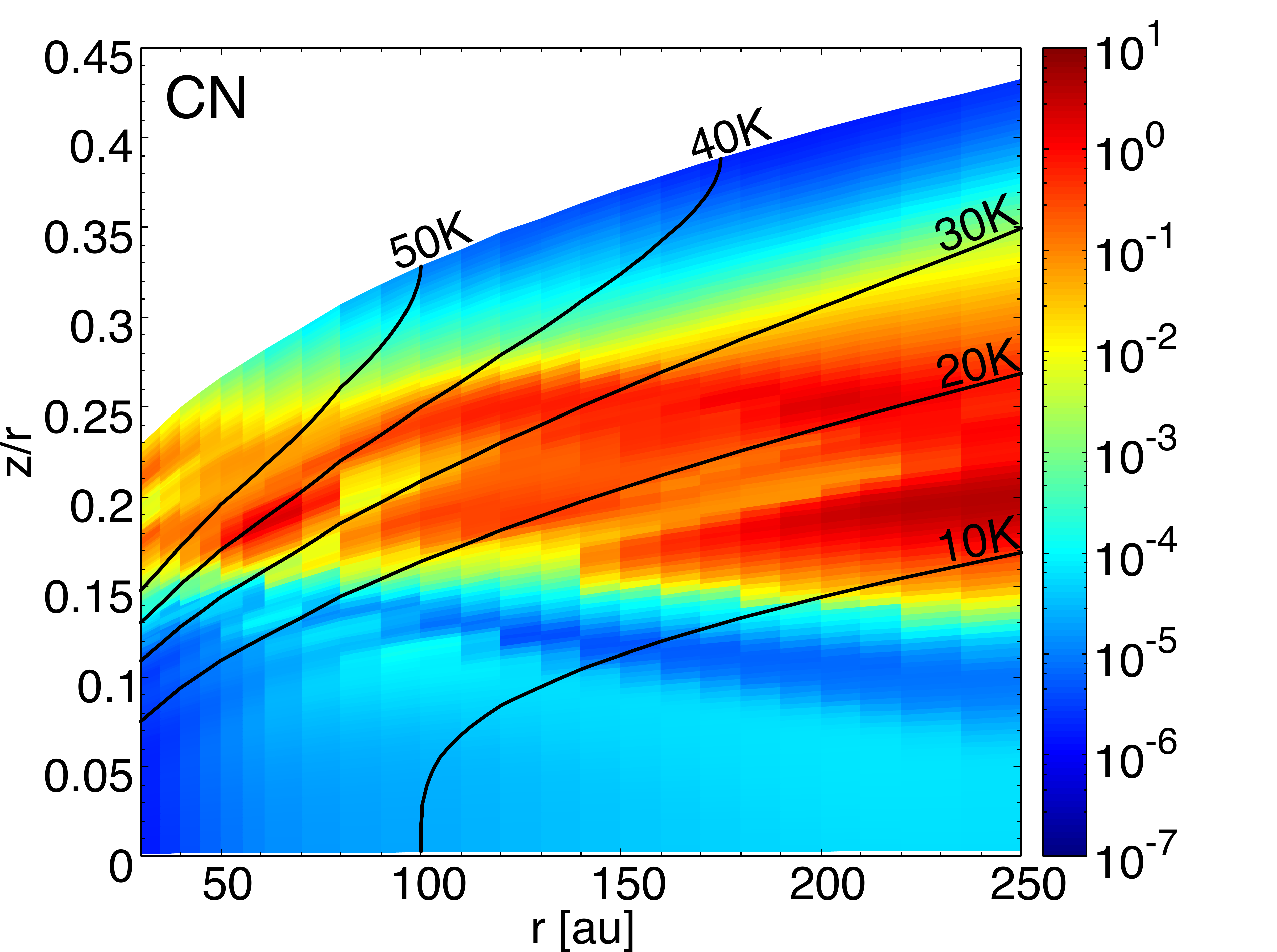}
   \subcaption{\shtg}   
\end{subfigure}
\begin{subfigure}{.33\linewidth}
  \centering
  \includegraphics[width=1.05\linewidth]{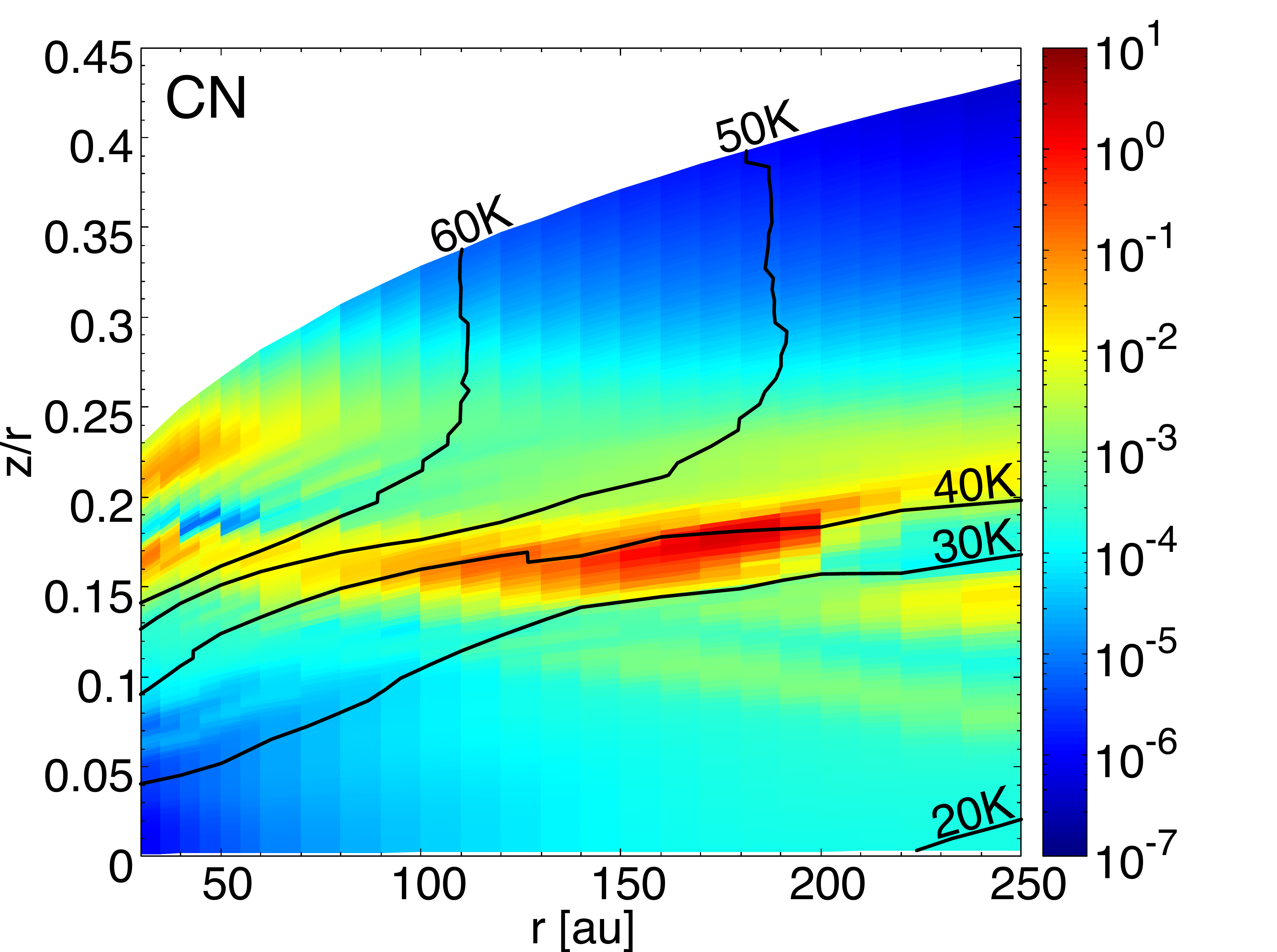}
   \subcaption{\shteff}   
\end{subfigure}
\begin{subfigure}{.33\linewidth}
  \centering
  \includegraphics[width=1.05\linewidth]{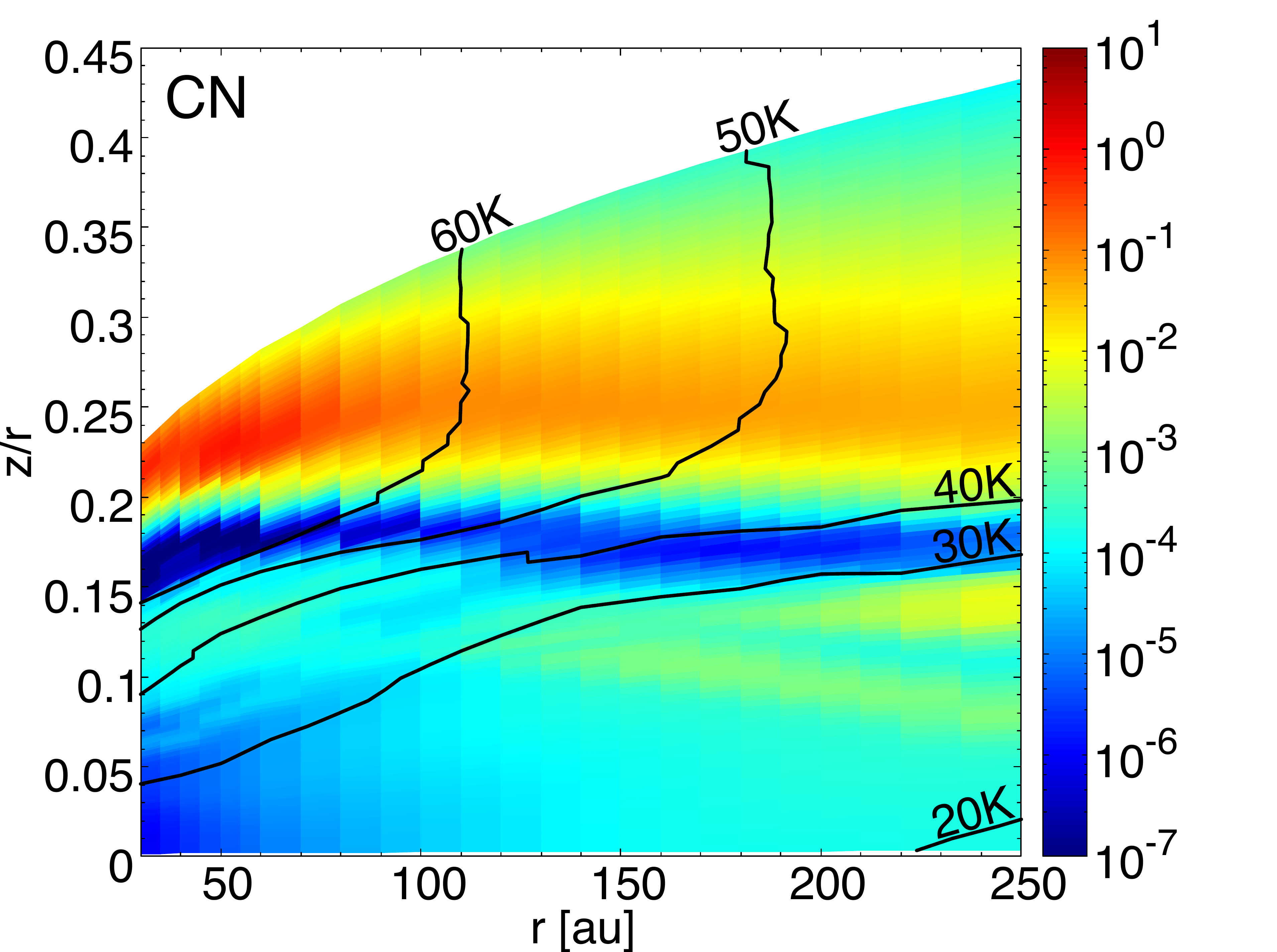}
   \subcaption{\shb} 
\end{subfigure}
\caption{Density [cm$^{-3}$] of H$_2$, CO, CS and CN in the gas-phase of the single-grain models in HUV regime. Left column is the \shtg model, middle one is \shteff and right one is \shb.  Black contours represent the dust temperature (T$_\mathrm{d}$ = T$_\mathrm{g}$ in the left column and T$_\mathrm{d}$ = T$_\mathrm{a}$ in the middle and right columns).}
\label{fig:s-maps-huv}
\end{figure*}


\begin{figure*}
\begin{subfigure}{.33\linewidth}
  \centering
  \includegraphics[width=1.05\linewidth]{figures/SINGLE/HUV_HL_Tg/maps/H2.pdf}
\end{subfigure}
\begin{subfigure}{.33\linewidth}
  \centering
  \includegraphics[width=1.05\linewidth]{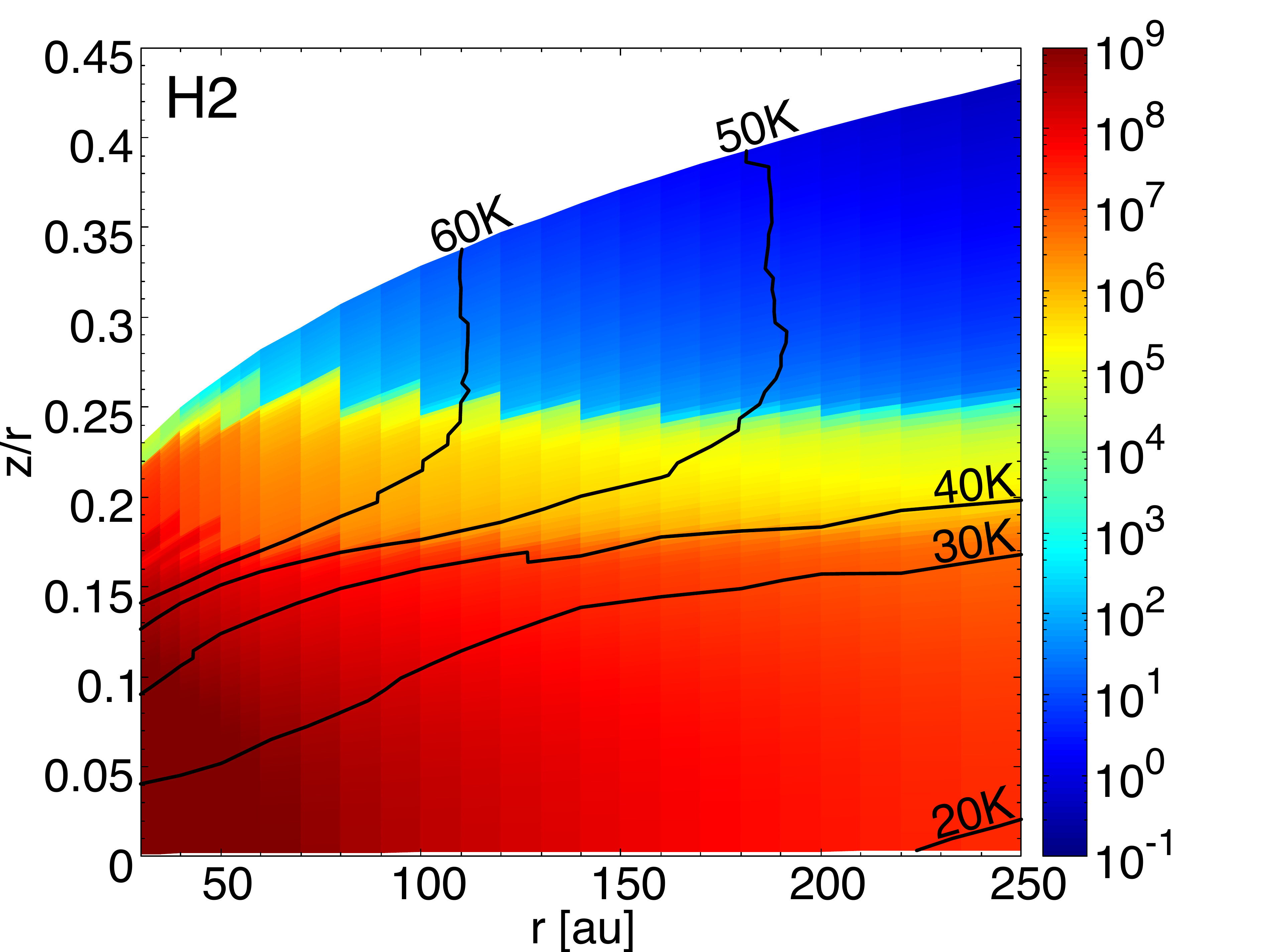}
\end{subfigure}
\begin{subfigure}{.33\linewidth}
  \centering
  \includegraphics[width=1.05\linewidth]{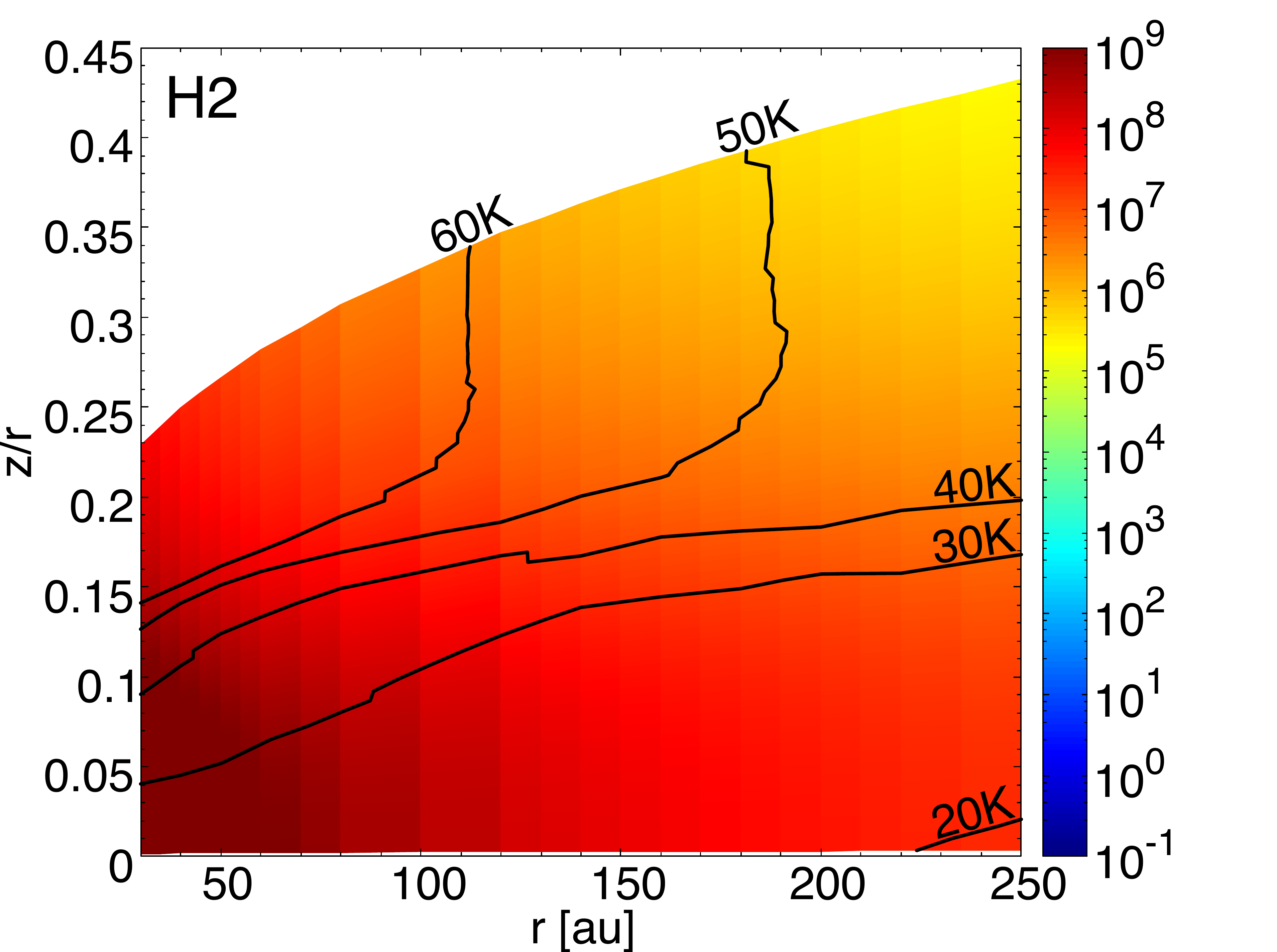} 
\end{subfigure}
\label{fig:m-mapsH2}

\begin{subfigure}{.33\linewidth}
  \centering
  \includegraphics[width=1.05\linewidth]{figures/SINGLE/HUV_HL_Tg/maps/CO.pdf}
\end{subfigure}
\begin{subfigure}{.33\linewidth}
  \centering
  \includegraphics[width=1.05\linewidth]{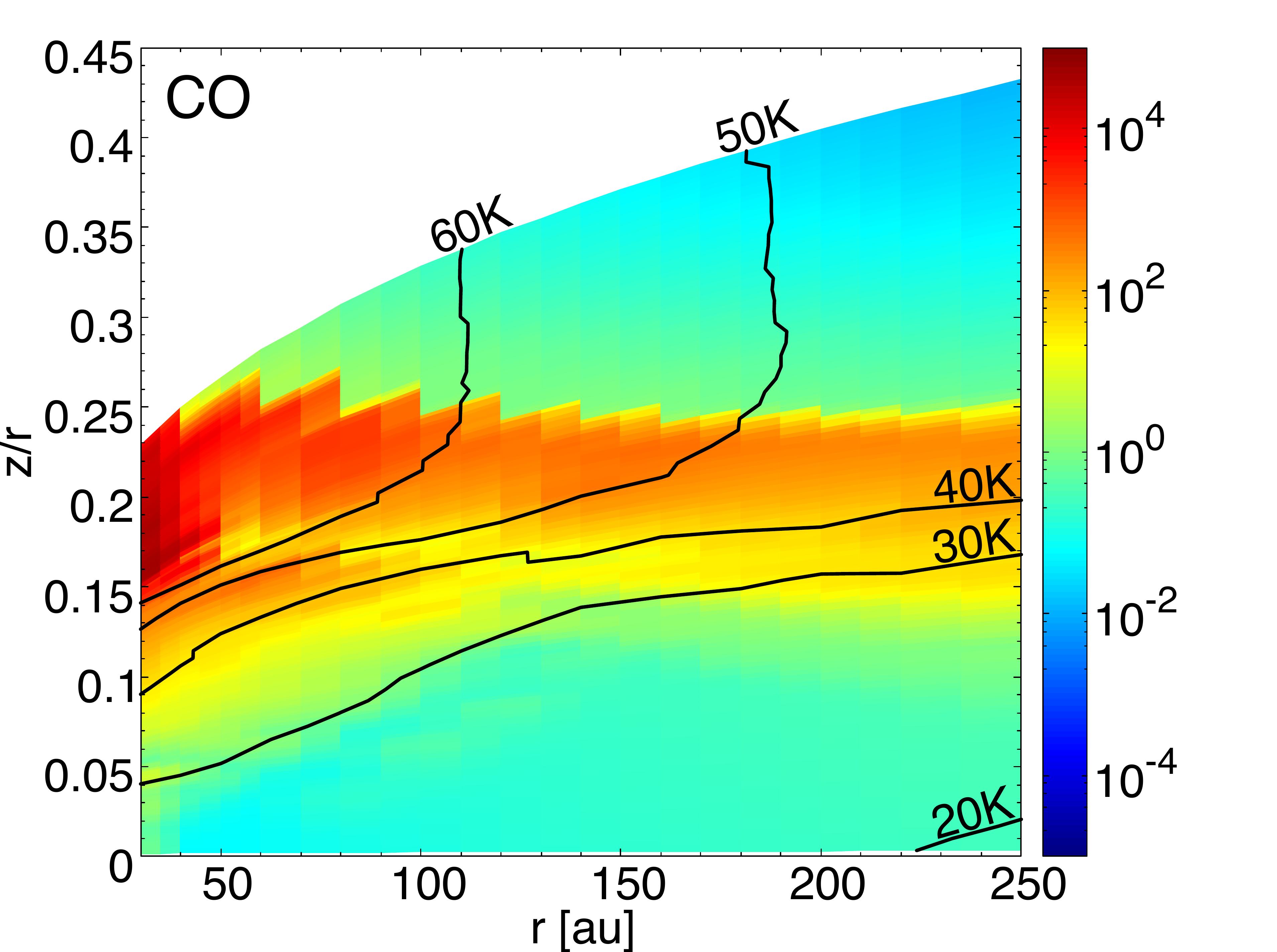}
\end{subfigure}
\begin{subfigure}{.33\linewidth}
  \centering
  \includegraphics[width=1.05\linewidth]{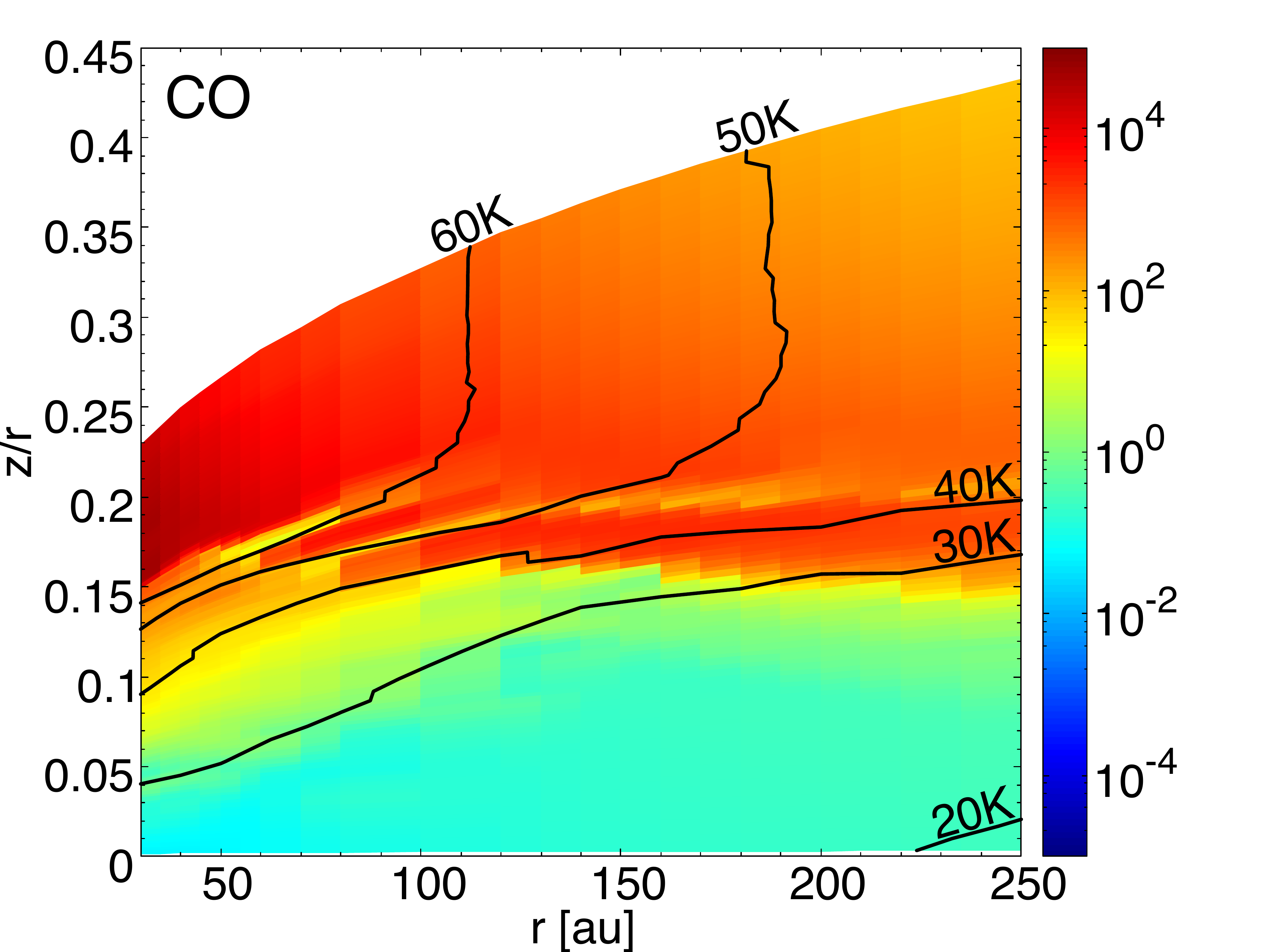} 
\end{subfigure}
\label{fig:m-mapsCO}

\begin{subfigure}{.33\linewidth}
  \centering
  \includegraphics[width=1.05\linewidth]{figures/SINGLE/HUV_HL_Tg/maps/CS.pdf}
\end{subfigure}   
\begin{subfigure}{.33\linewidth}
  \centering
  \includegraphics[width=1.05\linewidth]{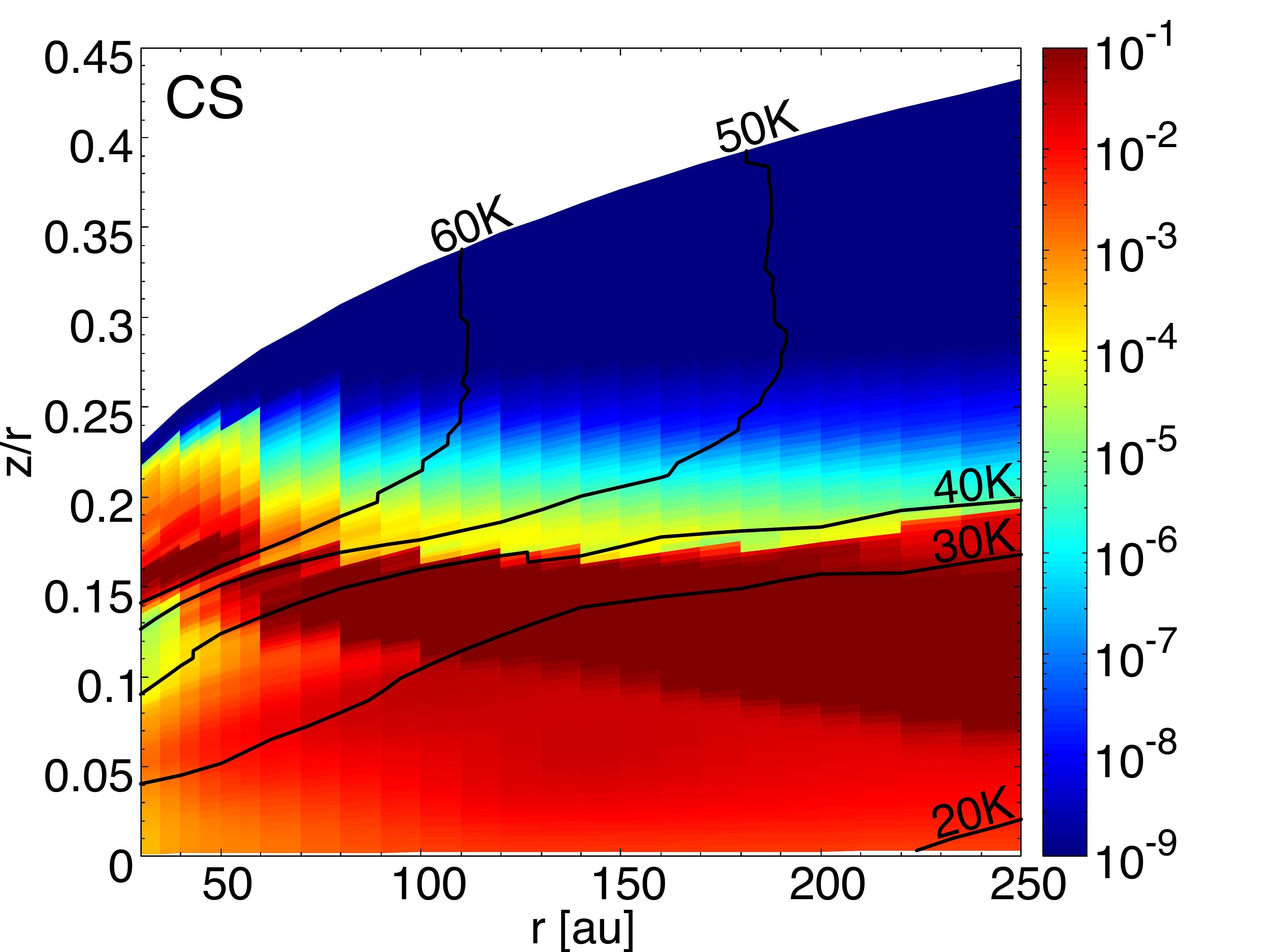}
\end{subfigure}
\begin{subfigure}{.33\linewidth}
  \centering
  \includegraphics[width=1.05\linewidth]{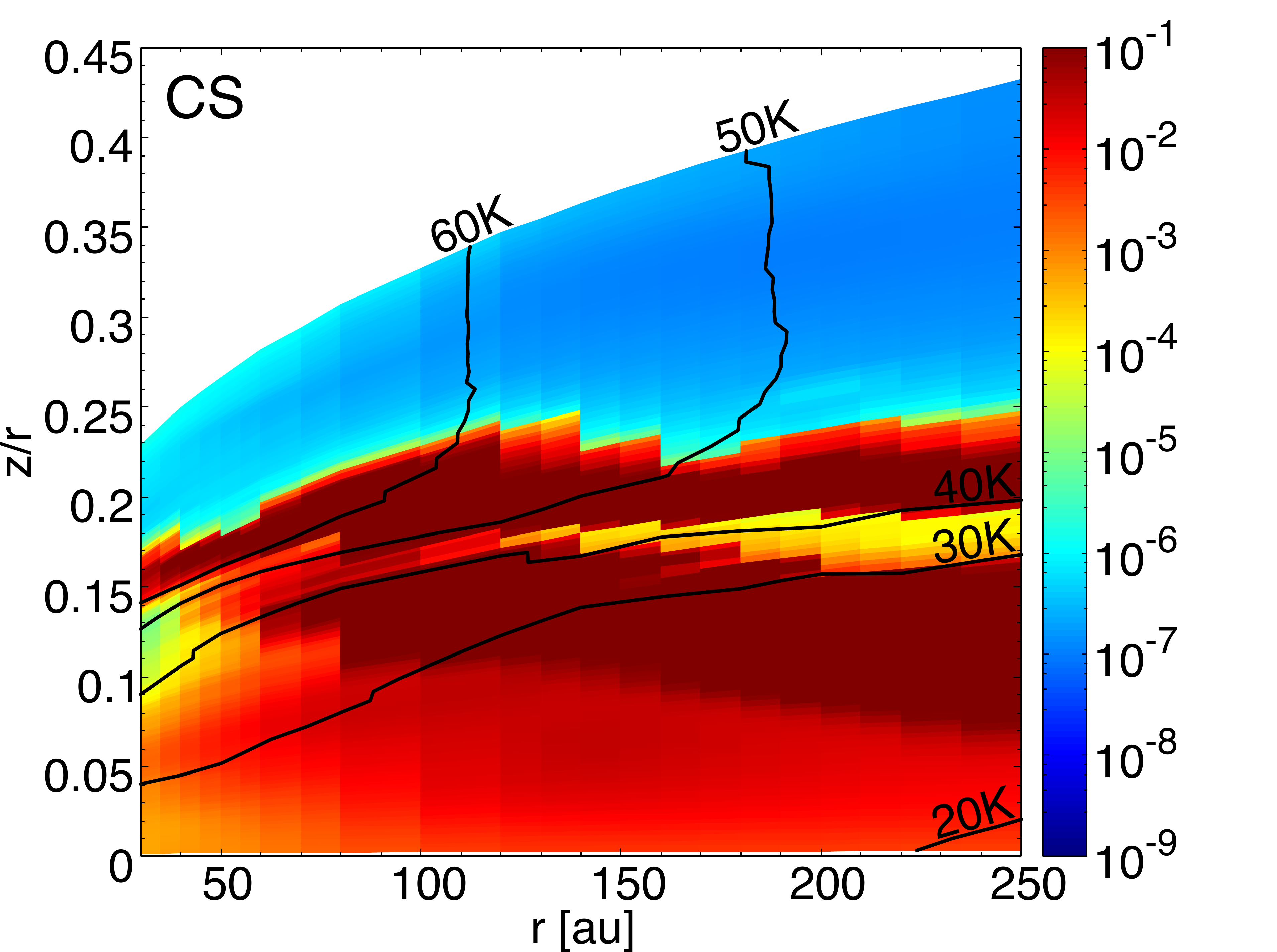}
\end{subfigure}
\label{fig:m-mapsCS}

\begin{subfigure}{.33\linewidth}
  \centering
  \includegraphics[width=1.05\linewidth]{figures/SINGLE/HUV_HL_Tg/maps/CN.pdf}
   \subcaption{\shtg}   
\end{subfigure}
\begin{subfigure}{.33\linewidth}
  \centering
  \includegraphics[width=1.05\linewidth]{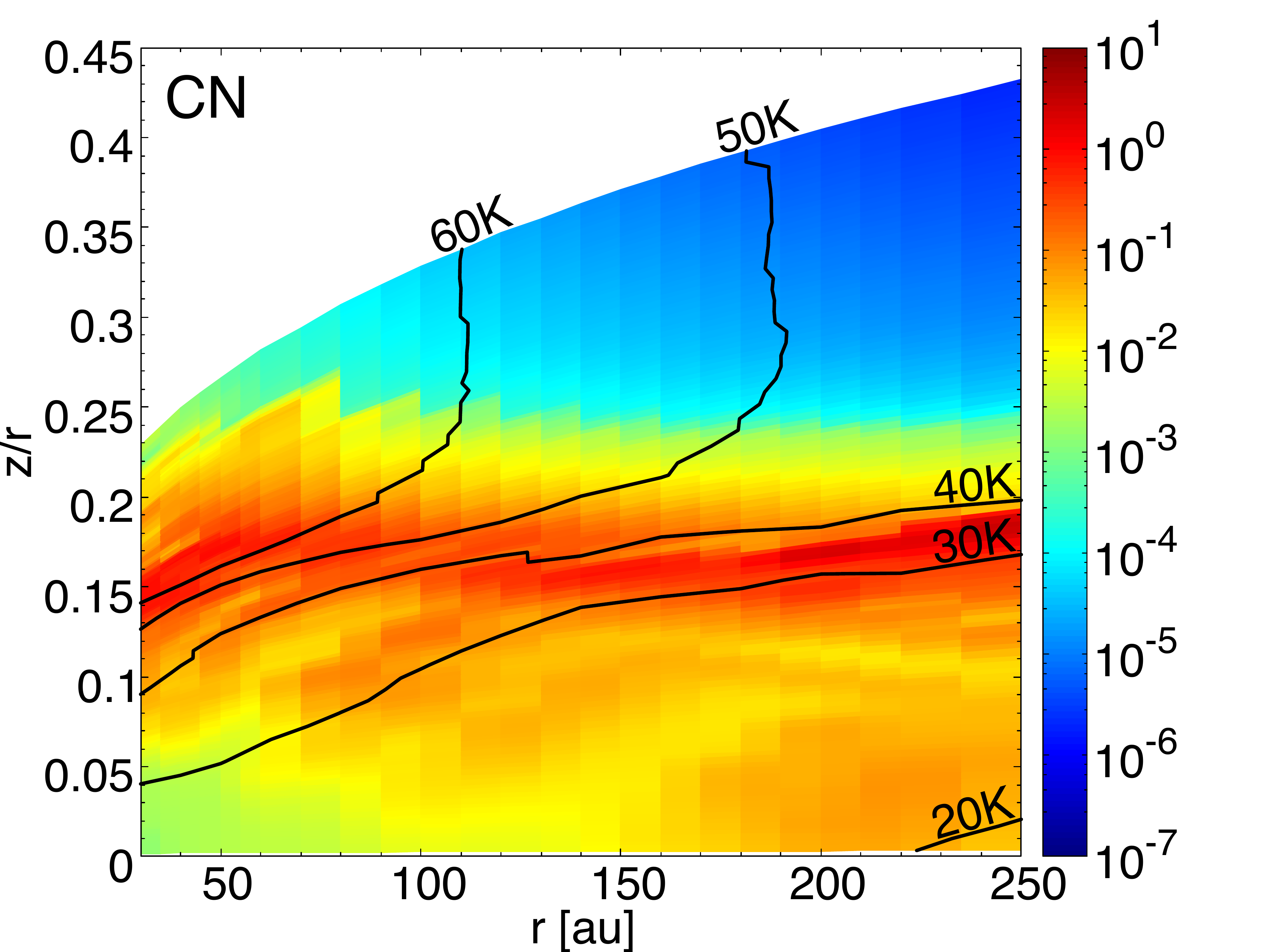}
   \subcaption{\mhlh}   
\end{subfigure}
\begin{subfigure}{.33\linewidth}
  \centering
  \includegraphics[width=1.05\linewidth]{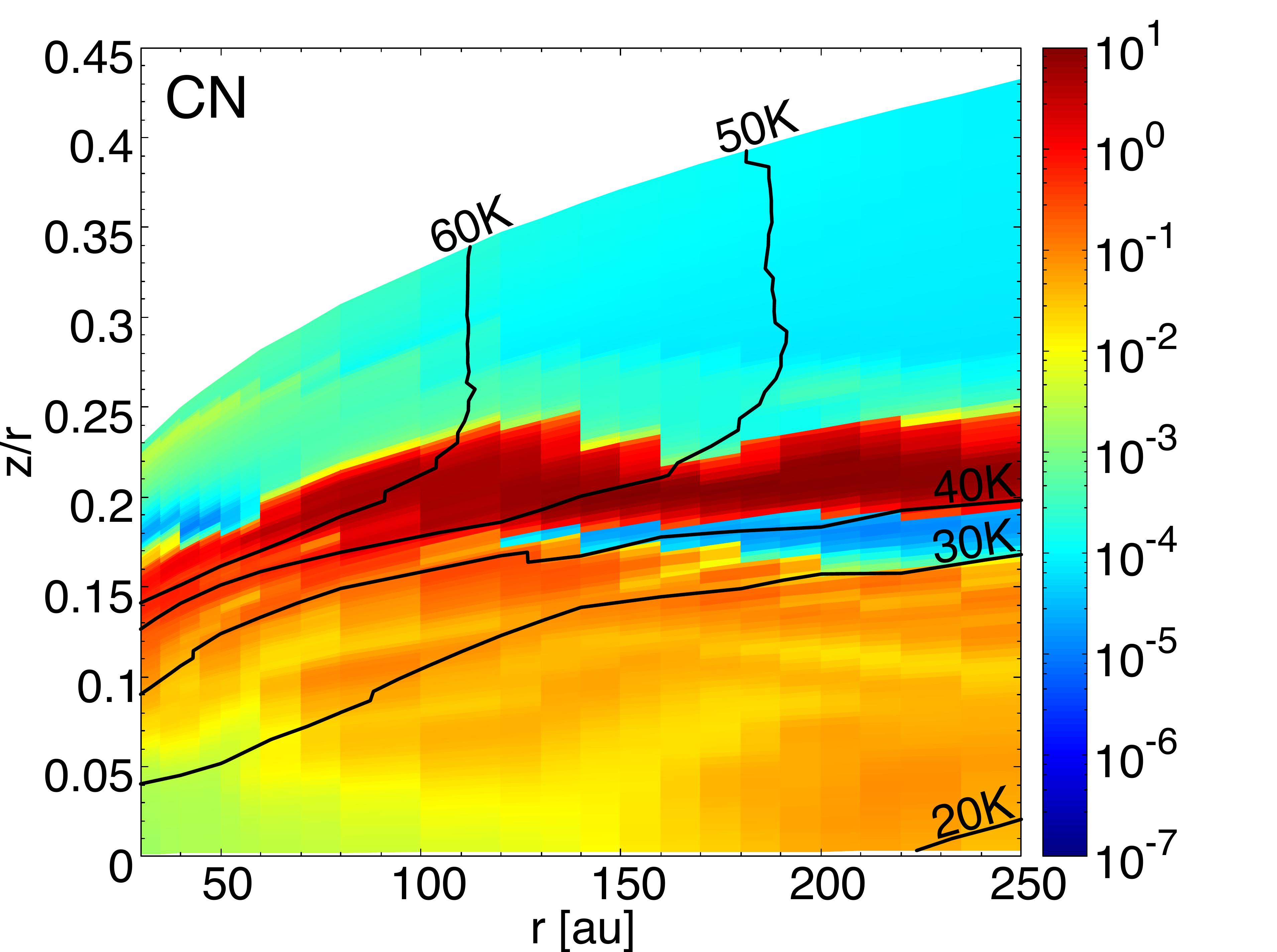}
   \subcaption{\mhb}   
\end{subfigure}
\label{fig:m-mapsCN}
\caption{Density [cm$^{-3}$] of H$_2$, CO, CS and CN in the gas-phase of \shtg (left column) and of the multi-grain models in HUV regime. Black contours represent the dust temperature (T$_\mathrm{d}$ = T$_\mathrm{g}$ in the left column and T$_\mathrm{d}$ = T$_\mathrm{a}$ in the middle and right columns).}
\label{fig:m-maps}
\end{figure*}

\subsubsection{Vertical variations}
    
 \paragraph{CO}
\revisionA{ Gas-phase CO abundance is both influenced by grain temperature 
and by  abundance of \hh. }
\revisionA{For multi-grain models, the flux penetration is larger than in 
single-grain models, so that CO column densities appear to be globally 
smaller than in single-grain models (Fig.\,\ref{fig:compare_high}). In 
\mhlh, the CO abundance remains low above $z \sim 3 H$ (see 
Fig.\,\ref{fig:m-100profile_high}). On the other hand, \mhb produces 
more \hh allowing CO to survive at higher altitudes. The resulting CO 
column density is of the same order than in \shteff and \shb. \revisionA{Although
this has little impact on the total column density, the abundance of CO at 
$z \leq 1 H$ is much higher} 
in multi-grain models than in single-grain ones. In single-grain 
models, the single grain temperature becomes low enough for CO to 
freeze out efficiently while in multi-grain models there is always a 
fraction of abundant small grains whose temperature remains high enough 
to prevent CO from being depleted as efficiently as in single-grain 
models.}

\paragraph{CN and Nitrogen bearing Species} 
For all models, the main reactions that create CN in the upper layers 
at 100 au are bimolecular reactions in the gas-phase: N + 
$\mathrm{C_2}$  $\rightarrow$ C + CN (69\%) and N + CH $\rightarrow$ H 
+ CN (22\%). Note that CH is mostly destroyed by the reaction H + CH 
$\rightarrow$ C + $\mathrm{H_2}$. Therefore, more \revisionA{atomic} H in 
the upper layers implies less CH available to form CN. The effect is 
clearly visible for the single-grain models 
(Fig.\ref{fig:s-100profile_high}) where \shb produces more CN above 
3.5H because $\mathrm{H_2}$ is more abundant compared to \shteff. The 
effect is less obvious in the multi-grain models 
(\ref{fig:m-100profile_high}). Below $z= 1.5\,H$, all the single-grain 
models exhibit roughly the same abundance, including \shtg, which 
suggests a low sensitivity of CN due to the dust temperature in this range 
of values.


 \paragraph{CS}   
The main formation process of CS in the upper 
layers is a sequence of a bimolecular reaction and a dissociative 
recombination reaction: \hh + $\mathrm{CS^+}$ $\rightarrow$ H + 
$\mathrm{HCS^+}$ and $\mathrm{HCS^+}$ + $\mathrm{e^-}$ $\rightarrow$ H + CS. 
The sequence shows that the production of CS is dependent on \hh. 
As a consequence, forcing the formation of \hh leads to the formation 
of CS, providing that the ionization fraction is the same (as it is the 
case here). 
 
For single-grain models, the grain temperature is higher in \shteff 
than in \shtg. The production of \hh being more efficient in the latter 
model at high elevation, the peak of CS in \shtg  (Fig.\,\ref{fig:s-100profile_high}) 
is broader ($\sim 2$ to $3\,H$). In the 
meantime, CS also tends to undergo sticking processes on surfaces 
higher in the disk, competing with its production in the gas-phase. 
This explains the dramatic drop of CS density below $2\,H$. As regards 
\shteff, the production of \hh starts to be efficient at a lower scale 
height ($\sim$ 2 versus $3\,H$). Then, CS abundance dramatically 
increases at $\sim 1.8\,H$ (because of the steep temperature slope) 
before rapidly decreasing as temperature is low enough for CS to 
deplete on the grains, resulting in a very narrow peak. In \shb, the 
production of \hh is efficient even in the upper layers and CS starts 
to form higher in the disk, explaining why the CS abundance is two 
orders of magnitude larger than in \shtg and \shteff at $4\,H$ 
(Fig.\ref{fig:s-100profile_high}). 
 
\revisionA{For multi-grain models, we find a \revisionA{larger} CS column 
density but the location of the peak of abundance is lower in altitude 
(Fig. \ref{fig:100cumul_huv} (c)) because of the relatively low visual 
extinction. Just as for the single-grain models, the surface density is 
larger in the case of \mhb because the formation of $\mathrm{H_2}$ is 
more efficient and the quantity of H lower. Below $z = 1.5\,H$, the 
difference in abundance is strictly dependent on $T_d$ and as expected, 
the hot-grain models (\shteff and \shb) produce more CS in the gas-phase 
than the cold-grain model \shtg, whereas all multi-grain models exhibit 
the same vertical profile.}

\revisionC{Concerning the intermediate models (Set-HUV-B14-T$\bm{\mathrm{_a}}$ and Set-HUV-LH-T$\bm{\mathrm{_a}}$), Fig.\,\ref{fig:interm-100profile_LH}  and
Fig.\,\ref{fig:interm-100profile_B14} show that above the altitude of $z \sim 2H$, these models are, as expected, very similar to what is observed for models \shteff because they share the same grain temperature for a number of grain sites which is of the same order. Beyond $z \sim 2H$,
the local gas densities for CO, CN and CS are closer to what is found for the multi-grain models. This is because in the intermediate models the number of grain sites around the midplane is lower than in the case of \shteff models and equal to those of multi-grain models.  Moreover, the amount of UV should be locally higher than in single grain models because of dust settling. Note that in all cases, at the midplane, the abundance of gas species is always very low, well below observational levels.}

 \subsection{\revisionA{Element} reservoirs: C, N, O in the cold midplane}  \label{subsec:reservoirs} 

 \begin{figure}
\begin{subfigure}{1\linewidth}
  \centering
 \includegraphics[width=1\linewidth]{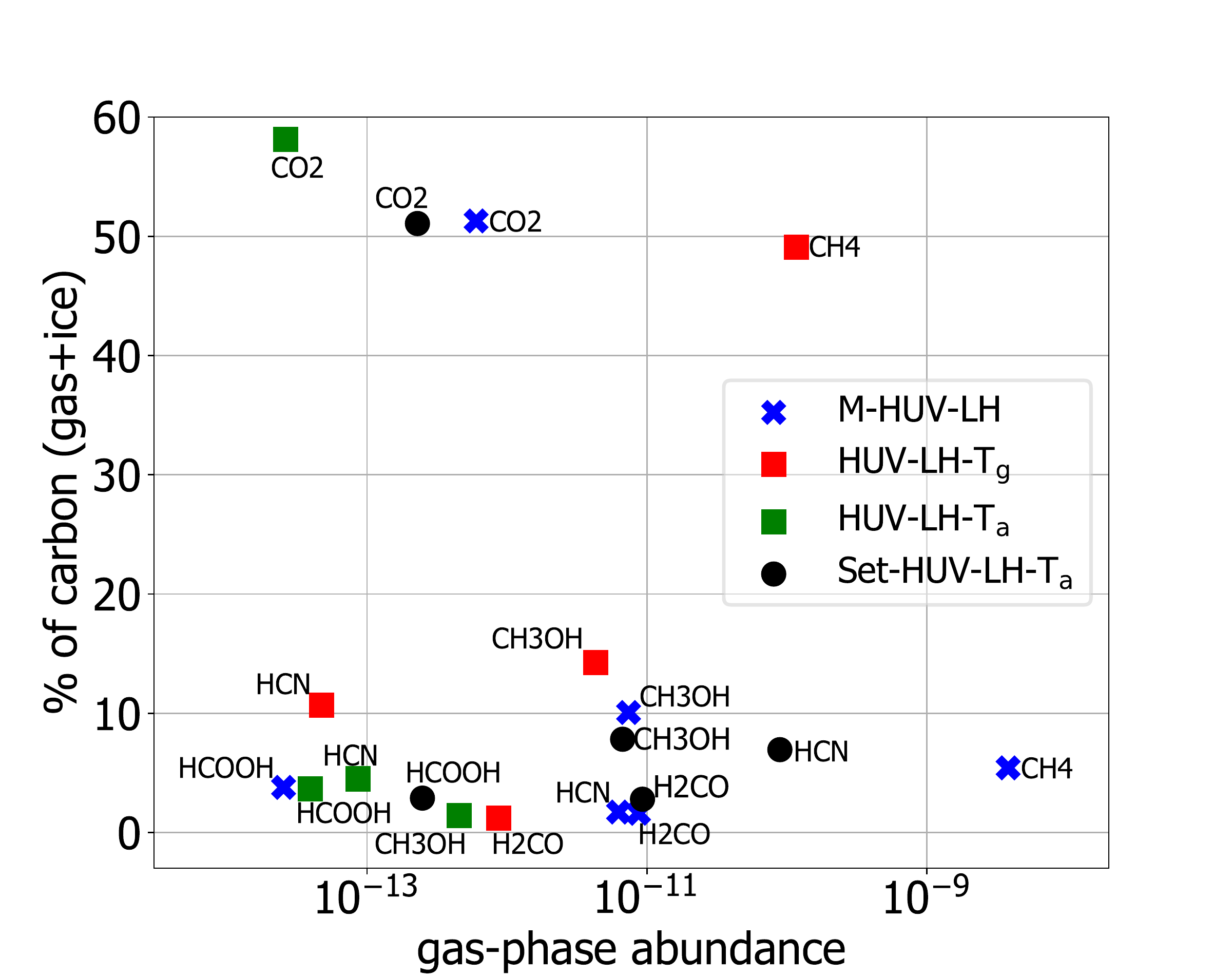}
 \label{fig:reservoirs_C}
\end{subfigure}
\begin{subfigure}{1\linewidth}
  \centering
  \includegraphics[width=1\linewidth]{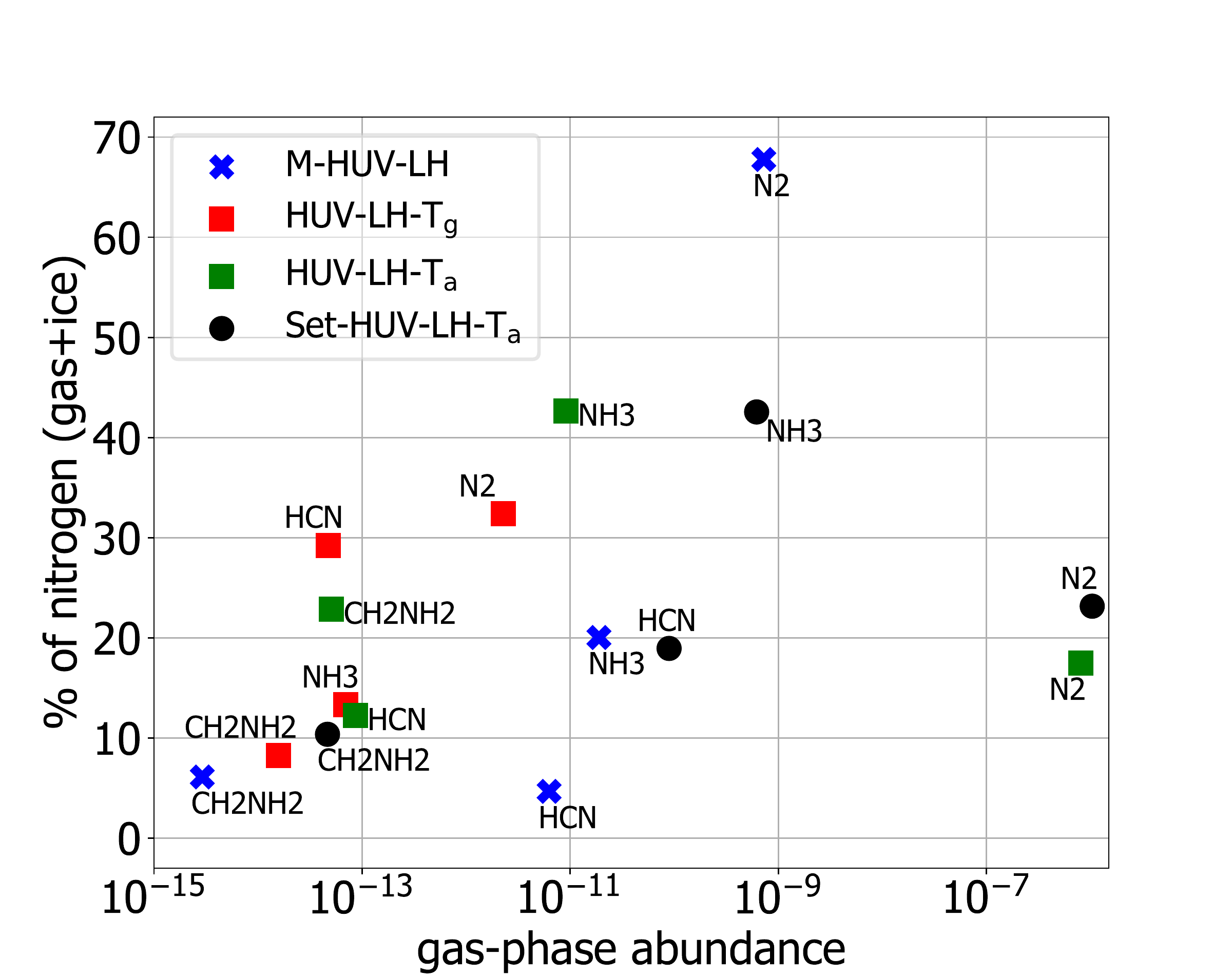}
  \label{fig:reservoirs_N}
\end{subfigure}
\begin{subfigure}{1\linewidth}
  \centering
 \includegraphics[width=1\linewidth]{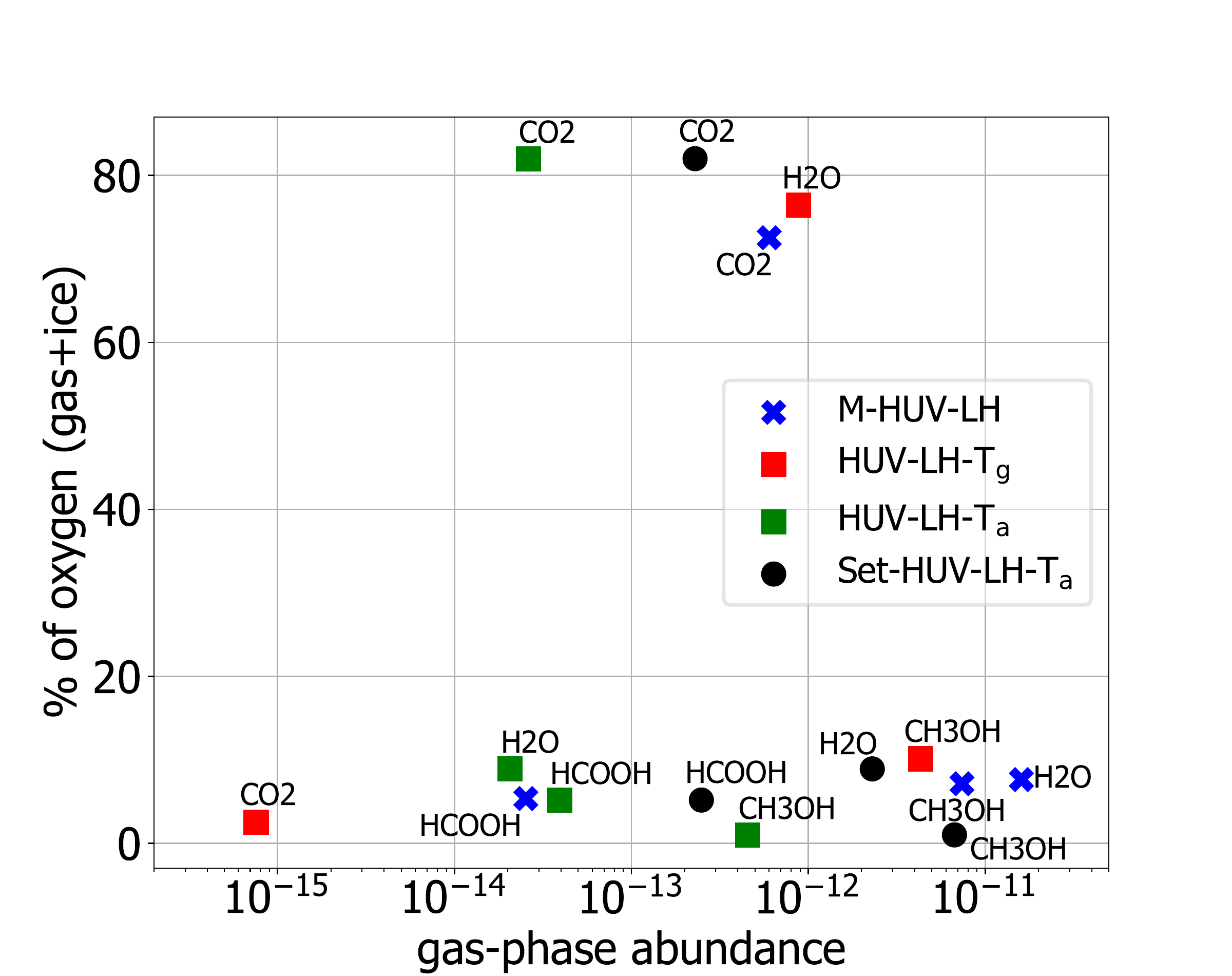}
 \label{fig:reservoirs_O}
\end{subfigure}
\caption{Species that contain at least 0.5\% of elemental carbon (top), nitrogen (middle) and oxygen (bottom) in the midplane at 100 au:  percentage of the carbon/nitrogen/oxygen locked in the species as a function of the gas-phase species abundance.}
 \label{fig:reservoirs}
\end{figure}

The relative importance of reservoirs appearing roughly constant along 
the distance from the star, we decide to study the reservoirs at 100 
au. All species that have been investigated represent at least 0.5 \% 
of the total quantity of carbon, nitrogen or oxygen. As previously 
seen, the chemistry below $z = 1.5\,H$ \revisionA{mostly depends on the 
grain temperature}. Therefore, it turns out to be sufficient to discuss 
\shtg, \shteff, \mhlh \, \revisionC{and Set-HUV-LH-T$\bm{\mathrm{_a}}$} only. 

\revisionA{\paragraph{Carbon bearing species}} 
Figure \ref{fig:reservoirs} (top) shows that frozen CO$_2$ is the main 
carrier of carbon, both in \shteff and \mhlh~ \revisionC{(and Set-HUV-LH-T$\bm{\mathrm{_a}}$)} 
with 58.1\% and 51.3\% of elemental carbon, respectively, while the quantity of CO$_2$ is 
negligible in comparison to the other species in \shtg. The dust temperature 
appears too high in these two models to allow for a large quantity of 
hydrogen to successively hydrogenate the frozen atomic oxygen to form 
water. Instead, the atomic oxygen easily diffuses on the surface and 
the relatively high temperature allows to overcome the activation 
barrier to rapidly form CO$_2$ by \revisionA{reacting with the frozen CO}. In 
\shtg, on the other hand, Fig.\,\ref{fig:reservoirs} (top) 
shows that the main carrier of carbon is CH$_4$ with 49.1\%.  In this 
colder dust temperature model, CO is initially formed in the gas before 
being efficiently and rapidly adsorbed onto the grain surfaces, just 
like in hot-grain models. However, once on the surface, CO is channeled 
to s-CH$_3$OH through hydrogenation sequences. The photodissociation of 
CH$_3$OH then leads to CH$_4$ formation.  Hence, in the midplane, 
disks with colder grains will produce more CH$_4$ while warmer disks will produce 
more CO$_2$. \revisionA{The other main carrier is the complex organic species 
CH$_3$OH, which holds about 10\% of carbon in \mhlh and \shtg.} \revisionC{Finally, it is important to mention that CO contributes as a reservoir of C for less than  $0.5$\%, in all models.}

\revisionA{\paragraph{Oxygen bearing species}}  CO$_2$ ice is the main 
carrier of oxygen in \shteff~\revisionC{(and Set-HUV-LH-T$\bm{\mathrm{_a}}$)} and \mhlh  with 82.0\% and 72.5\%, 
respectively (Fig. \ref{fig:reservoirs}, bottom). The temperature of 
the grains in these models is too high ($T_a$ = 24.6 K) to allow 
for a large quantity of atomic hydrogen to remain on the grains and 
hydrogenate \revisionA{the atomic oxygen to form water. As described above,
atomic oxygen easily diffuses and also combines with CO to form CO$_2$.}
In \shtg, H$_2$O is by far the largest carrier of oxygen with 
76.4\%. Disks with colder grains will 
produce more H$_2$O while disks with warmer grains will produce 
more CO$_2$. \revisionA{CH$_3$OH is, again, an important carrier and holds about 10\% of oxygen in \shtg.}
\revisionC{As for C, CO holds less than $0.5$\% of Oxygen, in all models.}

\paragraph{Nitrogen bearing species}  Nitrogen is mostly in the form of N$_2$ in \mhlh (67.8\%) and in \shtg (32.4\%) (Fig. \ref{fig:reservoirs}, middle). Moreover, in almost all models, the gas-phase abundance of N$_2$ is noticeably larger than the one of the other species. The reason for that is a combined effect of the faster conversion of atomic nitrogen into N$_2$ in the gas-phase compared to the depletion of N and the high grain temperature that prevents N$_2$ from depleting onto the surfaces. Indeed, The binding energy of N$_2$ ($E_{\mathrm{b}}(\mathrm{N}_2) = 1100$\,K on amorphous water ice surface) is smaller than most of the other molecules and slightly smaller than that of CO \citep{Minissale+etal_2016, Wakelam+etal_2017} so that N$_2$ is retained more efficiently in the gas. We note that, given the low dust temperature, the gas-phase abundance of N$_2$ in \shtg is the smallest of all models. In \shteff\  \revisionC{and in Set-HUV-LH-T$\bm{\mathrm{_a}}$}, \revisionC{more N$_2$ remains in the gas-phase because of the large dust temperatures.} On the grains, s-NH$_3$ is the main carrier of nitrogen with 42.7\% of the elemental nitrogen. Therefore N$_2$ is expected to be the carrier of nitrogen in the disk midplane but remains more in the gas-phase in hot-grain models. The other main carriers of nitrogen are HCN (29.2\% in \shtg, 12.3\% in \shteff) and CH$_2$NH$_2$ (22.9\% in \shteff, 8.3\% in \shtg). 

 
\subsection{Surface chemistry in the midplane} \label{subsec:surface}

\revisionA{One important conclusion of Section \ref{subsec:reservoirs} is 
that disks with colder grains will produce more s-CH$_4$ and s-H$_2$O in the midplane 
while disks with warmer grains will produce more s-CO$_2$. This clearly demonstrates 
the major role of the grain temperature on the surface chemistry in the plane. We 
investigate here more deeply the midplane surface chemistry of a few key molecules 
for the same four models i.e. \shtg, \shteff, \mhlh}\revisionC{, and Set-HUV-LH-T$_\mathrm{a}$}.
\revisionC{Another important conclusion is that our intermediate models, Set-HUV-LH-T$\bm{\mathrm{_a}}$ and Set-HUV-B14-T$\bm{\mathrm{_a}}$, 
are not able to reproduce the complex behavior of the gas-grain chemistry observed with the multi-grain model, since it happens to be closer to the single grain model  \shteff in terms of main reservoirs in the midplane. Small differences occur for relatively
minor constituents (s-CO, s-H$_2$CO, s-CH$_3$OH, s-HCN for example) because of the
different effective dust area in the two cases (Single vs Set), and hence different timescales for conversion on grain surfaces.}  
  
\begin{figure*}
\begin{subfigure}{.33\linewidth}
  \centering
  \includegraphics[width=1.05\linewidth]{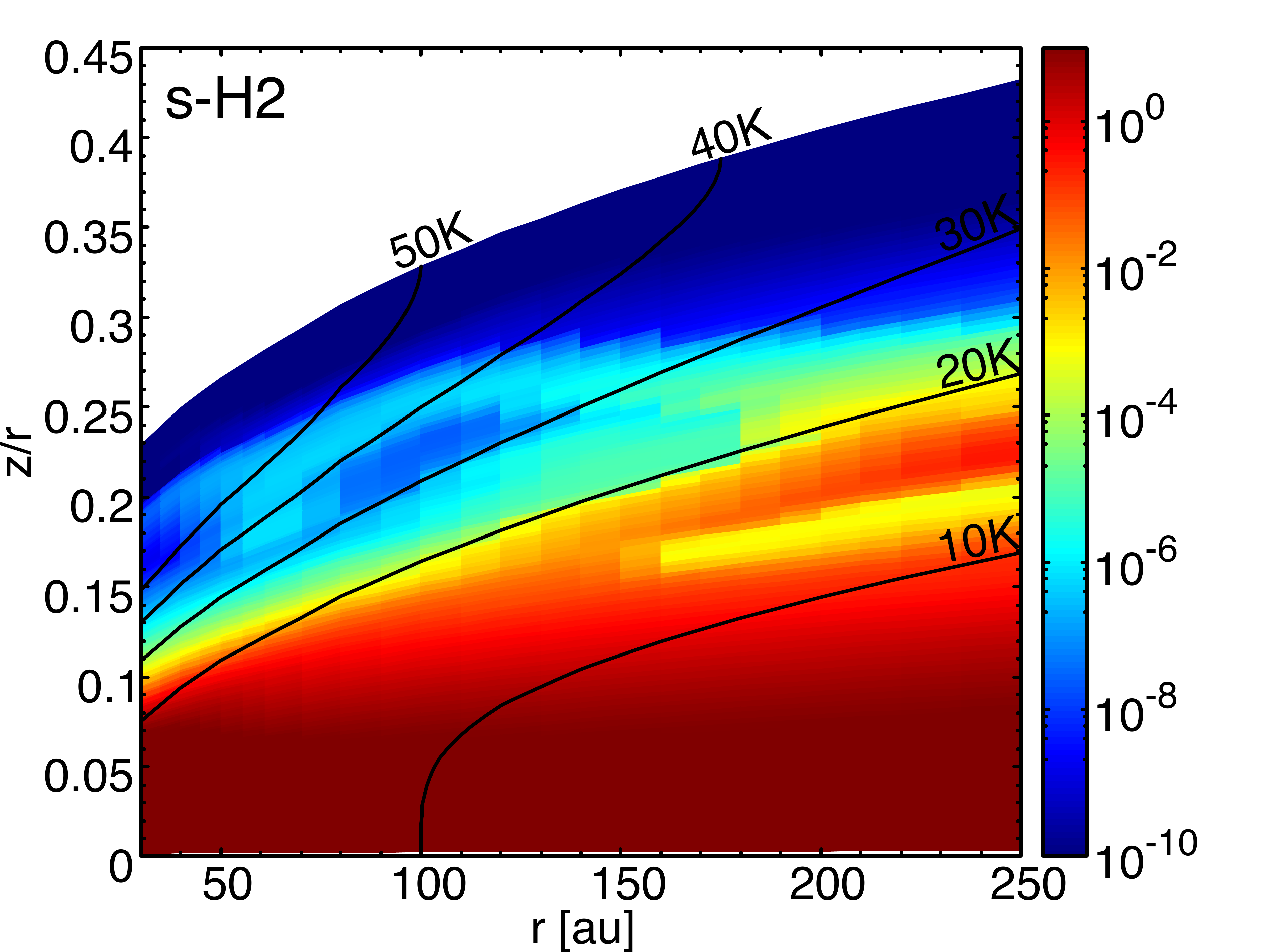}
\end{subfigure}
\begin{subfigure}{.33\linewidth}
  \centering
  \includegraphics[width=1.05\linewidth]{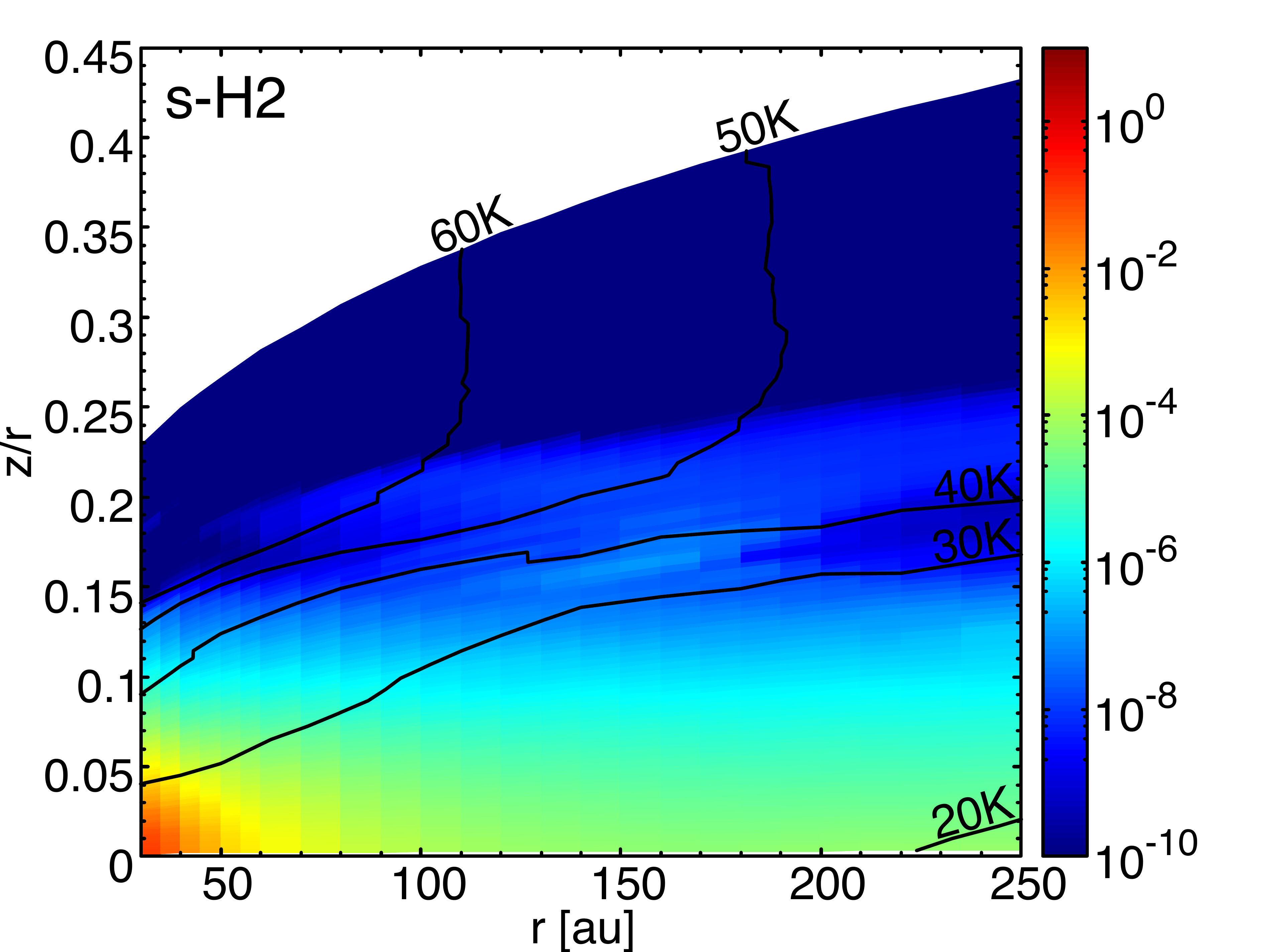}
\end{subfigure}
\begin{subfigure}{.33\linewidth}
  \centering
  \includegraphics[width=1.05\linewidth]{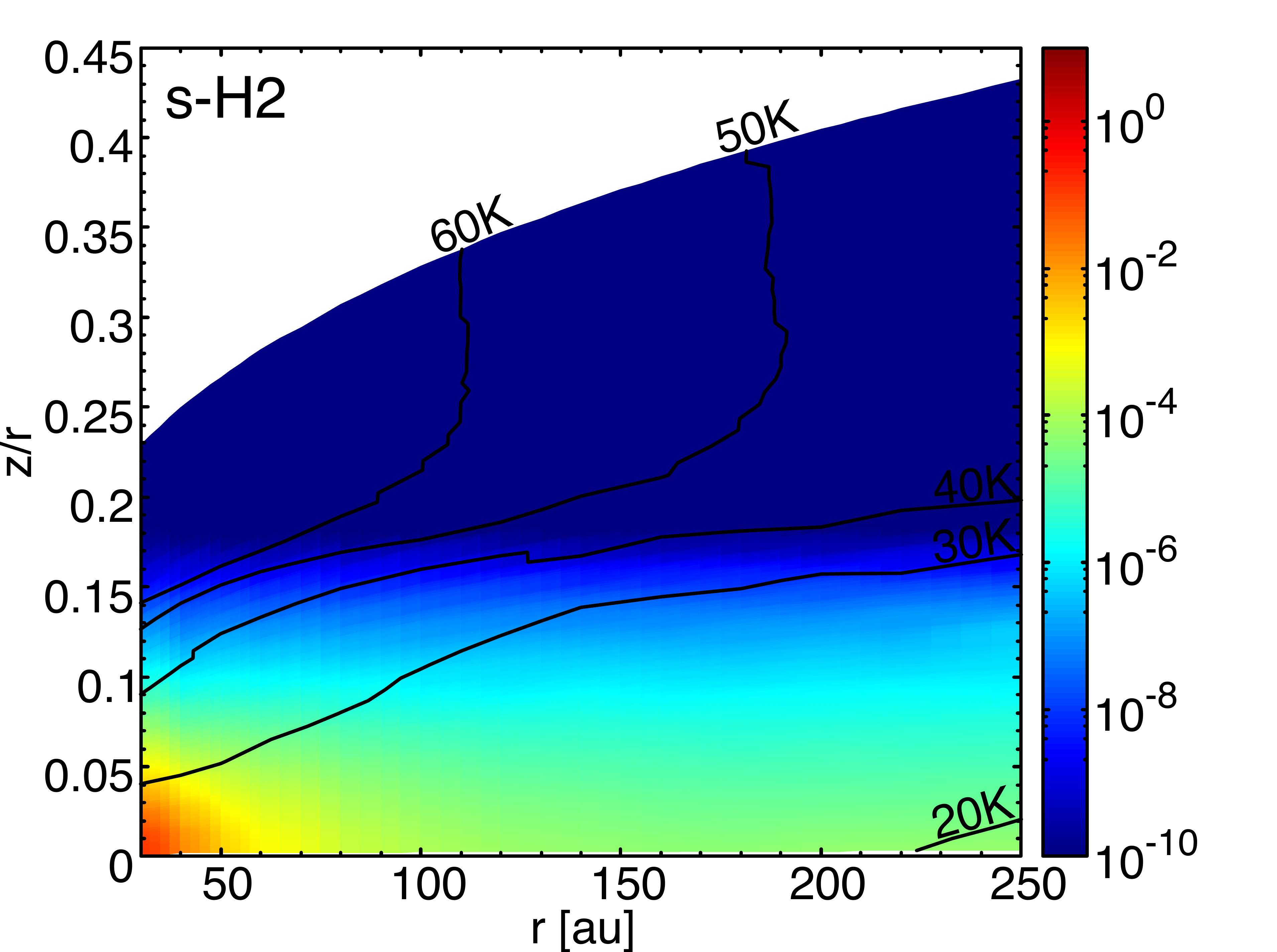}
\end{subfigure}

\begin{subfigure}{.33\linewidth}
  \centering
  \includegraphics[width=1.05\linewidth]{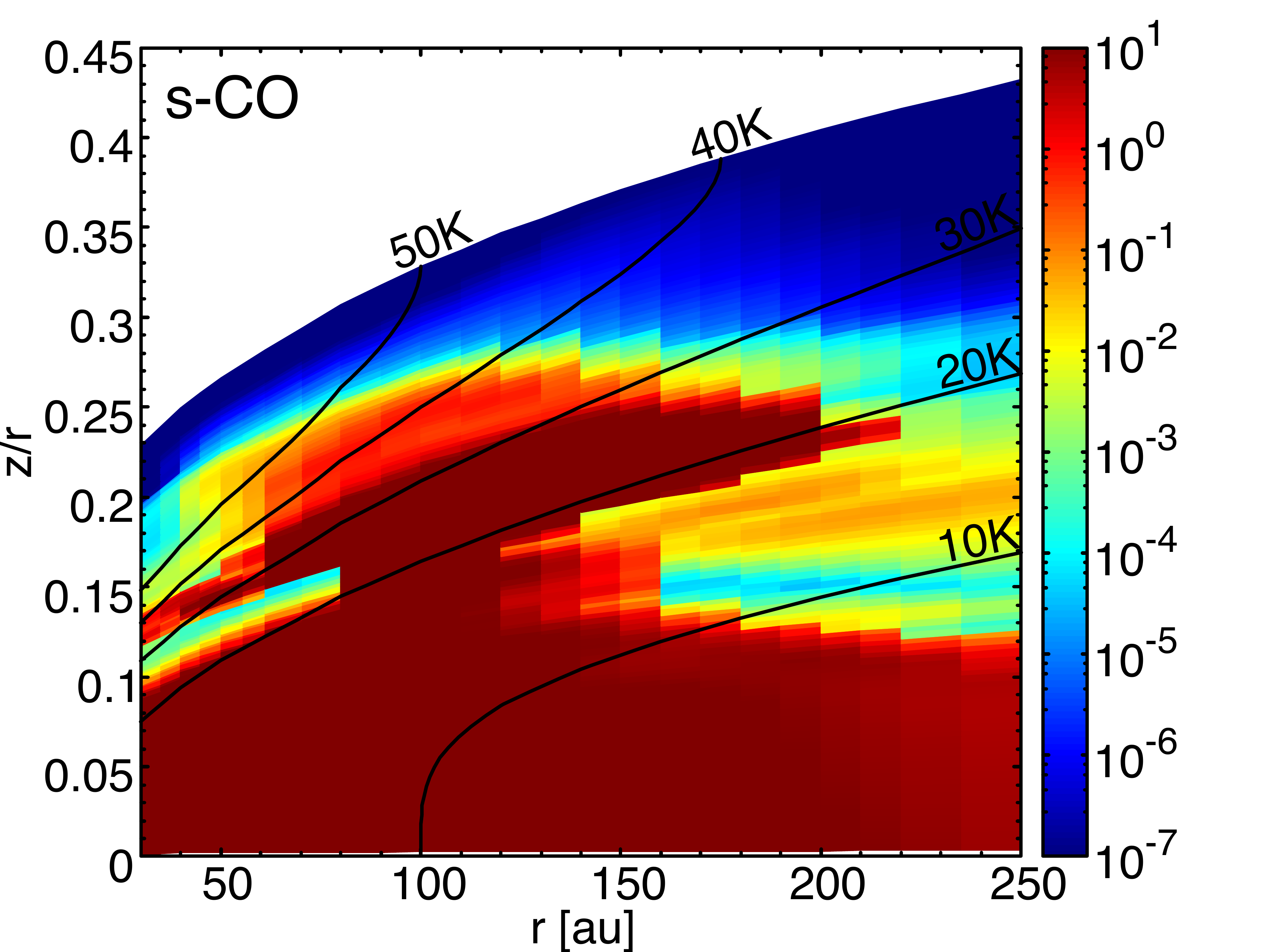}
\end{subfigure}
\begin{subfigure}{.33\linewidth}
  \centering
  \includegraphics[width=1.05\linewidth]{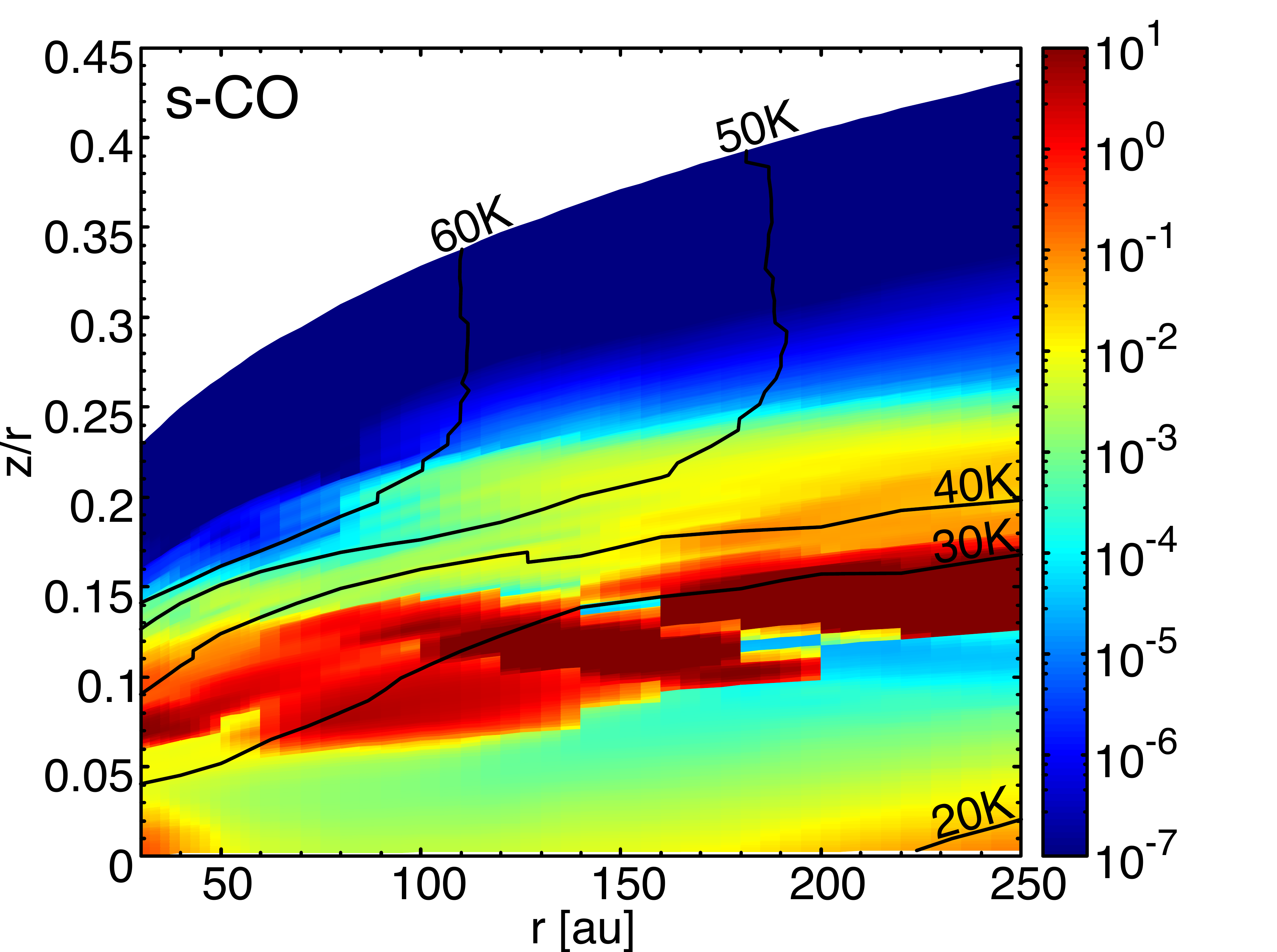}
\end{subfigure}
\begin{subfigure}{.33\linewidth}
  \centering
  \includegraphics[width=1.05\linewidth]{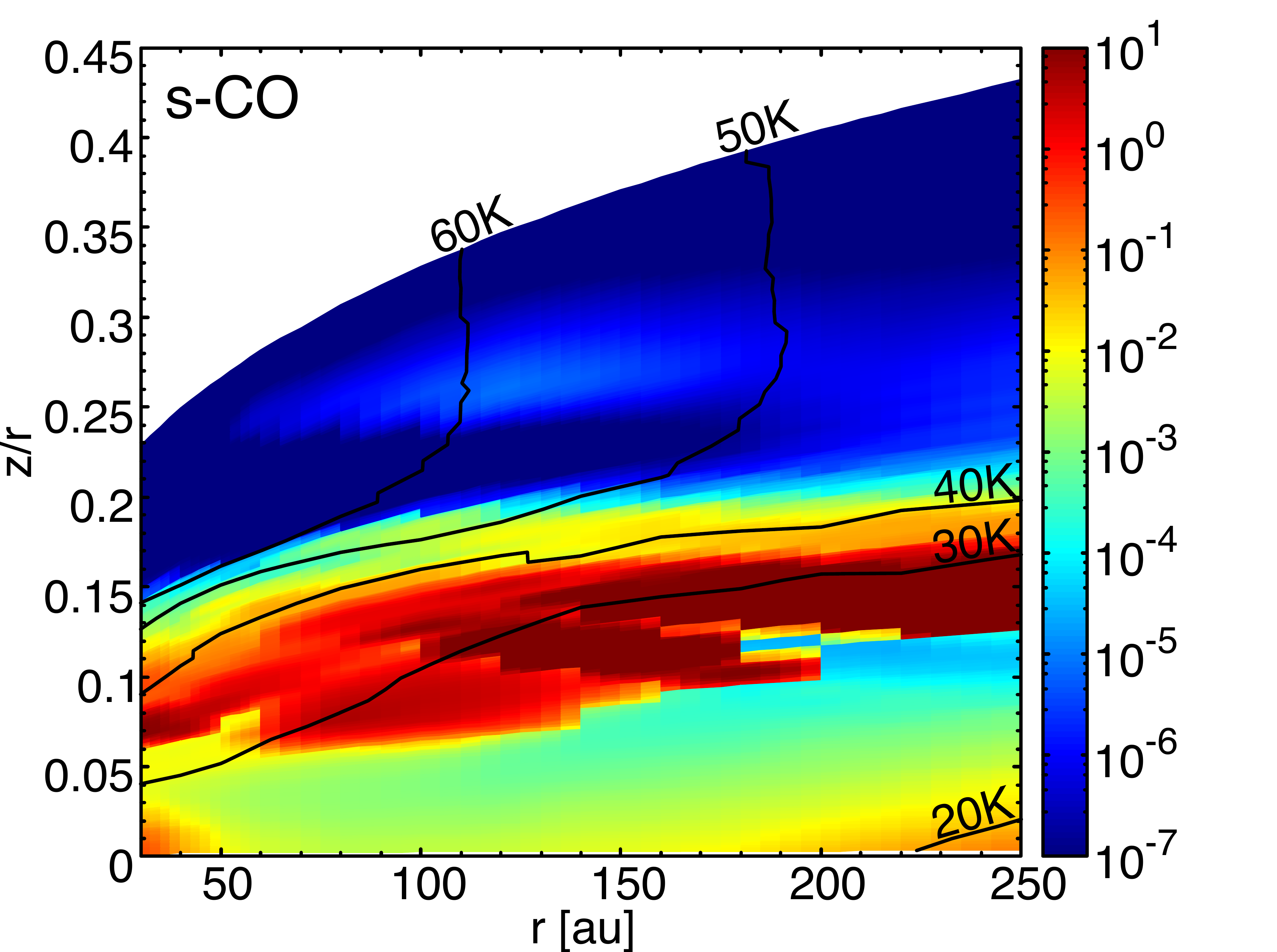}
\end{subfigure}

\begin{subfigure}{.33\linewidth}
  \centering
  \includegraphics[width=1.05\linewidth]{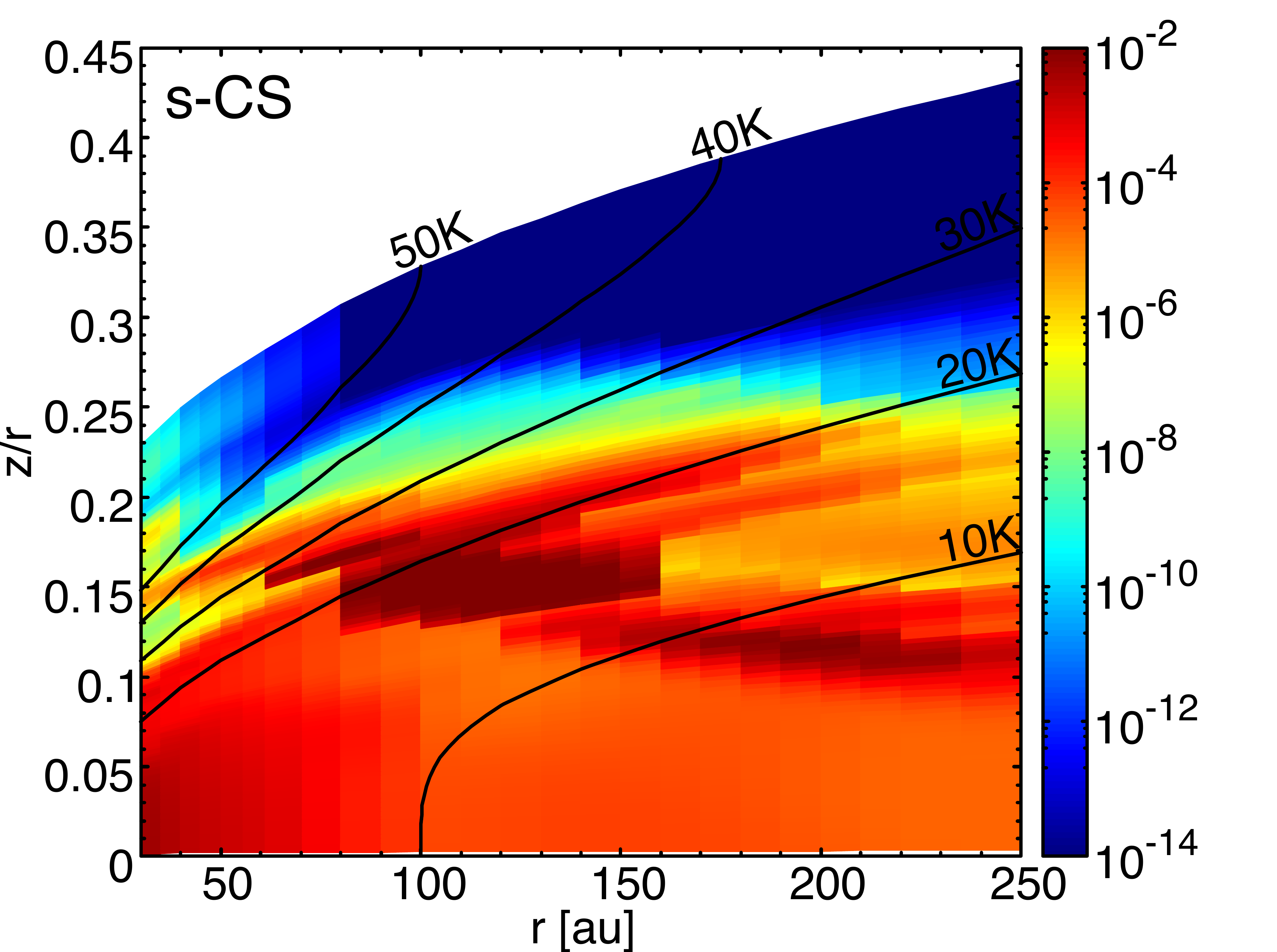}
\end{subfigure}
\begin{subfigure}{.33\linewidth}
  \centering
  \includegraphics[width=1.05\linewidth]{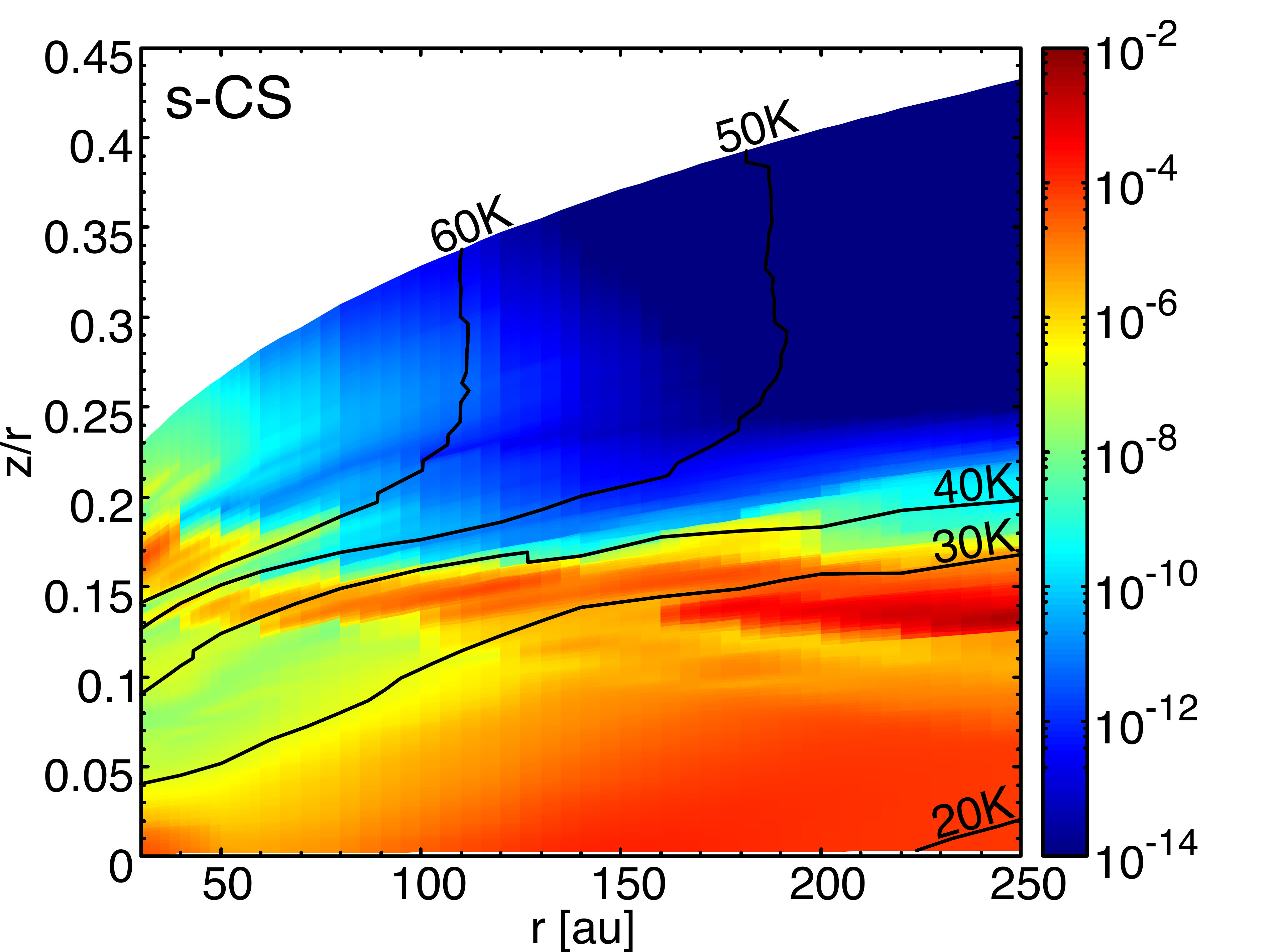}  
\end{subfigure}
\begin{subfigure}{.33\linewidth}
  \centering
  \includegraphics[width=1.05\linewidth]{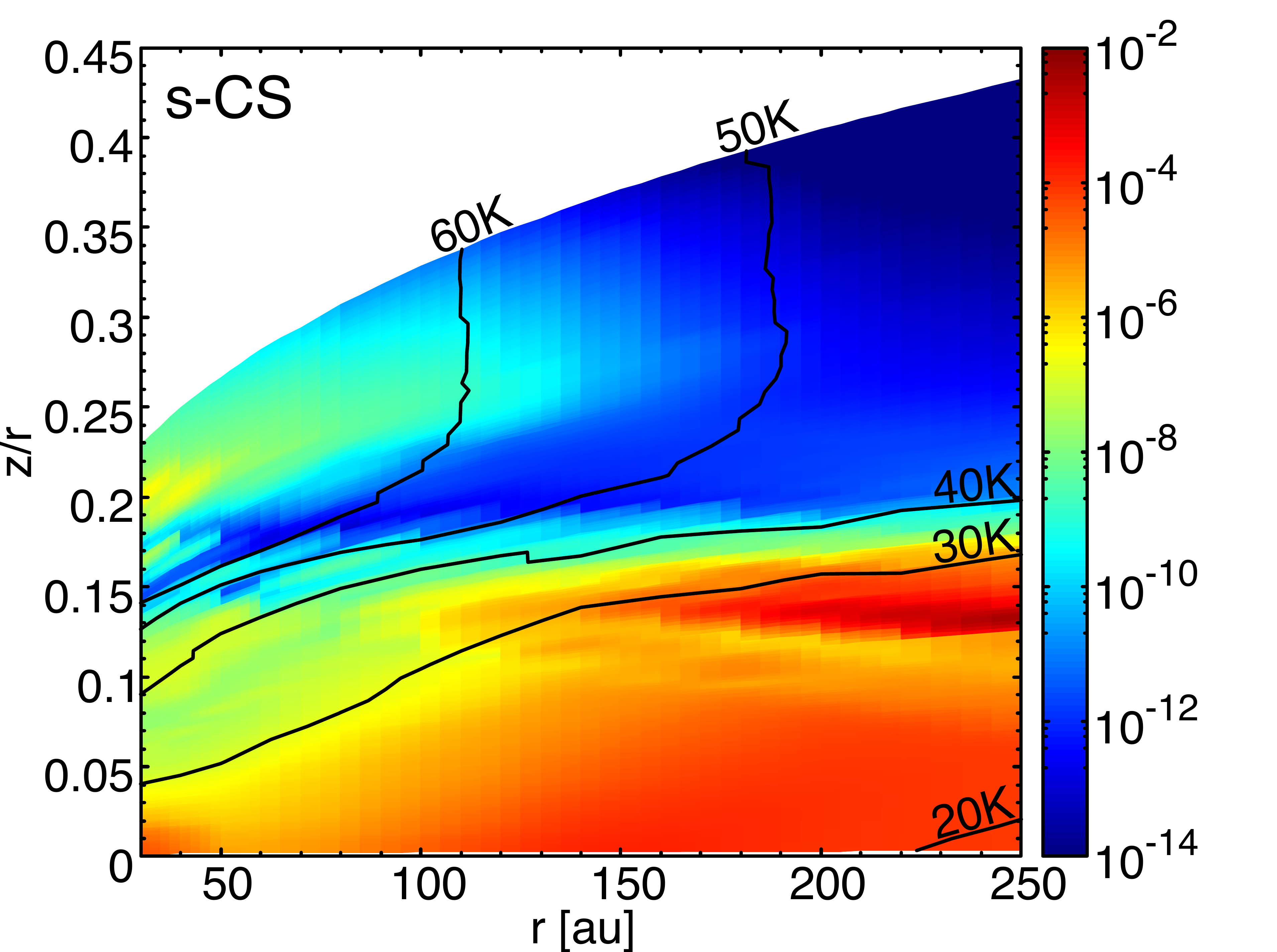}
\end{subfigure}

\begin{subfigure}{.33\linewidth}
  \centering
  \includegraphics[width=1.05\linewidth]{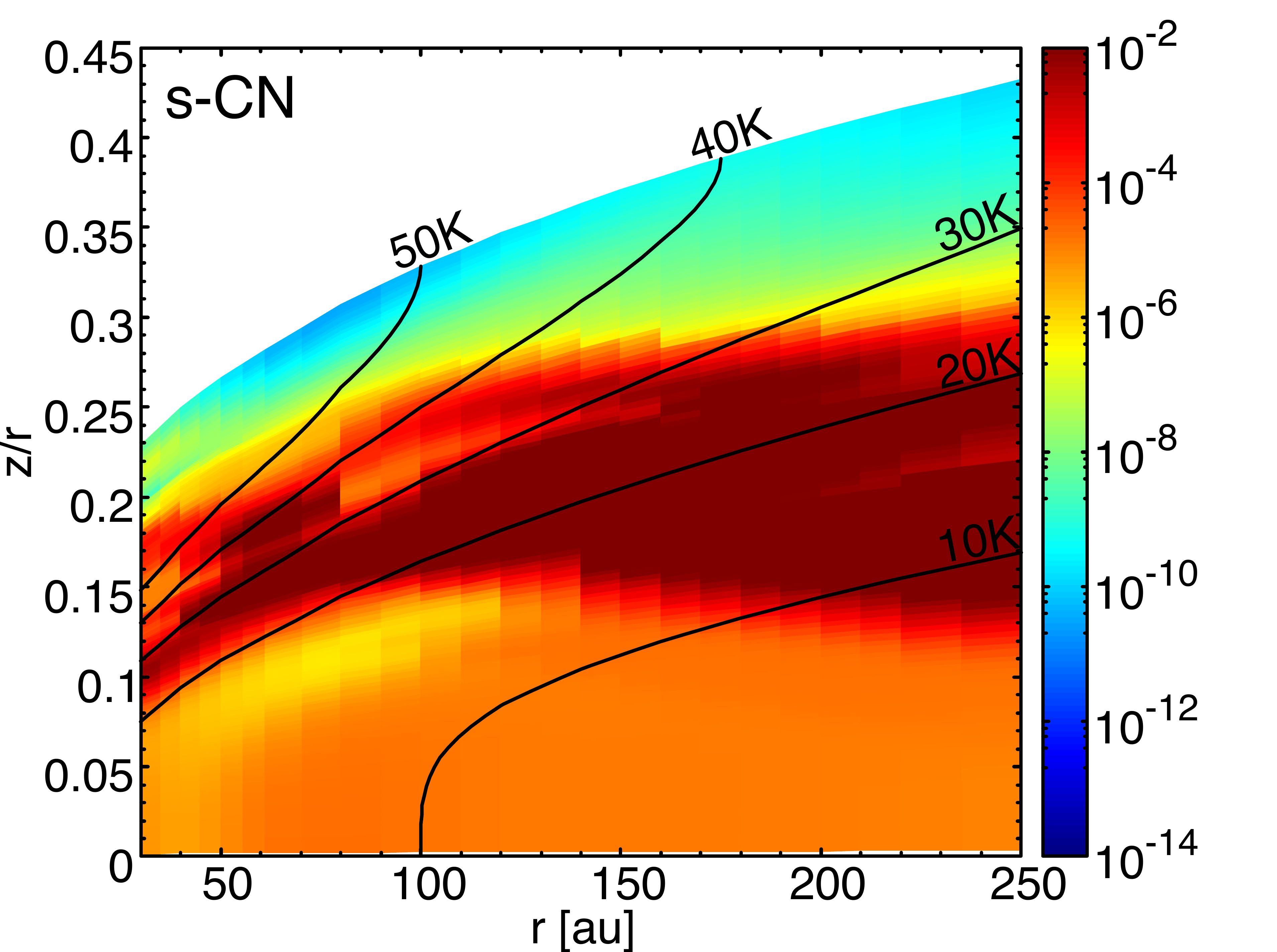}
   \subcaption{\shtg}   
\end{subfigure}
\begin{subfigure}{.33\linewidth}
  \centering
  \includegraphics[width=1.05\linewidth]{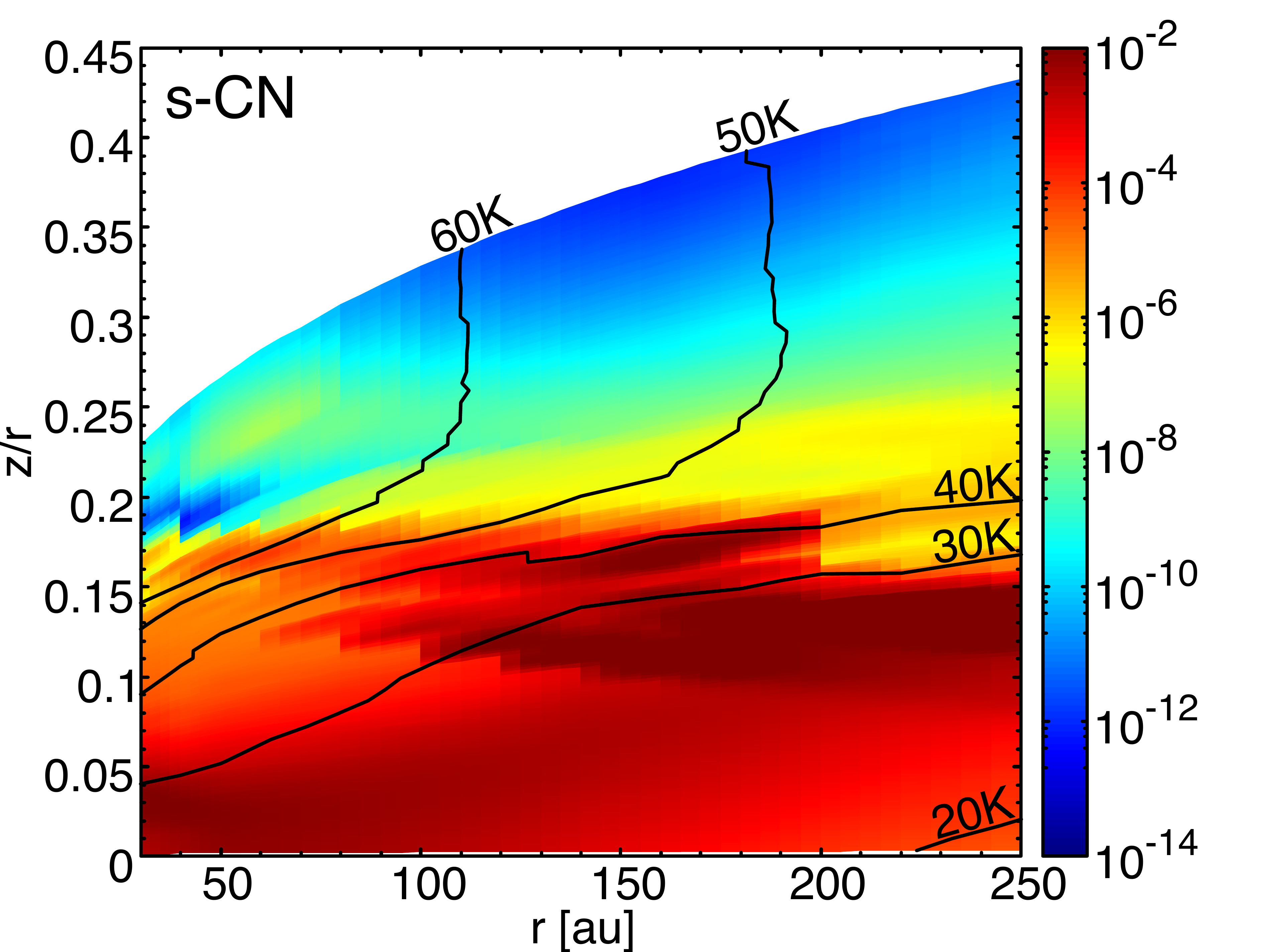}
   \subcaption{\shteff}   
\end{subfigure}
\begin{subfigure}{.33\linewidth}
  \centering
  \includegraphics[width=1.05\linewidth]{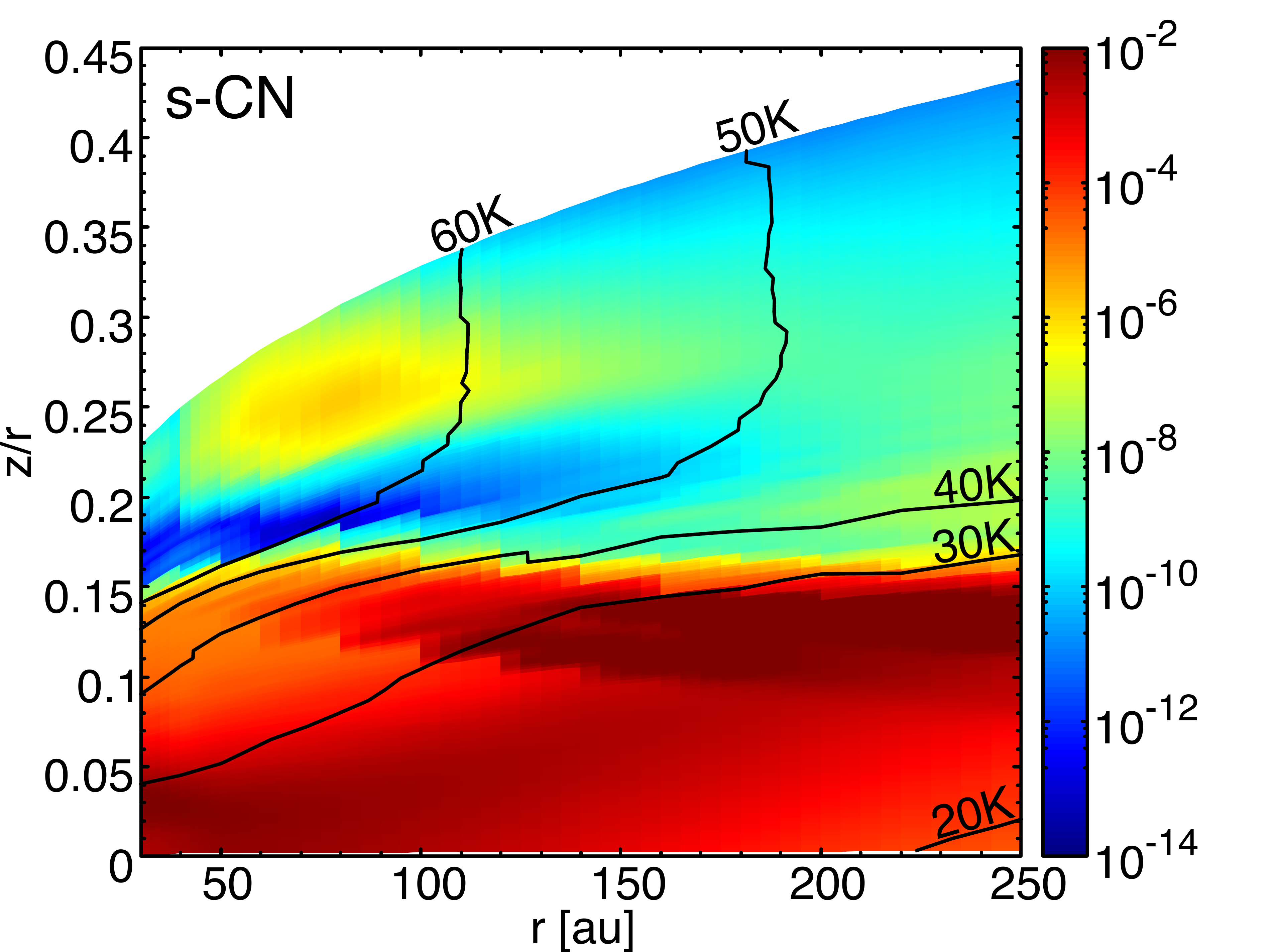}
   \subcaption{\shb} 
\end{subfigure}
\caption{Density [cm$^{-3}$] of H$_2$, CO, CS and CN in the gas-phase of 
the single-grain models in HUV regime. Left column is the \shtg model, 
middle one is \shteff and right one is \shb. Black contours represent the dust temperature (T$_\mathrm{d}$ = T$_\mathrm{g}$ in the left column and T$_\mathrm{d}$ = T$_\mathrm{a}$ in the middle and right columns).}
\label{fig:s-s-maps-huv}
\end{figure*}

\begin{figure*}
\begin{subfigure}{.33\linewidth}
  \centering
  \includegraphics[width=1.05\linewidth]{figures/SINGLE/HUV_HL_Tg/maps/JKH2.pdf}
\end{subfigure}
\begin{subfigure}{.33\linewidth}
  \centering
  \includegraphics[width=1.05\linewidth]{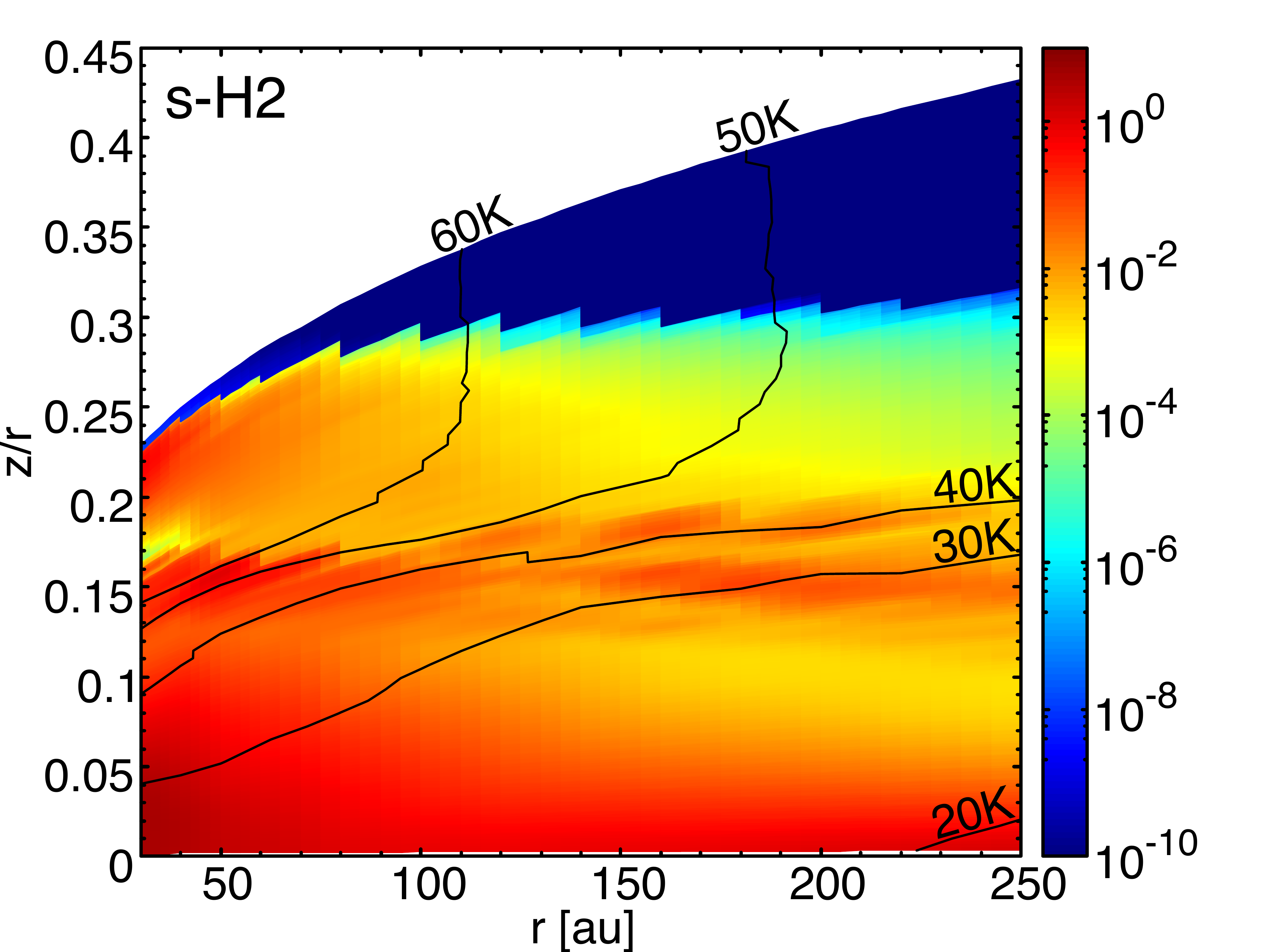}
\end{subfigure}
\begin{subfigure}{.33\linewidth}
  \centering
  \includegraphics[width=1.05\linewidth]{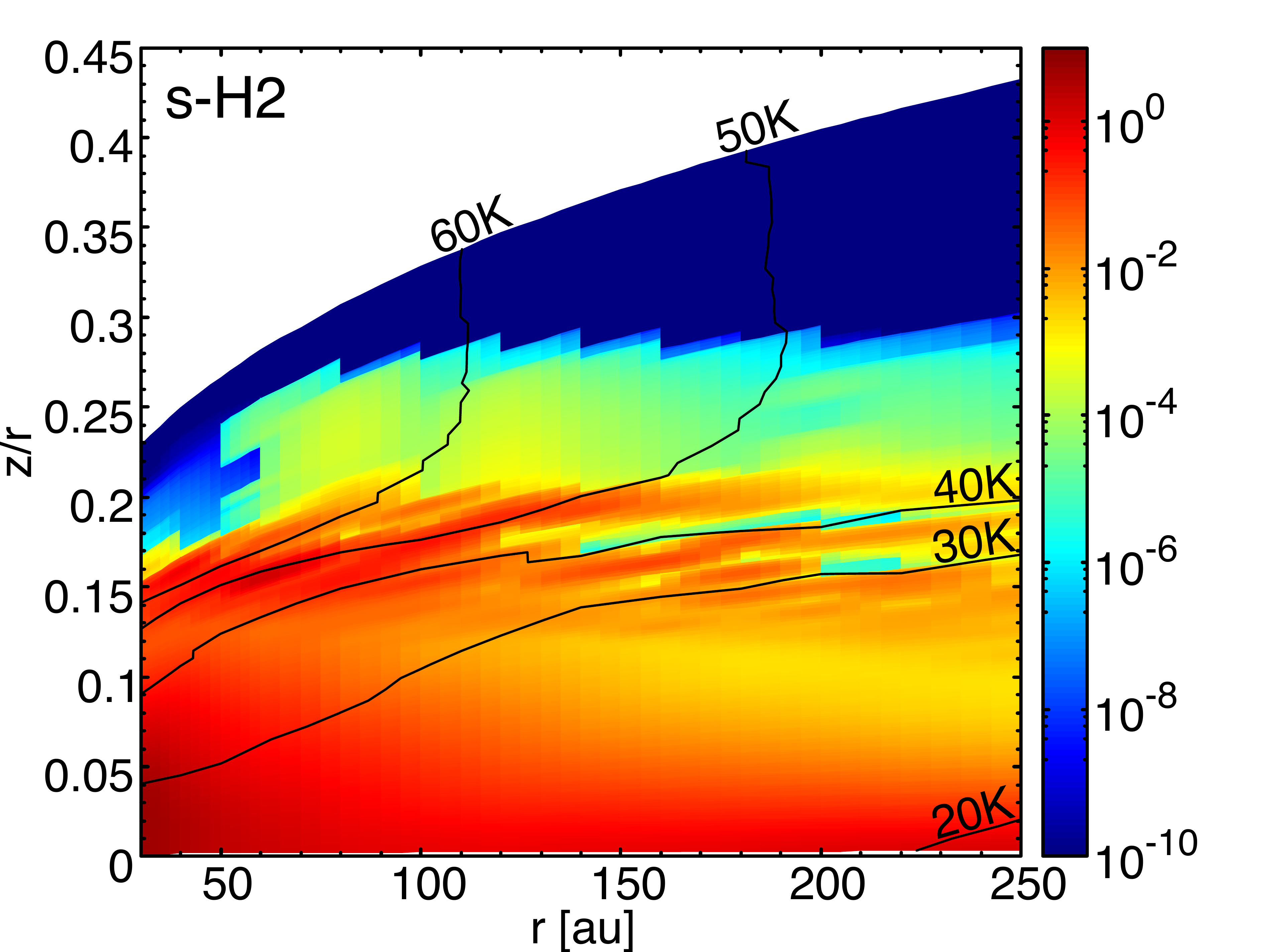} 
\end{subfigure}
\label{fig:m-surf-mapsH2}

\begin{subfigure}{.33\linewidth}
  \centering
  \includegraphics[width=1.05\linewidth]{figures/SINGLE/HUV_HL_Tg/maps/JKCO.pdf}
\end{subfigure}
\begin{subfigure}{.33\linewidth}
  \centering
  \includegraphics[width=1.05\linewidth]{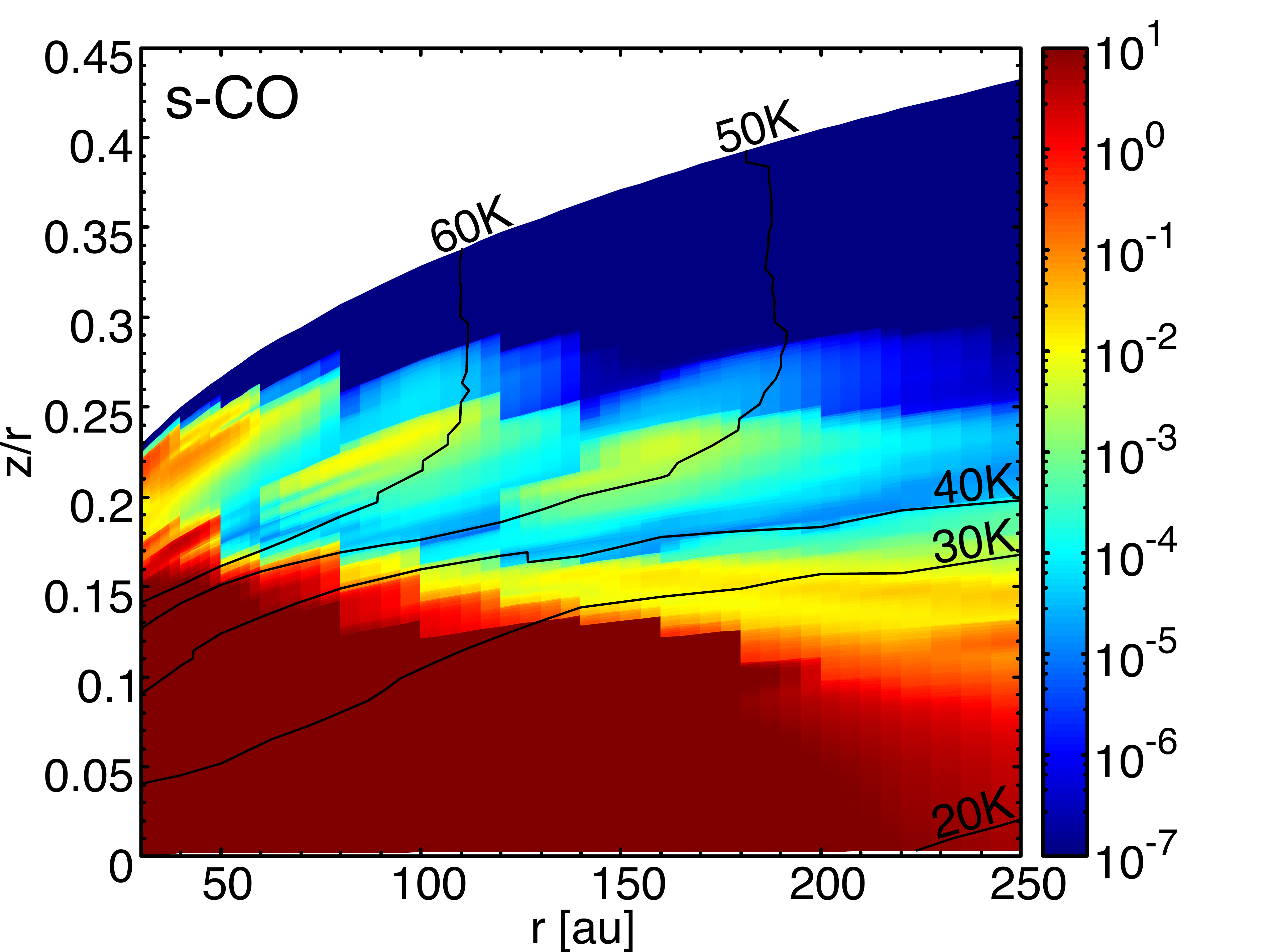}
\end{subfigure}
\begin{subfigure}{.33\linewidth}
  \centering
  \includegraphics[width=1.05\linewidth]{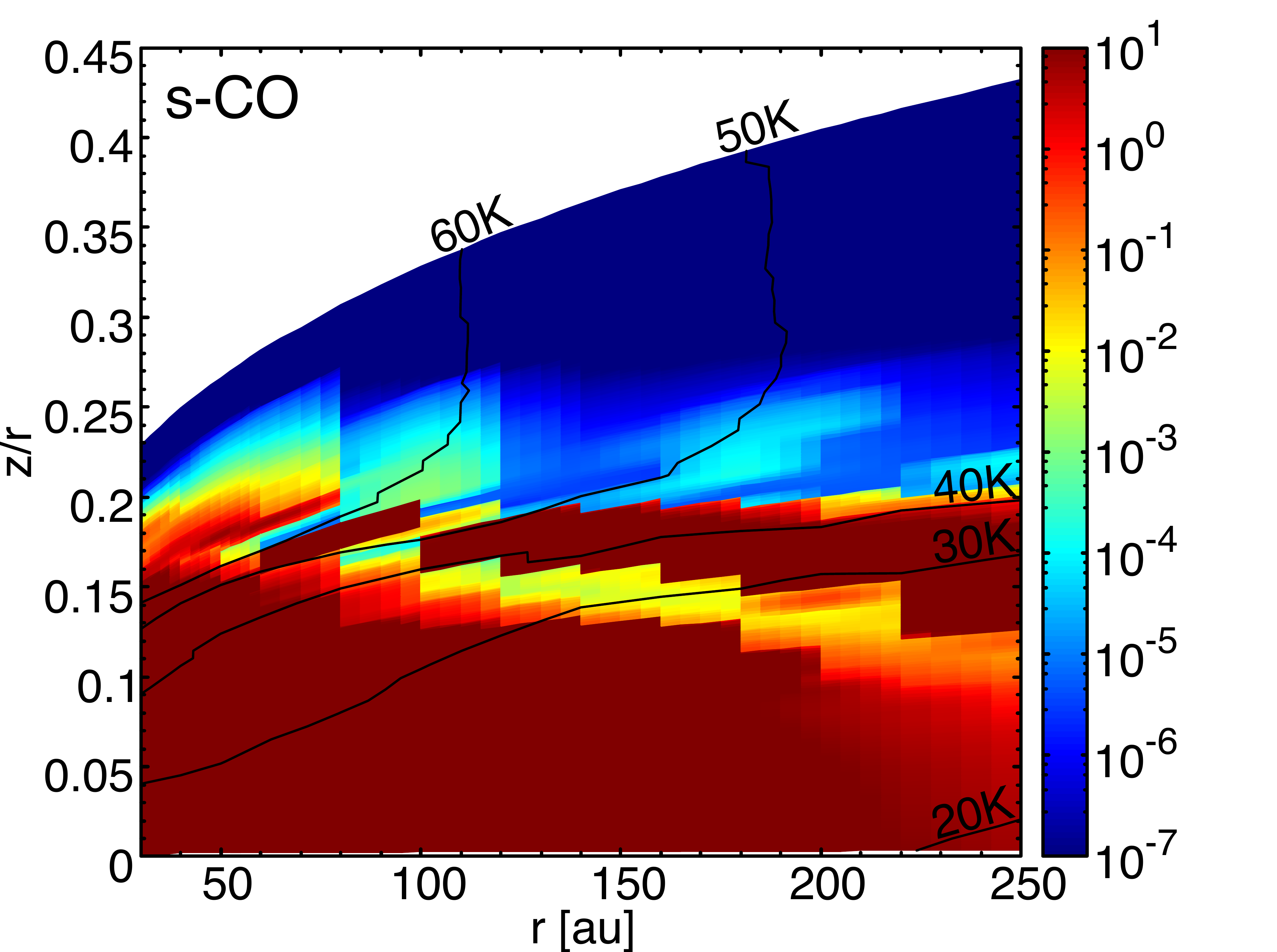} 
\end{subfigure}
\label{fig:m-surf-mapsCO}

\begin{subfigure}{.33\linewidth}
  \centering
  \includegraphics[width=1.05\linewidth]{figures/SINGLE/HUV_HL_Tg/maps/JKCS.pdf}
\end{subfigure}
\begin{subfigure}{.33\linewidth}
  \centering
  \includegraphics[width=1.05\linewidth]{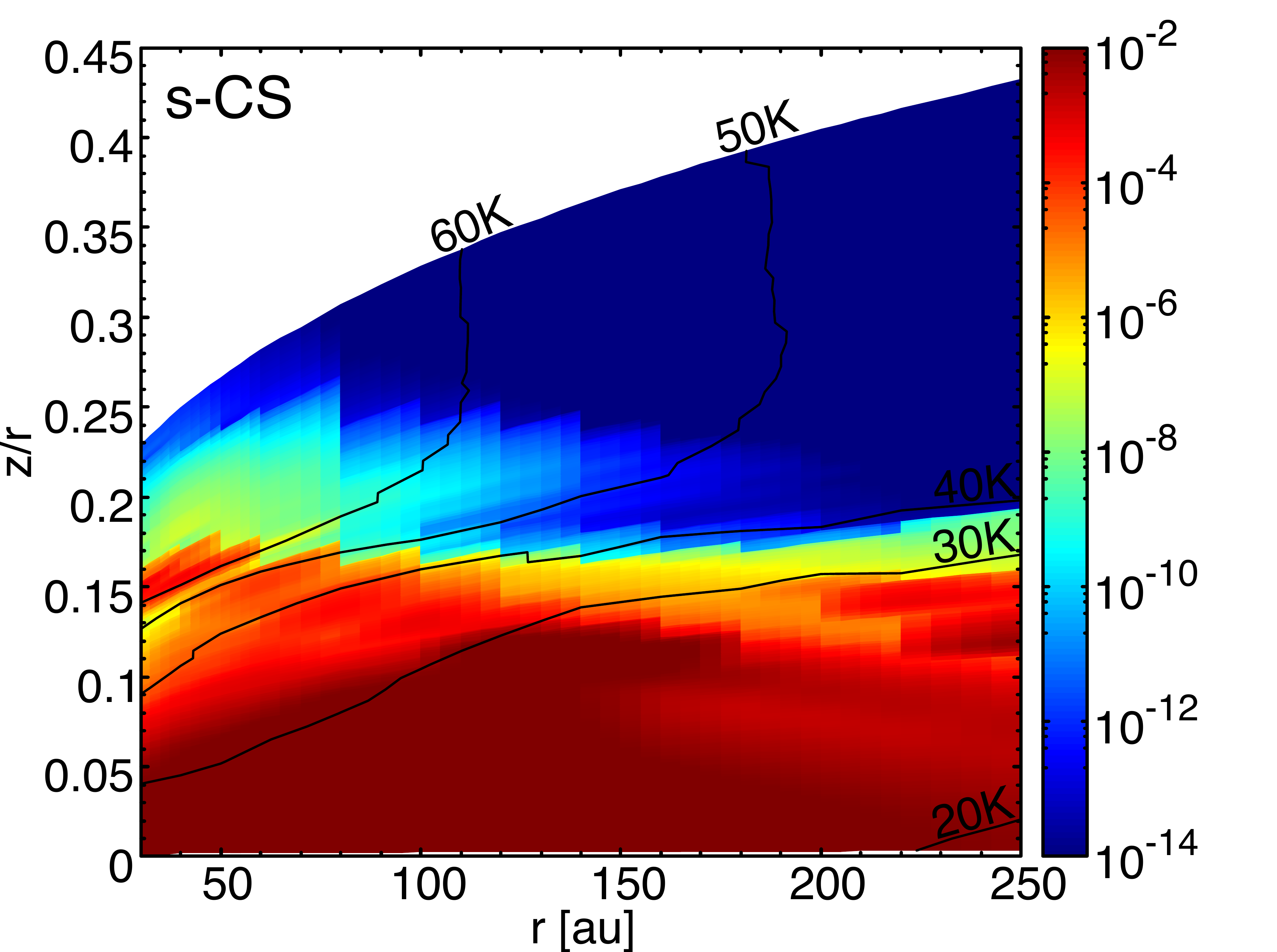}
\end{subfigure}
\begin{subfigure}{.33\linewidth}
  \centering
  \includegraphics[width=1.05\linewidth]{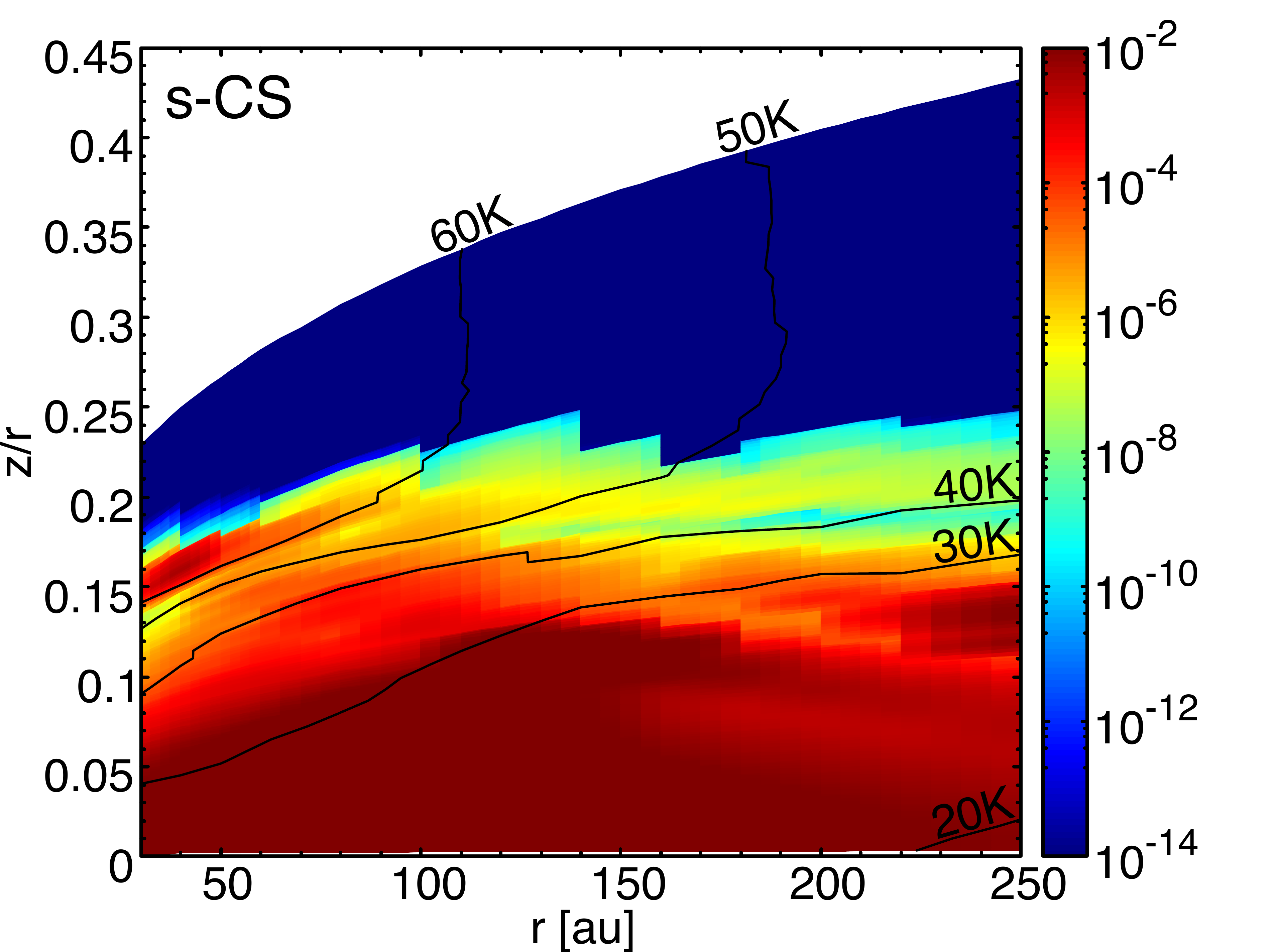}
\end{subfigure}
\label{fig:m-surf-mapsCS}

\begin{subfigure}{.33\linewidth}
  \centering
  \includegraphics[width=1.05\linewidth]{figures/SINGLE/HUV_HL_Tg/maps/JKCN.pdf}
   \subcaption{\shtg}   
\end{subfigure}
\begin{subfigure}{.33\linewidth}
  \centering
  \includegraphics[width=1.05\linewidth]{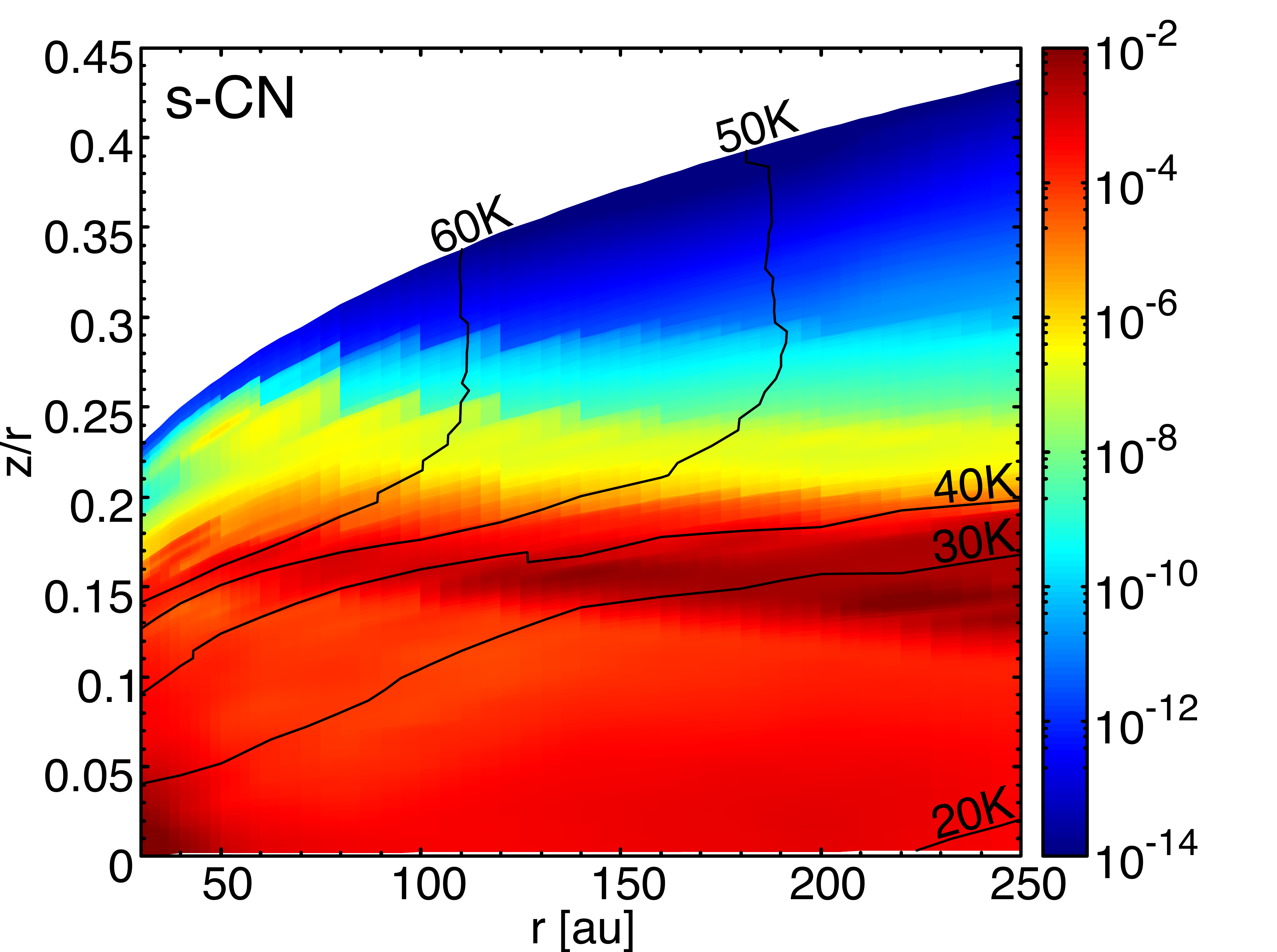}
   \subcaption{\mhlh}   
\end{subfigure}
\begin{subfigure}{.33\linewidth}
  \centering
  \includegraphics[width=1.05\linewidth]{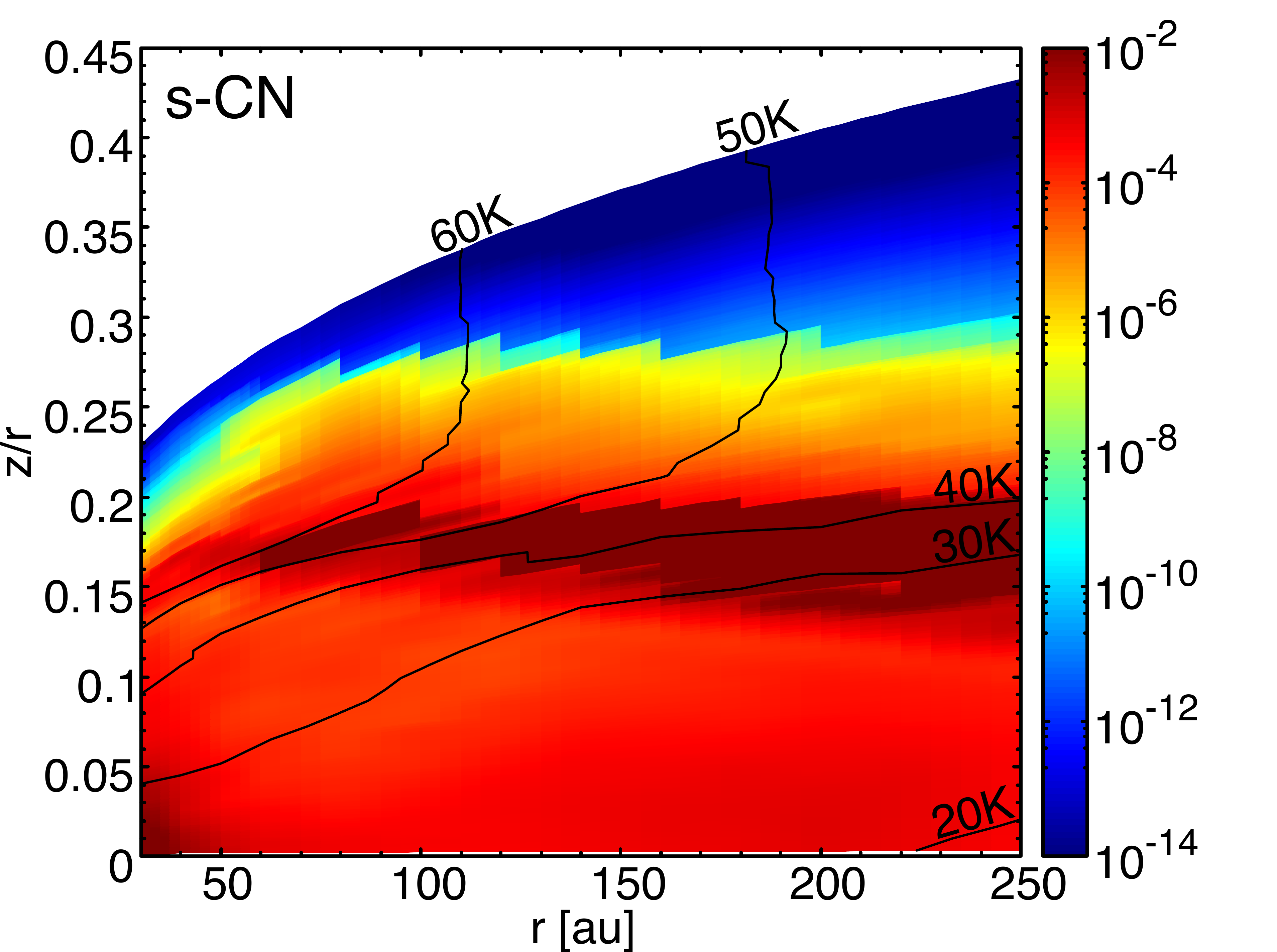}
   \subcaption{\mhb}   
\end{subfigure}
\label{fig:m-surf-mapsCN}
\caption{Density [cm$^{-3}$] of H$_2$, CO, CS and CN in the gas-phase of \shtg (left column) and of the multi-grain models in HUV regime. Black contours represent the dust temperature (T$_\mathrm{d}$ = T$_\mathrm{g}$ in the left column and T$_\mathrm{d}$ = T$_\mathrm{a}$ in the middle and right columns).}
\label{fig:m-s-maps-huv}
\end{figure*}

\revisionA{Figures \ref{fig:s-s-maps-huv} and \ref{fig:m-s-maps-huv} 
present maps of number density for s-H$_2$, s-CO, s-CS and s-CN locked on grain 
surfaces in the case of single-grain and multi-grain models, 
respectively. Figure \ref{fig:ab_surface} presents the surface abundance 
per total atomic hydrogen for various molecules as a function of the grain radius $a$ 
and their temperature in the midplane of the disk. The crosses, the 
square markers, and the round markers stand for the multi-grain model 
at 30 au, 100 au and 200 au respectively. The triangle pointing upward 
represents the surface abundance in the single-grain model \shtg and 
the triangle pointing downward represents the abundance in the 
single-grain model \shteff, both at 100 au. }
   
\subsubsection{Carbon and oxygen-bearing molecules: CO, CO$_2$ and CH$_4$} 
CO is formed in the gas-phase prior to being accreted on small grains 
during the first 10$^6$ yrs of the simulation. However, because of its 
small binding energy \citep[$E_{\mathrm{b}}$(CO) = 1300 K][]{Wakelam+etal_2017}, 
it is rapidly thermally desorbed from the hot ($T_g > 20$\,K)
smaller grains and gets accreted by large cold grains ($T_g < 20$\,K), \revisionA{where} 
it remains locked until the simulation is 
stopped. Considering this mechanism, CO is distributed according to the 
grain population and Fig.\,\ref{fig:ab_surface} (top left panel) shows 
two regimes of abundance, accounting for the desorption barrier, where CO ice tends to stay locked on grains of size $\gtrsim$ 
1 $\mu$m at all radii. 

For the single-grain models, the same mechanism is at play. However, 
rather than involving a distribution of s-CO on various grain 
populations, this simply results in a small 
abundance in 
the hot-grain model \shteff and a high abundance in the cold-grain model 
\shtg.

As for CO$_2$, its binding energy is relatively high 
\citep[$E_{\mathrm{b}}(\mathrm{CO}_2) = 2600$ K][]{Wakelam+etal_2017}. 
As a consequence the abundance of s-CO$_2$ depends very little on the 
dust temperature {and simply depends on the dust surface area instead}. s-CO$_2$ is efficiently formed 
through
\begin{gather}
\label{eq:CO2b}
\mathrm{s\mhyphen O + s\mhyphen CO \rightarrow s\mhyphen CO_2 \: (E_A = 1000\,K)}\\
\label{eq:CO2d}
\mathrm{s\mhyphen OH + s\mhyphen CO \rightarrow s\mhyphen CO_2 + s\mhyphen H}.
\end{gather}
\noindent Therefore, s-CO$_2$ is more abundant on the smaller grains because their 
temperature allows to overcome the activation barrier of reaction 
\ref{eq:CO2b} and the drop in abundance on large grains at 200 au (and 
further) originates from the grain temperature getting too low for 
reaction \ref{eq:CO2b} to be activated. Therefore, the outer disk forms 
fewer CO$_2$ ice. 

In cold-grain models, s-CH$_4$ is the largest carrier of carbon (see 
Section \ref{subsec:reservoirs}). As expected, Figure \ref{fig:ab_surface} 
(Bottom-left) shows that s-CH$_4$ is more abundant on cold grains ($T_g < 20$\,K). 
s-CH$_4$ is formed via the hydrogenation of s-CH$_3$. The latter is 
formed through the following two sequences:
\begin{gather}
\label{eqn:CH4a}
\mathrm{s\mhyphen CO + h\nu \rightarrow s\mhyphen C + s\mhyphen O} \\ 
\mathrm{s\mhyphen H_2 + s\mhyphen C \rightarrow s\mhyphen CH_2} \\
\mathrm{s\mhyphen H_2 + s\mhyphen CH_2 \rightarrow s\mhyphen CH_3 + s\mhyphen H}
\end{gather}
\noindent and
\begin{gather}
\label{eqn:CH4b}
\mathrm{s\mhyphen CH_3OH + h\nu \rightarrow s\mhyphen OH + s\mhyphen CH_3}. 
\end{gather}

\noindent In view of the above, s-CH$_3$OH is formed on the grains prior 
to being photodissociated (if the computation domain is sufficiently 
extended). Then the released atomic carbon (Reaction \ref{eqn:CH4a}) 
and s-CH$_3$ (Reaction \ref{eqn:CH4b}) lead to s-CH$_4$ formation. 
Consequently, the formation of s-CH$_4$ is more efficient on large cold 
grains since 1) CO must be adsorbed and 2) hydrogenation is needed to 
create s-CH$_3$OH. s-CH$_4$ ice is thus found to be more abundant on large 
grains and in the outer disk.
  
  \subsubsection{Sulfur-bearing species: H$_2$S} 
In the gas phase, H$_2$S has been identified in dense cloud cores, 
cometary comae and \citet{Phuong+etal_2018} reported the first 
detection of  H$_2$S in the cold and dense ring about the TTauri star 
GG Tau A. The low H$_2$S column densities observed in GG Tau A and in 
dense molecular clouds are assumed to be the result of a strong sulfur 
depletion \citep{Hudson+Gerakines_2018, Phuong+etal_2018}. 
 
We investigate here the formation routes of S-H bonds and more 
specifically the simple S-bearing molecule Hydrogen sulfide, H$_2$S, in 
the midplane icy mantles of our multi-grain models. 
Figure \ref{fig:ab_surface} (second row, right column) shows the abundance 
of s-H$_2$S as a function of the grain size. There is a strong 
anti-correlation with grain temperatures. The smallest s-H$_2$S 
abundances are located on the hottest grains ($T_d \sim 27$\,K) while the largest 
abundances are found on large colder grains ($T_d \lessapprox 15$\,K). This anti-correlation is 
observed at all radii. The abundances on the large grains ($> 10 \mathrm{\mu m}$) 
are independent of the radius. For the small grains, 
however, the inner disk exhibits larger abundances than the outer disk 
because of higher collision rates in the inner regions. The strong 
anti-correlation with temperature is directly related to the main 
formation pathways of H$_2$S that implies hydrogenation sequences:

\begin{gather}
\label{eqn:H2Sa}
\mathrm{s\mhyphen S \xrightarrow{s\mhyphen H} s\mhyphen HS \xrightarrow{s\mhyphen H} s\mhyphen H_2S} \\
\label{eqn:H2Sb}
\mathrm{s\mhyphen CH_3SH \xrightarrow{s\mhyphen H} s\mhyphen CH_3 + s\mhyphen H_2S}.
\end{gather}

\noindent We note that the channels to s-CH$_3$SH also involve 
successive hydrogenations (of s-CS, mainly). Reaction \ref{eqn:H2Sa} 
occurs on grains of all sizes but is more effective on the large cold 
grains ($T_d \lessapprox 15$\,K) and Reaction \ref{eqn:H2Sb} is only found to be effective on the 
colder grains.

\subsubsection{Nitrogen-bearing species: NH$_3$} 

\begin{figure*} 
  \centering
  \includegraphics[width=1.0\linewidth]{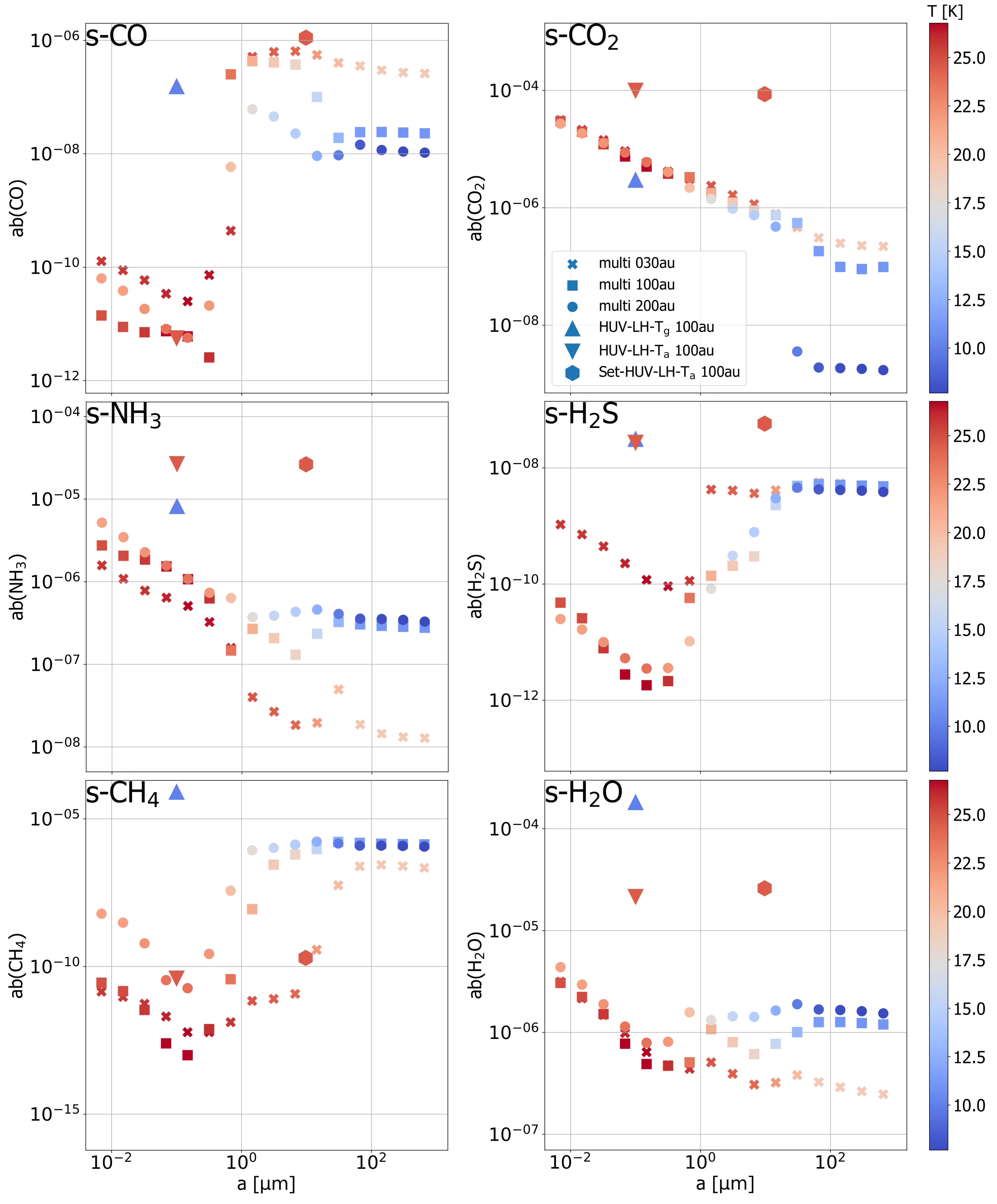}  
\caption{Grain surface abundance per total atomic hydrogen \revisionC{and per grain size bin} for various 
molecules at the final stage of integration (i.e. 5\,10$^6$ yrs) as a 
function of the grain radius $a$ in relation to the grain temperatures. 
Cross markers stand for the multi-grain model at 30 au, the square 
markers stand for the multi-grain model at 100 au and the round markers 
stand for 200 au. The triangles pointing upward represent the surface 
abundance in the single-grain model \shtg (grains of size 0.1 
$\mathrm{\mu m}$) and the triangle pointing downward represents the 
abundance in the single-grain model \shteff, both at 100 au.  \revisionC{Abundances
being per grain size bin, it is the sum of the abundances in the multi-grain models
that should be compared to the abundances in the single grain case.}}
\label{fig:ab_surface}
\end{figure*}

The main nitrogen-bearing icy species in the midplane are s-NH$_3$, s-HCN, 
s-N$_2$ and s-CN at all radii. In the multi-grain models, as seen in 
Fig.\,\ref{fig:ab_surface}, the abundance of s-NH$_3$ is inversely 
proportional to the grain size until $\approx 1 \mathrm{\mu m}$ where the 
abundance stops decreasing. The abundances of s-NH$_3$ are roughly 
the same at 100 and 200 au but not at 30 au. 
 
The main pathway to s-NH$_3$ originates from the accretion of N at the 
surface of the grains and its successive hydrogenations to form s-NH$_3$ 
\citep[see][]{Aikawa+etal_2015, Eistrup+etal_2018, Ruaud+Gorti_2019}. 
Such hydrogenation reactions are efficient on cold surfaces ($T_d \lessapprox 15$\,K), which 
explains why the abundance of s-NH$_3$ does not keep dropping 
when grain temperatures are low. As for the  warmer small grain surfaces, 
their high quantity implies a high collision rate with NH$_3$ which 
stays on their surface given its binding energy 
(E$_{\mathrm{b}}$(NH$_3$) = 5500 K). As the disk evolves, a large 
fraction of nitrogen-bearing species are transformed into complex 
molecules and other routes become effective such as the destructive 
hydrogenation of s-NH$_2$CO:
 
\begin{equation}
\label{eq:nh3}
	  \mathrm{s\mhyphen NH_2CO \xrightarrow{s\mhyphen H} s\mhyphen NH_3 + s\mhyphen CO}.
\end{equation}   

\subsection{Water} \label{subsec:water} 
  
The formation of water is key to explain the evolution of 
protoplanetary disks and formation of comets, both decisive elements to 
understand the delivery of water to planetary surfaces. Spectral lines 
of H$_2$O in the gas-phase of inner disk regions have now been widely 
detected in disks around TTauri stars using near- and mid-infrared 
\citep{Carr+Najita_2008, Pontoppidan+etal_2010} or far-infrared 
observations \citep{Riviere-Marichalar+etal_2012}. Cold water in the 
outer disk regions, on the other hand, is harder to detect due to low 
excitation lines. 

It is assumed \citep{Dominik+etal_2005, lecar+etal_2006, 
Podio+etal_2013, Du+Bergin_2014} that water on grain surfaces cannot 
exist inside the snowline ($T_{d} \gtrsim 150$\,K), which represents radii 
smaller than a few to ten au around T\,Tauri stars. The innermost radius 
computed by our models is 30 au, which is beyond the radial ice line of 
water. Therefore, we do not discuss the radial location of the water 
ice line. 

\subsubsection{Gas-phase}
\revisionA{Three main parameters control  the abundance of water in 
gas-phase, the photodesorption by FUV photons and cosmic rays, the 
dust temperature and the chemical reactions in the gas-phase, in 
particular: }  

\revisionA{ \begin{gather}
\label{eqn:h2od}
\mathrm{H_2 + OH^+ \rightarrow H + H_2O^+} \\ 
\mathrm{H_2O^+ + H_2 \rightarrow H + H_3O^+} \\
\mathrm{H_3O^+ + e^-  \rightarrow H + H_2O} 
\end{gather}}

\revisionA{Thus \hh and OH$^+$ are the main precursors for H$_2$O. Whether H$_2$ or OH$^+$ is the limiting factor is determined by this other sequence:}

\revisionA{ \begin{gather}
\label{eqn:oh}
\mathrm{H_2 + CR \rightarrow H_2^+ + e^-}\\
\label{eqn:oh2} 
\mathrm{H_2^+ + H_2 \rightarrow H_3^+ + H} \\
\label{eqn:oh3} 
\mathrm{H_3^+ + O  \rightarrow OH^+ + H_2}. 
\end{gather}}
			
\revisionA{\noindent These two sequences show that H$_2$O formation 
depends on \hh and OH$^+$ but the latter also depends on \hh. 
Therefore, this is the \hh abundance that sets the gas-phase formation 
of water and accounts for the difference between the models (not 
cosmic-rays or OH$^+$). However, the gas-phase formation of H$_2$O 
only represents a small fraction of all H$_2$O molecules formed as most 
H$_2$O are formed on grain surfaces. The abundance of gas-phase water 
is thus mainly governed by both the formation rate of icy water and by 
the photodesorption rate. Thermal-desorption is ineffective in most 
parts of the disk because the binding energy of water is high. The 
formation rate depends on the grain temperature while the 
photodesorption rate depends on the FUV flux and cosmic-rays. The 
analysis of the results herein (Figs.\,\ref{fig:s-water_maps} and 
\ref{fig:m-water_maps}) shows that the dust temperature plays the major 
role in the production of gas-phase water.} 

\revisionA{\paragraph{Single-grain models} The vertical snowline forms a 
clear boundary between high and low gas-phase H$_2$O abundance at $z/r 
\sim 0.15$ (Fig.\,\ref{fig:s-water_maps}) and we see that \shb produces 
the most distinct boundary of all three models. The snowline 
corresponds roughly to the abrupt drop in the UV flux as shown in 
Fig.\,\ref{fig:struct} (bottom-right panel). The cold-grain model \shtg 
produces substantially more water in the upper layers than the other 
single-grain models, 
because low dust temperature 
increase the hydrogenation rates on the grain surfaces. The 
photodesorption rates are approximately equal in the three models 
(same flux) but since \shtg produces a larger water abundance on grain 
surfaces the final gas-phase abundance is much larger. It results in a 
column density of water about a factor 10 larger in \shtg than in the 
other two models. On the other hand, the B14's prescription increases 
only marginally the column density of gas-phase water.} 
	
\revisionA{ \paragraph{Multi-grain models} The vertical snowline is far 
less apparent (Fig.\,\ref{fig:m-water_maps}, top row, middle and right 
column). Both models have approximately the same distribution of 
abundance. because both have the same dust temperature distribution and 
same UV. We notice, however, that B14's prescription in \mhb produces 
marginally more water in the upper layer of the disk than in 
\mhlh, due to the gas-phase formation of water via the two sequences 
presented above.} 	

\subsubsection{Water ice}

\begin{figure*}
\begin{subfigure}{.33\linewidth}
  \centering
  \includegraphics[width=1.05\linewidth]{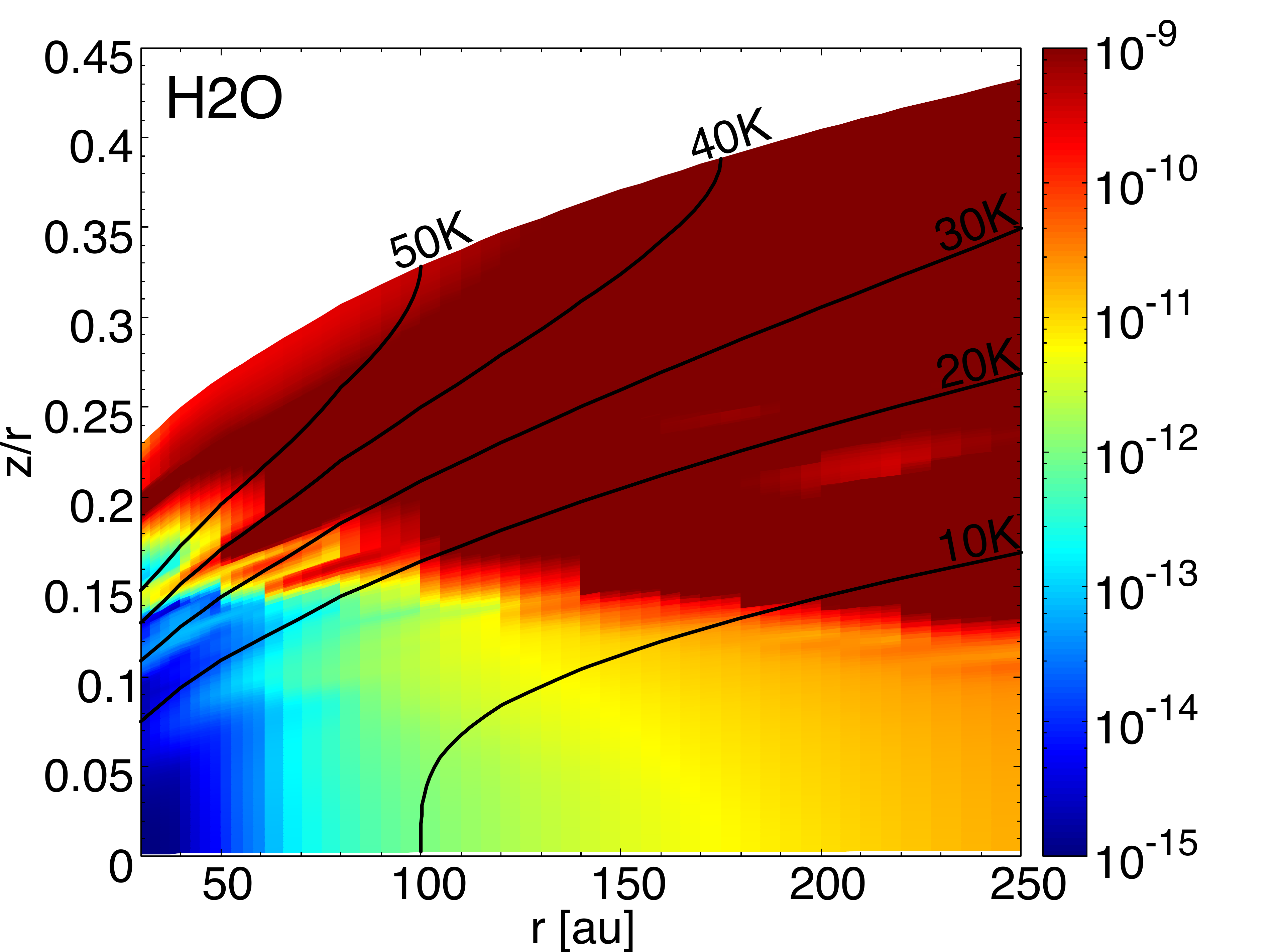}
\end{subfigure}
\begin{subfigure}{.33\linewidth}
  \centering
  \includegraphics[width=1.05\linewidth]{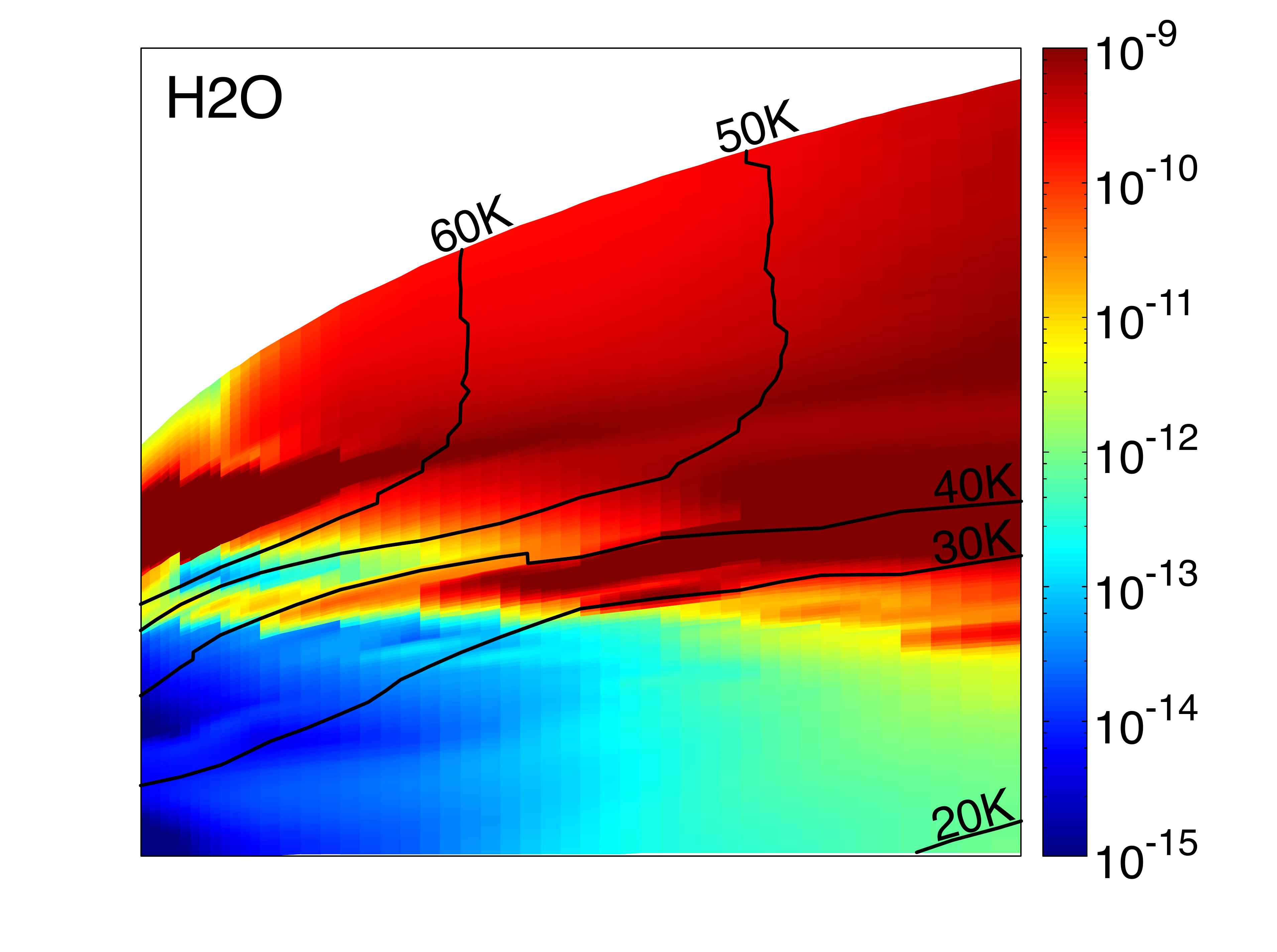}
 \end{subfigure}
\begin{subfigure}{.33\linewidth}
  \centering
  \includegraphics[width=1.05\linewidth]{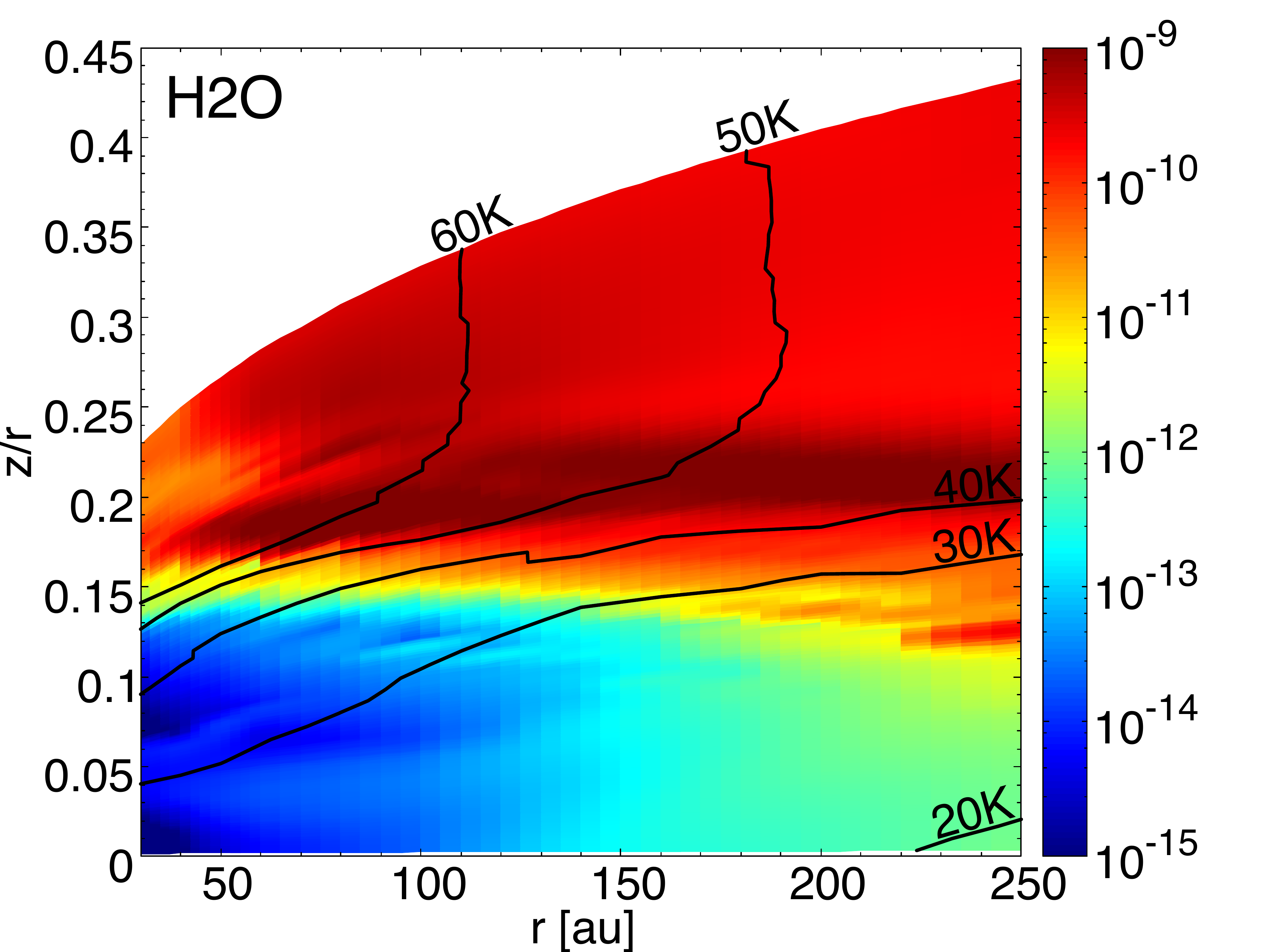}
\end{subfigure}
\begin{subfigure}{.33\linewidth}
  \centering
  \includegraphics[width=1.05\linewidth]{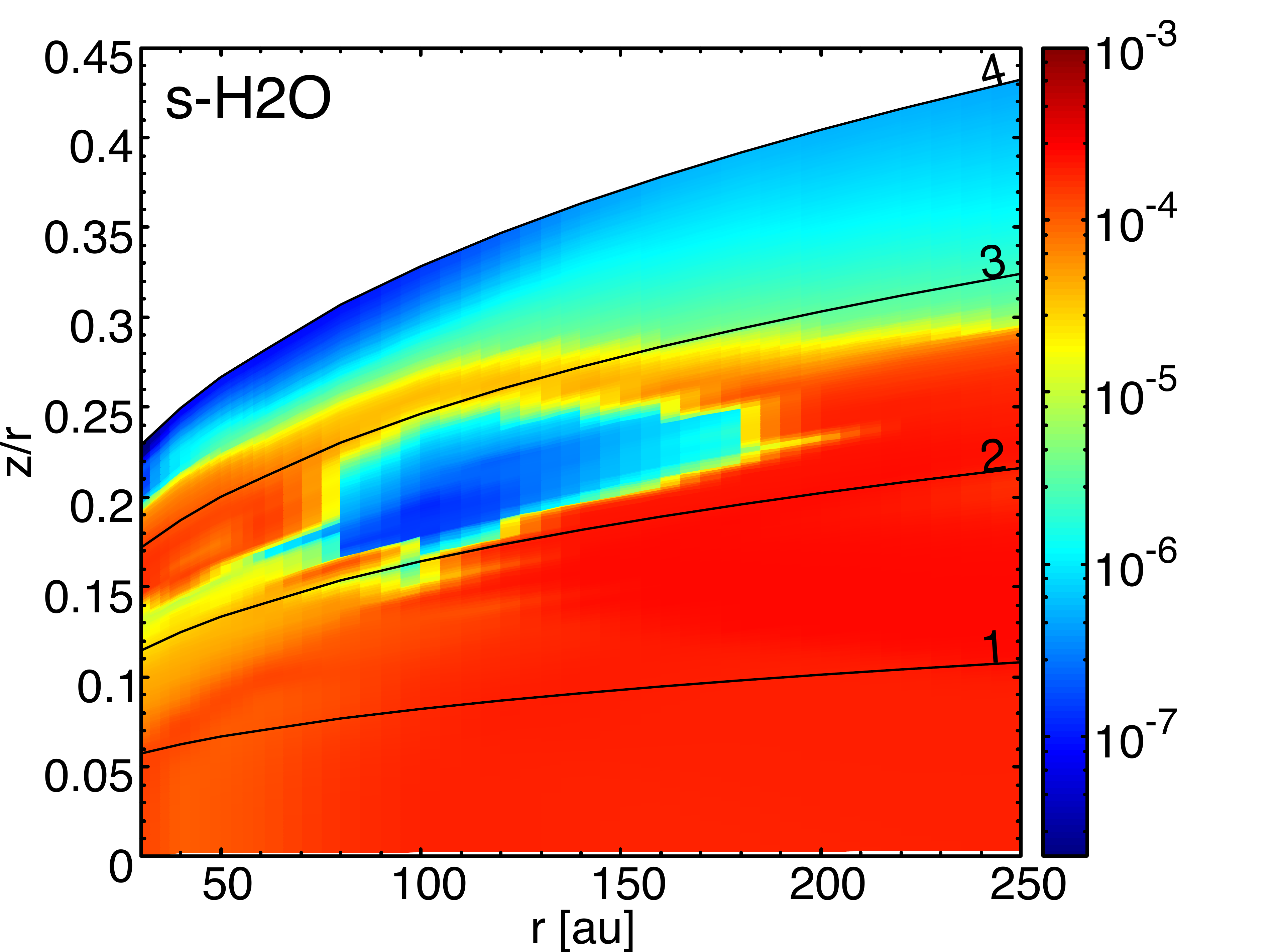}
  \subcaption{\shtg}   
\end{subfigure}
\begin{subfigure}{.33\linewidth}
  \centering
  \includegraphics[width=1.05\linewidth]{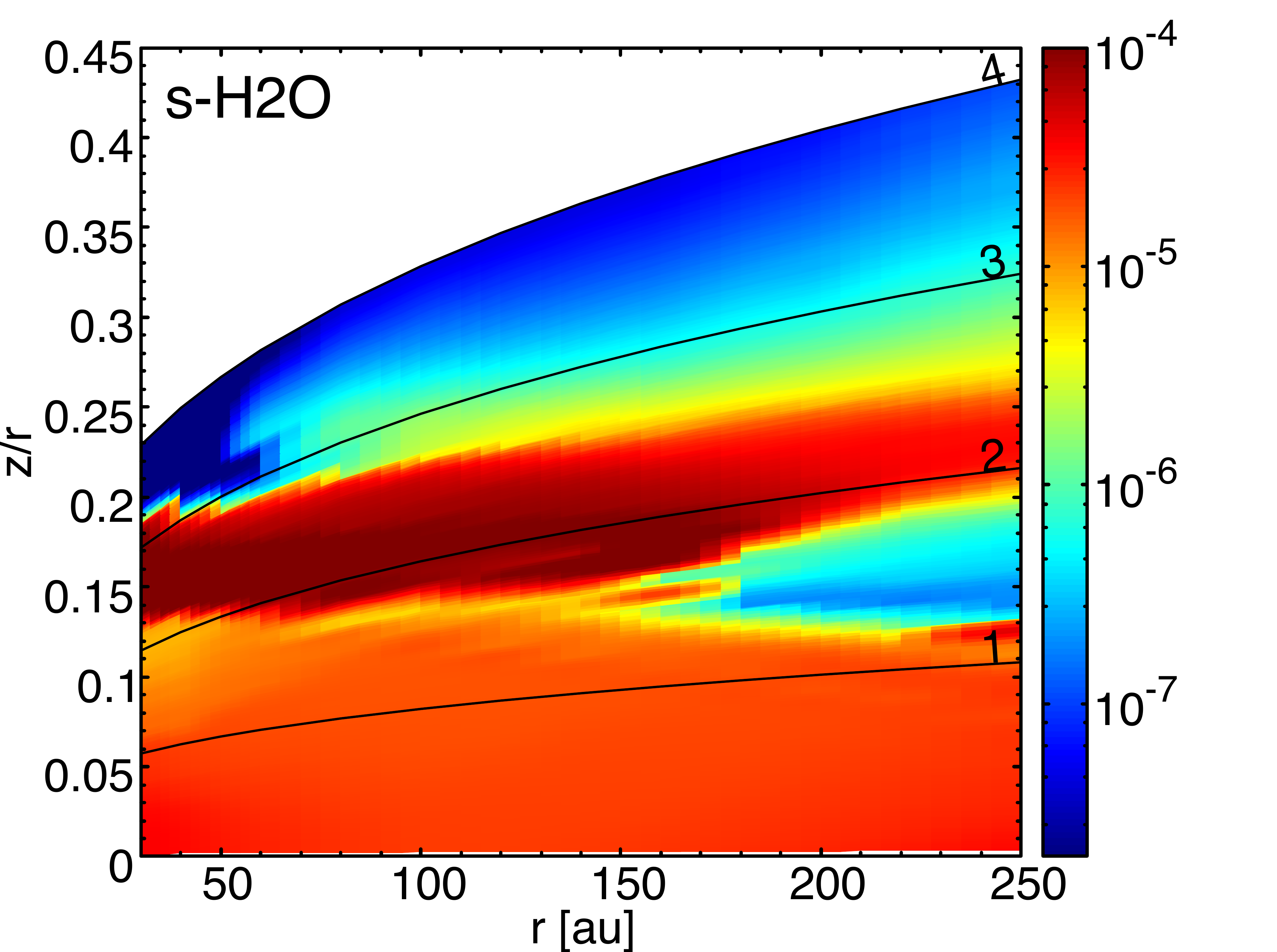}
  \subcaption{\shteff}   
\end{subfigure}
\begin{subfigure}{.33\linewidth}
  \centering
  \includegraphics[width=1.05\linewidth]{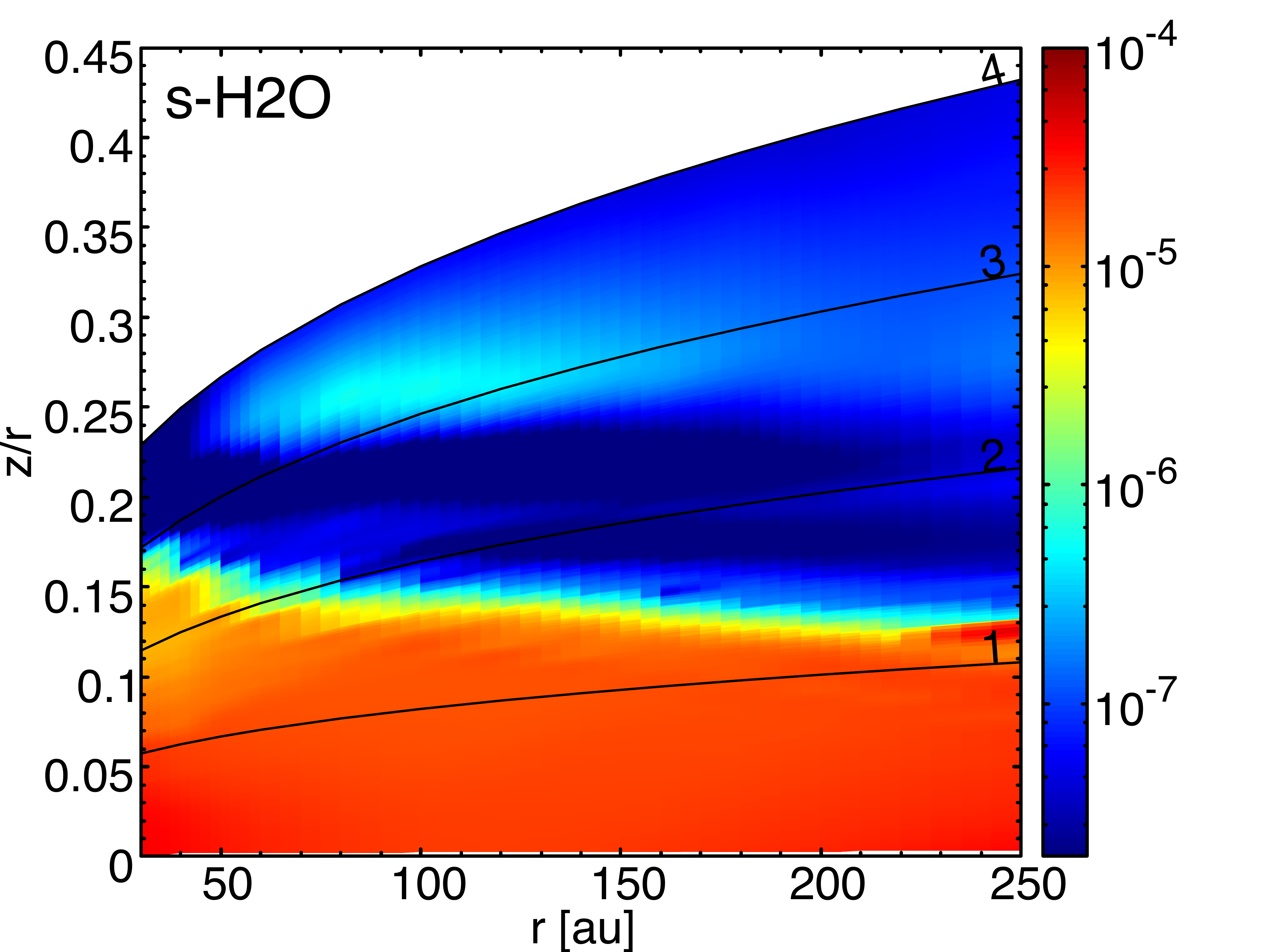}
  \subcaption{\shb}   
\end{subfigure}

\caption{Water abundance in the gas-phase (top) and on the grain surfaces (bottom) of the single-grain models. Left column shows the abundances in \shtg, middle in \shteff and right one in \shb. In the top row, the solid black lines show the dust temperature isocontours, in the bottom row they denote isocontours of scale heights.}
\label{fig:s-water_maps}
\end{figure*}

\revisionA{ \paragraph{Single-grain models} The vertical snowline location 
is clearly defined (see Fig.\,\ref{fig:s-water_maps}, bottom row), 
although we note the presence of a notable band around $z = 3\,H$ ($z = 
2\,H$) in \shtg  (resp. \shteff). The reason is the following activated 
hydrogenation reaction: }

\begin{gather}
\label{eqn:h2oc}
\mathrm{s\mhyphen H + s\mhyphen HOOH \rightarrow s\mhyphen H_2O + s\mhyphen OH \: (E_A = 1400\,K)}.
\end{gather}

\noindent Indeed, the band illustrates the location in the disk where 
this reaction is activated. Below the band, the reaction is not 
activated and water ice is mostly photodesorbed while above the band 
the photodesorption rates competes with the formation rate and the 
abundance drops. In \shteff (Fig. \ref{fig:s-water_maps}, middle 
column), Reaction \ref{eqn:h2oc} is activated at smaller altitudes in 
the disk since the dust temperature is globally higher, resulting in a 
peak in abundance at around 2 scale heights. In contrast, \shb does not 
exhibit such a peak near $z = 2\,H$ since a large fraction of atomic 
hydrogen on surfaces is used to create molecular hydrogen. This 
dramatically decreases the quantity of available H atoms necessary to 
form water. 

\begin{figure*}
\begin{subfigure}{.33\linewidth}
  \centering
  \includegraphics[width=1.05\linewidth]{figures/SINGLE/HUV_HL_Tg/maps/H2O_ab.pdf}
\end{subfigure}
\begin{subfigure}{.33\linewidth}
  \centering
  \includegraphics[width=1.05\linewidth]{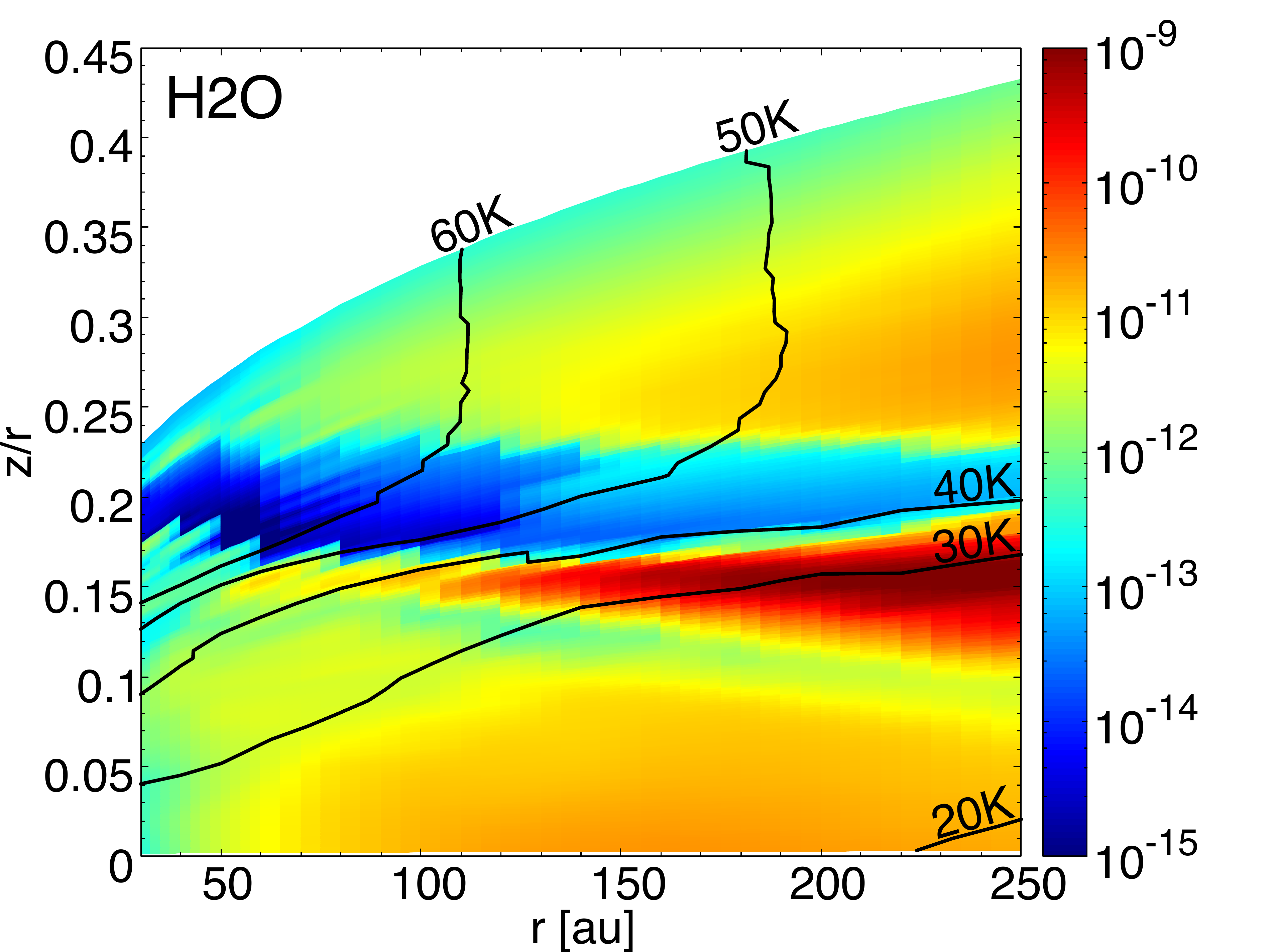}
\end{subfigure}
\begin{subfigure}{.33\linewidth}
  \centering
  \includegraphics[width=1.05\linewidth]{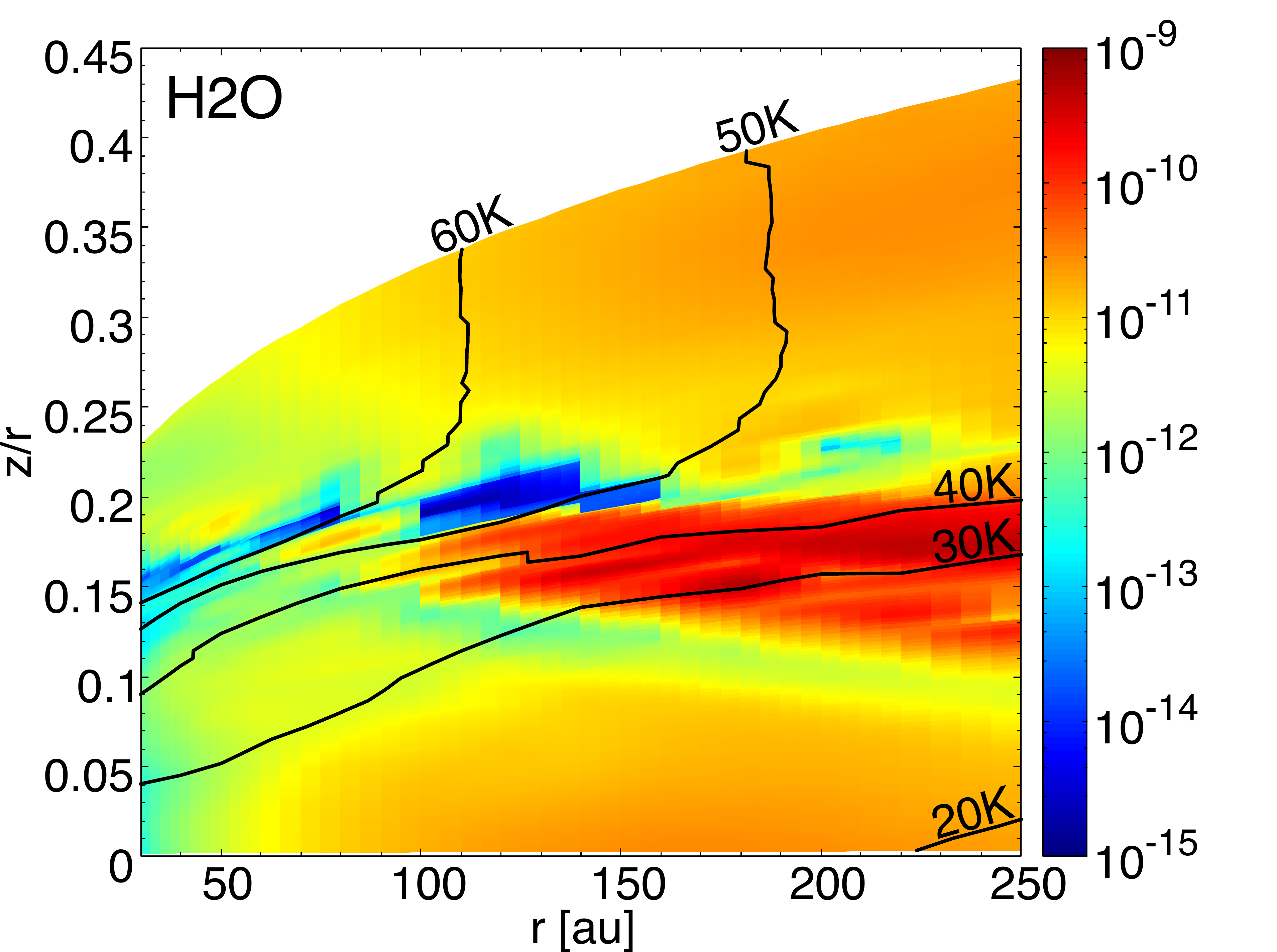}
 \end{subfigure}
 
 \begin{subfigure}{.33\linewidth}
  \centering
  \includegraphics[width=1.05\linewidth]{figures/SINGLE/HUV_HL_Tg/maps/sH2O_ab.pdf}
  \subcaption{\shtg}   
\end{subfigure}
 \begin{subfigure}{.33\linewidth}
  \centering
  \includegraphics[width=1.05\linewidth]{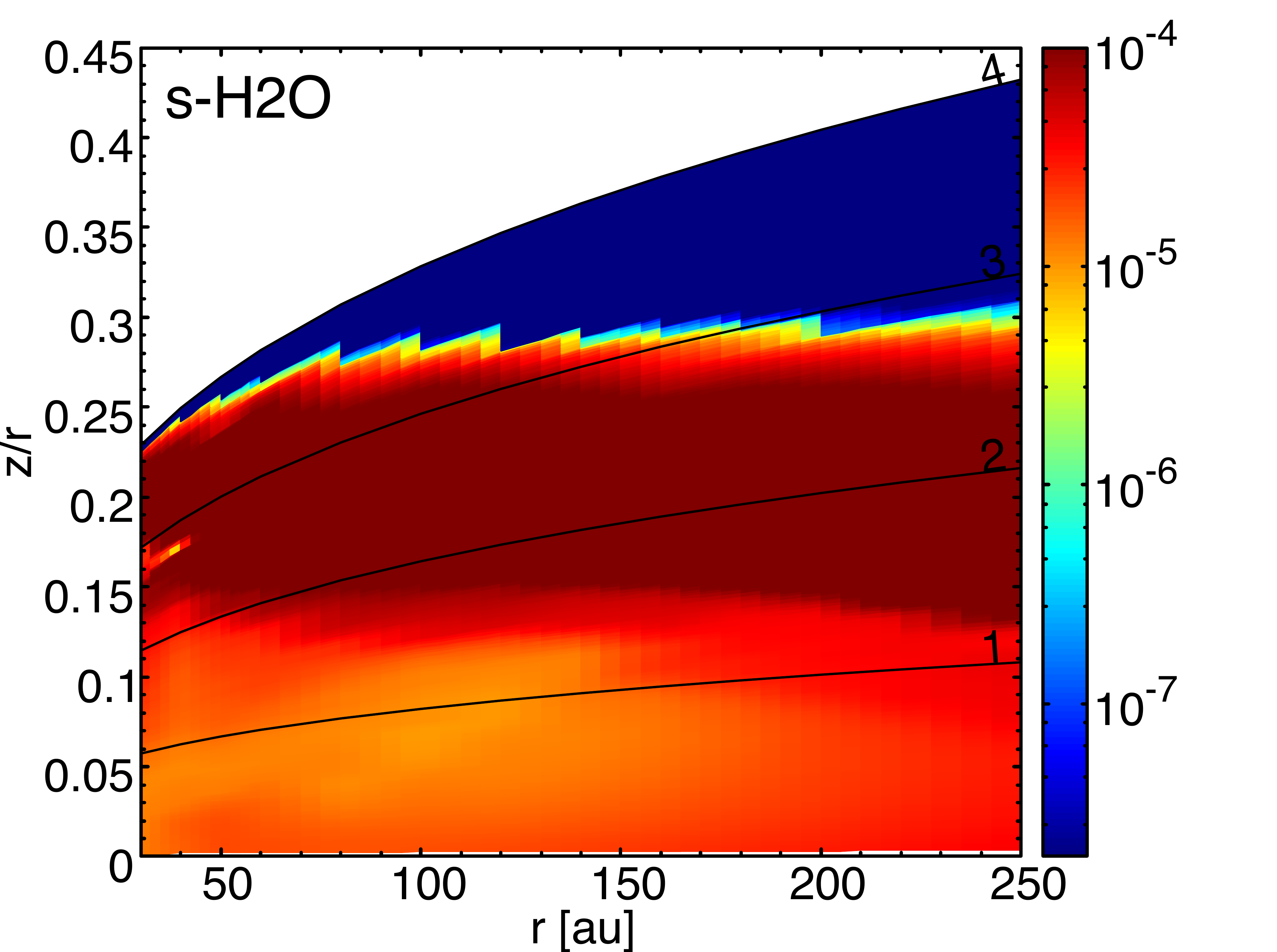}
  \subcaption{\mhlh}   
\end{subfigure}
\begin{subfigure}{.33\linewidth}
  \centering
  \includegraphics[width=1.05\linewidth]{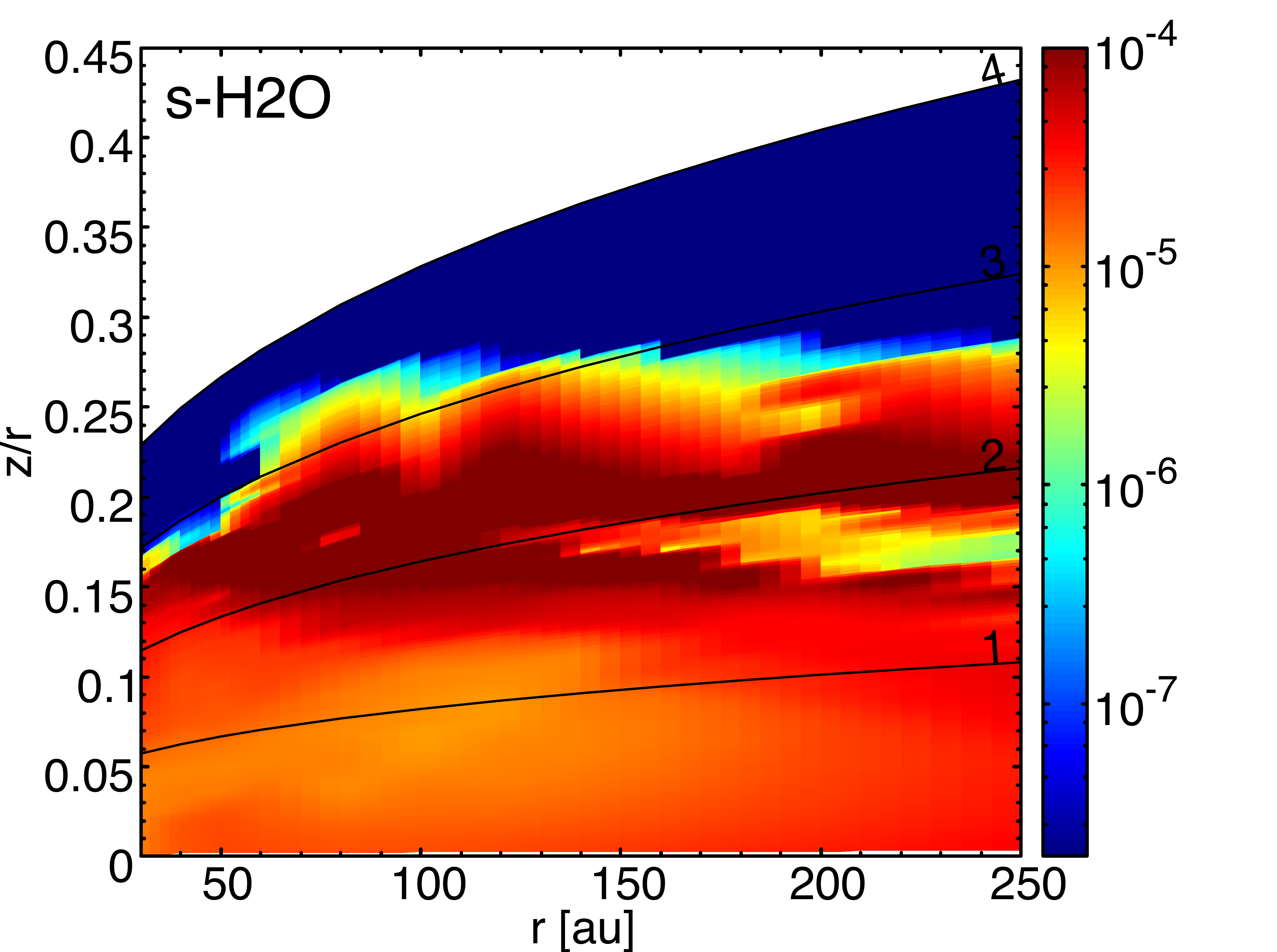}
  \subcaption{\mhb}
\end{subfigure}

\caption{Water abundance in the gas-phase (top) and on the grain surfaces (bottom). Left column shows the abundances in \shtg and middle and right column in the multi-grain models. In the top row, the solid black lines show the dust temperature isocontours, in the bottom row they denote isocontours of scale heights.}
\label{fig:m-water_maps}
\end{figure*}

\revisionA{ \paragraph{Multi-grain models} The same mechanisms are at work 
(see middle and right columns of Fig.\ref{fig:m-water_maps}). Yet, the 
dust temperature distribution involves that we find both hot (up to $\sim$ 95 K) and cold (as low as $\sim$ 7 K)
grains nearly everywhere in the disk. Consequently, Reaction 
\ref{eqn:h2oc} is activated at lower altitudes and the formation of 
s-H$_2$O through hydrogenation can occur at higher altitudes on cold 
medium- and large-sized grains so that the snowline is located at a 
higher altitude. }

\revisionA{\paragraph{Around the midplane} In all models, water is found 
in the form of ice around the midplane, where the desorption by UV 
flux and cosmic-rays is limited and where hydrogenation is effective. 
For single-grain models, the width of the region where water is on the 
icy mantles is larger in \shtg than in the two others because, again, 
the dust temperature is smaller. } 
As presented in Fig.\,\ref{fig:ab_surface} (bottom right panel), the abundance is roughly inversely proportional to the grain size at 30 au. The 
reason is the combination of the large binding energy of water and the 
high density of small grains. At 100 and 200 au, however, the abundance 
on large grains (size > 0.1 $\mu m$) is larger because the grain 
temperature is low enough to allow for effective hydrogenation.
  
  \subsection{COMs}  \label{subsec:coms} 
  
Complex organic molecules (COMs), supposedly constituting the crucial bond between simple ISM molecules and prebiotic chemistry, are hard to detect due to their low abundance and weak emission lines originating from their molecular complexity. This is especially true for objects with small spatial extensions like disks. Simple organic molecules have been regularly detected over the last decades such as H$_2$CO \citep{Dutrey+etal_1997, Aikawa+etal_2003, Oberg+etal_2011}, HC$_3$N \citep{Chapillon+etal_2012}, CH$_3$CN \citep{Oberg+etal_2015} and  CH$_3$OH \citep{Walsh+etal_2016}. We investigate here  the effectiveness of our models to synthesize COMs from surfaces or gas-phase when given sufficient time ($5\times10^6$ yrs).

 \begin{figure}
\includegraphics[width=\linewidth]{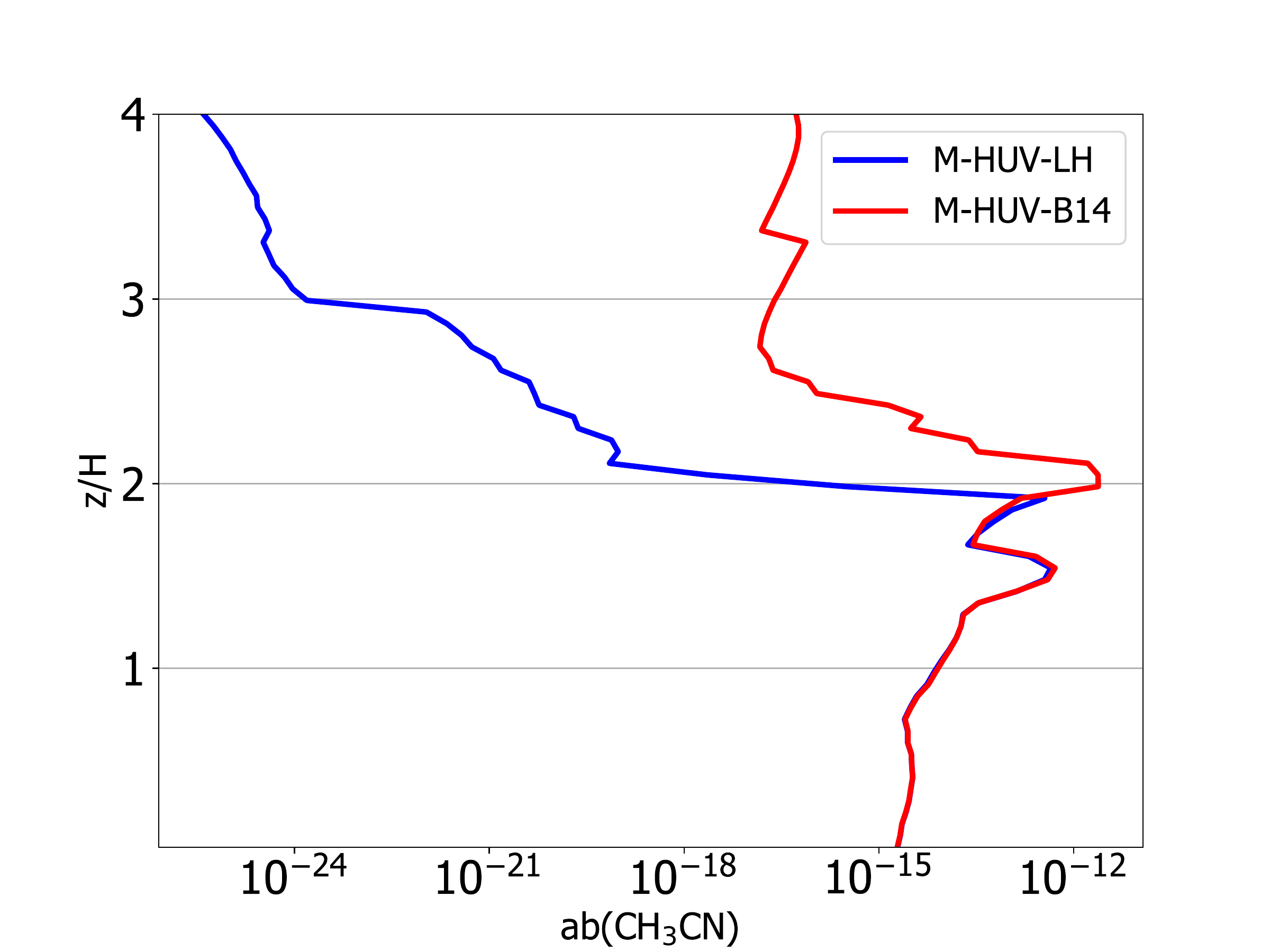} 
\caption{Altitude above the midplane as a function of the abundance relative to H of CH$_3$CN at 100 au. Blue line shows the abundance produced by model \mhlh, red line shows the abundance produced by model \mhb. We see that the abundance dramatically drops above $\sim$ 2 scale heights in \mhlh while \mhb keeps a rather high abundance at high altitudes. This is due to the available quantity of gas-phase H$_2$ in \mhb, which enhances the production of CH$_3^+$. 
\label{fig:methyl_compare}}
\end{figure}
 
  \subsubsection{Methyl Cyanide (CH$_3$CN)}
  
 
\revisionA{C-N bonded species are very important for the synthesis of 
Glycine \citep{Goldman+etal_2010}. \citet{Oberg+etal_2015} have 
observed methyl cyanide (CH$_3$CN) in the disk around the Herbig Ae star 
MWC 480, a star exhibiting a stronger UV flux and a warmer disk than a 
TTauri star. }
 
\revisionA{\paragraph{CH$_3$CN to HCN abundance ratio}
To facilitate comparisons,  we discuss the abundance of CH$_3$CN 
relative to HCN, as practiced in cometary studies 
\citep{Mumma+Charnley_2011, Oberg+etal_2015}. }The model that produces 
the largest CH$_3$CN/HCN ratio is the single-grain model \shtg with 
$\sim$ 14\% (vertically integrated) at 100, 120 and 140 au while the 
other single-grain models produce nearly more than $\sim$ 1\% at these 
radii. The outer regions (200-250 au) produce a larger ratio (up to 
9\%) than the inner disk regions in the warm grain single-models, 
whereas the peak is in the region around 100 au in \shtg. In 
multi-grain models, the gas-phase abundance ratio never exceeds 1\%. 
However, it never goes below 0.1\%, implying a rather uniform 
production of CH$_3$CN along the radius as compared to the single-grain 
models.  \mhb produces slightly more CH$_3$CN than \mhlh. \revisionA{ The 
model \shtg (with the coldest grains overall) is the one that globally 
produces the largest CH$_3$CN/HCN ratio.}
 
\revisionA{\paragraph{CH$_3$CN in gas-phase}
In the upper regions,} the main known pathways to form gas-phase 
CH$_3$CN are the reaction of HCN with CH$_3^+$ and the photodesorption 
of icy CH$_3$CN (thermal desorption is not effective because of the 
large binding energy of CH$_3$CN). Results show that the reaction 
between HCN and CH$_3^+$  is the most effective route in \mhb while 
only photodesorption is effective in \mhlh. The reason is 
straightforward since the production of CH$_3^+$ requires the presence 
of molecular hydrogen:
 
\begin{gather}
\label{eqn:ch3+}
\mathrm{C^+ \xrightarrow{H_2} CH_2^+ \xrightarrow{H_2} H + CH_3^+}
\end{gather}

\noindent and H$_2$ is much more abundant in \mhb in the upper layers. 
At 100 au, the abundance of CH$_3^+$ in \mhlh is very low ($\sim 
10^{-20}$) as compared to the abundance in \mhb ($\sim 10^{-10}$). It 
follows that, as seen in Fig.\,\ref{fig:methyl_compare}, \mhb produces 
much more CH$_3$CN than in \mhlh. Therefore, the presence of H$_2$ 
appears to be decisive for the production of gas-phase CH$_3$CN in the 
upper regions of the disk. However, this difference between the two 
models does not significantly impact the total column density of 
CH$_3$CN as most of it is produced near the midplane.
 
\revisionA{\paragraph{CH$_3$CN ices} Figure \ref{fig:surfdensmap}} 
(bottom) shows the abundance of s-CH$_3$CN ices in the midplane relative 
to the grain size as a function of the radius. s-CH$_3$CN mostly stays on 
warm small grains while there is a large drop in abundance around grain 
populations of size $\sim$ 0.317 $\mathrm{\mu m}$. The results show 
that the two main reactions that lead to s-CH$_3$CN are the 
adsorption of gas-phase CH$_3$CN onto the surfaces and the following 
surface reaction:

\begin{gather}
\label{eqn:ch3cn}
\mathrm{s\mhyphen CH_3 + s\mhyphen CN \rightarrow s\mhyphen CH_3CN}.
\end{gather}

\noindent \revisionA{s-CH$_3$CN is more abundant on small grains because of 
three main reasons.} The first reason is the large number density of 
small grains that implies large collision rates with 1) gas-phase 
produced CH$_3$CN and 2) with CH$_3$ that, despite a low binding 
energy, will briefly stick and quickly react with s-CN (see reaction 
\ref{eqn:ch3cn}). The second reason is the high binding energy of 
CH$_3$CN \citep[E$_\mathrm{b}$(CH$_3$CN) = 4680 K][]{Wakelam+etal_2017} 
and of the precursors (s-CN) so that they stay on small grain surfaces. 
Moreover, the temperature of small grains is sufficiently high to allow 
precursors to diffuse on the surface and increase reaction rates. The 
last reason is that even if s-CH$_3$CN ices can be formed on large grain 
surfaces, photodesorption rates are large enough for CH$_3$CN to be 
desorbed from large grains and seed small grains as the disk evolves.

\subsubsection{Methanol and formaldehyde} 

 
\begin{figure}
\includegraphics[width=\linewidth]{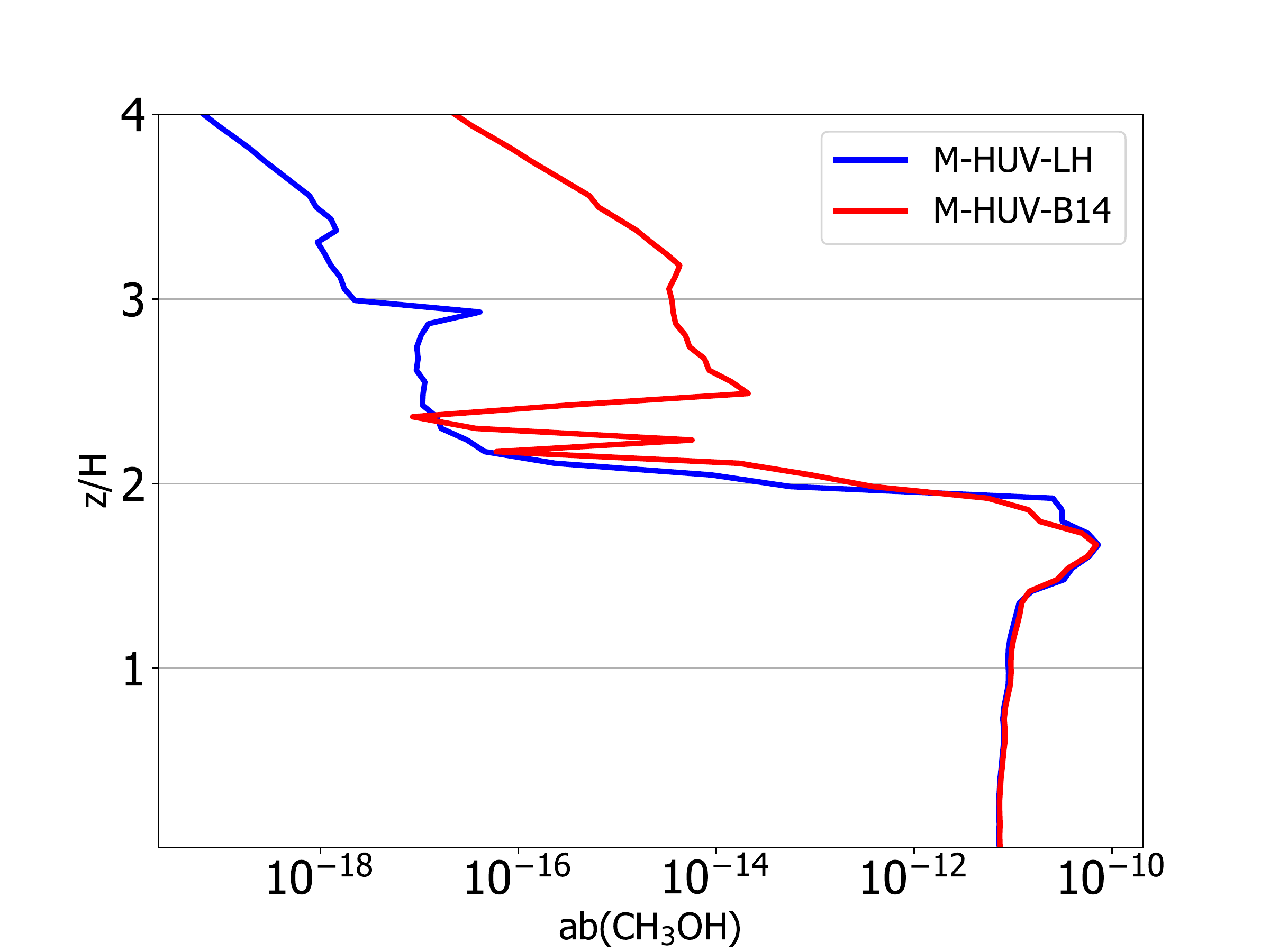} 
\caption{Altitude above the midplane as a function of the abundance relative to H of CH$_3$OH at 100 au. Blue line shows the abundance produced by model \mhlh, red line shows the abundance produced by model \mhb. We see that the abundance dramatically drops at $\sim$ 2 scale heights in both models, although \mhb maintains a larger abundance than \mhlh at high altitudes. 
\label{fig:CH3OH_compare}}
\end{figure}

\paragraph{CH$_3$OH in gas-phase} For multi-grain models, 
Fig.\,\ref{fig:CH3OH_compare} shows that the vertical profile of 
gas-phase CH$_3$OH is divided into two layers separated by an abrupt 
drop in abundance at $\sim 2\,H$, which corresponds roughly to the dust 
temperature transition (Fig. \ref{fig:weighted_100}). There is a peak 
in gas-phase CH$_3$OH at $\sim 1.8\,H$ in both models. This corresponds 
to the location where dust temperature is low enough for adsorbed CO to 
be hydrogenated and where the UV flux is high enough to efficiently 
photodissociate a significant fraction of newly formed methanol. The 
lower layer ($ < 2\,H$) contains most of CH$_3$OH column density. In the 
upper layer, \mhb produces more CH$_3$OH than \mhlh.  
   
\paragraph{CH$_3$OH to H$_2$CO abundance ratio} In multi-grain models, 
the ratio CH$_3$OH/H$_2$CO increases by a factor of $\sim$ 2 to 3 in 
the outermost region ($\geq$ 200 au) between $10^6$ and $5\times10^6$ 
years while it remains stable in the inner region (< 100 au). In \mhb, 
after $10^6$ years, the averaged ratio in the region 80 - 120 au is 
equal to 1.3. This appears to be consistent with the observed ratio in 
TW Hya \citep{Walsh+etal_2016, Carney+etal_2019}. In \mhlh, the 
averaged ratio is equal to 2.4. The difference is due to a more 
efficient pathway to form H$_2$CO in the upper disk regions of \mhb. 
Gas-phase H$_2$CO is efficiently formed through the following sequence:   
\begin{gather}
\label{eqn:h2co}
\mathrm{H_2 +CH_3^+ \rightarrow CH_5+} \\
\mathrm{CH_5^+ + e^- \rightarrow H + H + CH_3/H_2 + CH_3} \\ 
\mathrm{O + CH_3 \rightarrow H + H_2CO}.
\end{gather}
  
\noindent 
This shows that molecular hydrogen is a key element in the formation 
process of gas-phase formaldehyde and \mhb provides much more H$_2$ 
than \mhlh. 

\paragraph{Grain size dependence of surface chemistry for  CH$_3$OH}

\begin{figure*}
\begin{subfigure}{.49\linewidth}
  \centering
  \includegraphics[width=1.0\linewidth]{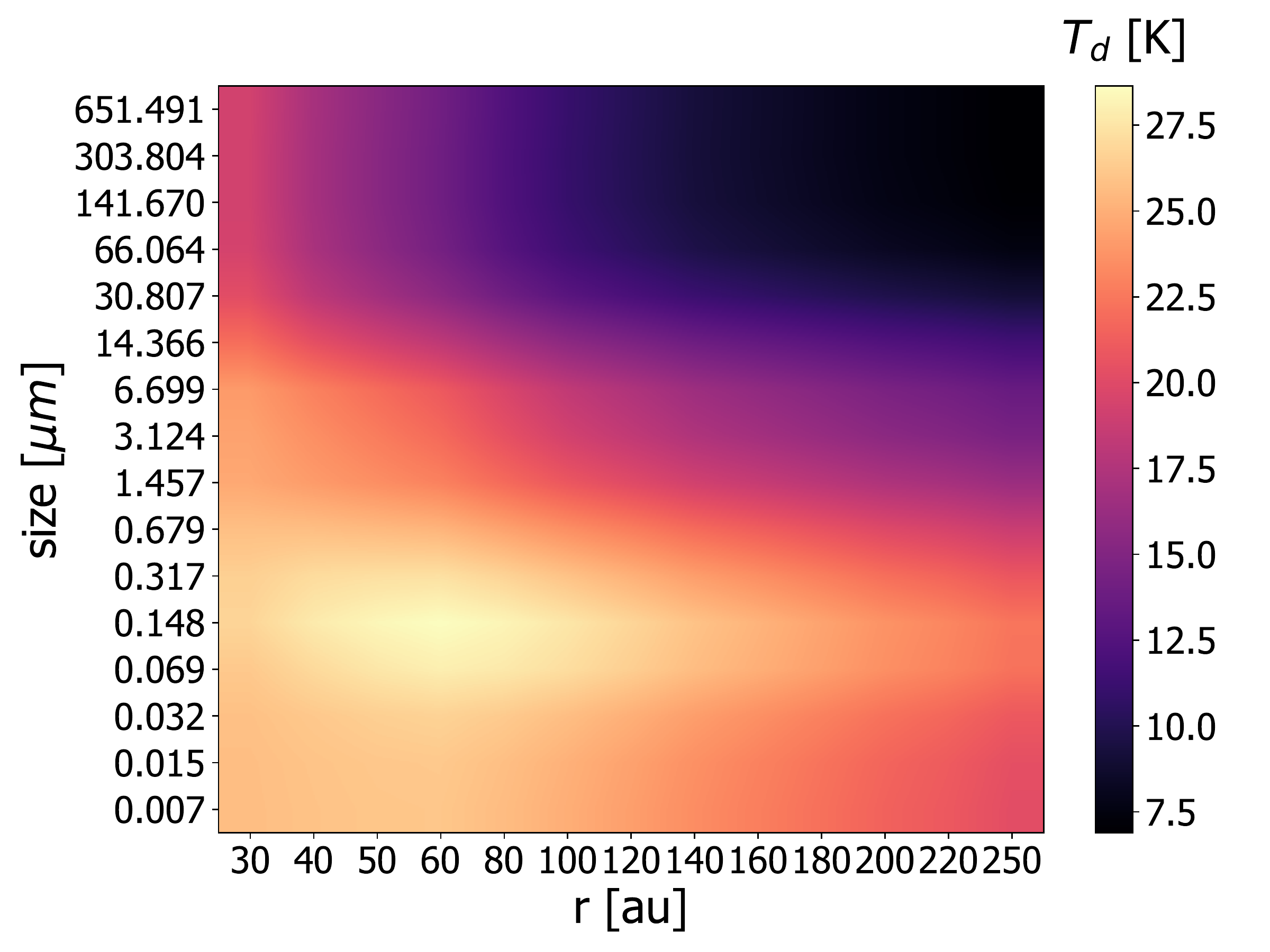}  
\end{subfigure}
\begin{subfigure}{.49\linewidth}
  \centering
  \includegraphics[width=1.0\linewidth]{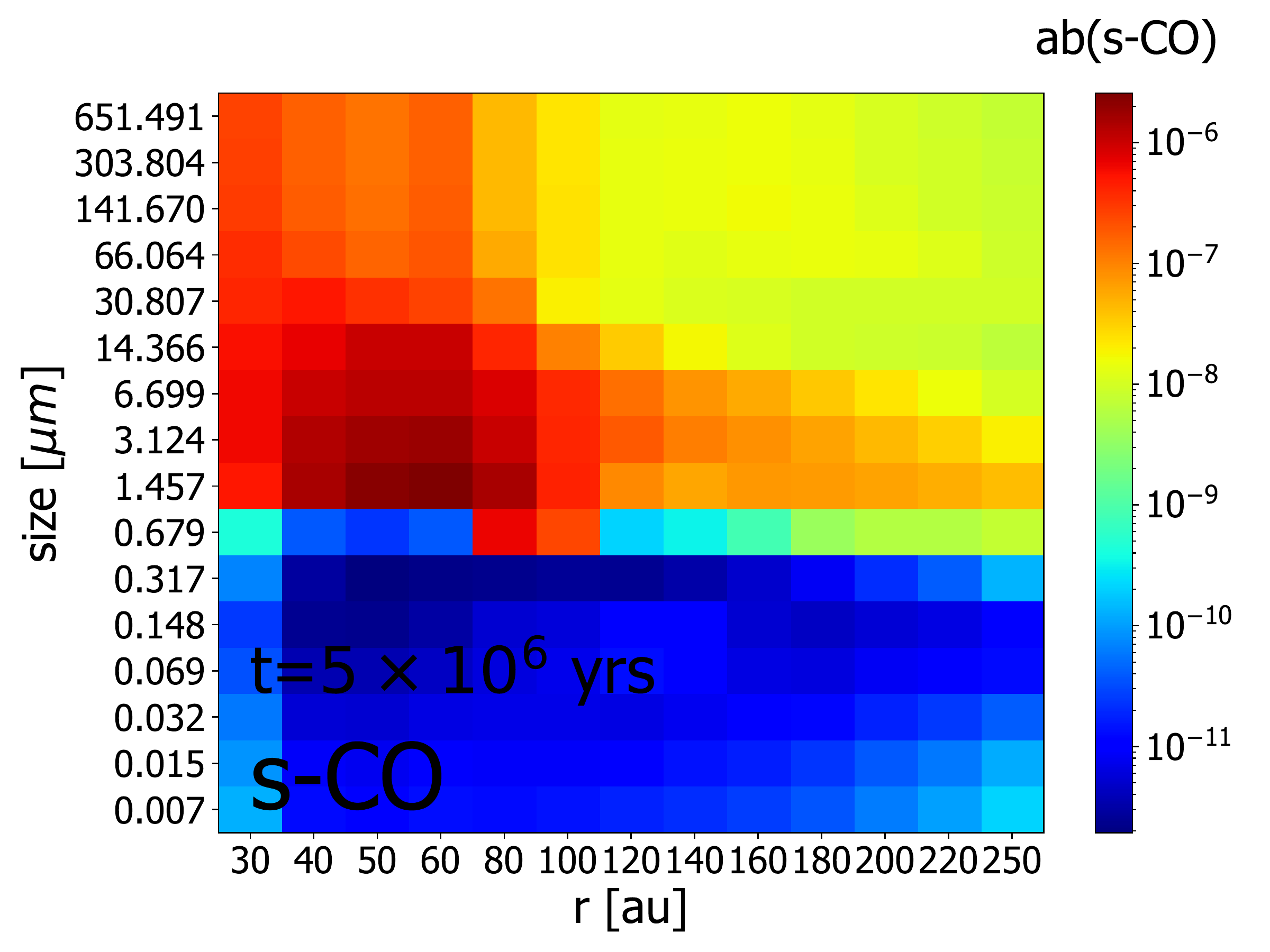}
\end{subfigure}
\begin{subfigure}{.49\linewidth}
  \centering
  \includegraphics[width=1.0\linewidth]{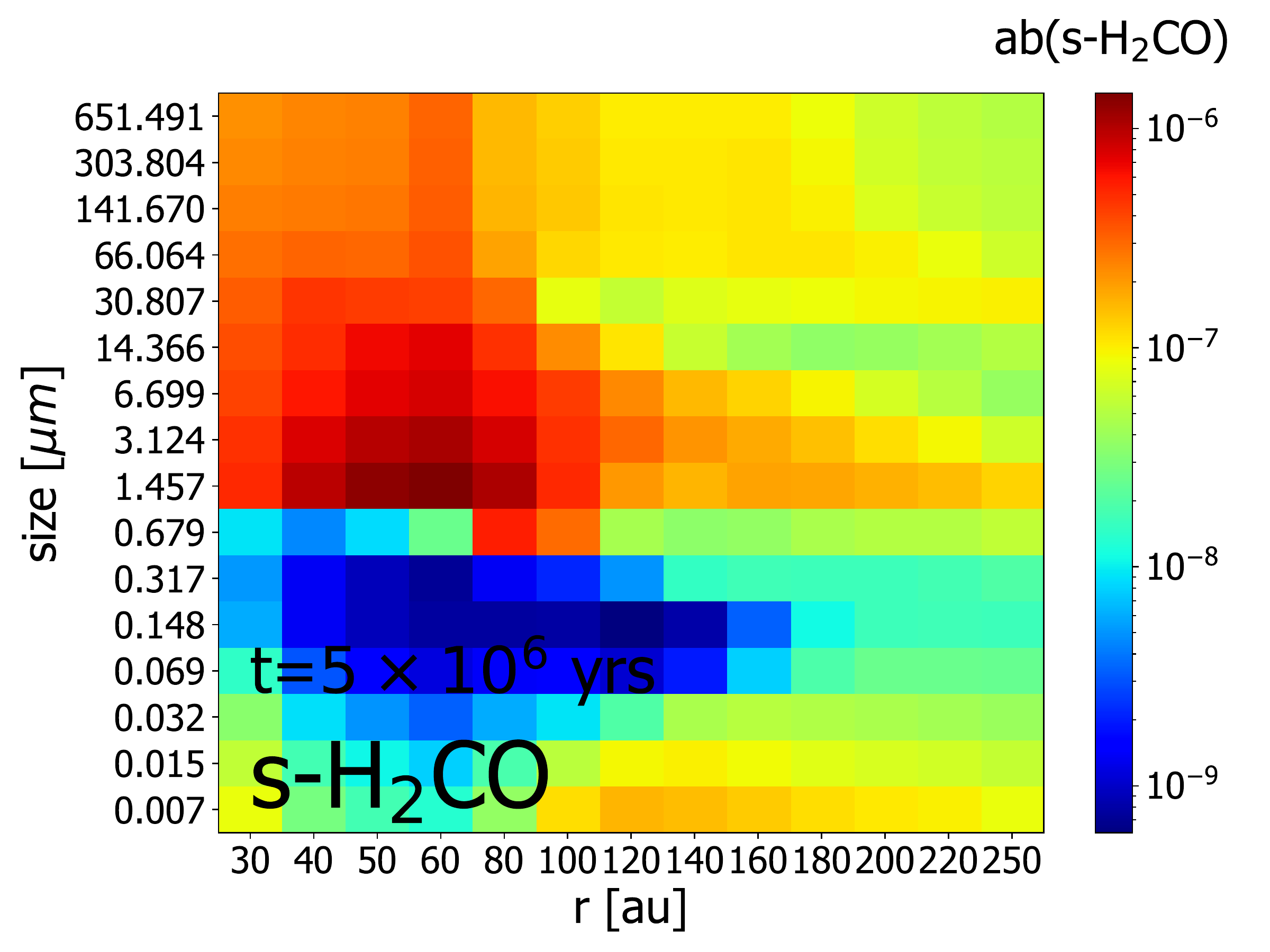}
\end{subfigure}
\begin{subfigure}{.49\linewidth}
  \centering
  \includegraphics[width=1.0\linewidth]{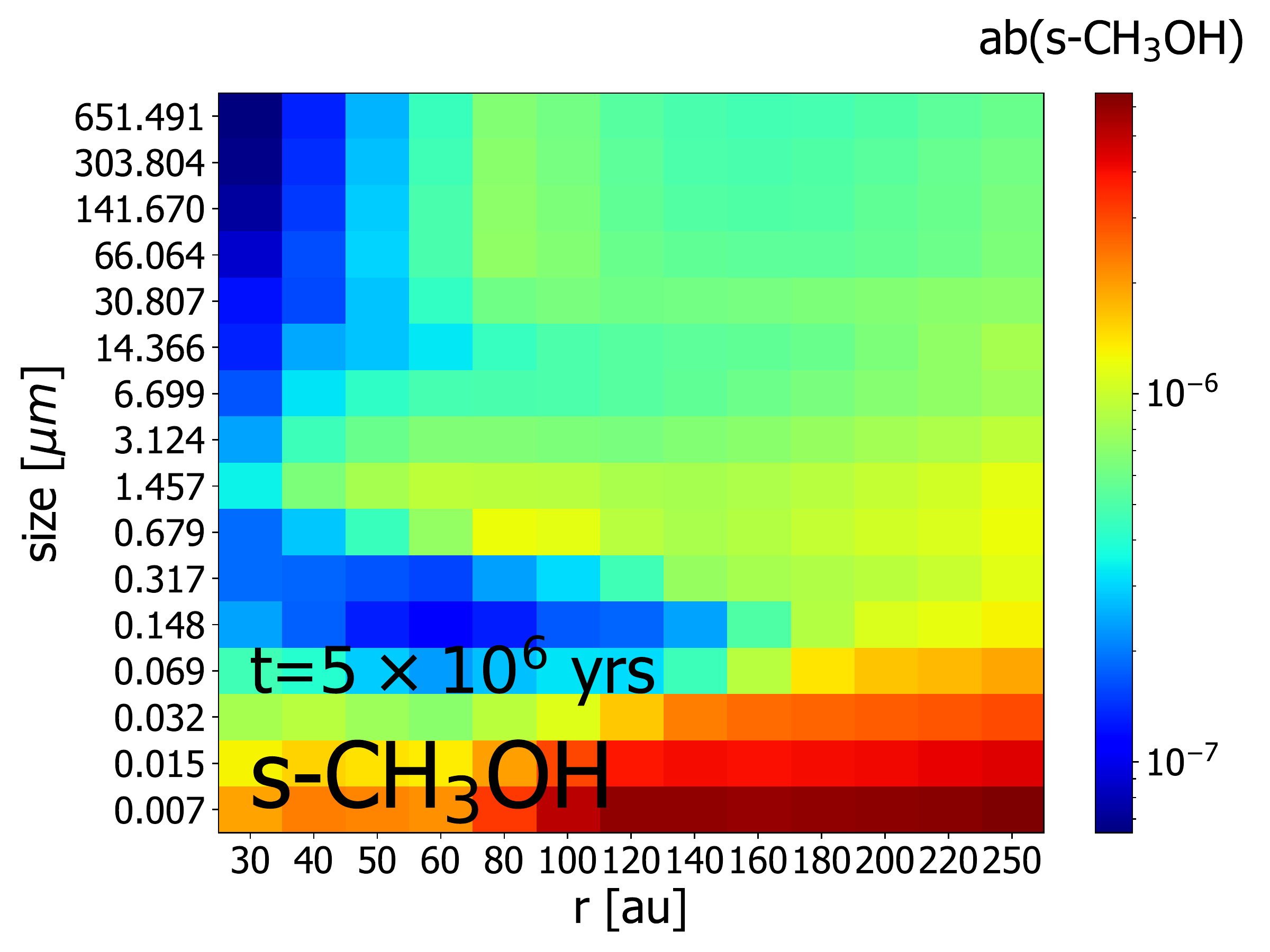}  
\end{subfigure}
\begin{subfigure}{.49\linewidth}
  \centering
  \includegraphics[width=1.0\linewidth]{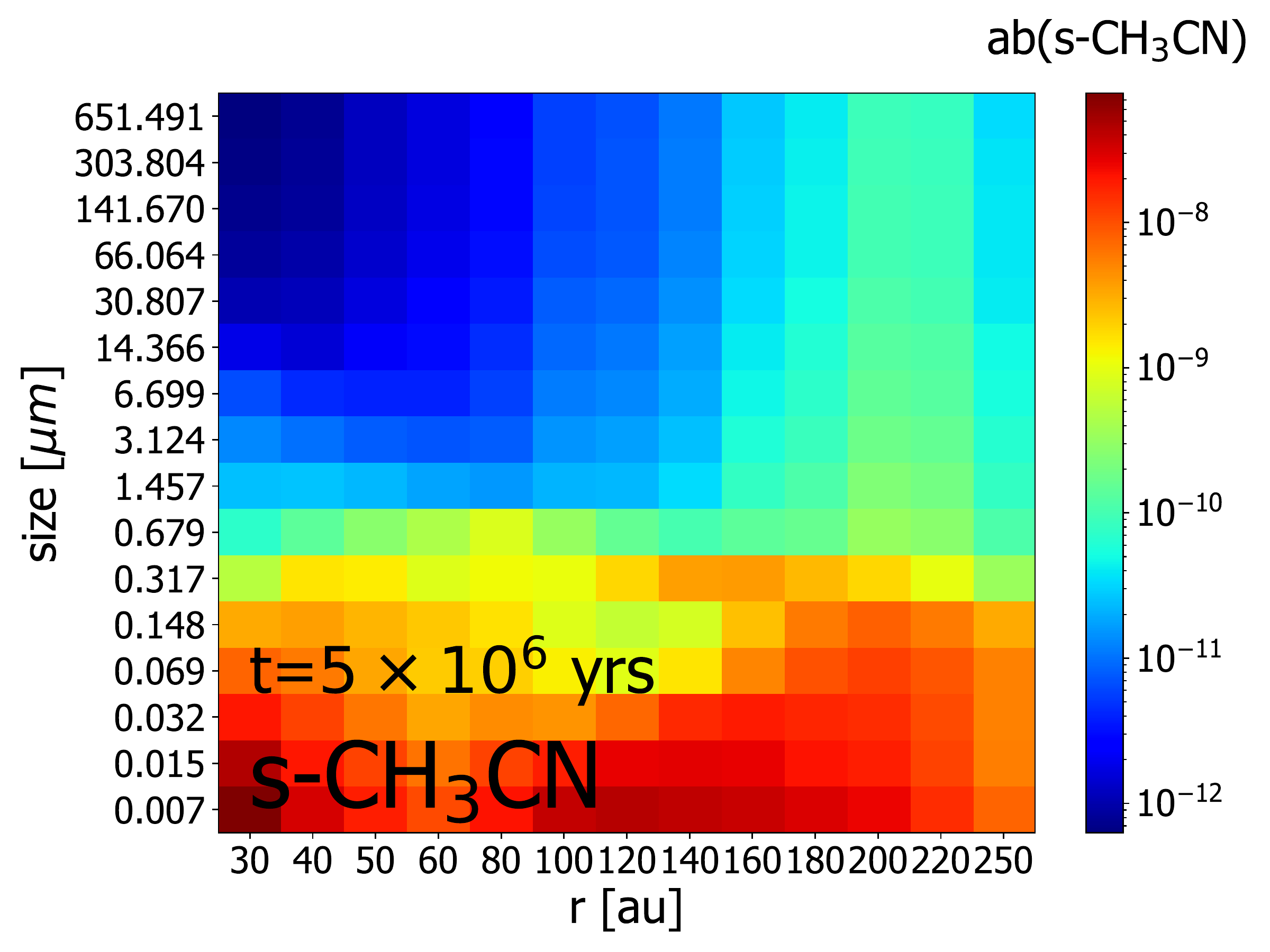}
\end{subfigure}

\caption{multi-grain models: Maps of ice abundance relative to H in the 
midplane, with grain sizes as a function of the radius. Top left panel 
is the distribution in grain temperature. We see that the hottest 
region is located at around 60 au from the star on grains of size 0.148 
$\mathrm{\mu m}$.}
\label{fig:surfdensmap}
\end{figure*}

\revisionA{Figure \ref{fig:surfdensmap} gives a set of maps of ice abundance 
in the midplane with grain sizes as a function of the computed radii. 
Top left panel is the grain temperature which allows to analyze the 
surface abundance in view of the surface temperature. We have same 
results for both multi-grain models. On icy mantles, both s-H$_2$CO and 
s-CH$_3$OH are formed via the same hydrogenation sequence:}
 
\revisionA{
\begin{gather}
\label{eqn:hydro}
\mathrm{s\mhyphen CO \xrightarrow{H} s\mhyphen HCO \xrightarrow{H} s\mhyphen H_2CO \xrightarrow{H} s\mhyphen CH_3O \xrightarrow{H} s\mhyphen CH_3OH}.
\end{gather}
}
\noindent However, we note that H$_2$CO is mainly formed in the 
gas-phase. As seen in Fig.\,\ref{fig:surfdensmap} (top right), CO tends 
to be efficiently adsorbed onto cold surfaces (T$_d$ $\lessapprox$ 17 K). The map of H$_2$CO 
(bottom left) grossly follows the same distribution although the trend 
is globally less pronounced. CH$_3$OH, on the other hand, exhibits a 
totally different distribution and tend to concentrate on small grain 
surfaces (bottom right). The main reason comes from the fact that the 
binding energy increases as the molecules become more complex 
\citep[e.g., $E_{\mathrm{b}}$(CO) $\sim$1200 K, 
$E_{\mathrm{b}}$(H$_2$CO) $\sim$3200 K, $E_{\mathrm{b}}$(CH$_3$OH) 
$\sim$5500 K][]{Minissale+etal_2016, Noble+etal_2012, 
Collings+etal_2004} and that the hydrogenation of s-H$_2$CO (which leads 
to s-CH$_3$OH) has an activation barrier and cannot efficiently occur on 
cold grain surfaces.
  
\subsection{Impact for planet formation}
 
Planet formation is strongly linked to grain growth and dust disk evolution. Our study reveals some interesting trends which should influence it. 
 
For a T\,Tauri disk of a few Myr, radial changes of the millimeter 
spectral index \citep[e.g.][]{Birnstiel+etal_2010} reveal that, after 
settling, large grains have drifted toward the central star. 
Typically, grains larger than a few  0.1- 1 mm are located inside a 
radius of 50-80 au. Beyond this, dust disks have a spectral index which 
is similar to that observed in the ISM, suggesting a similar size 
distribution, even if settling still occurs with larger grains residing 
onto the midplane. As a consequence, toward the midplane, one 
expects to observe different icy layers on grains in the inner (large 
cold grains) and outer disk (warmer smaller grains). We have seen in 
the previous section that  s-CH$_4$ and s-H$_2$O are preferentially located 
on large grains while s-CO$_2$ remains on smaller ones (Fig.\,\ref{fig:ab_surface}). 
This implies also a different chemical coupling 
between the gas and solid phase. Moreover, we have seen that complex 
molecules tend to stick and remain on small warmer grains ($\leq$ $\sim$ 
0.1 $\mathrm{\mu m}$), which are less likely to drift toward the inner 
disk. The outer disk (r $\geq$ 80 au) is therefore assumed to support 
rich chemistry on small grains which are probable precursors to 
cometary nuclei. On the other hand, as larger grains drift efficiently 
toward the inner region, their surface temperature is likely to vary 
during the journey which can allow for the growing chemical complexity 
to occur until the less refractory molecules become vaporized when 
sufficiently close to the star. Therefore, rich chemistry of the inner 
disk region can be assumed to come from the gas-phase which is later 
accreted by planetary embryos, while rich chemistry of the outer disk 
region occurs on the small grain surfaces and then seeds the inner 
region through planet-comet collisions in later phases of the disk. 

\revisionA{Grain-grain collisions may modify this picture.  The characteristic timescale (the time for a large grain to encounter
small grains of equivalent area) around 100 au is of order $10^4$--$10^5$ years. However, while large grains may accrete the smaller ones, and
inherit part of their surface composition while growing, small grains are re-created by disruptive events that may severely affect
their surface layer and ice mantles. The outcome of the surface state during such grain-grain collisions remains to be studied.} 

\revisionA{Nevertheless, in all cases, the grain size distribution, because it results in a spread of dust temperature, allows for a
faster depletion of C and O, that are converted to s-CO$_2$ on the colder grains in the disk midplane, but also a more 
active surface chemistry leading to formation of COMs on the warmer grains in the upper layer. Such a behavior cannot
be reproduced by a single grain temperature chemical model, because the formation processes are highly non-linear, 
since the mobility and evaporation efficiency depend exponentially on dust temperature. }

\revisionA{
The range in  dust temperatures also spreads out spatially the snowlines 
that become fuzzy. The spatial extent of these ``snow bands'' will depend
on the specific dust properties and disk mass. The temperature spread is
larger for less massive disks.} %

As disks evolve, the amount of dust (and gas) decreases until they 
reach the optically thin, debris disk phase. A less massive dust disk 
necessarily implies a warmer dust disk, which should affect the 
composition of the icy layers at the grain surface in evolved disks. 
Recent surveys of 5-40\,Myr old disks suggest the existence of an 
intermediate phase (at least for Herbig Ae stars) between debris and 
protoplanetary disks \citep[e.g.][]{Kospal+etal_2013, 
Pericaud+etal_2017}. The so-called hybrid disks share dust properties 
similar to those observed around debris disks, while the amount of gas 
is still significant, and a larger than standard gas-to-dust-ratio is 
observed. For instance, in HD141569, \citet{Difolco+etal_2020} observed 
a CO surface density which is only a factor 10 lower than in a typical 
protoplanetary HAe disk, whereas the amount of millimeter grains is a 
factor 50 times smaller than in typical young disks. The reduced amount 
of grains, and in some cases at least possible changes in the grain 
size distribution \citep[e.g.][]{hughes+etal_2018}, should affect the 
grain surface chemistry during the late phases of disk evolution. More 
simulations of older, gas-rich disks with a small amount of dust and a 
larger gas-to-dust ratio are needed to quantify the impact on chemistry 
and planet formation.


\section{Summary}. 
\label{sec:concl}
We presented here a new chemical model of protoplanetary disks that 
consistently accounts for the impact of grain sizes on the dust 
temperatures and UV penetration. Representative results were obtained 
for a typical disk around T Tauri stars, using parameters derived from 
the Flying Saucer observations \citep{Dutrey+etal_2017}. We coupled the chemical code NAUTILUS and 
radiative transfer code POLARIS to use a consistent distribution of grain sizes and temperatures, accounting for size-dependent dust settling. We implemented in NAUTILUS a new method to account 
for self and mutual shielding of molecules in photo processes, method 
that accounts for the frequency dependence of photorates.
\revisionA{
The outputs of  multi-grain models are compared to those of 
single grain size, using two different assumptions for the single-grain 
temperature: equal to the (assumed) gas temperature, or to the area-weighted 
mean temperature of the multi-grain models. }

\begin{itemize}
\item We identify the formation of H$_2$ through Langmuir-Hinshelwood 
mechanism as inefficient for a grain size distribution, and the amount 
of atomic H as critical for the chemistry of the upper layers (2-4\,$H$). 
The parametric method of \cite{Bron+etal_2014}, that accounts for 
temperature fluctuations for small grains, appears more appropriate to 
produce enough H$_2$.
\item \revisionA{Comparisons between the models reveal differences of more than 1-2 orders of magnitudes} 
for some abundant gas-phase species. Most of these differences arise
from above $1.5 H$, despite the larger UV penetration in multi-grain
models.
\item The location of the H$_2$O vertical 
snowline is partly regulated by the UV penetration but also by the 
amount of H$_2$. In the upper disk layers, the amount of H$_2$ also 
regulates the formation of H$_2$CO, although this molecule is mostly 
formed, like CH$_3$OH, on grain surfaces. 
\item CH$_3$CN, like CH$_3$OH to a lesser extent, is preferentially 
formed on warmer grains that allow for a larger diffusivity of 
precursors on the grain surface. 
\item At 100\,au, in the midplane, disks with colder dust produce more 
H$_2$O and CH$_4$ while those with warmer dust  produce more CO$_2$ on 
grain surfaces. 
\item As soon as CO is locked onto large (hence colder) grains 
($a >1\mu$m), it remains trapped until the end of the simulated evolution.
\item \revisionA{The spread of temperatures available with grain size distributions
simultaneously provides surface chemistry for cold and warm grains 
that cannot be mimicked by a unique dust temperature. In particular,
it allows depletion of C and O to proceed efficiently (by conversion
to CO$_2$ or CH$_4$ and H$_2$O on colder grains), while the COMs
production is boosted on the warmer grains. }
\item Using a single grain size that locally mimicks grain growth and dust settling fails to reproduce the complex gas-grain chemistry resulting from grain temperature depending on grain size.
\item \revisionA{Another direct consequence of having different dust temperatures 
at the same location is a spatial spread-out of the snowlines.} \revisionC{High angular observations
would be required to test this prediction.}

\end{itemize}
 
Our results clearly show that the chemistry of protoplanetary disks 
cannot simply be represented by a single grain size and temperature.

Finally, our disk model assumes a relatively high dust mass (i.e. an
optically thick dust disk). An optically thinner dust disk 
would impact the UV penetration and dust temperature,  
changing the chemistry at play even on the midplane. This remains to 
be investigated to determine how it can affect the grain surfaces and 
then the composition of planetesimals and embryos.

\begin{acknowledgements}
%
We thank E.Bron for fruitful discussions about the H$_2$ formation in PDRs and M.Ruaud for
a detailed reading of the article. We also thank an anonymous referee who helped to improve the paper.
This work was supported by the ``Programme National de Physique Chimie du Milieu Interstellaire'' (PCMI)
from INSU/CNRS. J.K. acknowledges support from the DFG grants WO 857/13-1, WO 857/15-1, and WO 857/17-1.
%
\end{acknowledgements}

\bibliography{ref}

\bibliographystyle{aa}

\begin{appendix} 
\appendix

\section{Formal description of the model}
\label{app:model}

Unless otherwise specified, we use subscript $g$ for gas quantities and subscript $d$ for dust.
Subscript $0$ is used for reference values.

\subsection{Gas disk structure} 
\label{app:gas_param}

We describe here the gas temperature and disk structure.

We neglect gas pressure and disk self-gravity, and thus assume that
the disk is in Keplerian rotation around the host star:
\begin{equation}
\label{eq:kepler}
	v_{g}(r) = \sqrt{\frac{GM_*}{r}}  \iff \Omega_{g}(r) = \sqrt{\frac{GM_*}{r^3}}
\end{equation}
where $r$ is the radius.

The gas surface density $\Sigma_g(r)$ also follows a power law:
\begin{equation}
\label{eq:surdensg}
	\Sigma_g(r) = \Sigma_{g,0} \bigg(\frac{r}{R_0}\bigg)^{-p}.
\end{equation}
and, by integration, the gas surface density $\Sigma_{g,0}$ at 
the reference radius $R_0$ for $p\neq 2$ is related to the gas disk mass by:
\begin{equation}
\Sigma_{g,0}  = \frac{M_g R_0^{-p} \left( 2-p\right)}{2 \pi \left( R_\mathrm{out}^{2-p} - R_\mathrm{in}^{2-p} \right)},
\end{equation}	
\noindent where $R_\mathrm{in}$ and $R_\mathrm{out}$ are the inner and outer radii of the disk.

The gas temperature is not self-consistently derived from the dust temperature distribution but imposed by 
analytical laws.
The kinetic temperature $T_k$ in the midplane is given by
\begin{equation}
\label{eq:Tmiddef}
       T_{k}(r,z=0) = T_{\mathrm{mid},0} \bigg(\frac{r}{R_0}\bigg)^{-q}
\end{equation}
Following \citet{Dartois+etal_2003}, we allow for a warmer disk atmosphere
using the formulation of \citet{Williams+Best_2014}

\begin{equation}
\label{eq:verticalT}
       T_\mathrm{g}(r, z) = T_\mathrm{mid}(r) + (T_\mathrm{atm}(r) - T_\mathrm{mid}(r)) \sin\bigg({\frac{\pi z}{2 z_\mathrm{atm}}}\bigg)^{2\sigma}
\end{equation}
where $\sigma$ is the siffness of the vertical temperature profile and 
$z_\mathrm{atm}$ is the altitude at the upper boundary of our disk 
model (4 scale heights $H$), meaning that the temperature above $4\,H$ is constant. The 
atmosphere temperature is also given by a power law with the same 
exponent as that of the midplane temperature:
\begin{equation}
\label{eq:Tatmdef}
       T_\mathrm{atm}(r) = T_{\mathrm{atm},0} \bigg(\frac{r}{R_0}\bigg)^{-q}
\end{equation}

\paragraph{Isothermal case:} If the gas temperature is vertically isothermal, we can derive the vertical density structure by assuming hydrostatic 
equilibrium at the midplane temperature. Consequently the gas vertical mass distribution follows a Gaussian profile :
\begin{equation}
\label{eq:rhog}
\rho_g(r,z) =  \rho_{g,mid}(r) \exp\left(- \frac{z^2}{2\,H_\mathrm{g}\!\left(r\right)^2}\right)
\end{equation}
where $\mathrm{\rho_{g,mid}(r)}$ is the density in the midplane, that
is related to the surface density by 
\begin{equation}
\label{eq:int}
	\rho_{g,mid}(r) = \frac{\Sigma_g(r)}{\sqrt{2\pi}H_g(r)}
\end{equation}
The scale height $H_g(r)$ is given by 
\begin{equation}
\label{eq:Hg2}
	H_{g}(r) = \frac{C_s(r)}{\Omega_{g}(r)}
\end{equation} 
and using the definition of the temperature law from Eq.\ref{eq:Tmiddef} 
\begin{equation}
\label{eq:Href}
	H_{g,0} = \frac{C_s\left(R_0\right)}{\Omega \left(R_0\right)} = \sqrt{\frac{k_\mathrm{b} T_\mathrm{mid,0} R_0^3}
  {\mu_\mathrm{m}\,m_H\,G M_*} }       
\end{equation}
where $\mu_\mathrm{m}$ is the mean molecular weight, and $m_\mathrm{H}$ the 
Hydrogen mass. The temperature and velocity being power laws, the scale
height $H(r)$ also follows a power law:
\begin{equation}
\label{eq:H}
       H_g(r) = H_{g,0}\bigg(\frac{r}{R_0}\bigg)^{\frac{3}{2} - \frac{q}{2}}
\end{equation}

\paragraph{Vertical temperature gradient:}
With a vertical temperature gradient, the vertical gas profile deviates from a Gaussian profile. 
In Nautilus, we compute the gas density for each vertical cell $z_i$ at a given $r$ by solving out
locally the equation as in \citet{Hersant+etal_2009}:
\begin{equation}
\label{eq:not-iso}
ln(\rho_g(z_i)= ln(\rho_g(z_i-1)) - (\Omega^2\frac{\mu m_H}{k_BT_g(z_i)}+(ln(T_g(z_i))-ln(T_g(z_{i-i})))
\end{equation} 


\subsection{Dust distribution and properties} 
\label{app:dust}

\begin{figure}
\includegraphics[width=\linewidth]{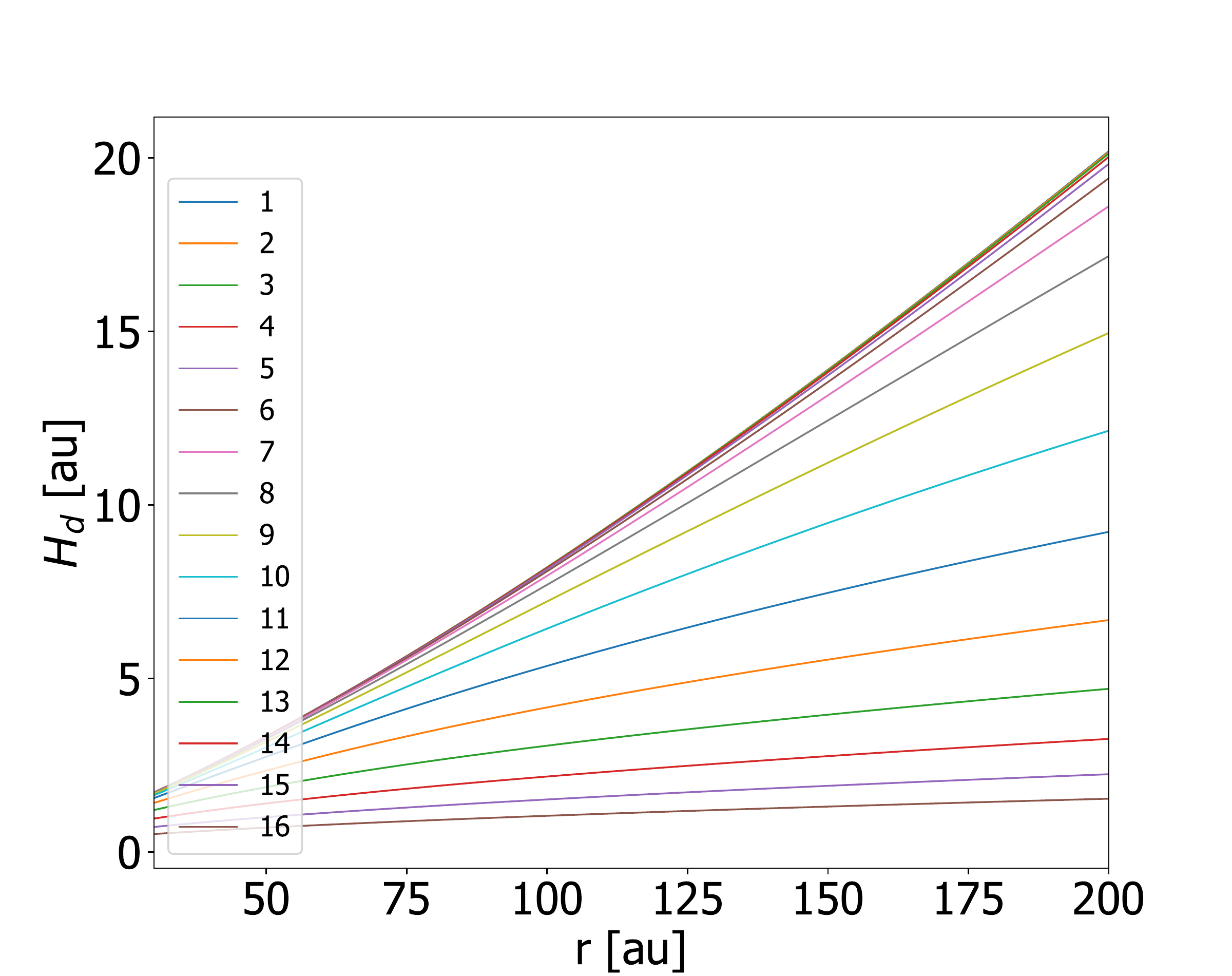}
\caption{Scale heights of the grain distribution for the 16 different grain sizes as a function of 
the radius. The grain averaged size interval ranges from 0.007 to 651 $\mathrm{\mu m}$. Small grains scale heights (size < 0.317 $\mathrm{\mu m}$)
practically follow the gas scale height. For larger grains the settling factor 
(Eq.\,\ref{eq:settling-factor}) becomes significant and the dust scale height 
decreases.
\label{fig:hd}}
\end{figure}

\subsubsection{Dust material} 
We assume that the grains are made of an homogeneous mixture of a 
population of silicate grains (62.5\%) and graphite grains (37.5\%).
This composition is used both for radiative transfer and chemical modeling. This leads to a material specific mass 
density for dust grains in all this study equal to $\mathrm{\rho_m = 2.5}$ $\mathrm{g.cm^{-3}}$.
The grain temperatures are defined for each grain sizes using radiative 
transfer simulations as described in Section \ref{sec:RT_sim}.

\subsubsection{Size distribution} 
\label{app:distrib}
We consider a size dependent grain population. Overall, this grain population is derived
assuming no radial drift, but accounting for dust settling as a function of height.
The surface density of a grain population of size $a$ is thus given by
\begin{equation}
\label{eq:sigma_d}
	\sigma_d(r, a) = \sigma_{d,0}(a)\bigg(\frac{r}{R_0}\bigg)^{-p}.
\end{equation}
where $\sigma_{d,0}(a)$ is the surface density of grains of size $\mathrm{a}$ at reference radius. Eq. \ref{eq:sigma_d} implies that the total dust surface density is:
\begin{equation}
\label{eq:Sigma_d}
	\Sigma_{d}(r) = \int_a \sigma_{d}(r, a)da,
\end{equation}
which we associate to the overall (vertically integrated, equivalently disk averaged) 
dust-to-gas mass ratio $\zeta$ through 
$\Sigma_d(r) = \zeta \Sigma_g(r)$. The mass fraction of grains of size $[a,a+da]$ is given by
\begin{equation}
\label{eq:fraction}
	x_a da = \frac{m(a)dn(a)}{\int_{a_{min}}^{a_{max}}m(a)dn(a)} = \frac{m(a)dn(a)}{m_{dust}}
\end{equation}
where the size distribution in number $dn(a)$ is assumed to follow a power law
of exponent $d$:
\begin{equation}
\label{eq:da}
dn(a) = C a^{-d} da 
\end{equation}
where $C$ is a normalization constant. $C$ is related to the
dust mass by simply integrating over the size as follows: 
\begin{equation}
\label{eq:totalmassdust}
	m_{dust} = \int_{a_\mathrm{min}}^{a_\mathrm{max}}m(a)dn(a) =  \int_{a_\mathrm{min}}^{a_\mathrm{max}} \frac{4\pi}{3}\rho_{m} C a^{3-d}da
\end{equation}
where $\mathrm{\rho_m}$ is the specific mass density defined in the previous section. Using Eq.\ref{eq:fraction} and the definition
of $\mathrm{\zeta}$ yields
\begin{equation}
\label{eq:sigma_frac}
	\sigma_d(r, a)  = x_a \zeta \Sigma_g(r) = \frac{\zeta \Sigma_g(r) a^{3-d}}{\int_{a_\mathrm{min}}^{a_\mathrm{max}} a^{3-d}da}
\end{equation}

For radiative transfer and chemistry simulations the grain size 
distribution is discretized into several logarithmically distributed 
subranges with relative dust masses of the i$^\mathrm{th}$ grain size interval:
\begin{equation}\label{eq:interval}
\frac{m_{d,i}}{m_d} = \frac{\int_{a_{\mathrm{min},i}}^{a_{\mathrm{max},i}} \frac{4}{3} \pi \rho_\mathrm{grain} a^3 n(a) \mathrm{d}a}{\int_{a_{\mathrm{min}}}^{a_{\mathrm{max}}} \frac{4}{3} \pi \rho_\mathrm{grain} a^3 n(a) \mathrm{d}a} = \frac{a_{\mathrm{max},i}^{4-d}-a_{\mathrm{min},i}^{4-d}}{a_{\mathrm{max}}^{4-d}-a_{\mathrm{min}}^{4-d}}.
\end{equation}
\revisionA{The relative dust mass per bin and the relative area per bin 
(which also quantifies the number of chemical sites) are given in Table \ref{tab:grains}. }

\begin{table}
\caption{Overview of the grain size intervals \label{tab:grains}}
\centering
\begin{tabular}{p{0.07\linewidth} c p{0.2\linewidth} c p{0.2\linewidth}c p{0.2\linewidth}c}
\hline
\noalign{\smallskip}
\revisionA{bin} & \revisionA{averaged size} ($\mathrm{\mu m}$) & \revisionA{relative dust mass} & \revisionA{relative area}\\
\noalign{\smallskip}
\hline	
\noalign{\smallskip}
1 &	 0.007 	& 0.10 \%& 30.5 \%	\\
2 &	 0.015  	& 0.15 \%& 20.8 \%	\\
3 &	 0.032 	& 0.22 \%& 14.2 \%	\\
4 &	 0.069 	& 0.33 \%& 9.73 \%	\\
5 &	 0.15		& 0.48 \%& 6.67 \%	\\
6 &	 0.32 	& 0.70 \%& 4.59 \%	\\
7 &	 0.68 	& 1.03 \%& 3.19 \%	\\
8 &	 1.46 	& 1.50 \%& 2.25 \%	\\
9 &	 3.13 	& 2.20 \%& 1.64 \%	\\
10 &	 6.69 	& 3.22 \%& 1.25 \%	\\
11 &	 14.4 	& 4.72 \%& 1.03 \%	\\
12 &	 31.0 	& 6.91 \%& 0.90 \%	\\
13 &	 66.0 	& 10.1 \%& 0.84 \%	\\
14 &	 142 		& 14.8 \%& 0.80 \%	\\
15 &	 304 		& 21.7 \%& 0.78 \%	\\
16 &	 651 		& 31.8 \%& 0.76 \%	\\
\end{tabular}
\end{table}

\subsubsection{Dust settling}
\label{app:settle} 
\begin{figure}
\includegraphics[width=\linewidth]{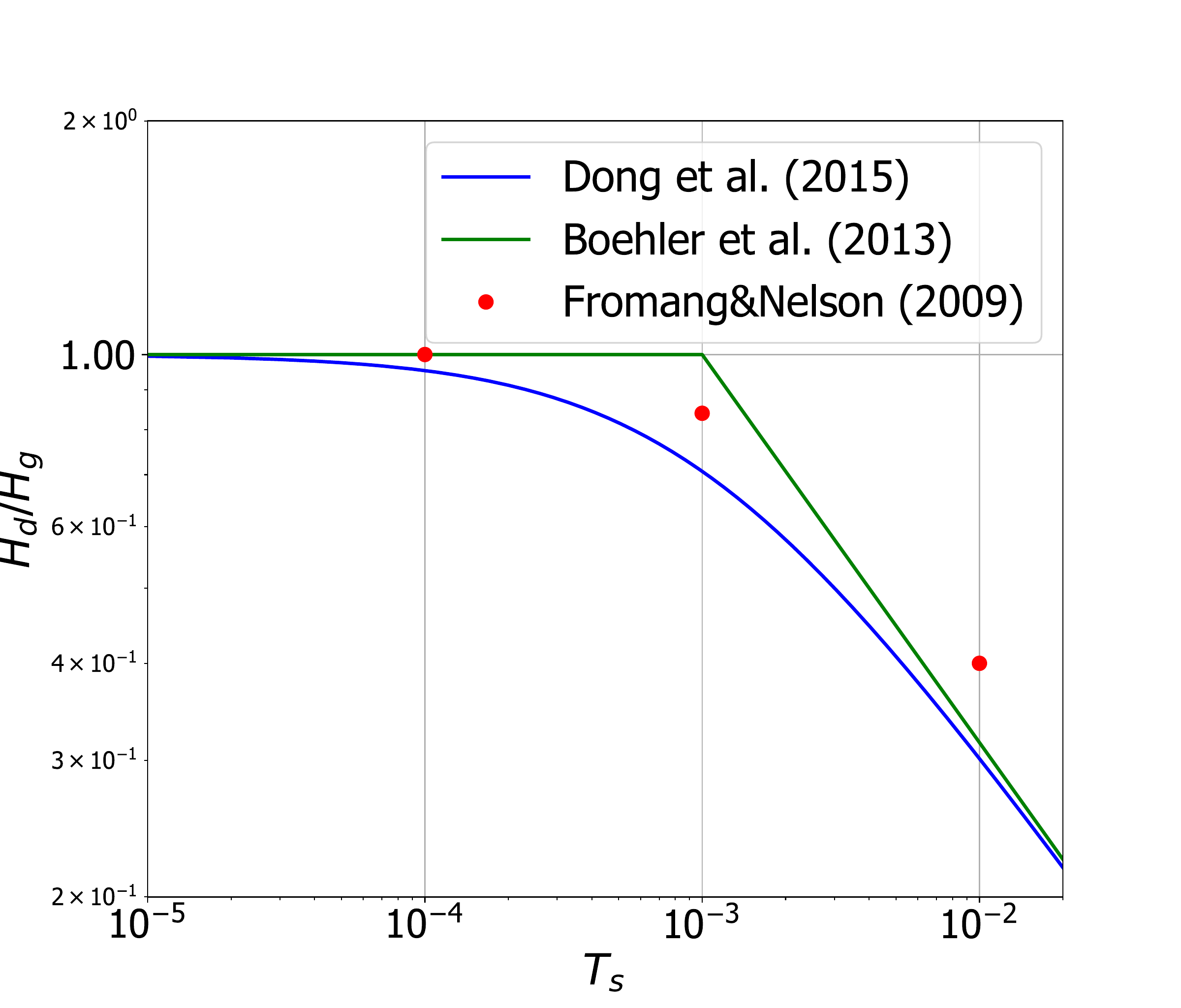} 
\caption{Settling factor as a function of the dimensionless stopping 
time at $R_0$. In green the model given by \citet{Boehler+etal_2013}, 
the red dots show the simulation values of \citet{Fromang+Nelson_2009} 
and the blue curve is the model of \citet{Dong+etal_2015} that we use.
\label{fig:settling_factor}}
\end{figure}

To account for size-dependent dust settling, we assume that the vertical profile
for a given grain size has \revisionA{a Gaussian shape:}
\begin{equation}
\label{eq:rhod}
\rho_d(r,z, a) =  \frac{\sigma_d(r, a)}{\sqrt{2\pi}H_d(r, a)} \exp\left(- \frac{z^2}{2\,H_{d}(r, a)^2}\right).
\end{equation}
Because dust grains do not react to pressure forces in the same
way as gas, the dust scale height will differ from that of the gas. 
Following \cite{Boehler+etal_2013}, we define the  
settling factor, that is the ratio between $H_d(r,a)$ and \revisionA{the gas scale height
near the midplane}  $H_g(r)$
\begin{equation}
\label{eq:settling-factor} 
	s(a,r) = \frac{H_{d}(a,r)}{H_g(r)} 
\end{equation}
that can be related to the dust properties through the stopping time.
The stopping time of a dust particle $\tau_s$ 
is the characteristic time for a dust particle initially at rest to 
reach the local gas velocity. We use the dimensionless 
stopping time $T_s(r, z)$ defined as the product of 
$\tau_s$ and the Keplerian angular momentum 
$\Omega(r)$ which is a way to compare the stopping time to 
the dynamical time in the disk. \citet{Garaud+etal_2004} showed that 
dust friction with gas in protoplanetary disks is well described 
by the Epstein regime, in which the dimensionless stopping time 
is given by:
\begin{equation}
\label{eq:stoppingtime}
	T_s(r, z) = \tau_s\Omega_g(r) = \frac{a \rho_{m}}{\rho_{g}(r, z)C_s}\Omega(r) 
\end{equation}
The stopping time depends linearly on the grain size.
Using Eq.\,\ref{eq:Hg2} we can write:
\begin{equation}
\label{eq:stoppingtime2}
	T_s(r, z) = \frac{a \rho_{m}}{\rho_{g}(r, z)H_g(r)} =  \frac{\sqrt{2\pi} a \rho_m}{\Sigma_g(r)} \exp\left(- \frac{z^2}{2 H_g(r)^2}\right).
\end{equation}
The settling factor is given by the approximation of \citet{Dong+etal_2015}:
\begin{equation}
\label{eq:settling-dong} 
	s(a,r) = \frac{H_{d}(a,r)}{H_g(r)} = \frac{1}{\sqrt{1 + T_{s,mid}\frac{S_c}{\alpha}}} 
\end{equation}
where $\alpha$ is the $\alpha$ viscosity coefficient,
\begin{equation}
\label{eq:stoppingtime3} 
T_{s,mid} = \frac{\sqrt{2\pi} a \rho_m}{\Sigma_g(r)}
\end{equation}
is the dimensionless stopping time in the midplane, and ${S_c}$ is 
the Schmidt number. The Schmidt number is a dimensionless number 
defined by the ratio between the turbulent viscosity $\nu_t$ and 
the turbulent diffusion $D_t$:
\begin{equation}
\label{eq:Sc} 
	S_c = \frac{\nu_t}{D_t}. 
\end{equation}
\autoref{fig:settling_factor} shows the settling factor at the reference 
radius as defined in \citet{Dong+etal_2015} compared to the asymptotic 
behavior used by \citet{Boehler+etal_2013} and the values  from the 
numerical simulations of  \citet{Fromang+Nelson_2009}.

\revisionC{Table\,\ref{tab:grains-cs} provides the total dust cross-section at 50, 100, and 200 au from the star, defined by the term $n_d(r, z=0, a) \pi a^2$.}

\begin{table}
\caption{\revisionC{Total dust cross-sections for each size interval in the midplane for three radii of the multi-grain model.} \label{tab:grains-cs}}
\centering
\begin{tabular}{l c c c}
\hline
\multicolumn{4}{r}{total cross-section (cm$^{-1}$)} \\
\cline{2-4} \bigstrut
size ($\mathrm{\mu m}$)   & 50 au  & 100 au & 200 au \\
\hline \bigstrut
 0.007 	& 3.39(-14) & 4.86(-15) & 6.99(-16)	\\
 0.015  & 2.31(-14) & 3.32(-15) & 4.78(-16)	\\
 0.032 	& 1.58(-14) & 2.27(-15) & 3.27(-16)	\\
 0.069 	& 1.08(-14) & 1.55(-15) & 2.24(-16)	\\
 0.15   & 7.38(-15) & 1.06(-15) & 1.55(-16)	\\
 0.32 	& 5.05(-15) & 7.32(-16) & 1.08(-16)	\\
 0.68 	& 3.47(-15) & 5.08(-16) & 7.69(-17)	\\
 1.46 	& 2.40(-15) & 3.59(-16) & 5.69(-17)	\\
 3.13 	& 1.68(-15) & 2.61(-16) & 4.46(-17)	\\
 6.69 	& 1.21(-15) & 2.00(-16) & 3.75(-17)	\\
 14.4 	& 9.06(-16) & 1.64(-16) & 3.37(-17)	\\
 31.0 	& 7.24(-16) & 1.44(-16) & 3.17(-17)	\\
 66.0 	& 6.21(-16) & 1.34(-16) & 3.06(-17)	\\
 142 	& 5.65(-16) & 1.28(-16) & 2.98(-17)	\\
 304 	& 5.37(-16) & 1.25(-16) & 2.90(-17)	\\
 651 	& 5.21(-16) & 1.22(-16) & 2.76(-17)	\\
\hline
\end{tabular}
\end{table}

\section{Self and Mutual Shielding}

\begin{table}
\caption{Adopted elemental initial abundances relative to H. \label{tab:init_ab}}
\centering
\begin{tabular}{c c c}
\hline
\noalign{\smallskip}
\textbf{Element} & \textbf{abundance (relative to H)} & \textbf{mass (amu)} \\
\noalign{\smallskip}
\hline
\noalign{\smallskip}	
He &	 9.0(-2)  & 4.00	\\
C &	 1.7(-4) & 12.00	\\
N &	 6.2(-5) & 14.00\\
O &	 2.4(-4) & 16.00 \\
Si &	 8.0(-9) & 28.00	\\
S &	 8.0(-8) & 32.00	\\
Fe &	 3.0(-9) & 56.00	\\
Na &	 2.0(-9) & 23.00	\\
Mg &	  7.0(-9) & 24.00	\\
Cl &	 1.0(-9) & 35.00	\\
P &	 2.0(-10) & 31.00	\\
F &	 6.7(-9) & 19.00	\\

\end{tabular}
\end{table}

\begin{figure}
    \includegraphics[width=\linewidth]{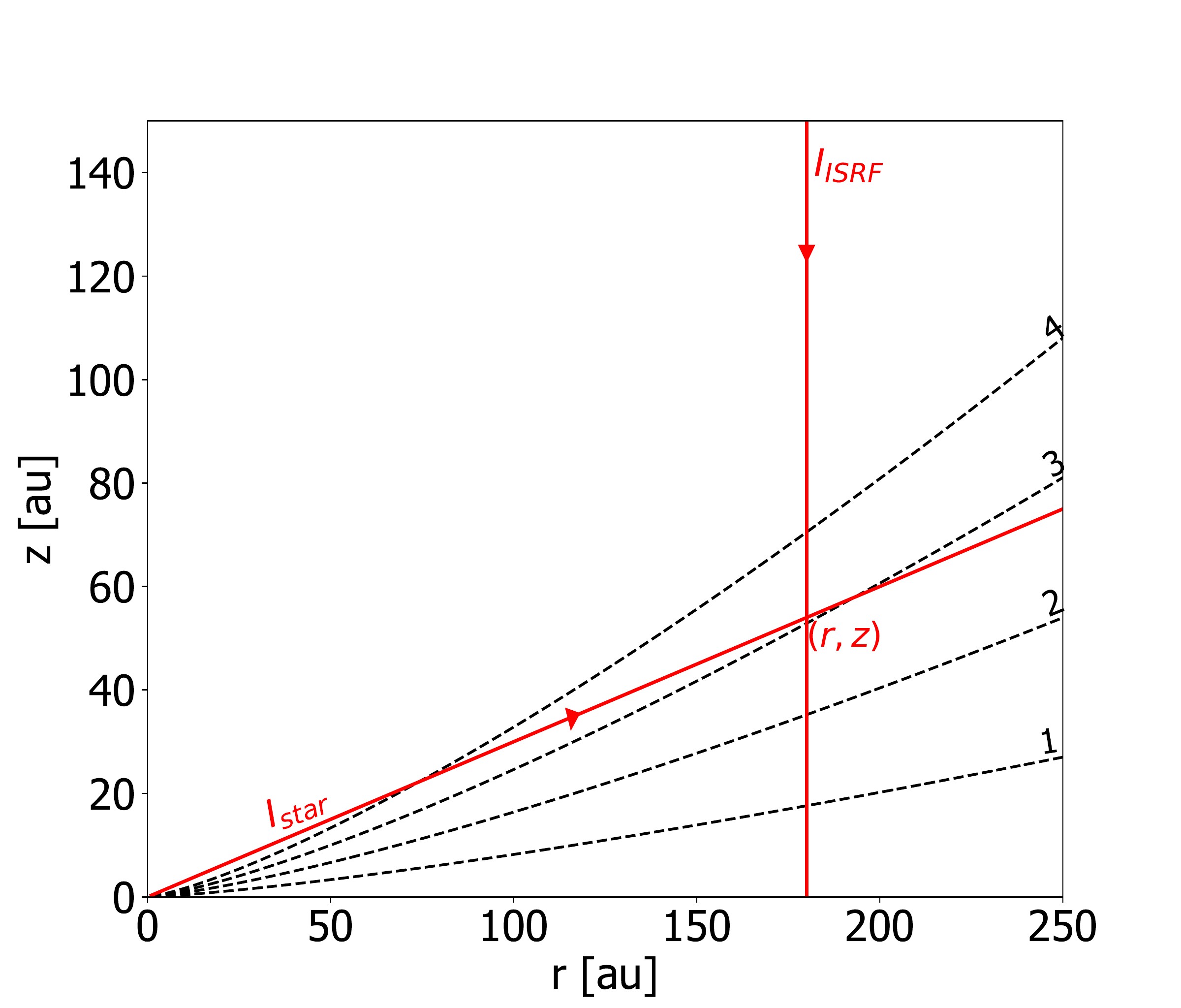} 
\caption{\label{fig:field} Elevation above the midplane as a function of the radius. The dotted black lines are scale heights of the gas (1 $\mathrm{H}$, 2 $\mathrm{H}$, 3 $\mathrm{H}$ and 4 $\mathrm{H}$). The red lines are radiation emitted from the central star and from the interstellar radiation field (ISRF). The coordinates $(r,z)$ represent a point at where both fields cross.}
\end{figure}

\label{app:self}

\revisionA{The local UV flux in the disk is not a simple sum of the attenuated stellar and interstellar
fields because of scattering by dust grains. Scattering can be significant near the disk midplane, 
where the optical depths are large}.
However, molecular shielding is most important only in the upper layers, when dust optical depth
remains low while molecular optical depth can be significant. 

Thus, at first order,
to evaluate the impact of self-shielding, we neglect scattering in our analysis by 
writing in a simple form the local flux $I$ in the cell of coordinates $(r, z)$ as follows:
\begin{equation}
\label{eq:local1}
	I(\lambda, r, z) \approx I_{ISRF}e^{-\tau^V} + I_*(r, z)e^{-\tau^R}
\end{equation}
where $I_{ISRF}$ is the incident flux from the ISRF, $I_*(r, z)$ is the spatially diluted, unattenuated,
incident flux from the central star, $\tau^V$ is the vertical opacity 
generated by the disk matter above the local point $(r, z)$ and 
$\tau^R$ is the radial opacity generated by the matter between the 
central star and the local point.  As the opacities are a contribution 
of both dust and gas, we write:
\begin{align}
\label{eqn:tau1}
\begin{split}
 \tau^V =   \tau^V_m + \tau^V_{d}
\\
 \tau^R=   \tau^R_m + \tau^R_{d}
\end{split}
\end{align}
where $m$ stands for molecules and $d$ for dust. So the local expression becomes:
\begin{equation}
\label{eq:local2}
	I(\lambda, r, z) \approx   I_{ISRF}e^{-(\tau^V_d + \tau^V_m)} + I_{*}e^{-(\tau^R_d + \tau^R_m)}
\end{equation}
\revisionA{The local flux weighted by the dust only $I_d$ can also be approximated by:}  
\begin{equation}
\label{eq:local_d}
	I_d(\lambda, r, z) \approx   I_{ISRF}e^{-\tau^V_d} + I_{*}e^{-\tau^R_d}
\end{equation}

\revisionA{POLARIS  treats consistently scattering by dust and this simple 
expression was checked to be a reasonable approximation in the upper layers where molecular shielding is important.}

At a given wavelength, the molecular opacity is defined as follows:
\begin{equation}
\label{eq:tau_m}
	\tau_m(\lambda, r, z) =  \sum_X \int_{l=f(r,z)} n(X, r, z) \sigma(X, \lambda) dl
\end{equation}
where $X$ includes all molecules for which we have cross-sections (including H$_2$O),
$n(X, r, 
z)$ is the number density [$\mathrm{cm^{-3}}$] of species $X$ at 
coordinates $(r, z)$ and $\sigma(X, \lambda)$ is the sum of the 
absorption, dissociation and ionization cross-sections of species $X$ 
at  wavelength $\lambda$. Depending on whether the radiation is from 
the interstellar medium or from the central star, the line $l$ over 
which we integrate is either the altitude $z$ above the cell of 
coordinates $(r,z)$ or the radial distance $\sqrt{r^2 + z^2}$ between 
the central star and the considered cell, respectively.

\begin{figure}
\includegraphics[width=\linewidth]{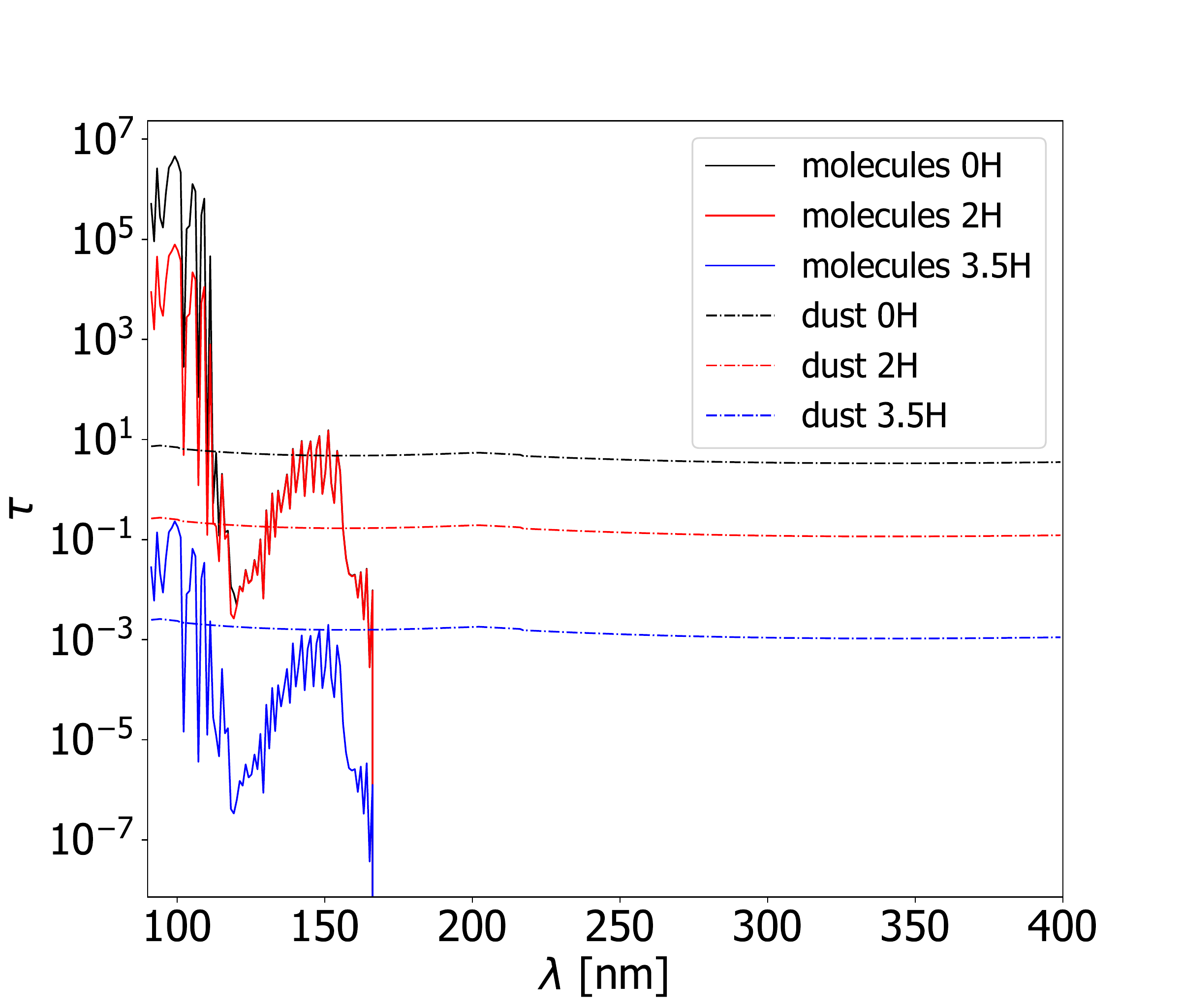}
\caption{\label{fig:opacities} Vertical optical depth as a function of 
wavelengths given at different scale heights $H\mathrm{_g}$ (0 
$\mathrm{H}$, 2 $\mathrm{H}$, 4 $\mathrm{H}$) of a disk at 100 
$\mathrm{au}$ away from the star. Solid lines are molecular opacities, 
dotted lines are dust opacities. }
\end{figure}

\autoref{fig:opacities} shows the vertical opacities generated both by 
the dust (dotted lines) and the molecules (solid lines). The opacities 
are given for different altitudes at 100 
$\mathrm{au}$. 
H$_2$, CO and N$_2$ are the main contributors to the line opacity.

As shown in Figs. \ref{fig:opacities}, 
at all wavelengths 
where H$_2$ contributes to the attenuation ($\lambda \lesssim 115$ 
$\mathrm{nm}$), its opacity is in all parts of the disk always 
substantially larger than the dust opacity.
 
\revisionA{While calculating the vertical molecular opacity $\tau^V_m$ is trivial, 
characterizing the radial molecular opacity $\tau^R _m$ is more 
difficult as it couples different radii and thus their chemical 
evolution together. Given the significant computing time involved in the multi-grain
model, we show below that we can reasonably simplify the problem 
through a simple approximation.} 

Self or mutual shielding is only relevant for high molecular opacities, 
because frequencies where molecular shielding occurs cover a limited 
fraction of the UV domain. In regard to \autoref{fig:field}, we introduce 
the ratio between the radial and vertical molecular optical depths as 
follows:

\begin{equation}
\label{eq:rv}
	RV = \frac{\tau^R_m}{\tau^V_m}
\end{equation}
and  Eq. \ref{eq:local2} can be rewritten as:
\begin{equation}
\label{eq:local3}
	I(\lambda, r, z) \approx  e^{-\tau^V_m} \bigg( I_{ISRF}e^{-\tau^V_d} + I_{*} e^{-\tau^R_d} e^{-\tau^V_m(RV - 1)} \bigg)
\end{equation}
Strictly speaking, $RV$ is position and frequency dependent, since the 
radial and vertical distribution of molecules differ. 

\revisionA{The geometry shown in Figure \ref{fig:field} indicates that $RV$ should be $\geq 1$ and we approximate Eq. \ref{eq:local3} by:}
\begin{equation}
\label{eq:approx1}
	I(\lambda, r, z) \approx  e^{-\tau^V_m} \bigg( I_{ISRF}e^{-\tau^V_d} + I_{*} e^{-\tau^R_d}\bigg) \approx e^{-\tau^V_m} I_d(\lambda, r, z)
\end{equation}

\revisionA{This approximation may underestimate the molecular shielding 
when $\tau^V_m$ is $\lesssim 1$, while $\tau^R_m$ is substantially 
larger than 1, impacting thus the details of the H$_2$ formation layer 
and more generally the physico-chemistry in the upper disk atmosphere. 
Studying the disk upper layer is out of the scope of our paper (and 
would require the use of a PDR code). In fact, our approximation 
becomes reasonable as soon as dust attenuation is significant, a 
situation which happens at $z/H < 3.5-3.8 $ in our dust disk model 
(which has a relatively high dust mass), and does not affect our 
results on the molecular layer and below.}

In summary, everywhere in the disk, self and mutual shielding are 
considered by applying the attenuation by the vertical opacity due to 
molecules to the radiation field computed with dust only by the POLARIS 
code:
\begin{equation}
\label{app:eq:uvfield}
	I(\lambda, r, z) =  e^{-\tau^V_m} I_d(\lambda, r, z)
\end{equation}
where $I_d(\lambda, r, z)$ explicitly includes the impact of dust
scattering that is handled by POLARIS. 

\revisionA{It is important to mention that compared to empirical, pre-computed factors as a function of visual 
extinction and H$_2$ column densities (or also CO and N$_2$ for self-shielding),
this approach has the advantage to take consistently the dust extinction
into account for the specific dust properties and distribution.}

\begin{figure}
\includegraphics[width=\linewidth]{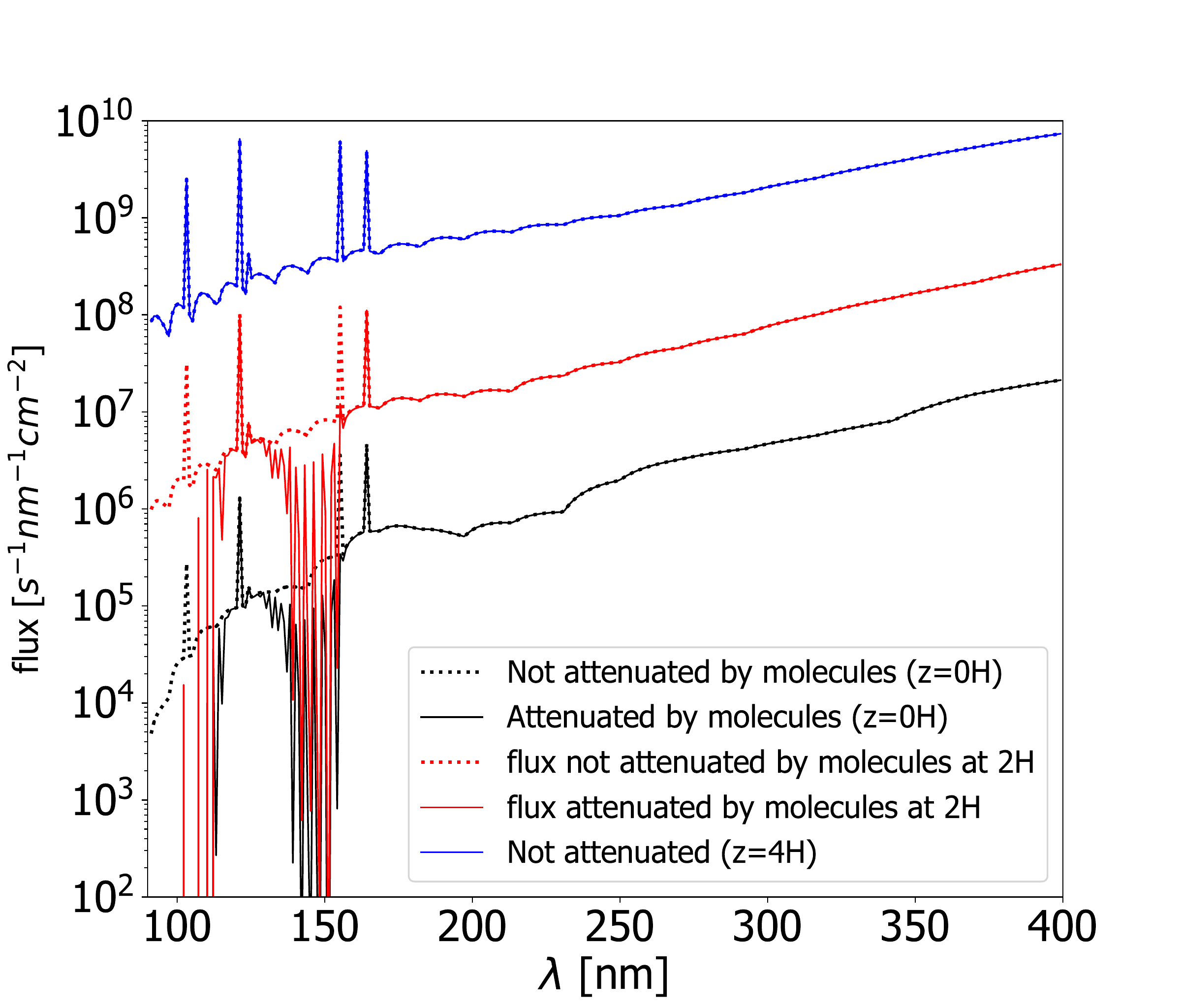}
\caption{\label{fig:flux_approx} Flux as a function of the wavelength 
in the midplane (black lines), at $2\,H$ (red lines) and at $4\,H$ (blue 
lines) at 100 $\mathrm{au}$. \revisionA{The unattenuated UV spectrum is 
derived from \cite{France+etal_2014}} \revisionA{with spectral resampling preserving
the equivalent widths of the lines.}. }
\end{figure}

Consequently, we use Eq. \ref{app:eq:uvfield} to compute the photorates and 
the vertical molecular opacities are computed using Eq. \ref{eq:tau_m} 
in all our simulations. \autoref{fig:flux_approx} gives the flux as a 
function of the wavelength in the midplane and in the upper atmosphere 
(4 $H$) at 100 $\mathrm{au}$ for the disk whose parameters are given in 
\autoref{tab:param}. It shows a comparison between the flux as given by 
the radiative transfer simulation (dotted lines) and the flux as given by  
Eq. \ref{eq:uvfield} which takes into account the molecular 
contribution to the opacity (solid line). The attenuation by spectral lines, especially 
those of H$_2$, is clearly visible. The contribution of the molecules to the opacity occurs in 
a limited range of the UV domain. 
Although our chemical code (Section \ref{sec:chemistry}) computes other 
molecular species that contribute to the UV opacity, their abundance 
relative to H$_2$ remains small and limit their contribution. 

\section{HUV cases}
\label{app:huv}

\begin{figure*}
\begin{subfigure}{.32\linewidth}
  \centering
  \includegraphics[width=1.00\linewidth]{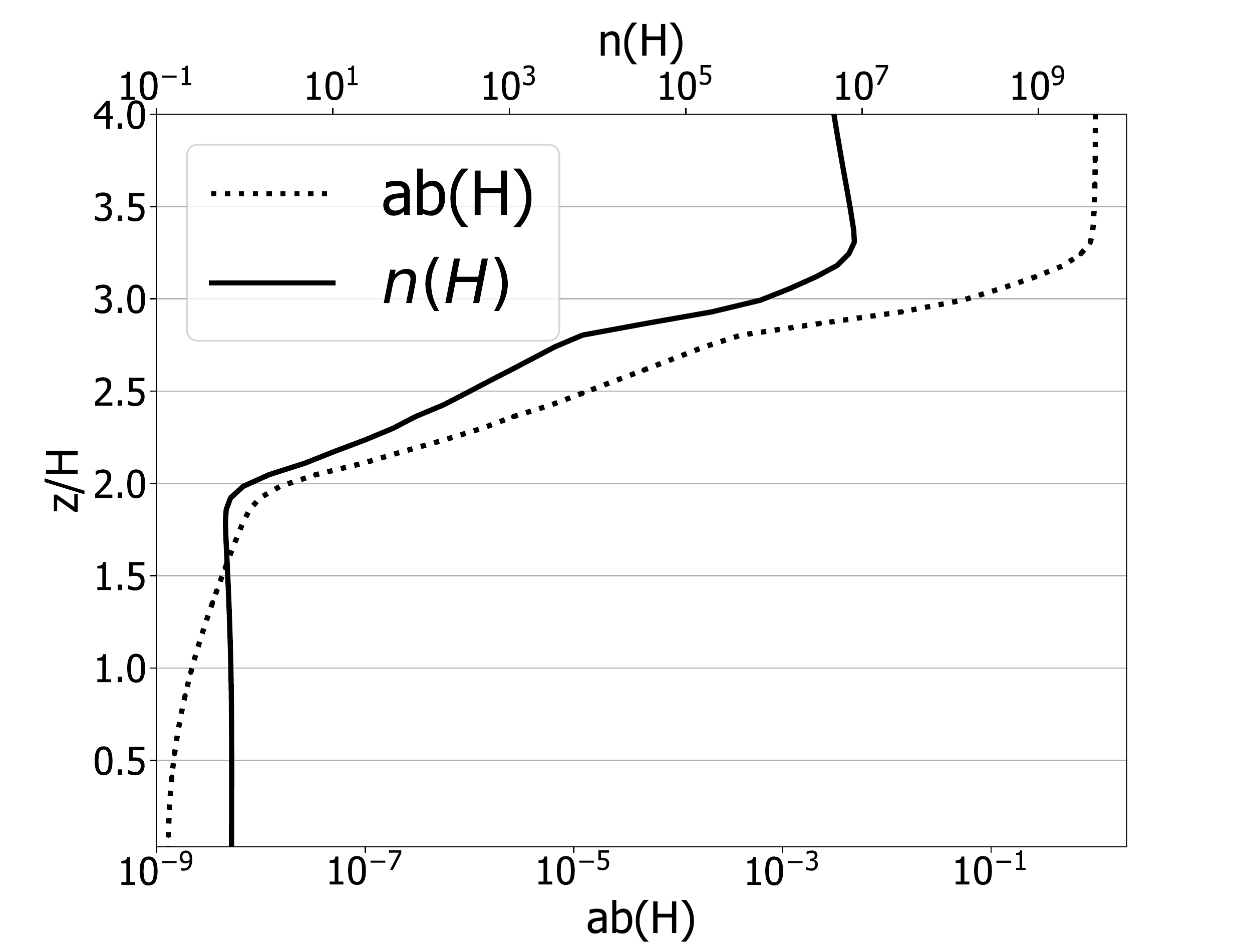}
\end{subfigure}
\begin{subfigure}{.32\linewidth}
  \centering
  \includegraphics[width=1.00\linewidth]{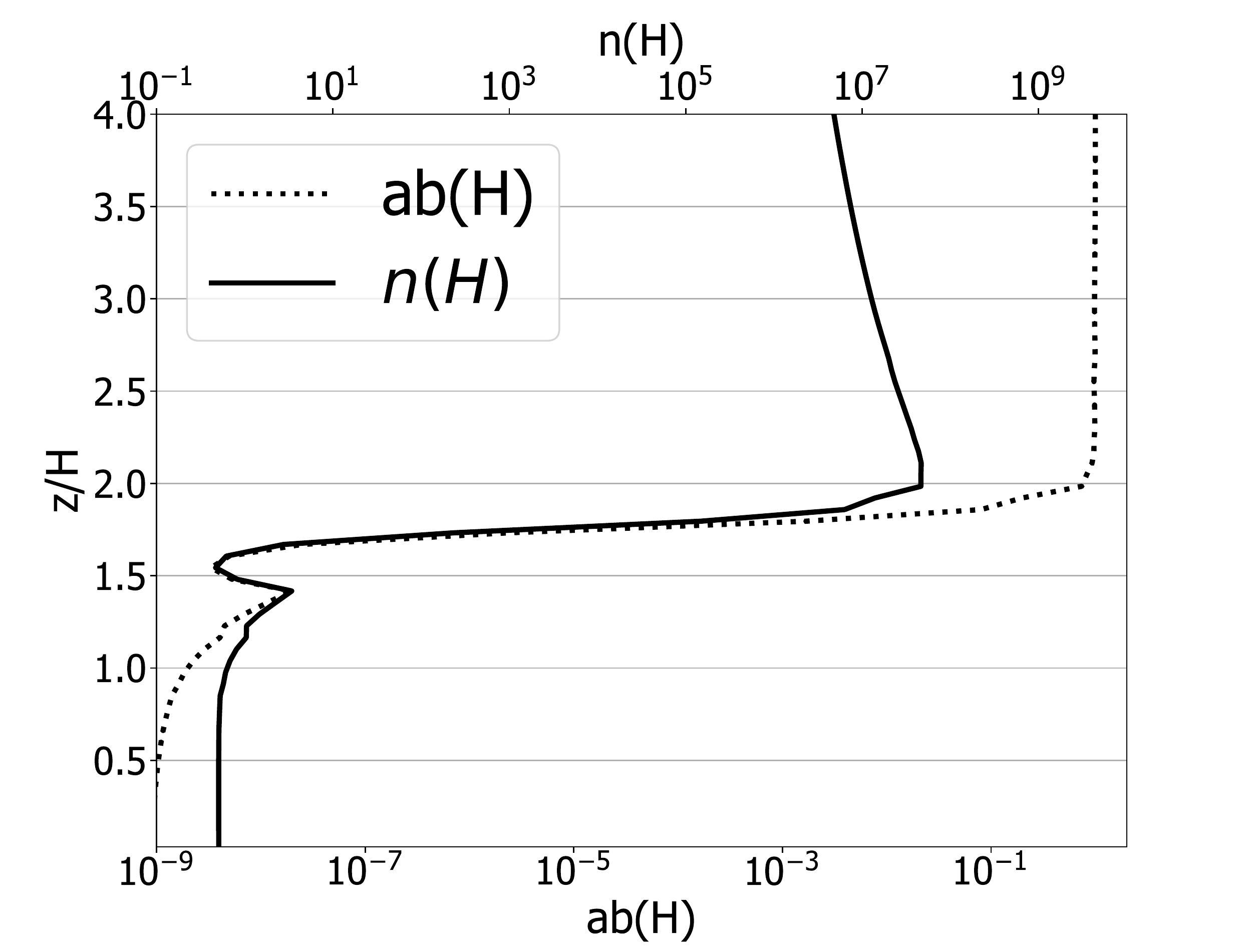}
\end{subfigure}
\begin{subfigure}{.32\linewidth}
  \centering
  \includegraphics[width=1.00\linewidth]{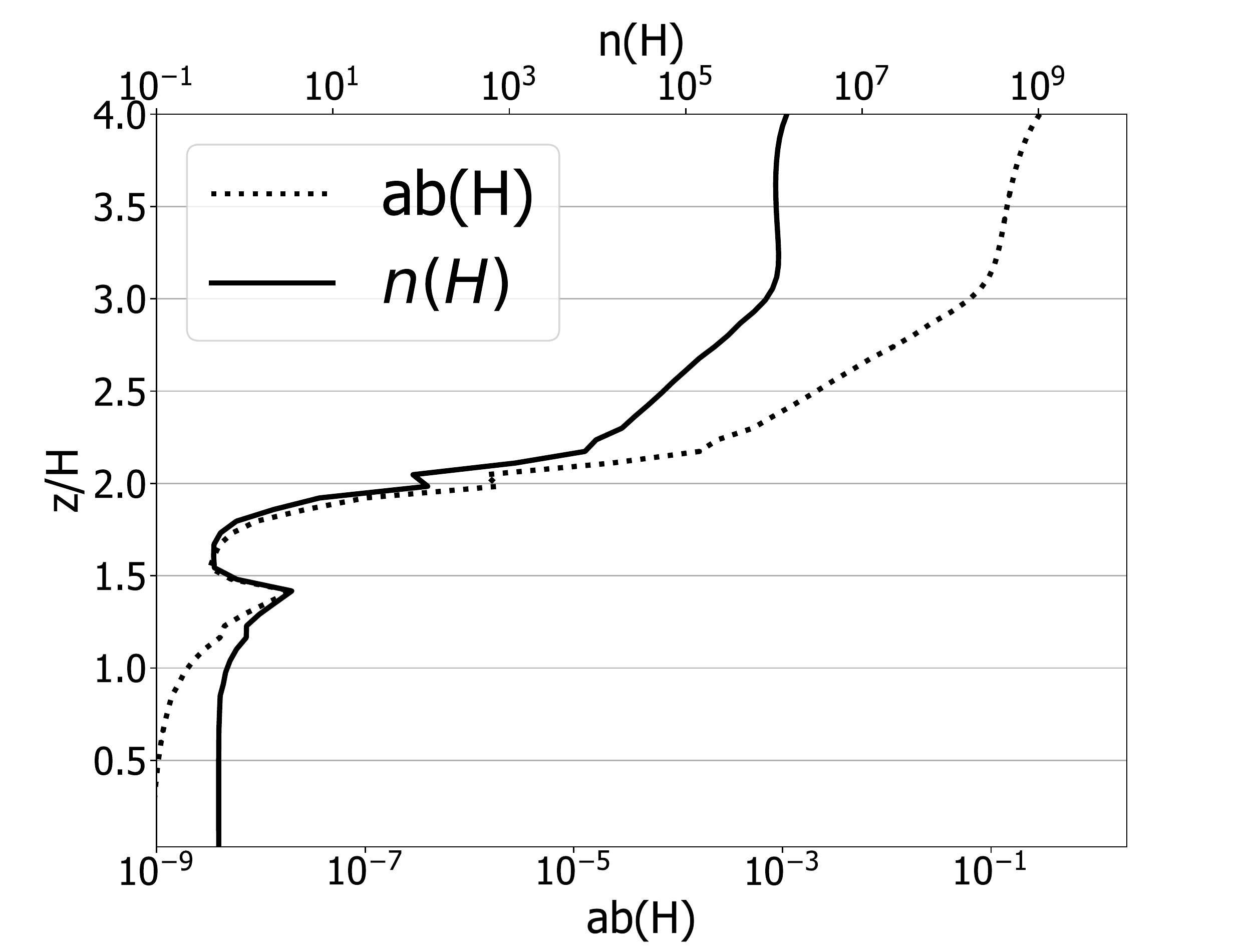}
\end{subfigure}

\begin{subfigure}{.32\linewidth}
  \centering
  \includegraphics[width=1.00\linewidth]{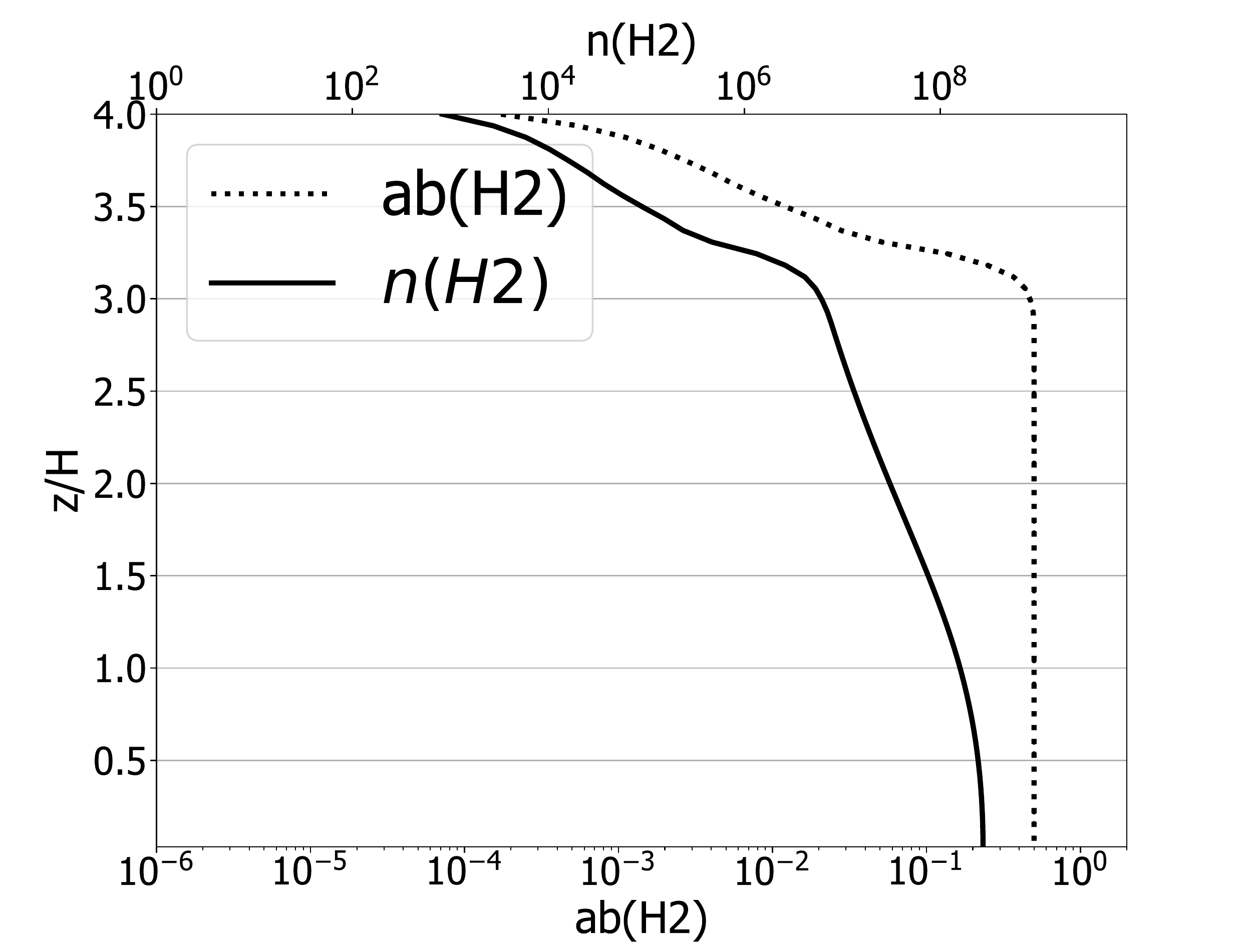}
\end{subfigure}
\begin{subfigure}{.32\linewidth}
  \centering
  \includegraphics[width=1.00\linewidth]{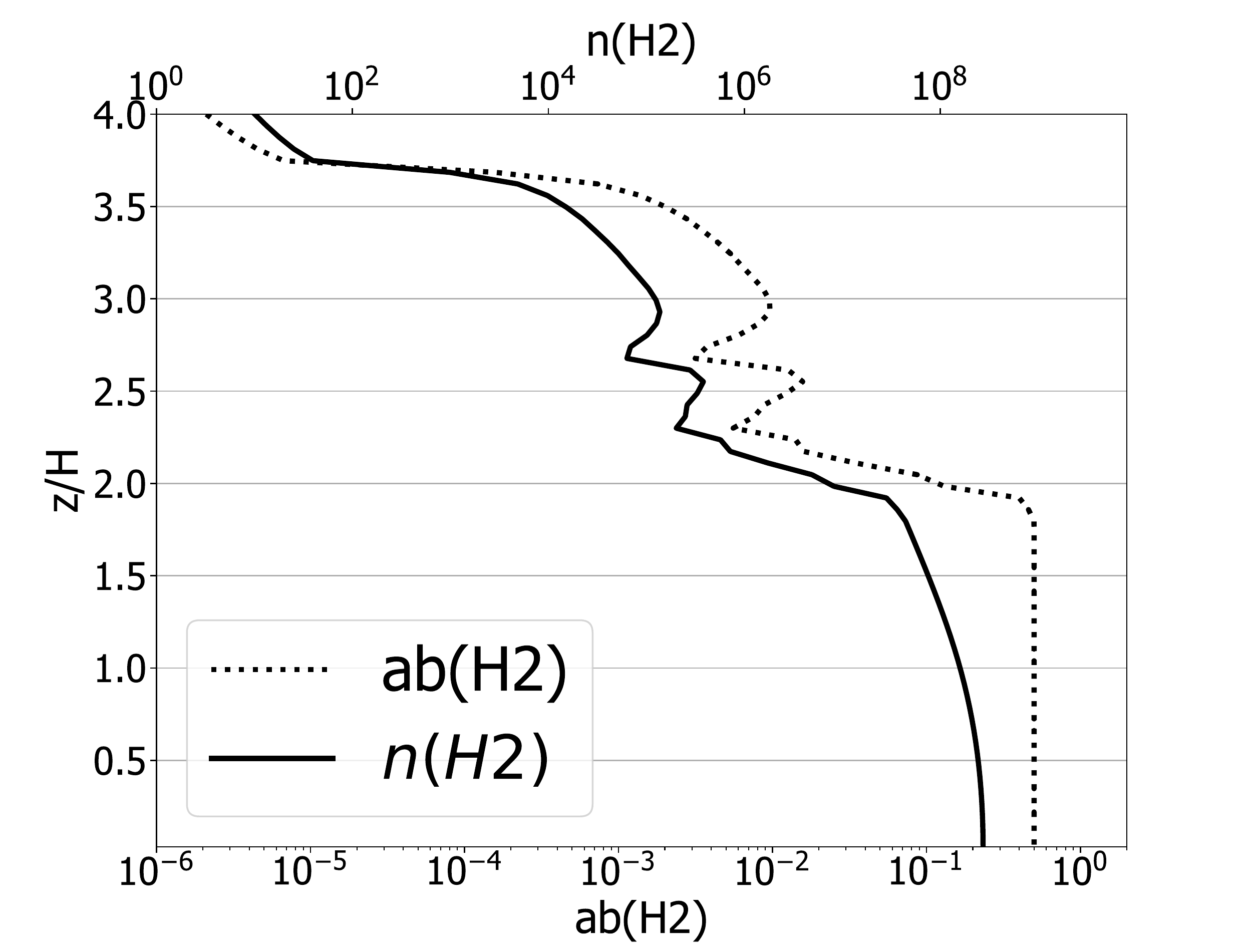}
\end{subfigure}
\begin{subfigure}{.32\linewidth}
  \centering
  \includegraphics[width=1.00\linewidth]{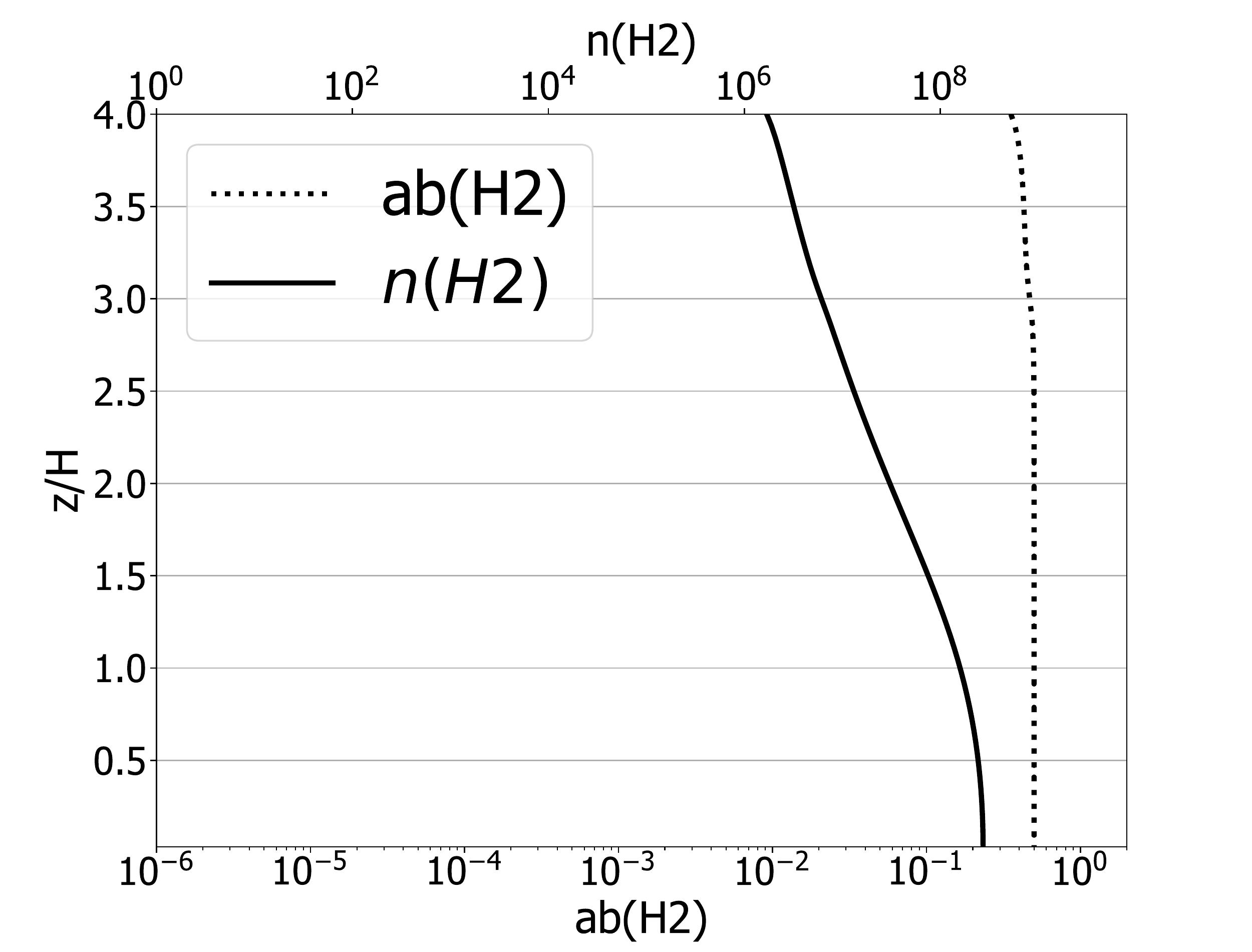}
\end{subfigure}

\begin{subfigure}{.32\linewidth}
  \centering
  \includegraphics[width=1.00\linewidth]{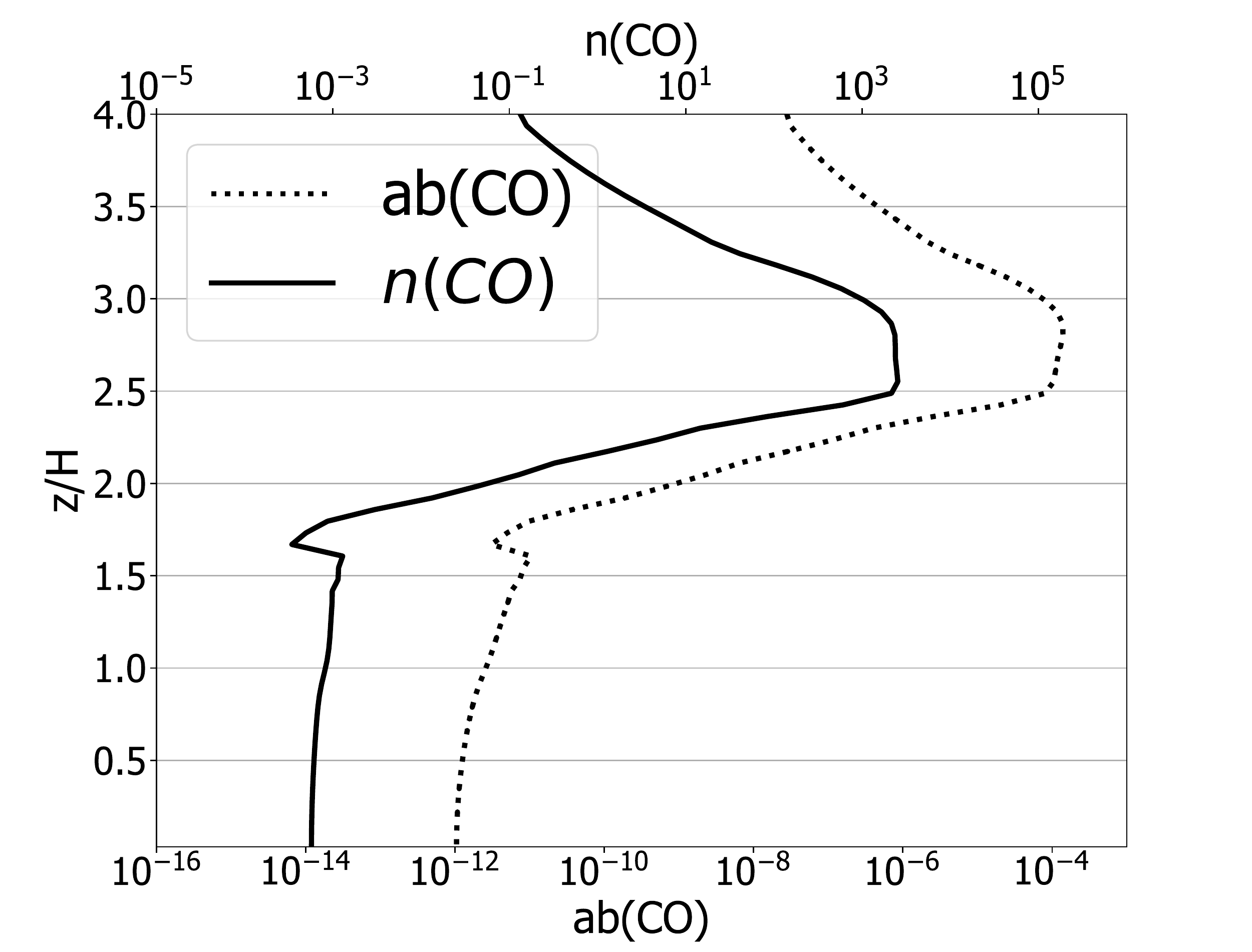}
\end{subfigure}
\begin{subfigure}{.32\linewidth}
  \centering
  \includegraphics[width=1.00\linewidth]{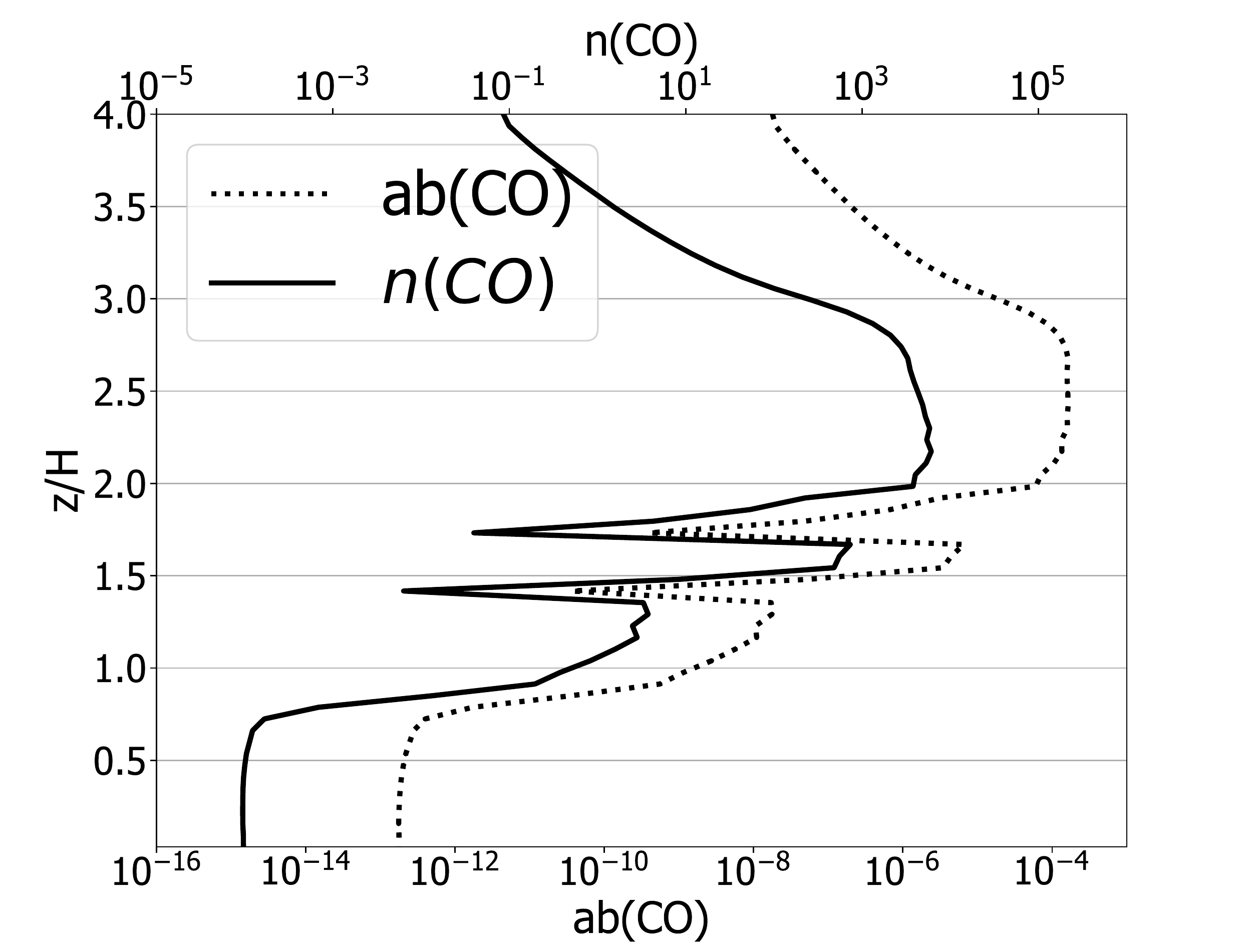}  
\end{subfigure}
\begin{subfigure}{.32\linewidth}
  \centering
  \includegraphics[width=1.00\linewidth]{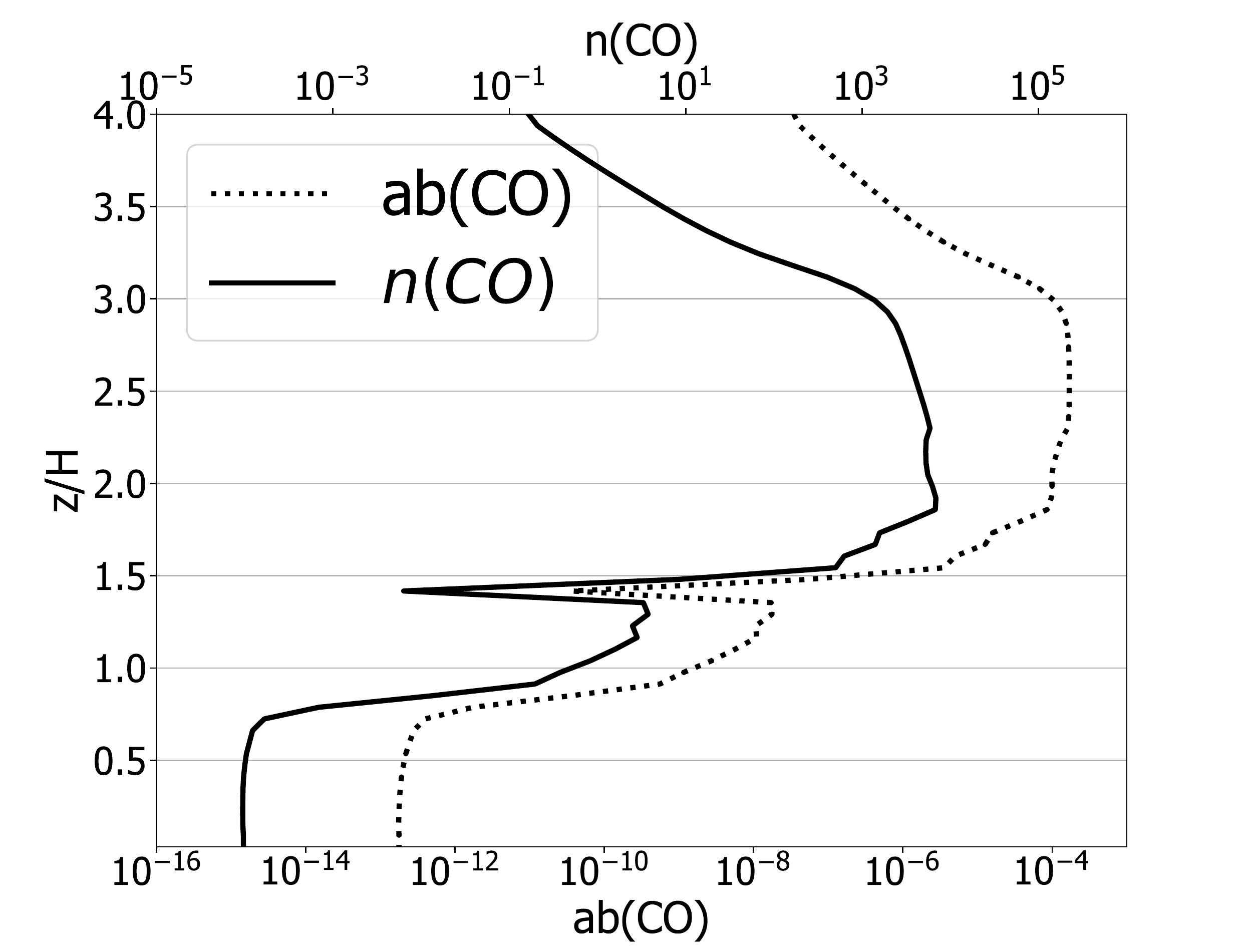}
\end{subfigure}

\begin{subfigure}{.32\linewidth}
  \centering
  \includegraphics[width=1.00\linewidth]{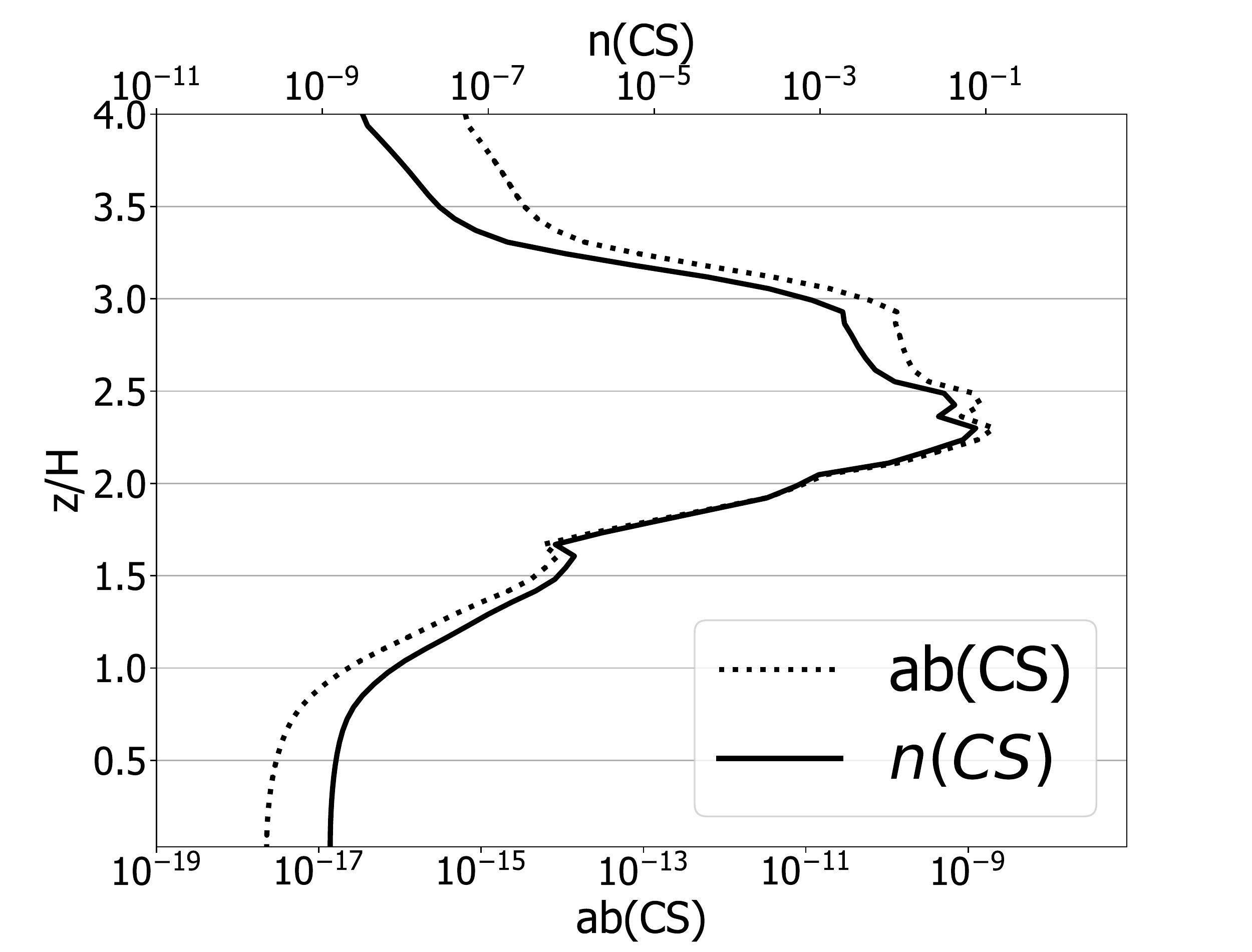}
\end{subfigure}
\begin{subfigure}{.32\linewidth}
  \centering
  \includegraphics[width=1.00\linewidth]{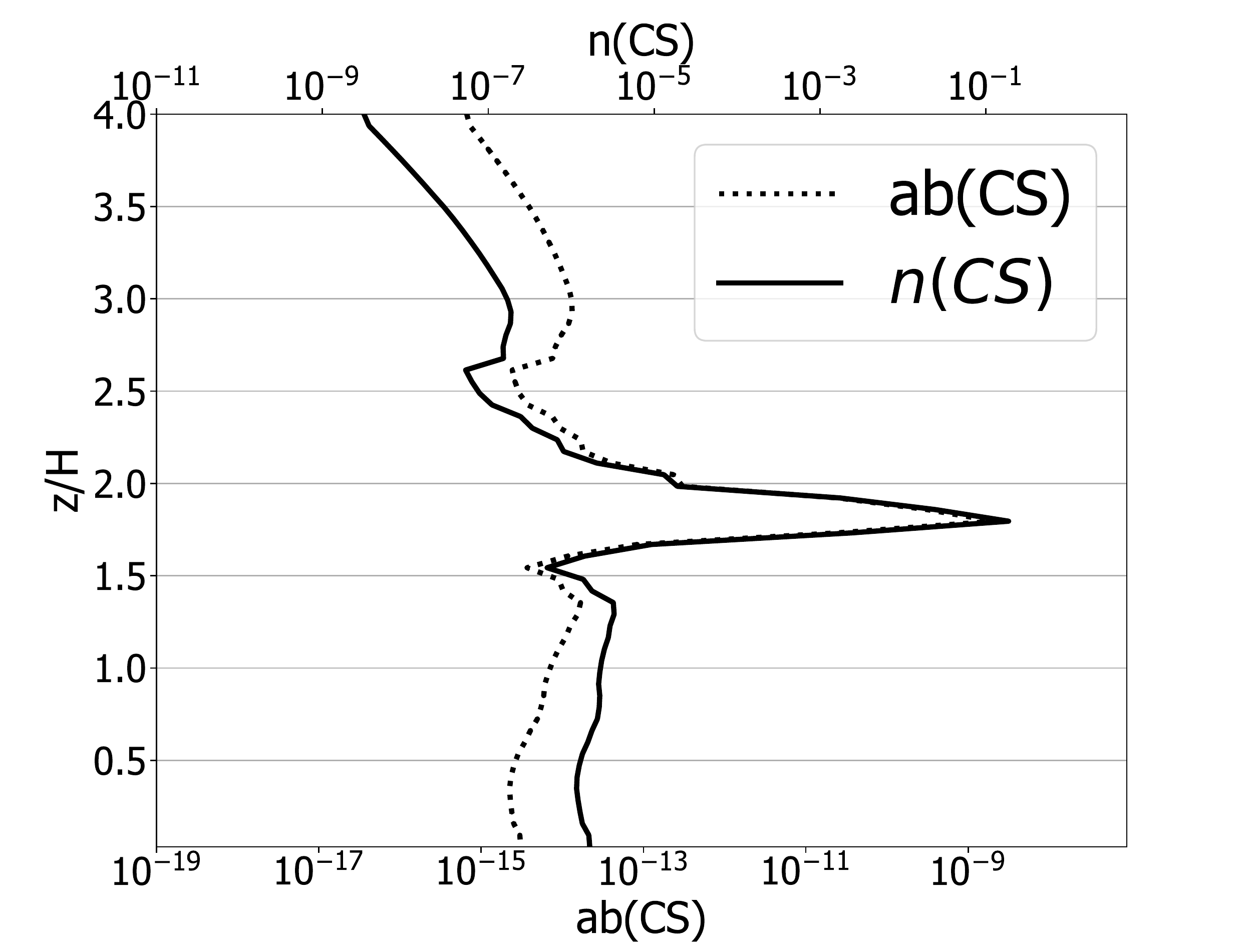}
\end{subfigure}
\begin{subfigure}{.32\linewidth}
  \centering
  \includegraphics[width=1.00\linewidth]{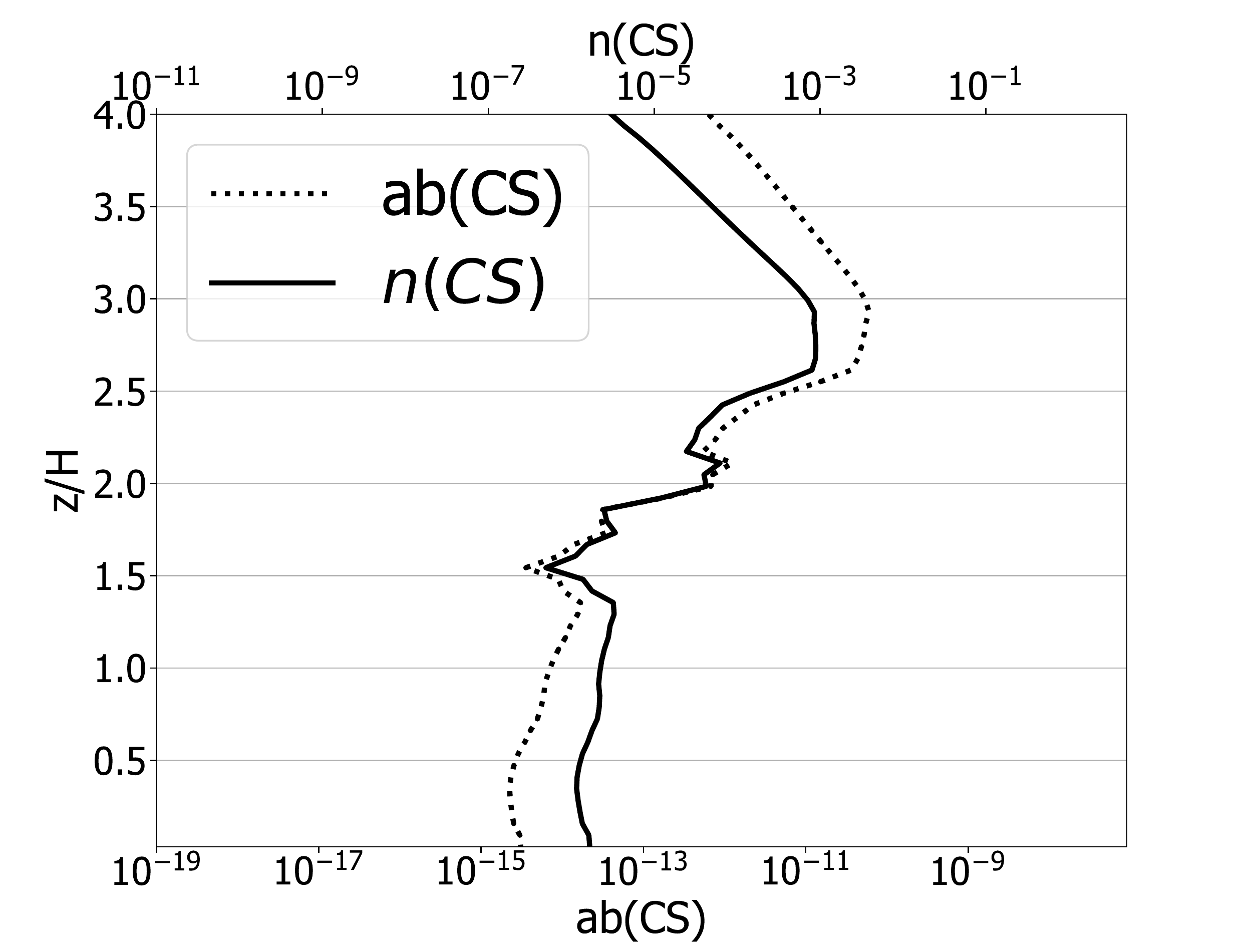}
\end{subfigure}

\begin{subfigure}{.32\linewidth}
  \centering
  \includegraphics[width=1.00\linewidth]{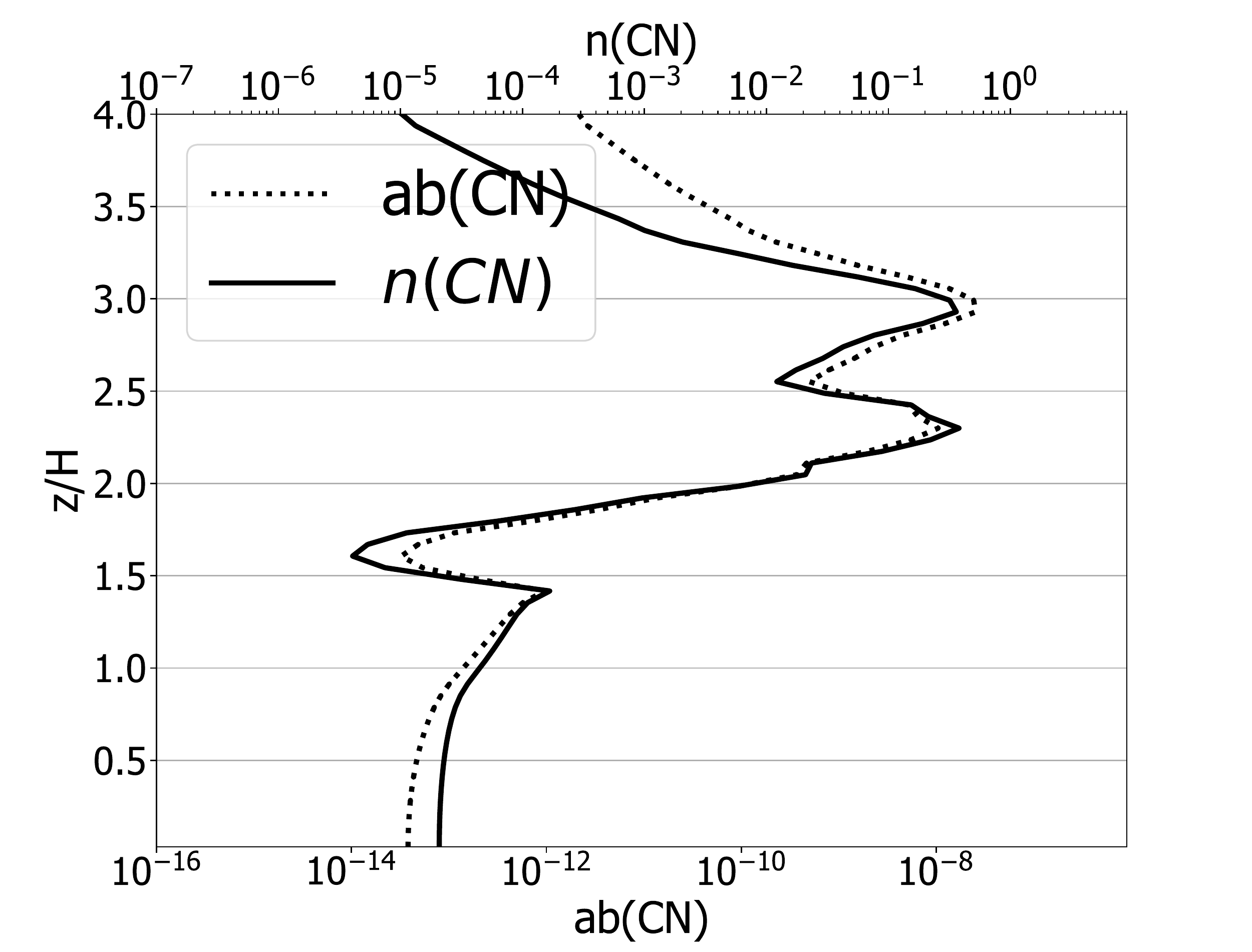}
   \subcaption{HUV-LH-$\mathrm{T_{g}}$}   
\end{subfigure}
\begin{subfigure}{.32\linewidth}
  \centering
  \includegraphics[width=1.00\linewidth]{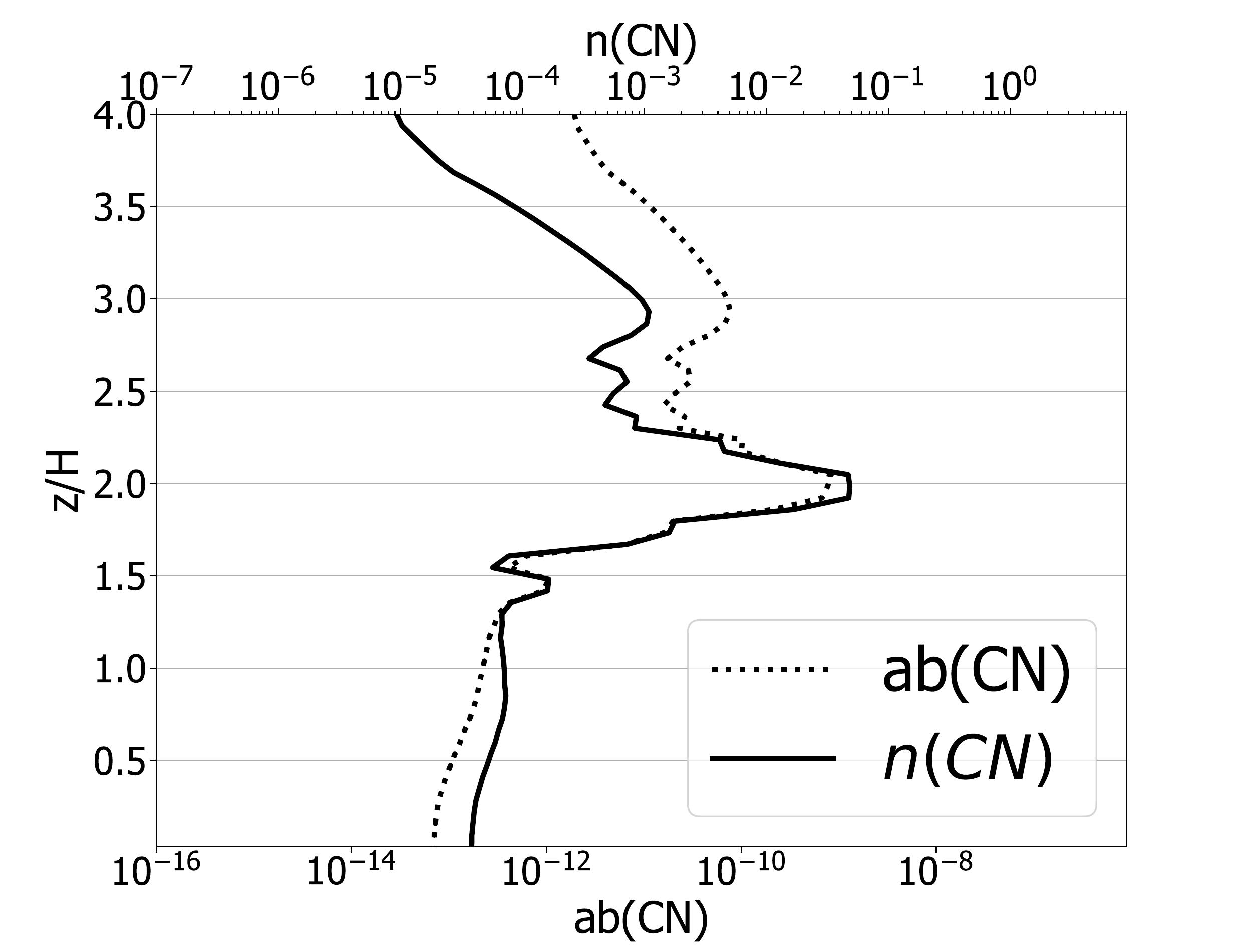}
   \subcaption{HUV-LH-$\mathrm{T_{a}}$}   
\end{subfigure}
\begin{subfigure}{.32\linewidth}
  \centering
  \includegraphics[width=1.00\linewidth]{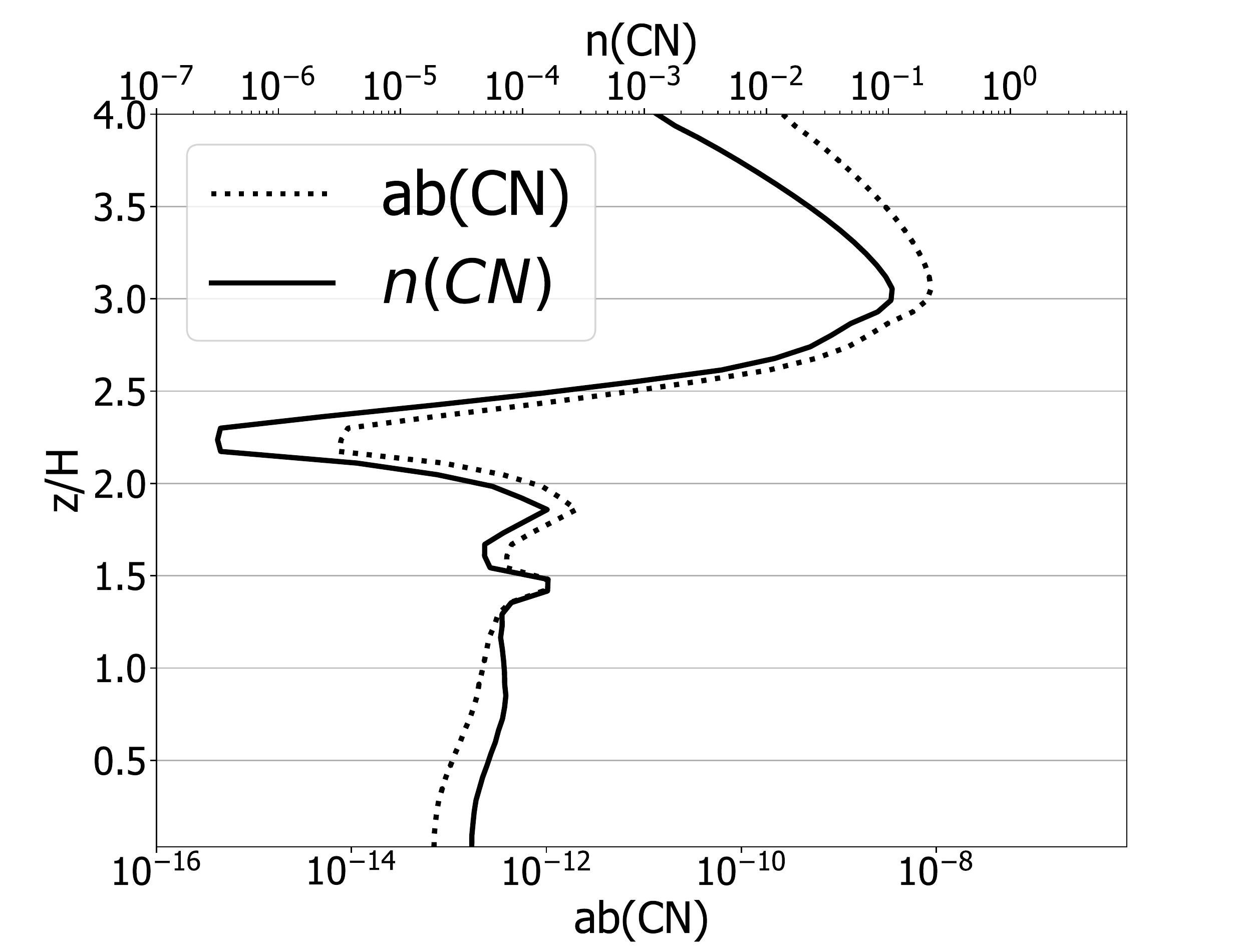}
   \subcaption{HUV-B14-$\mathrm{T_{a}}$} 
\end{subfigure}

\caption{Vertical profiles of H, $\mathrm{H_2}$, CO, CS and CN at 100 au from the star of the HUV single-grain models. The dotted line is the abundance relative to H and the solid line is the density [$cm^{-3}$].}
\label{fig:s-100profile_high}
\end{figure*}

\begin{figure*}
\begin{subfigure}{.32\linewidth}
  \centering
  \includegraphics[width=1.00\linewidth]{figures/SINGLE/HUV_HL_Tg/100AU/H_HUV_HL_Tg.pdf}
\end{subfigure}
\begin{subfigure}{.32\linewidth}
  \centering
  \includegraphics[width=1.00\linewidth]{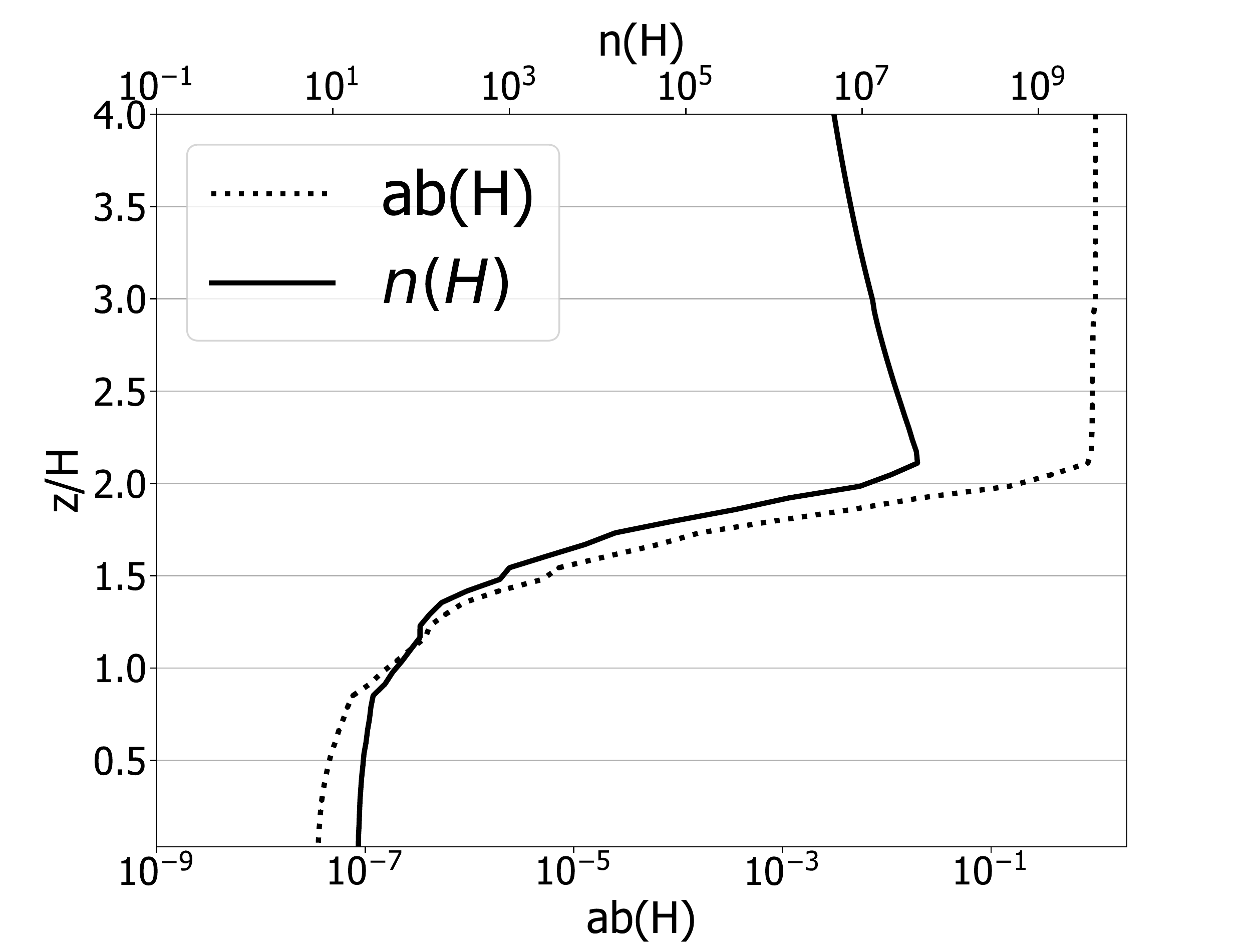}
\end{subfigure}
\begin{subfigure}{.32\linewidth}
  \centering
  \includegraphics[width=1.00\linewidth]{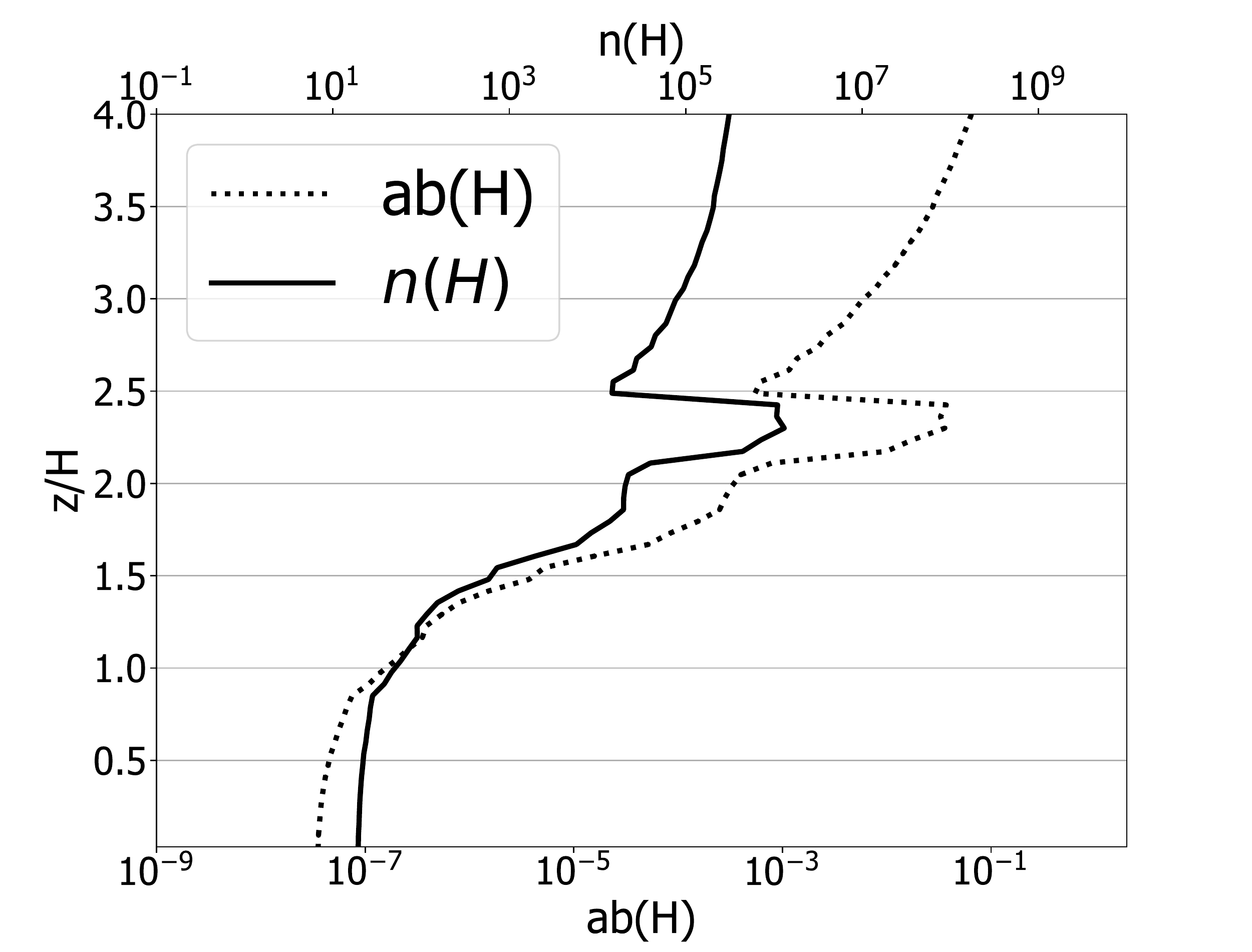}
\end{subfigure}\\

\begin{subfigure}{.32\linewidth}
  \centering
  \includegraphics[width=1.00\linewidth]{figures/SINGLE/HUV_HL_Tg/100AU/H2_HUV_HL_Tg.pdf}
\end{subfigure}
\begin{subfigure}{.32\linewidth}
  \centering
  \includegraphics[width=1.00\linewidth]{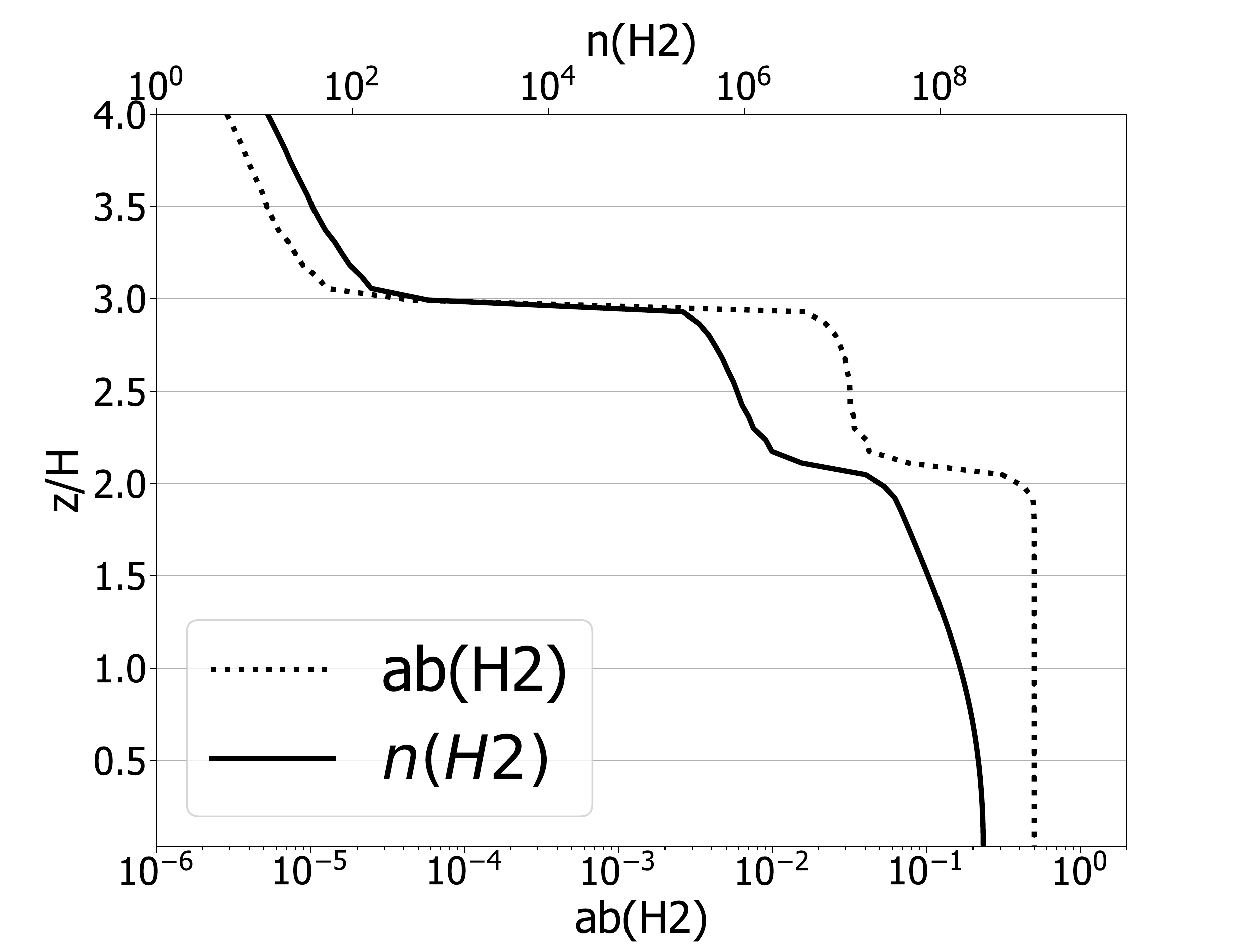}
\end{subfigure}
\begin{subfigure}{.32\linewidth}
  \centering
  \includegraphics[width=1.00\linewidth]{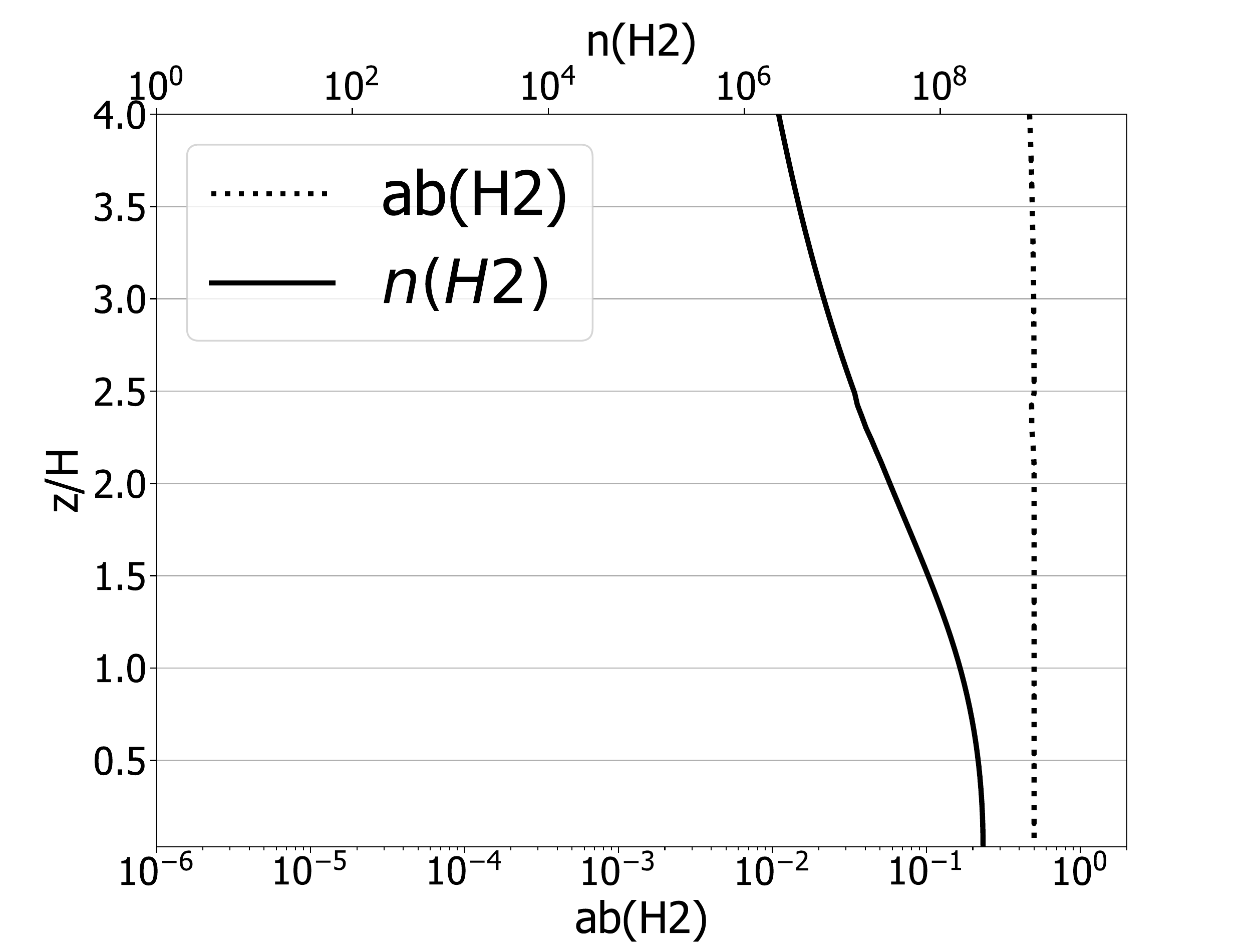}
\end{subfigure}\\

\begin{subfigure}{.32\linewidth}
  \centering
  \includegraphics[width=1.00\linewidth]{figures/SINGLE/HUV_HL_Tg/100AU/CO_HUV_HL_Tg.pdf}
\end{subfigure}
\begin{subfigure}{.32\linewidth}
  \centering
  \includegraphics[width=1.00\linewidth]{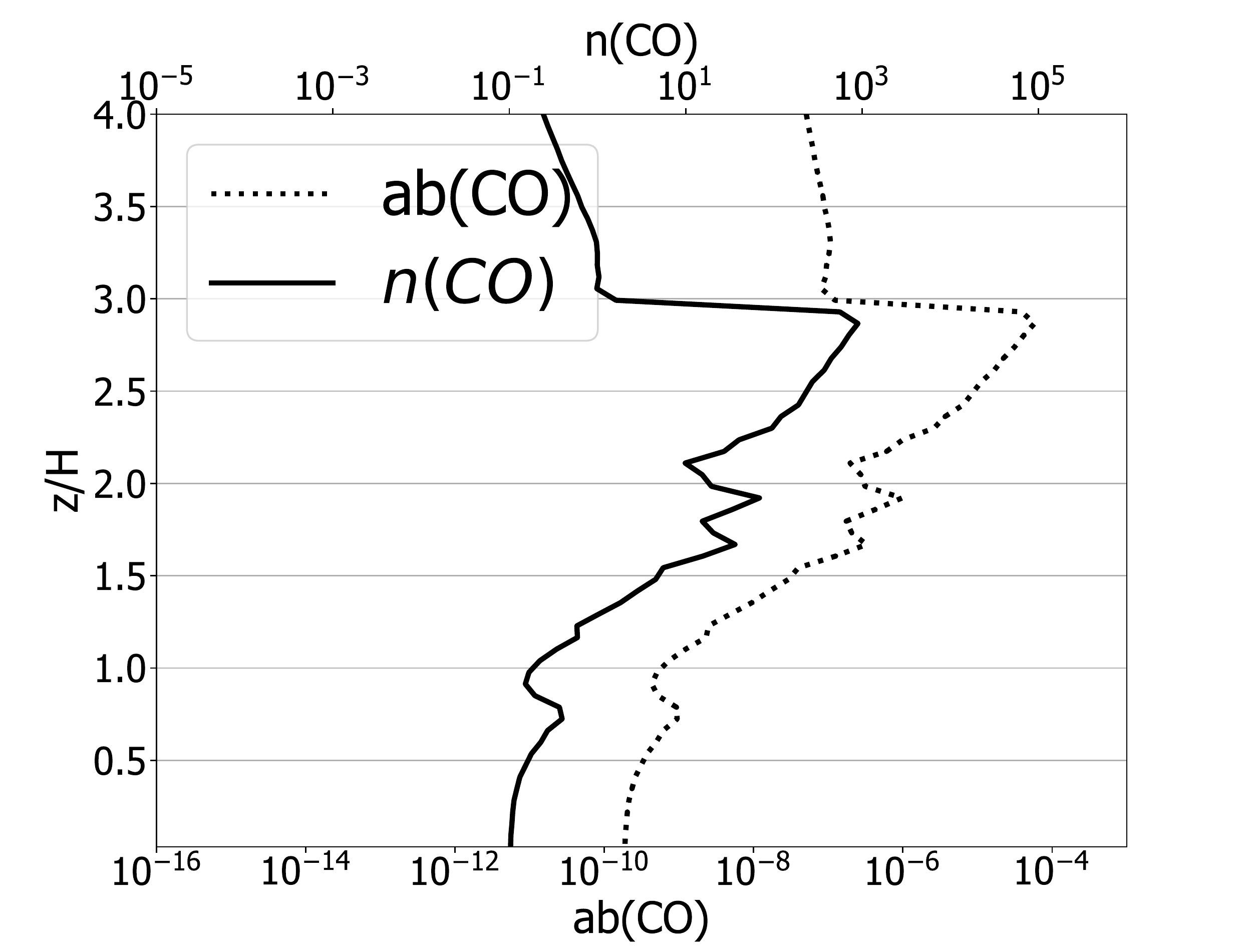}
\end{subfigure}
\begin{subfigure}{.32\linewidth}
  \centering
  \includegraphics[width=1.00\linewidth]{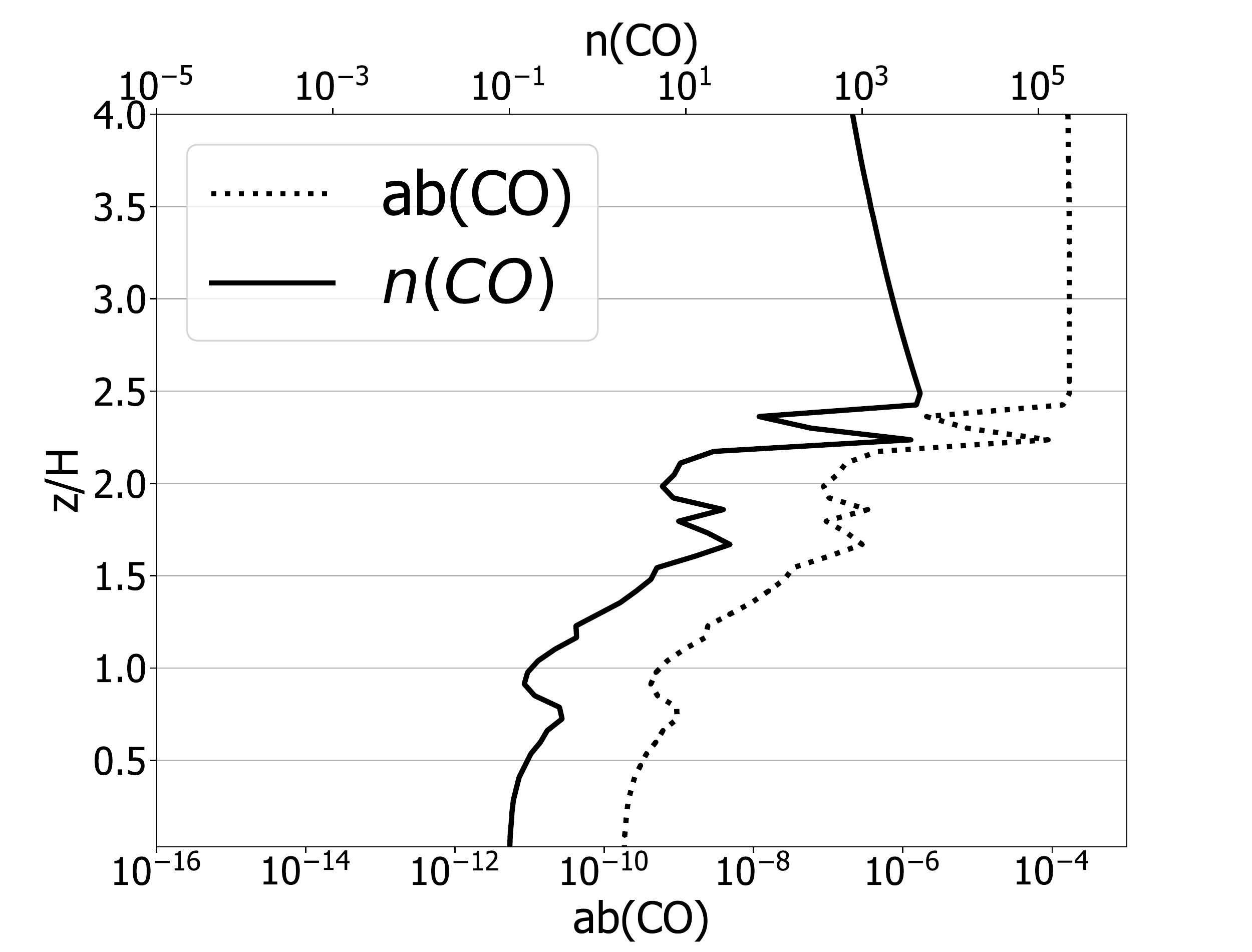}
\end{subfigure}\\

\begin{subfigure}{.32\linewidth}
  \centering
  \includegraphics[width=1.00\linewidth]{figures/SINGLE/HUV_HL_Tg/100AU/CS_HUV_HL_Tg.pdf}
\end{subfigure}
\begin{subfigure}{.32\linewidth}
  \centering
  \includegraphics[width=1.00\linewidth]{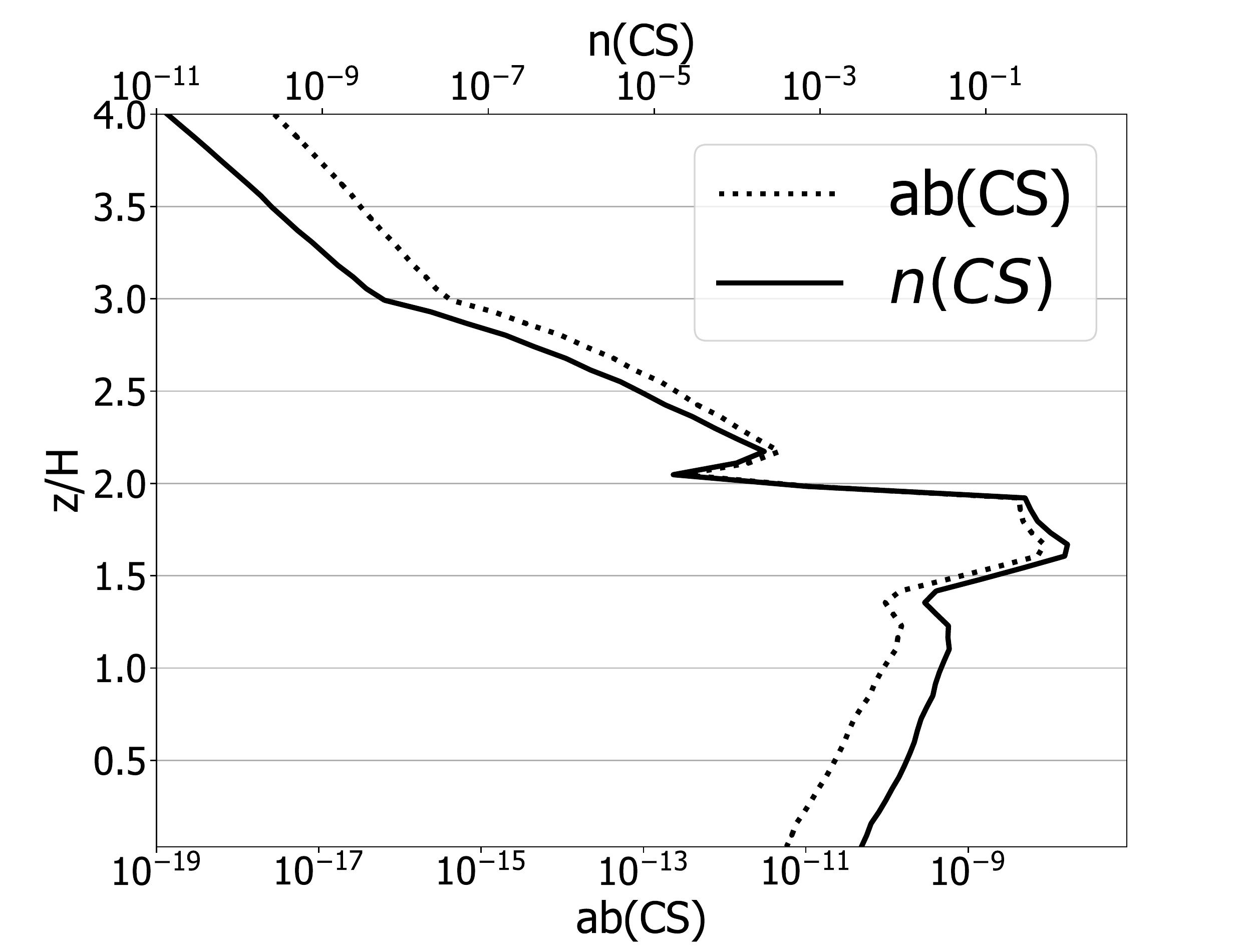}
\end{subfigure}
\begin{subfigure}{.32\linewidth}
  \centering
  \includegraphics[width=1.00\linewidth]{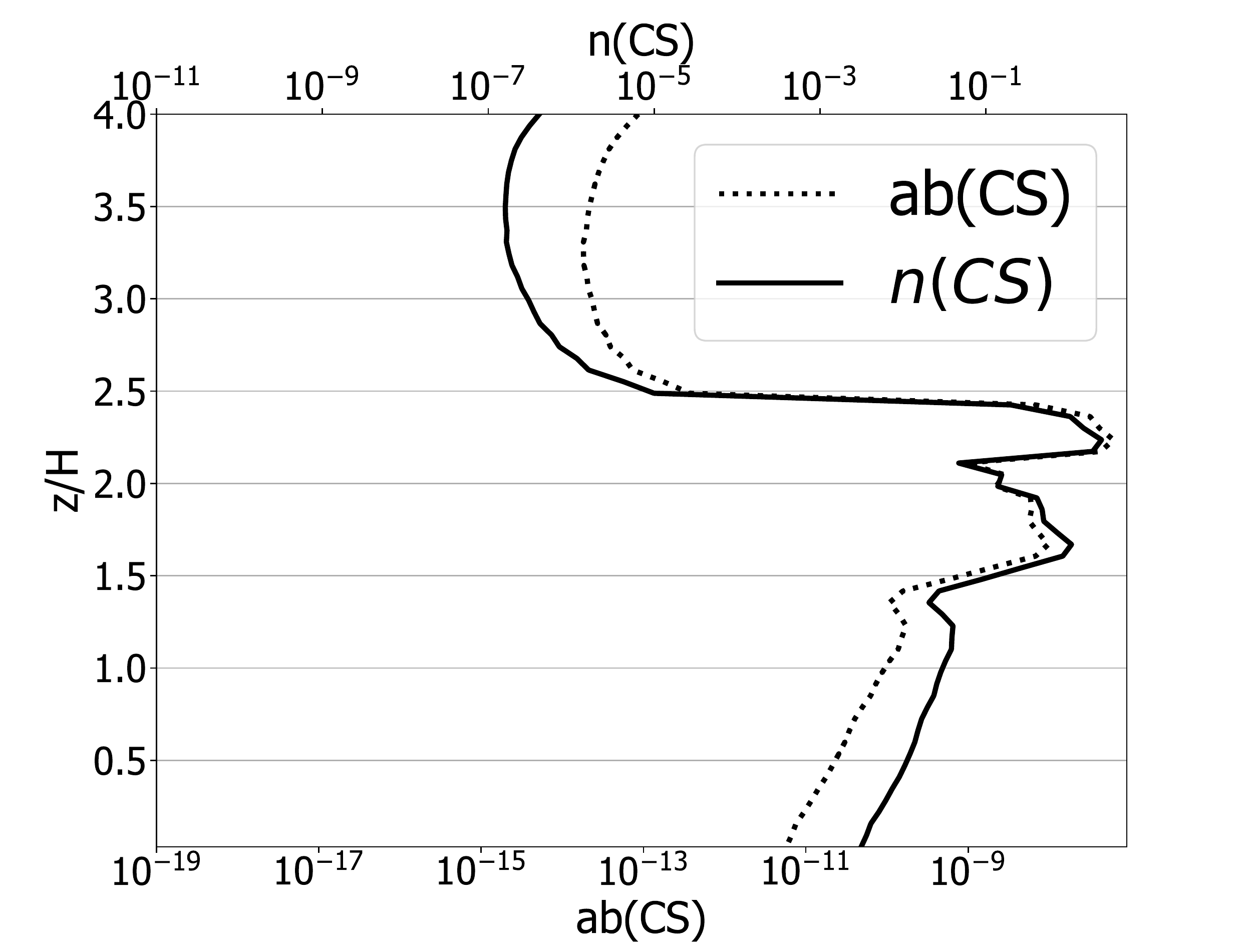}
\end{subfigure}\\

\begin{subfigure}{.32\linewidth}
  \centering
  \includegraphics[width=1.00\linewidth]{figures/SINGLE/HUV_HL_Tg/100AU/CN_HUV_HL_Tg.pdf}
   \subcaption{HUV-LH-$\mathrm{T_{g}}$}   
\end{subfigure}
\begin{subfigure}{.32\linewidth}
  \centering
  \includegraphics[width=1.00\linewidth]{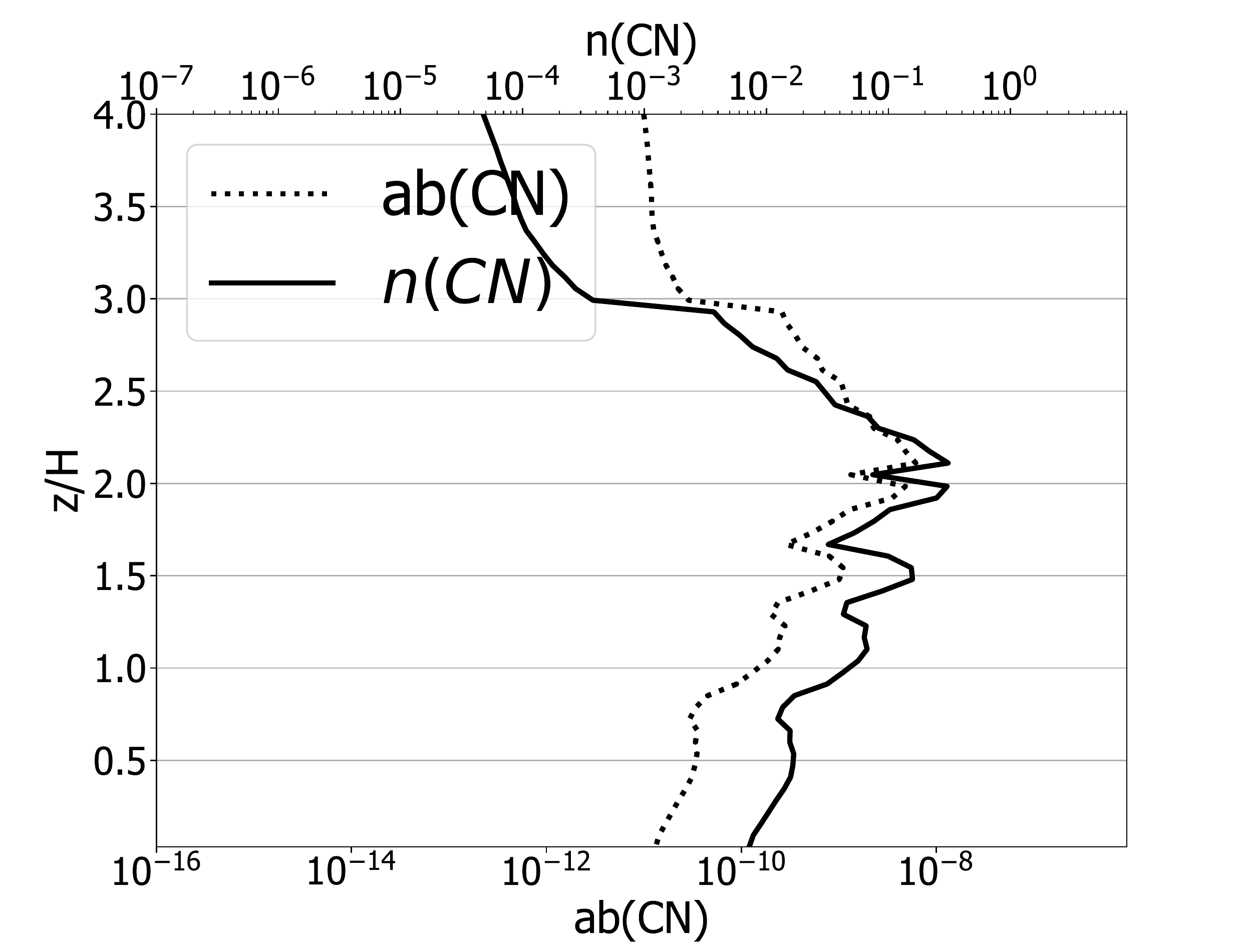}
   \subcaption{M-HUV-LH}   
\end{subfigure}
\begin{subfigure}{.32\linewidth}
  \centering
  \includegraphics[width=1.00\linewidth]{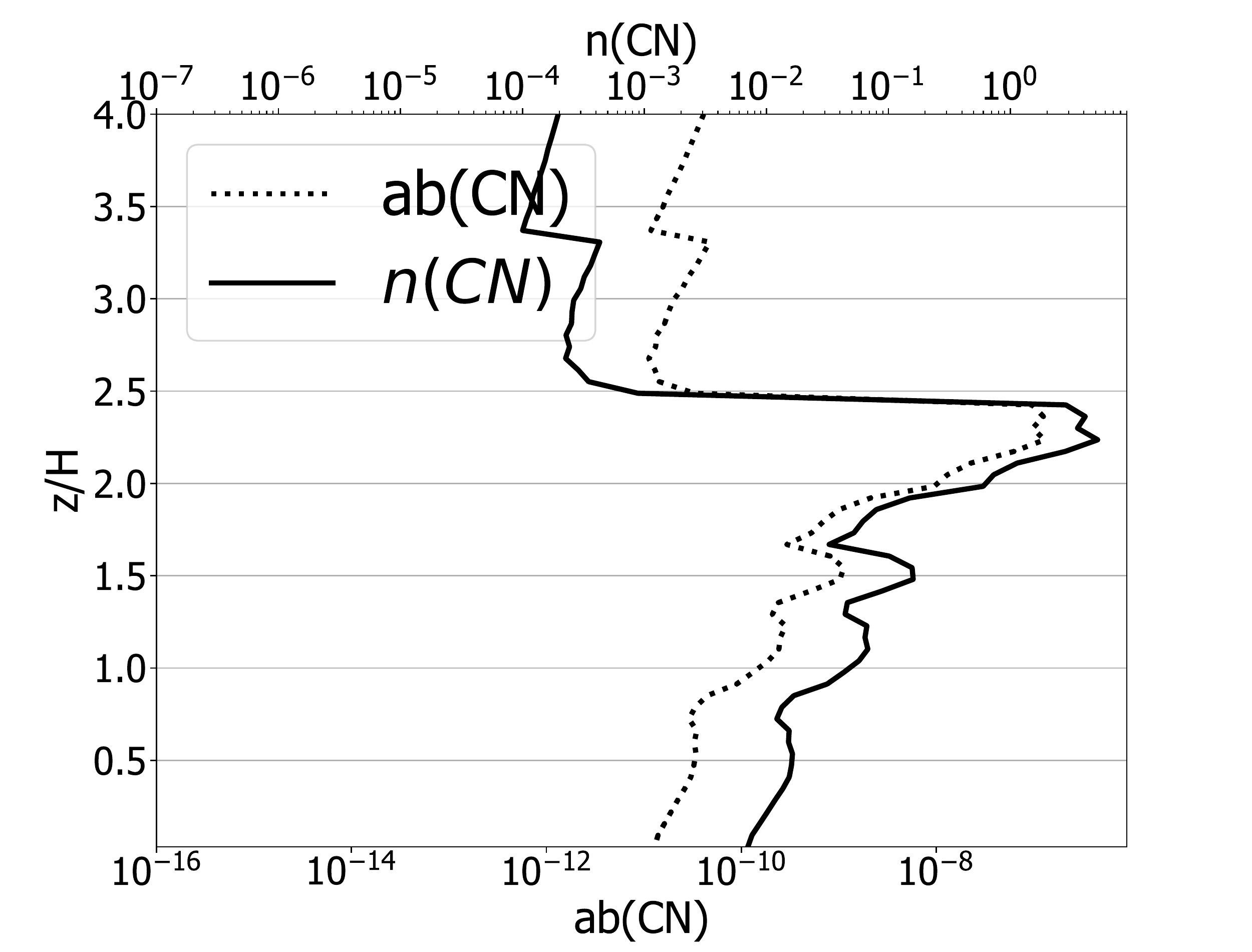}
   \subcaption{M-HUV-B14}   
\end{subfigure}\\

\caption{Vertical profiles of H, $\mathrm{H_2}$, CO, CS and CN at 100 au from the star of the HUV single-grain model \shtg on the left column, and multi-grain models on middle and right columns. The dotted line is the abundance relative to H and the solid line is the density [$cm^{-3}$].}
\label{fig:m-100profile_high}
\end{figure*}

In Figs.\ref{fig:s-100profile_high} and \ref{fig:m-100profile_high} we 
present the vertical profiles at 100 au of the fractional abundances 
and number densities of gas-phase H, H$_2$, CO, CS and CN in HUV 
single-grain models and multi-grain models, respectively.

\subsection{\revisionC{Intermediate models}}\label{app:interm}

\revisionC{The intermediate models have physical conditions identical to that of the multi-grain models,
but the dust grain distribution is represented using (locally) a single grain size, determined
so that its effective area and mass is identical to those of dust grains at the same location
in multi-grain models. The temperature of this equivalent grain is the area-weighted temperature T$_\mathrm{a}$. These models thus exhibit dust settling as multi-grain models. They also
have a larger UV penetration than in single-models, similar, though not identical to 
the UV penetration in  multi-grain models.}

\begin{figure*}

\begin{subfigure}{.32\linewidth}
  \centering
  \includegraphics[width=1.00\linewidth]{figures/SINGLE/HUV_HL_Teff/100AU/H_HUV_HL_Teff.pdf}
\end{subfigure}
\begin{subfigure}{.32\linewidth}
  \centering
  \includegraphics[width=1.00\linewidth]{figures/MULTI/HUV_LH/100AU/H_HUV_LH.pdf}
\end{subfigure}
\begin{subfigure}{.32\linewidth}
  \centering
  \includegraphics[width=1.00\linewidth]{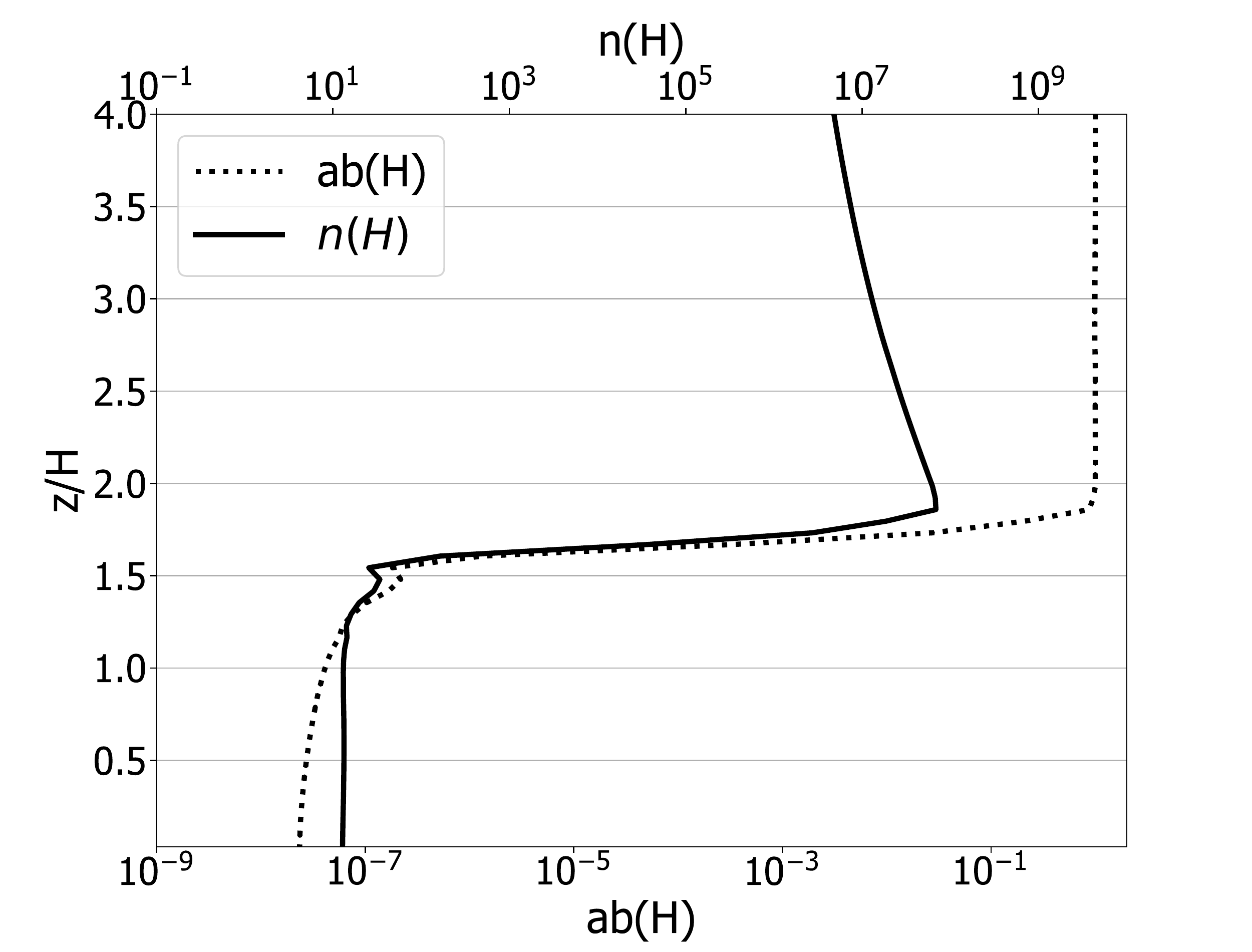}
\end{subfigure}

\begin{subfigure}{.32\linewidth}
  \centering
  \includegraphics[width=1.00\linewidth]{figures/SINGLE/HUV_HL_Teff/100AU/H2_HUV_HL_Teff.pdf}
\end{subfigure}
\begin{subfigure}{.32\linewidth}
  \centering
  \includegraphics[width=1.00\linewidth]{figures/MULTI/HUV_LH/100AU/H2_HUV_LH.pdf}
\end{subfigure}
\begin{subfigure}{.32\linewidth}
  \centering
  \includegraphics[width=1.00\linewidth]{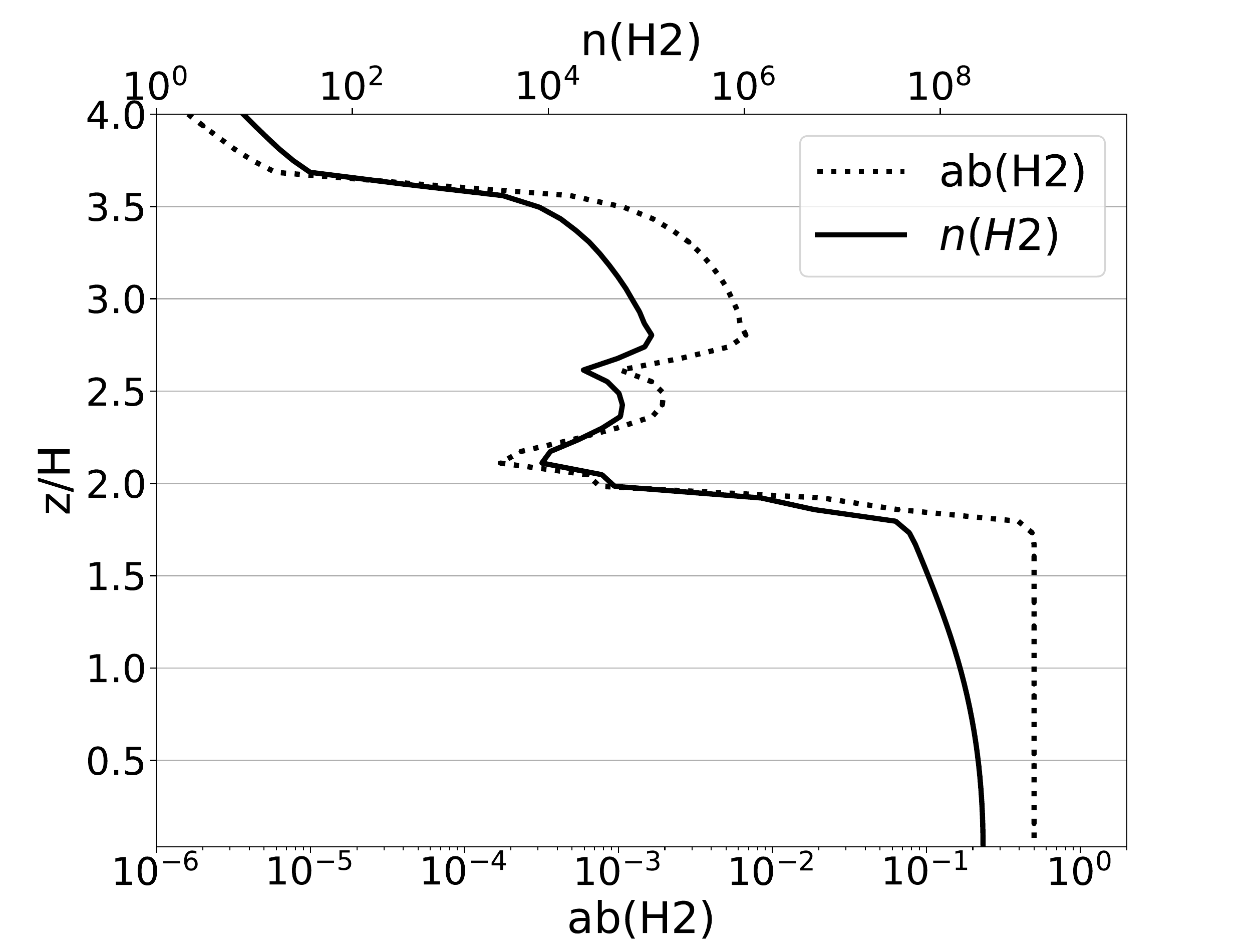}
\end{subfigure}

\begin{subfigure}{.32\linewidth}
  \centering
  \includegraphics[width=1.00\linewidth]{figures/SINGLE/HUV_HL_Teff/100AU/CO_HUV_HL_Teff.pdf}
\end{subfigure}
\begin{subfigure}{.32\linewidth}
  \centering
  \includegraphics[width=1.00\linewidth]{figures/MULTI/HUV_LH/100AU/CO_HUV_LH.pdf}  
\end{subfigure}
\begin{subfigure}{.32\linewidth}
  \centering
  \includegraphics[width=1.00\linewidth]{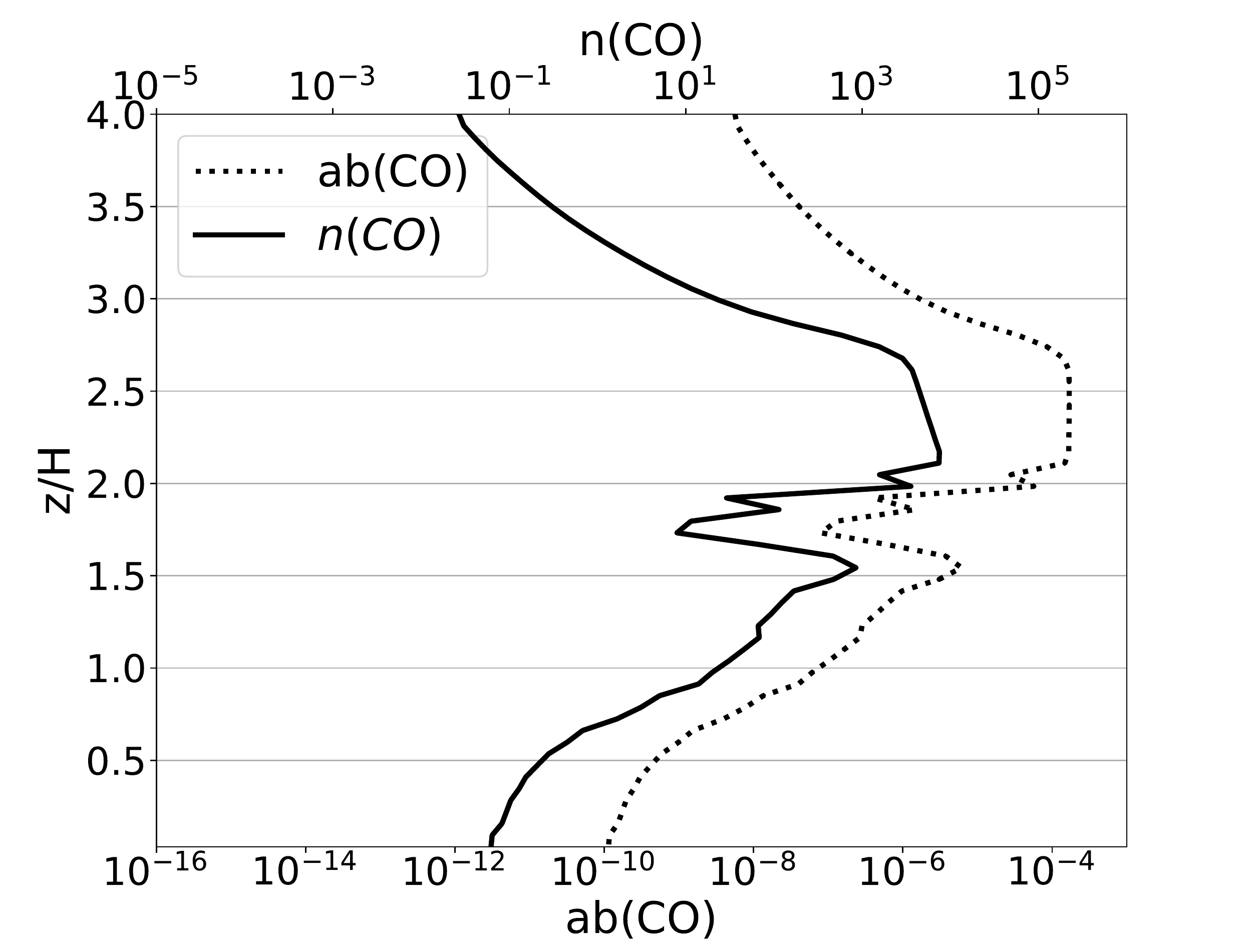}
\end{subfigure}

\begin{subfigure}{.32\linewidth}
  \centering
  \includegraphics[width=1.00\linewidth]{figures/SINGLE/HUV_HL_Teff/100AU/CS_HUV_HL_Teff.pdf}
\end{subfigure}
\begin{subfigure}{.32\linewidth}
  \centering
  \includegraphics[width=1.00\linewidth]{figures/MULTI/HUV_LH/100AU/CS_HUV_LH.pdf}
\end{subfigure}
\begin{subfigure}{.32\linewidth}
  \centering
  \includegraphics[width=1.00\linewidth]{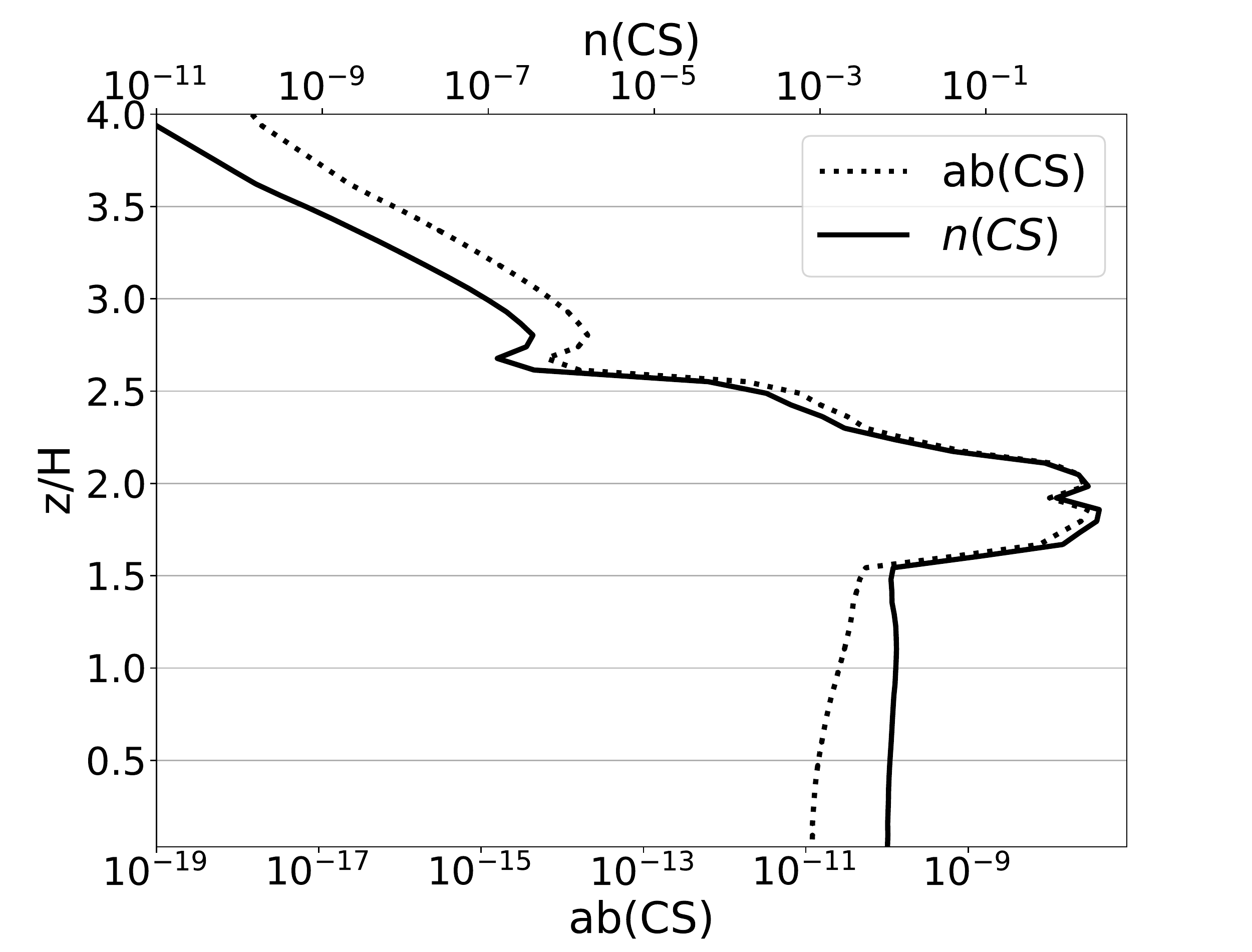}
\end{subfigure}

\begin{subfigure}{.32\linewidth}
  \centering
  \includegraphics[width=1.00\linewidth]{figures/SINGLE/HUV_HL_Teff/100AU/CN_HUV_HL_Teff.pdf}
   \subcaption{HUV-LH-$\mathrm{T_{a}}$}   
\end{subfigure}
\begin{subfigure}{.32\linewidth}
  \centering
  \includegraphics[width=1.00\linewidth]{figures/MULTI/HUV_LH/100AU/CN_HUV_LH.pdf}
   \subcaption{M-HUV-LH}   
\end{subfigure}
\begin{subfigure}{.32\linewidth}
  \centering
  \includegraphics[width=1.00\linewidth]{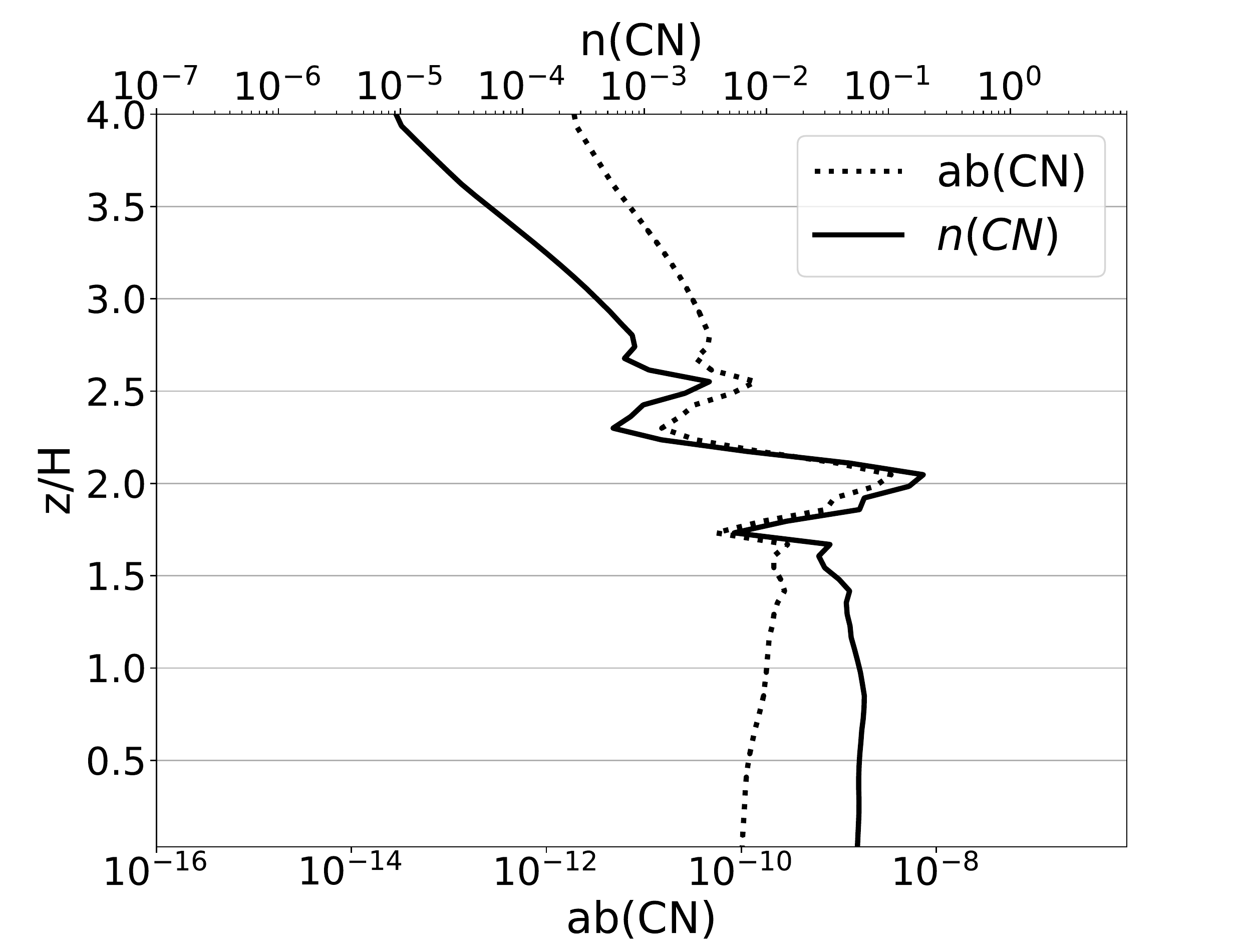}
   \subcaption{Set-HUV-LH-$\mathrm{T_{a}}$} 
\end{subfigure}

\caption{{Vertical profiles of H, $\mathrm{H_2}$, CO, CS and CN at 100 au from the star of the single-grain model HUV-LH-$\mathrm{T_{a}}$ (left), the multi-grain model M-HUV-LH (middle), and the intermediate model Set-HUV-LH-$\mathrm{T_{a}}$ (right). The dotted line is the abundance relative to H and the solid line is the density [$cm^{-3}$].}}
\label{fig:interm-100profile_LH}
\end{figure*}

\begin{figure*}
\begin{subfigure}{.32\linewidth}
  \centering
  \includegraphics[width=1.00\linewidth]{figures/SINGLE/HUV_B14_Teff/100AU/H_HUV_B14_Teff.pdf}
\end{subfigure}
\begin{subfigure}{.32\linewidth}
  \centering
  \includegraphics[width=1.00\linewidth]{figures/MULTI/HUV_B14/100AU/H_HUV_B14.pdf}
\end{subfigure}
\begin{subfigure}{.32\linewidth}
  \centering
  \includegraphics[width=1.00\linewidth]{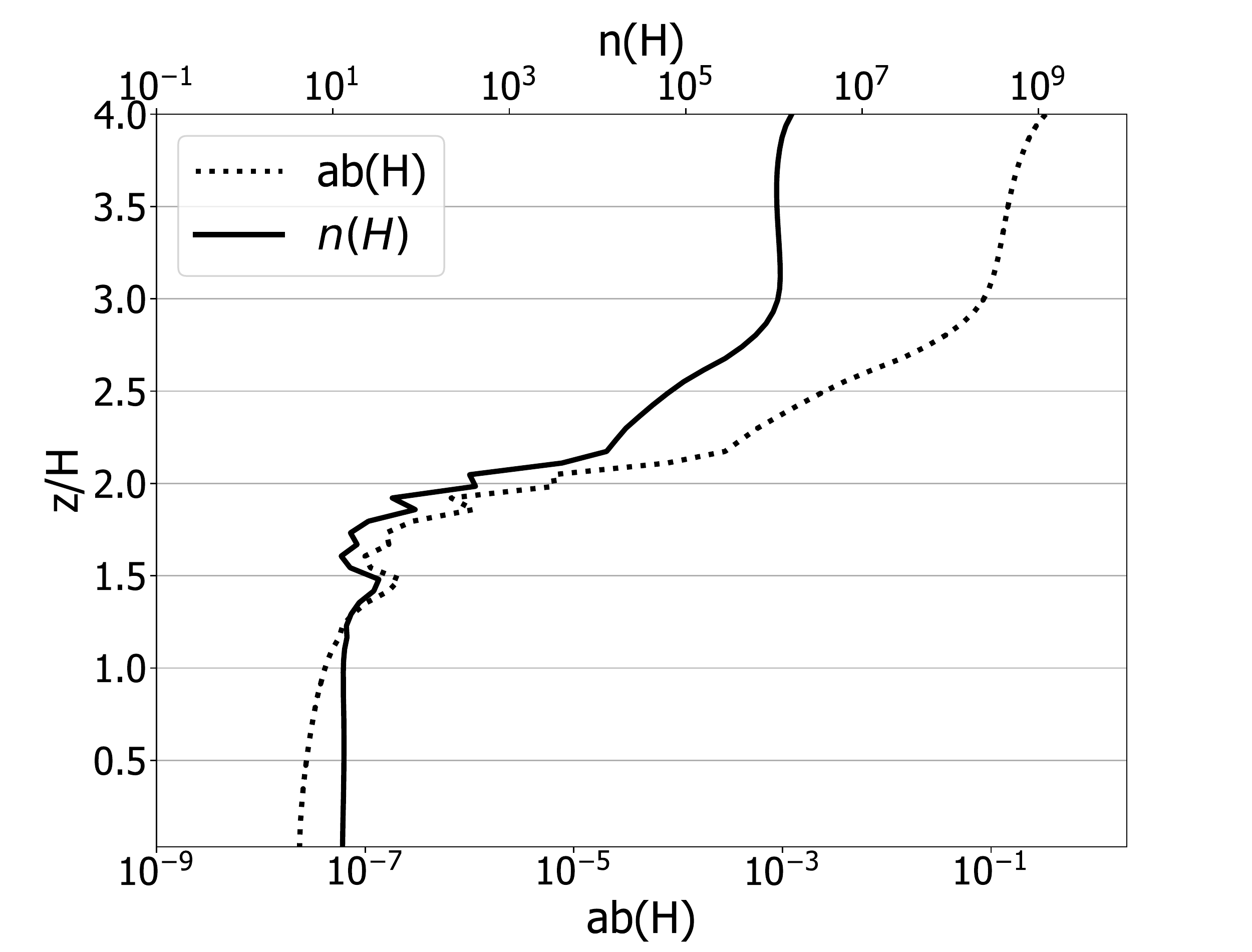}
\end{subfigure}

\begin{subfigure}{.32\linewidth}
  \centering
  \includegraphics[width=1.00\linewidth]{figures/SINGLE/HUV_B14_Teff/100AU/H2_HUV_B14_Teff.pdf}
\end{subfigure}
\begin{subfigure}{.32\linewidth}
  \centering
  \includegraphics[width=1.00\linewidth]{figures/MULTI/HUV_B14/100AU/H2_HUV_B14.pdf}
\end{subfigure}
\begin{subfigure}{.32\linewidth}
  \centering
  \includegraphics[width=1.00\linewidth]{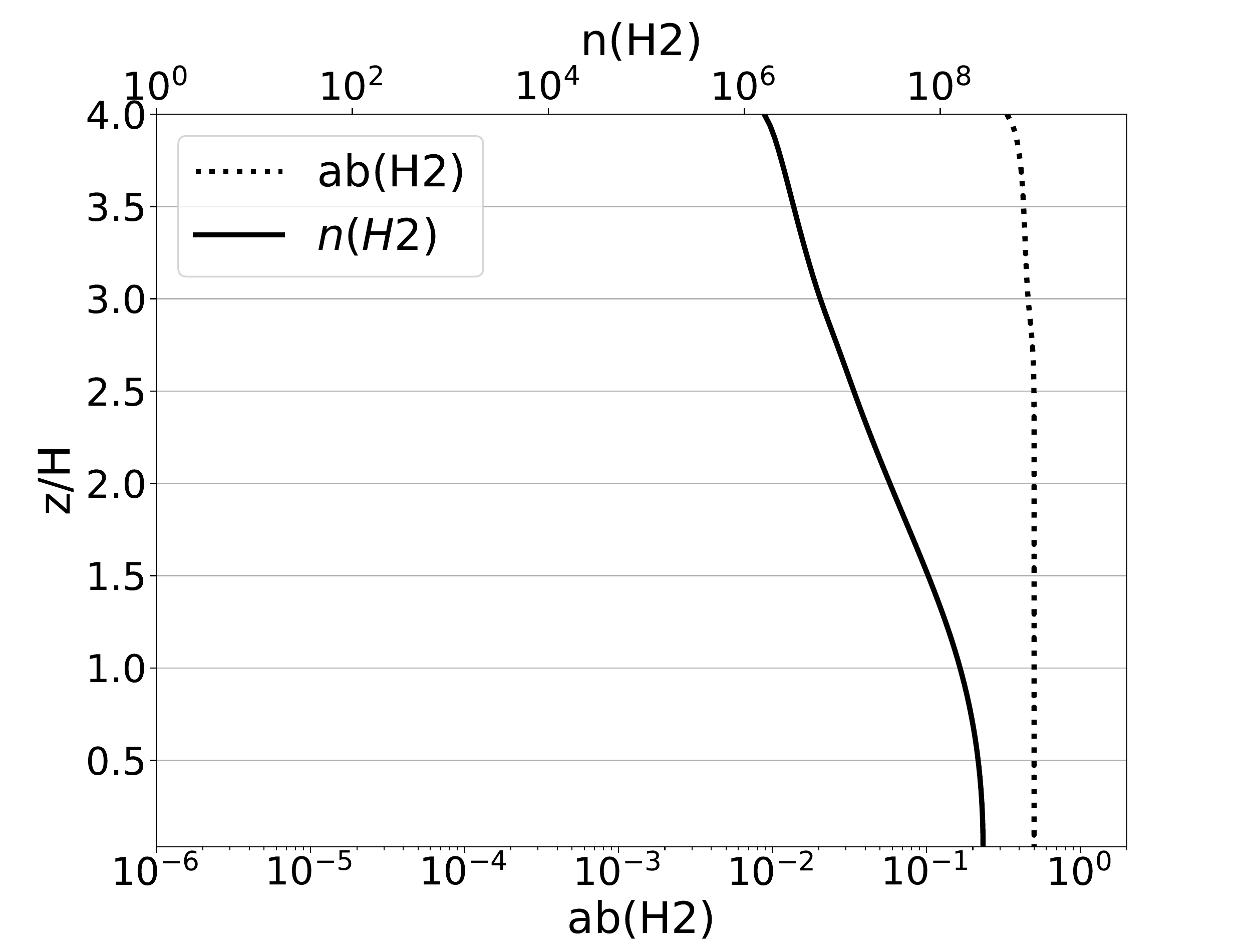}
\end{subfigure}

\begin{subfigure}{.32\linewidth}
  \centering
  \includegraphics[width=1.00\linewidth]{figures/SINGLE/HUV_B14_Teff/100AU/CO_HUV_B14_Teff.pdf}
\end{subfigure}
\begin{subfigure}{.32\linewidth}
  \centering
  \includegraphics[width=1.00\linewidth]{figures/MULTI/HUV_B14/100AU/CO_HUV_B14.pdf}  
\end{subfigure}
\begin{subfigure}{.32\linewidth}
  \centering
  \includegraphics[width=1.00\linewidth]{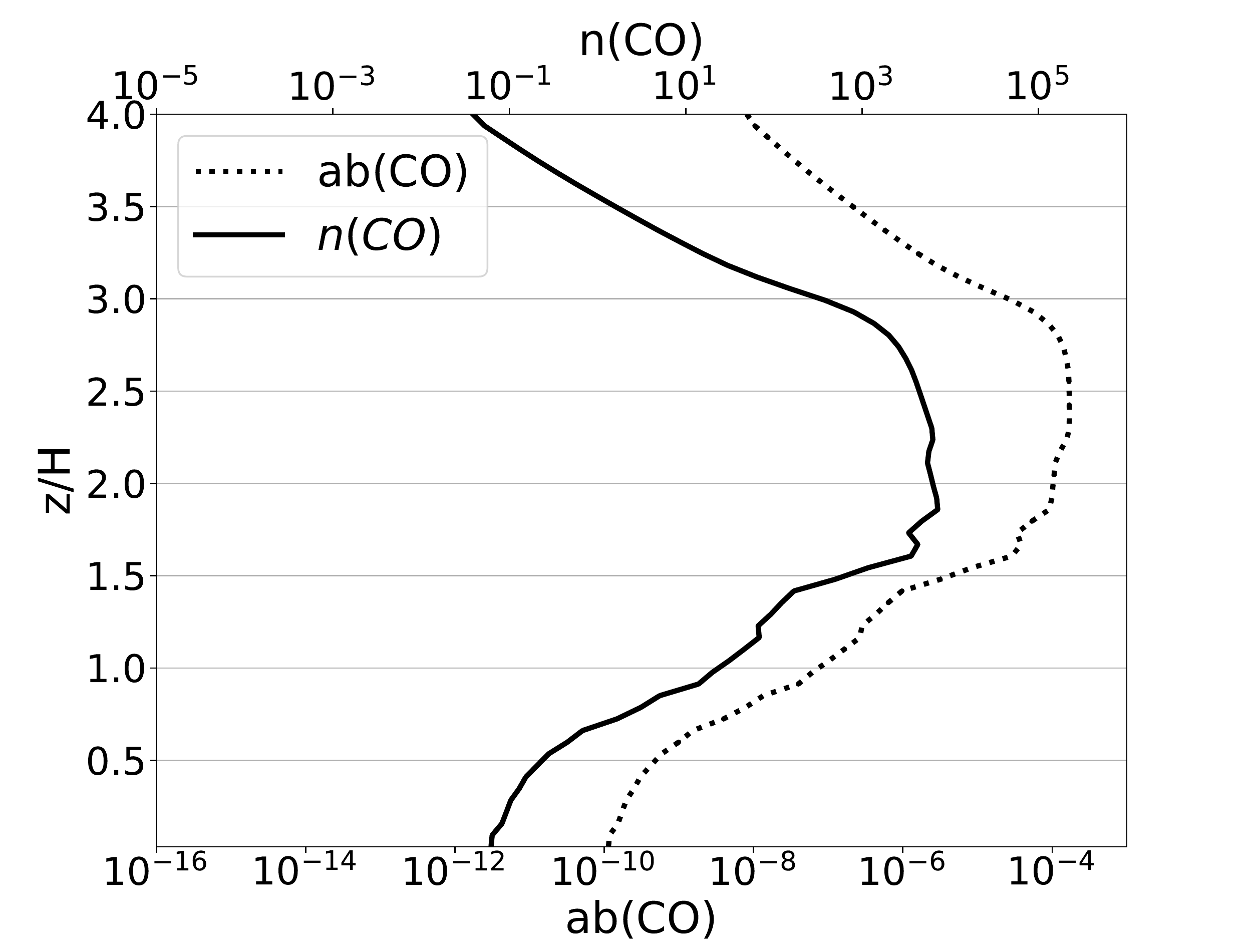}
\end{subfigure}

\begin{subfigure}{.32\linewidth}
  \centering
  \includegraphics[width=1.00\linewidth]{figures/SINGLE/HUV_B14_Teff/100AU/CS_HUV_B14_Teff.pdf}
\end{subfigure}
\begin{subfigure}{.32\linewidth}
  \centering
  \includegraphics[width=1.00\linewidth]{figures/MULTI/HUV_B14/100AU/CS_HUV_B14.pdf}
\end{subfigure}
\begin{subfigure}{.32\linewidth}
  \centering
  \includegraphics[width=1.00\linewidth]{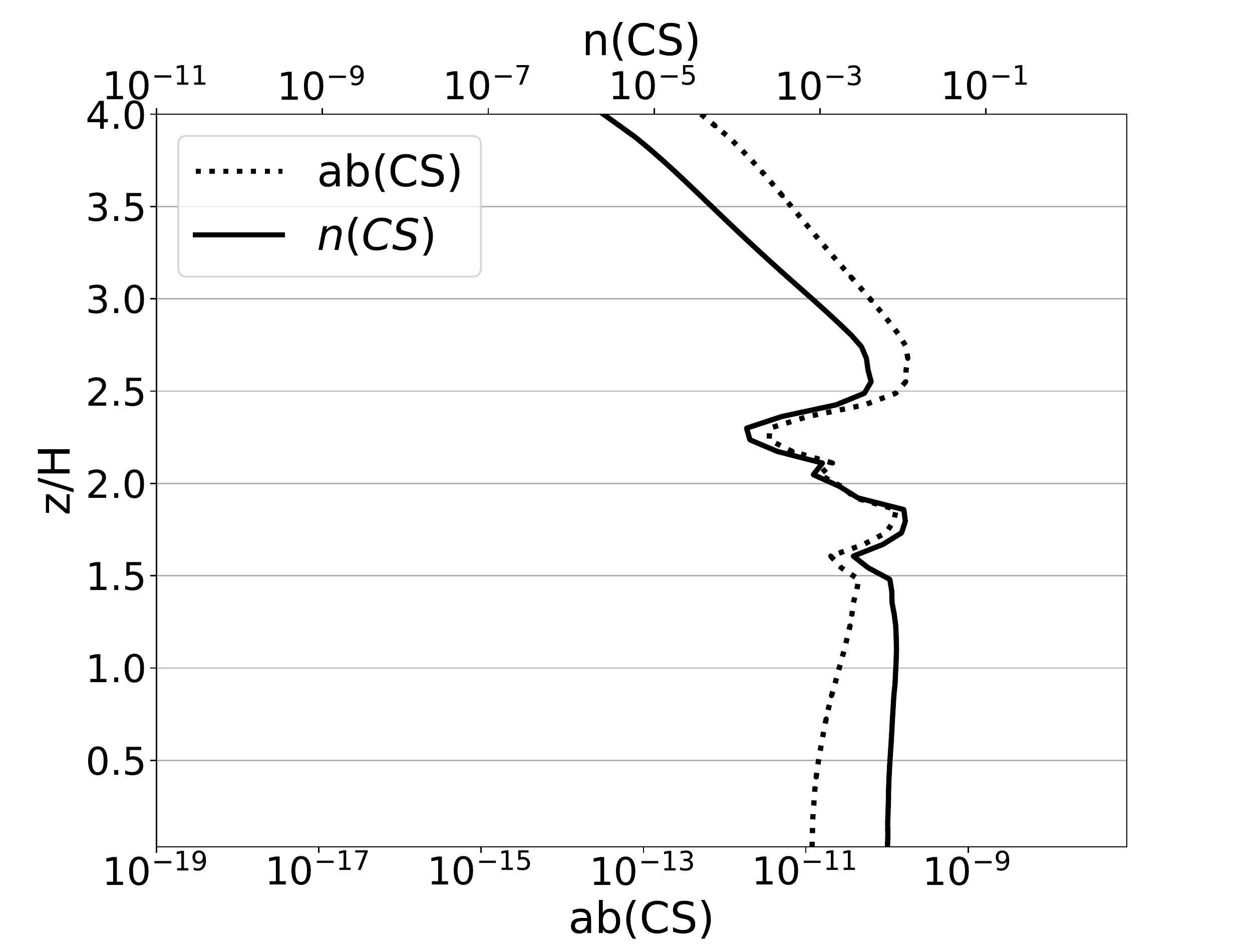}
\end{subfigure}

\begin{subfigure}{.32\linewidth}
  \centering
  \includegraphics[width=1.00\linewidth]{figures/SINGLE/HUV_B14_Teff/100AU/CN_HUV_B14_Teff.pdf}
   \subcaption{HUV-B14-$\mathrm{T_{a}}$}   
\end{subfigure}
\begin{subfigure}{.32\linewidth}
  \centering
  \includegraphics[width=1.00\linewidth]{figures/MULTI/HUV_B14/100AU/CN_HUV_B14.pdf}
   \subcaption{M-HUV-B14}   
\end{subfigure}
\begin{subfigure}{.32\linewidth}
  \centering
  \includegraphics[width=1.00\linewidth]{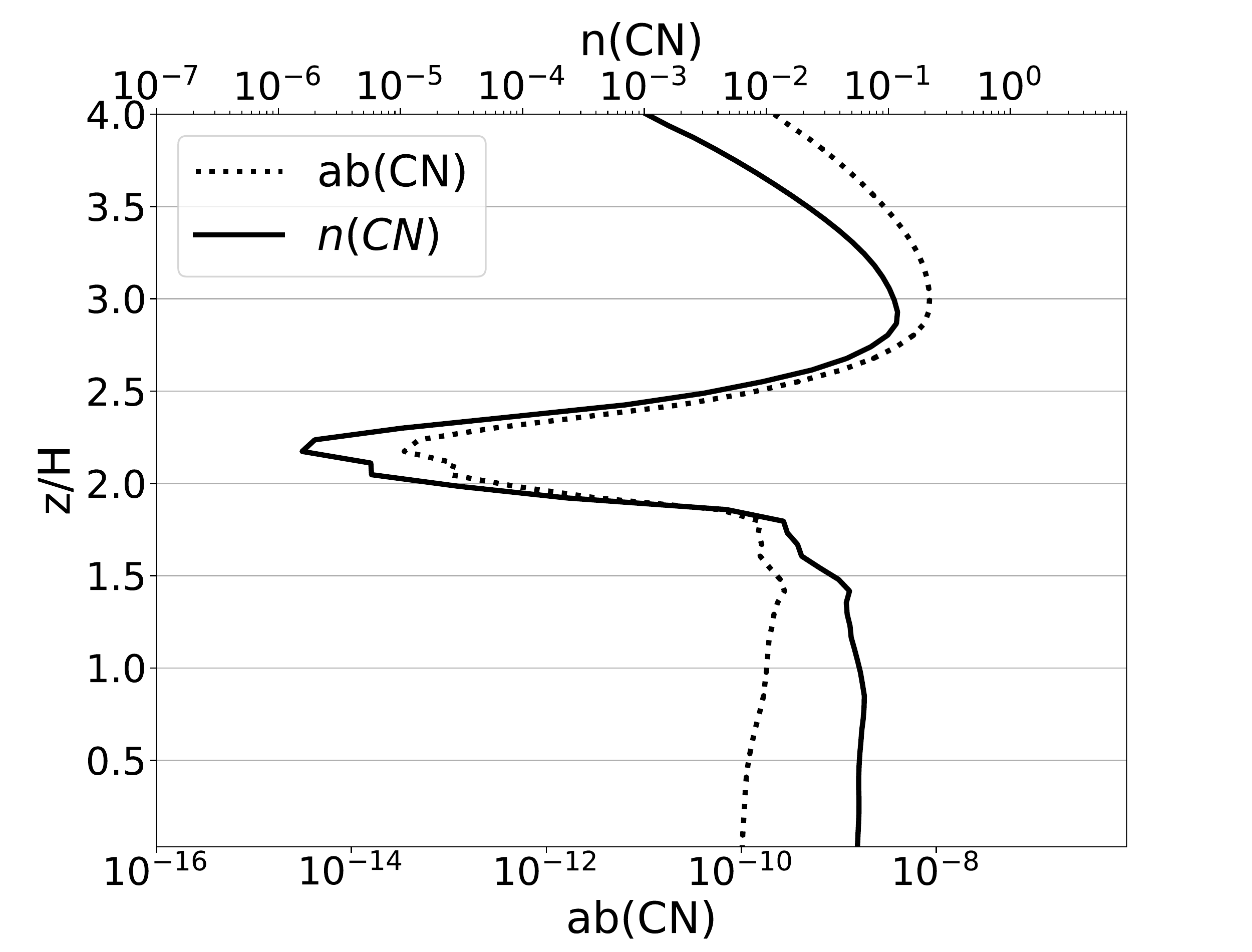}
   \subcaption{Set-HUV-B14-$\mathrm{T_{a}}$} 
\end{subfigure}

\caption{{Vertical profiles of H, $\mathrm{H_2}$, CO, CS and CN at 100 au from the star of the single-grain model HUV-B14-$\mathrm{T_{a}}$ (left), the multi-grain model M-HUV-B14 (middle), and the intermediate model Set-HUV-B14-$\mathrm{T_{a}}$ (right). The dotted line is the abundance relative to H and the solid line is the density [$cm^{-3}$].}}
\label{fig:interm-100profile_B14}
\end{figure*}

\revisionC{For simplicity, we only compute the chemistry for the HUV flux at 50, 100 and 200 au. 
Results are presented in Figures.\,\ref{fig:interm-100profile_LH}, \ref{fig:interm-100profile_B14} and have been added in Figures\,\ref{fig:compare_high}, \ref{fig:reservoirs} and \ref{fig:ab_surface} for discussion.}

\clearpage

%

\section{LUV cases}\label{app:luv}
\begin{figure*}
\begin{subfigure}{.48\linewidth}
  \centering
  \includegraphics[width=1.00\linewidth]{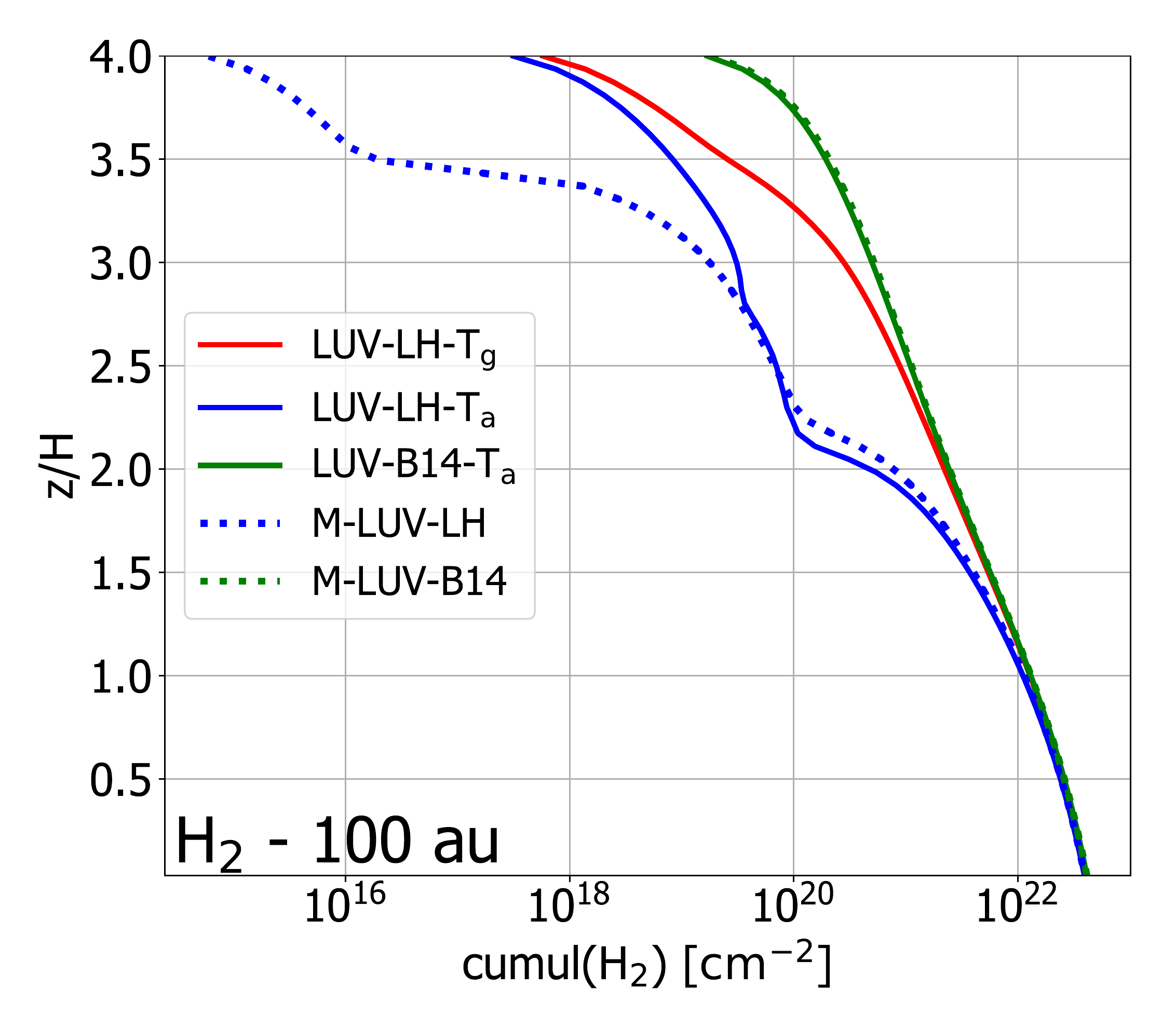}
   \subcaption{H2}   
\end{subfigure}
\begin{subfigure}{.48\linewidth}
  \centering
  \includegraphics[width=1.00\linewidth]{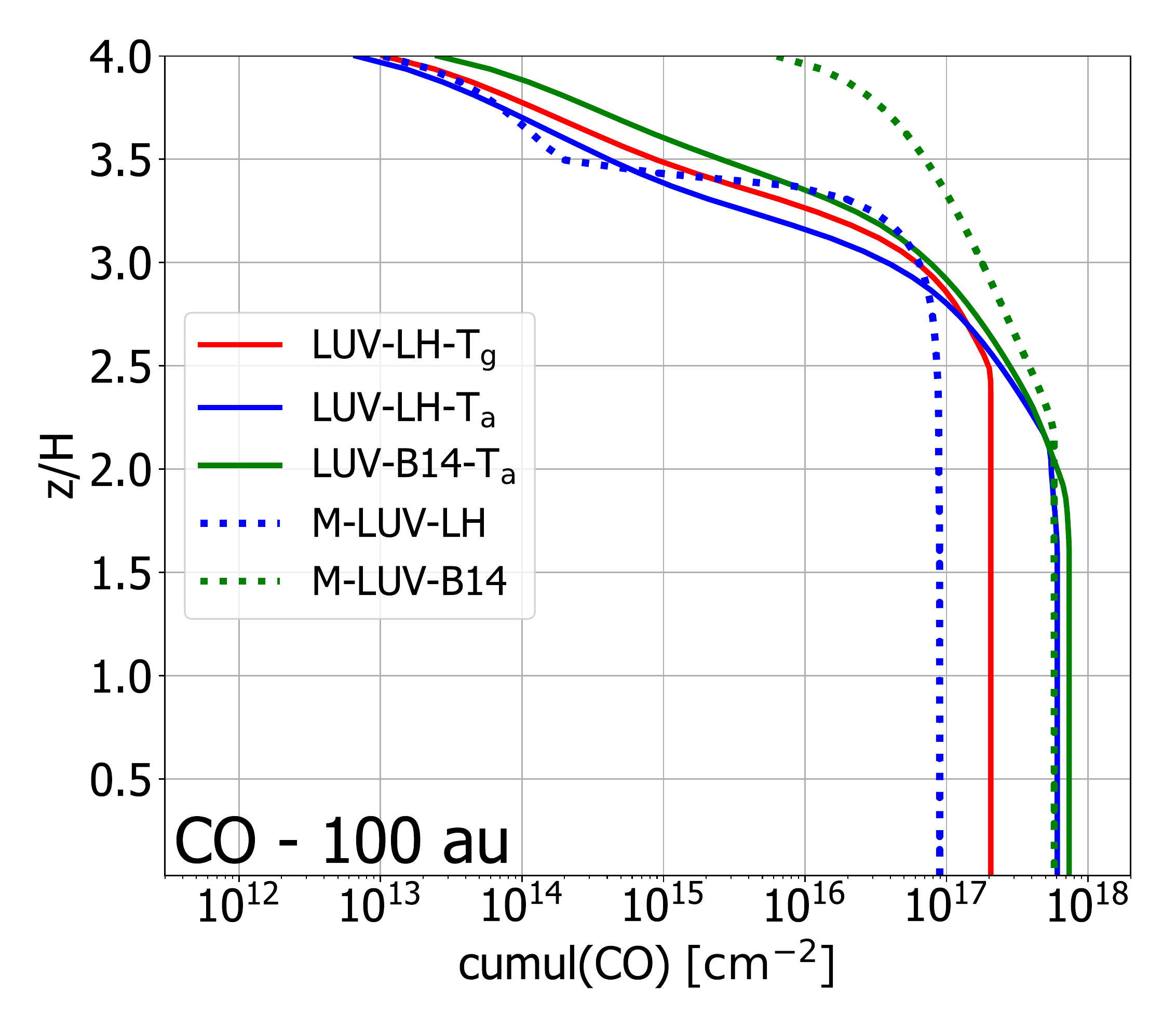}
   \subcaption{CO}   
\end{subfigure}\\
\begin{subfigure}{.48\linewidth}
  \centering
  \includegraphics[width=1.00\linewidth]{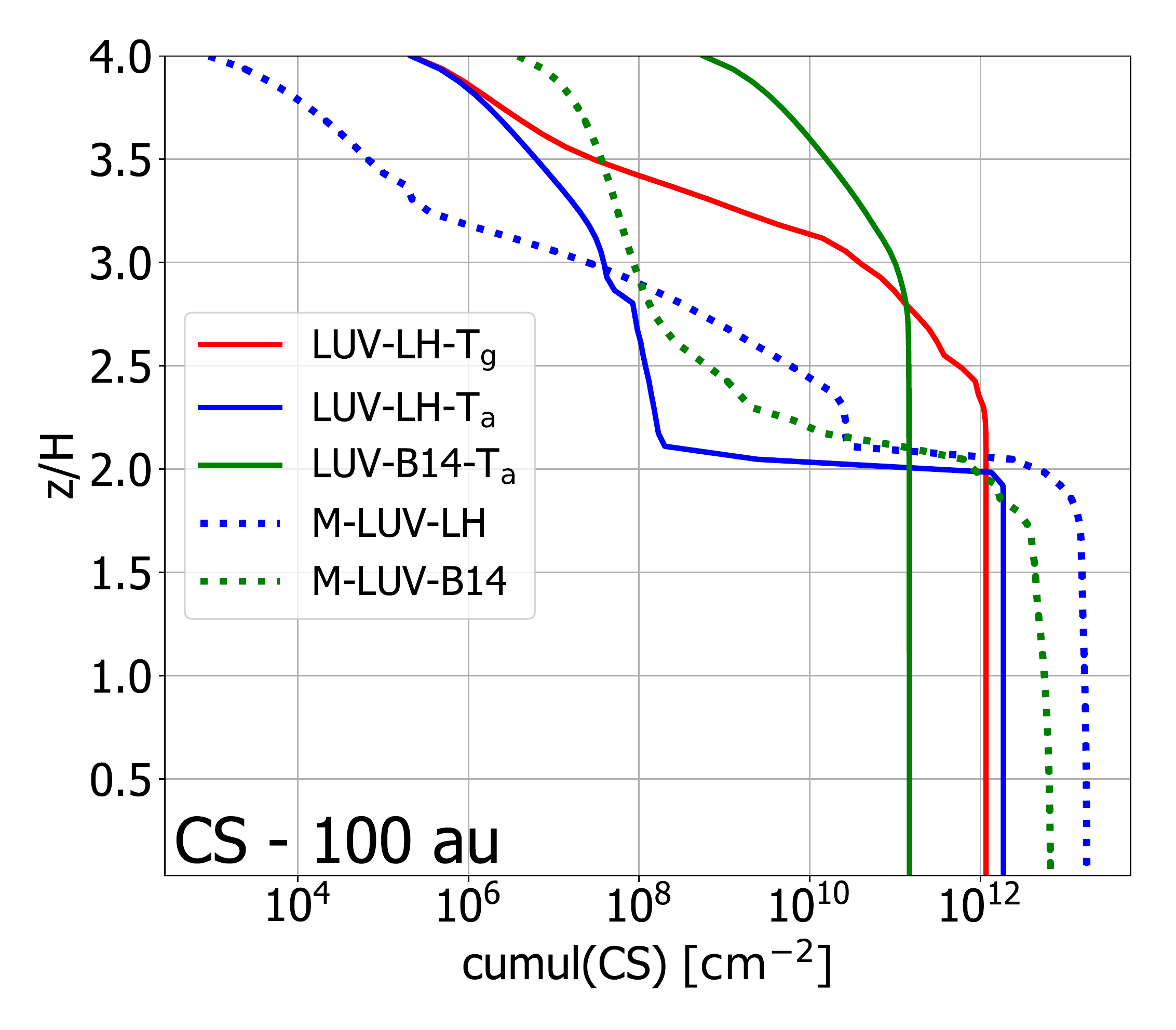}
   \subcaption{CS} 
\end{subfigure}
\begin{subfigure}{.48\linewidth}
  \centering
  \includegraphics[width=1.00\linewidth]{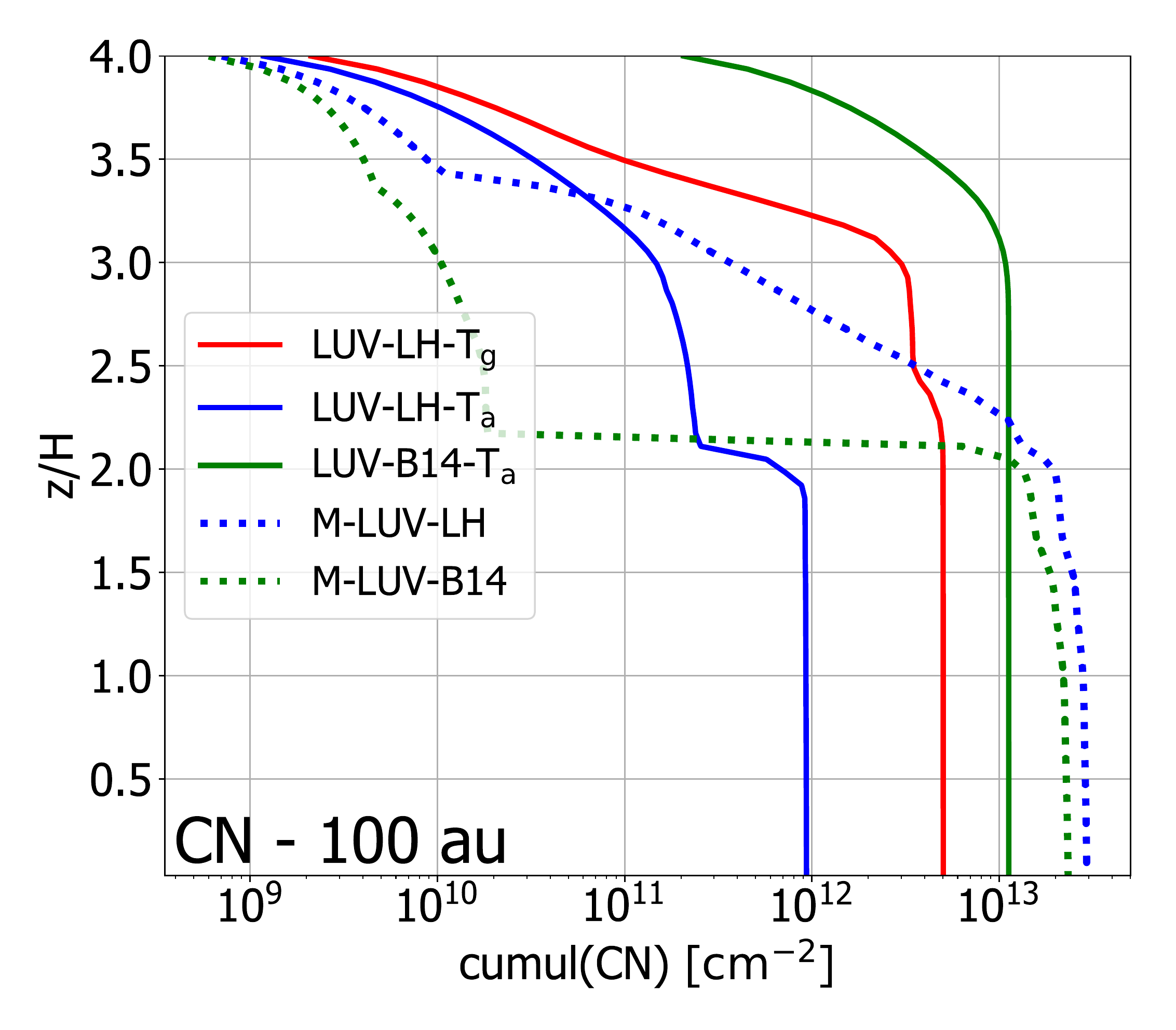}
   \subcaption{CN} 
\end{subfigure}

\caption{Vertical cumulative surface density [$\mathrm{cm^{-2}}$] of \hh, CO, CS and CN  at 100 au from the star of the LUV models.}
\label{fig:100cumul}
\end{figure*}

In \autoref{fig:100cumul} we present vertical cumulative column density 
of H$_2$, CO, CS and CN at 100 au from the LUV models. The profiles are 
very similar to that of the HUV models below $z \sim 2 H$ (see 
Fig.\,\ref{fig:100cumul_huv}). We shall note a few noticeable 
differences. The model \mllh produces a larger column density of CS 
than \mhlh. As for CN, it is clear that \mhb produces a larger column 
density than \mhlh. In the case of LUV multi-grain models, the 
difference is far less pronounced and \mllh produces a slightly larger 
column density than \mlb. These different results are mostly due to the 
lower UV penetration in the case of the LUV models than in the HUV 
models.  

\begin{figure*}
\begin{subfigure}{.33\linewidth}
  \centering
  \includegraphics[width=1.05\linewidth]{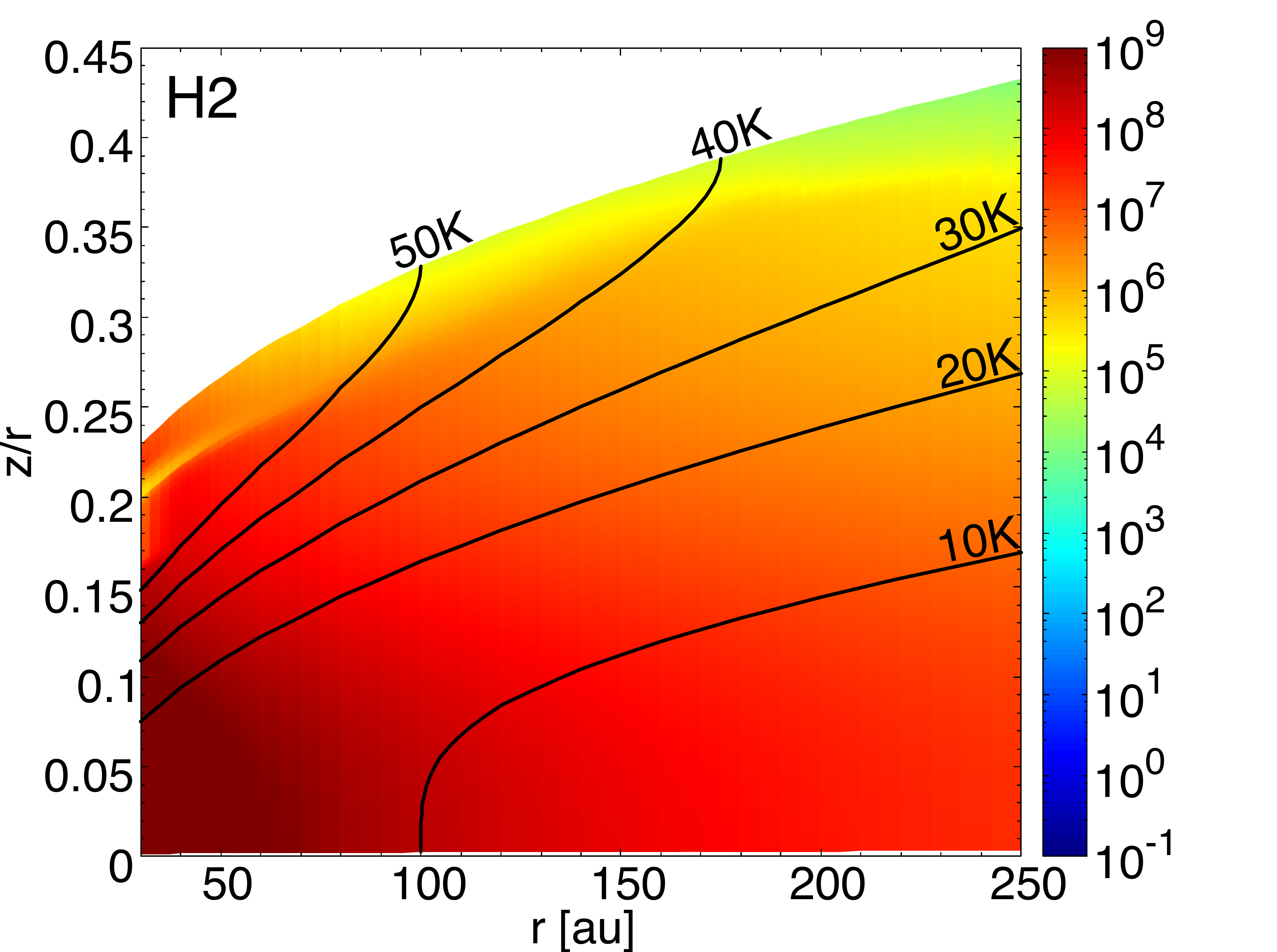} 
\end{subfigure}
\begin{subfigure}{.33\linewidth}
  \centering
  \includegraphics[width=1.05\linewidth]{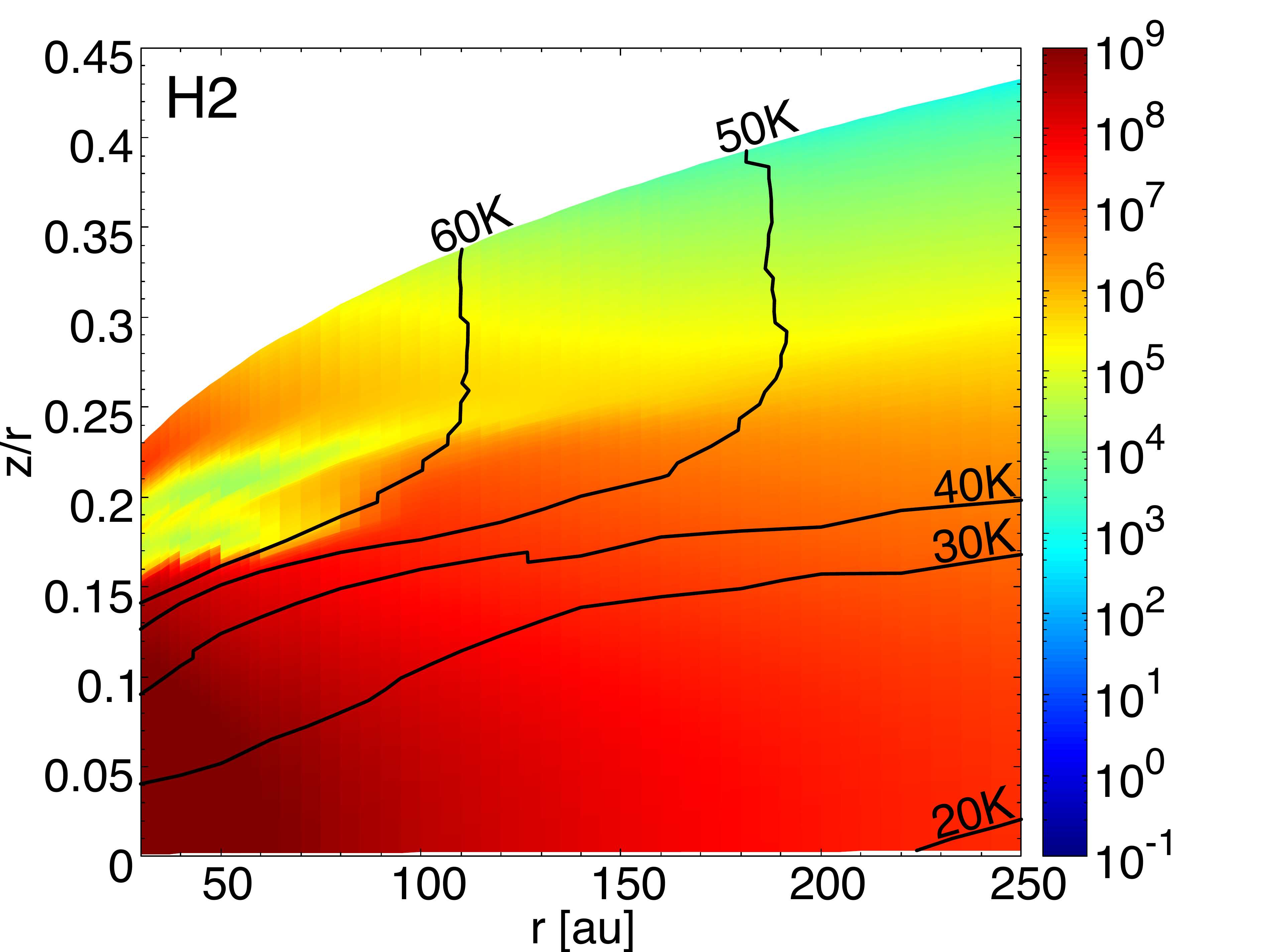} 
\end{subfigure}
\begin{subfigure}{.33\linewidth}
  \centering
  \includegraphics[width=1.05\linewidth]{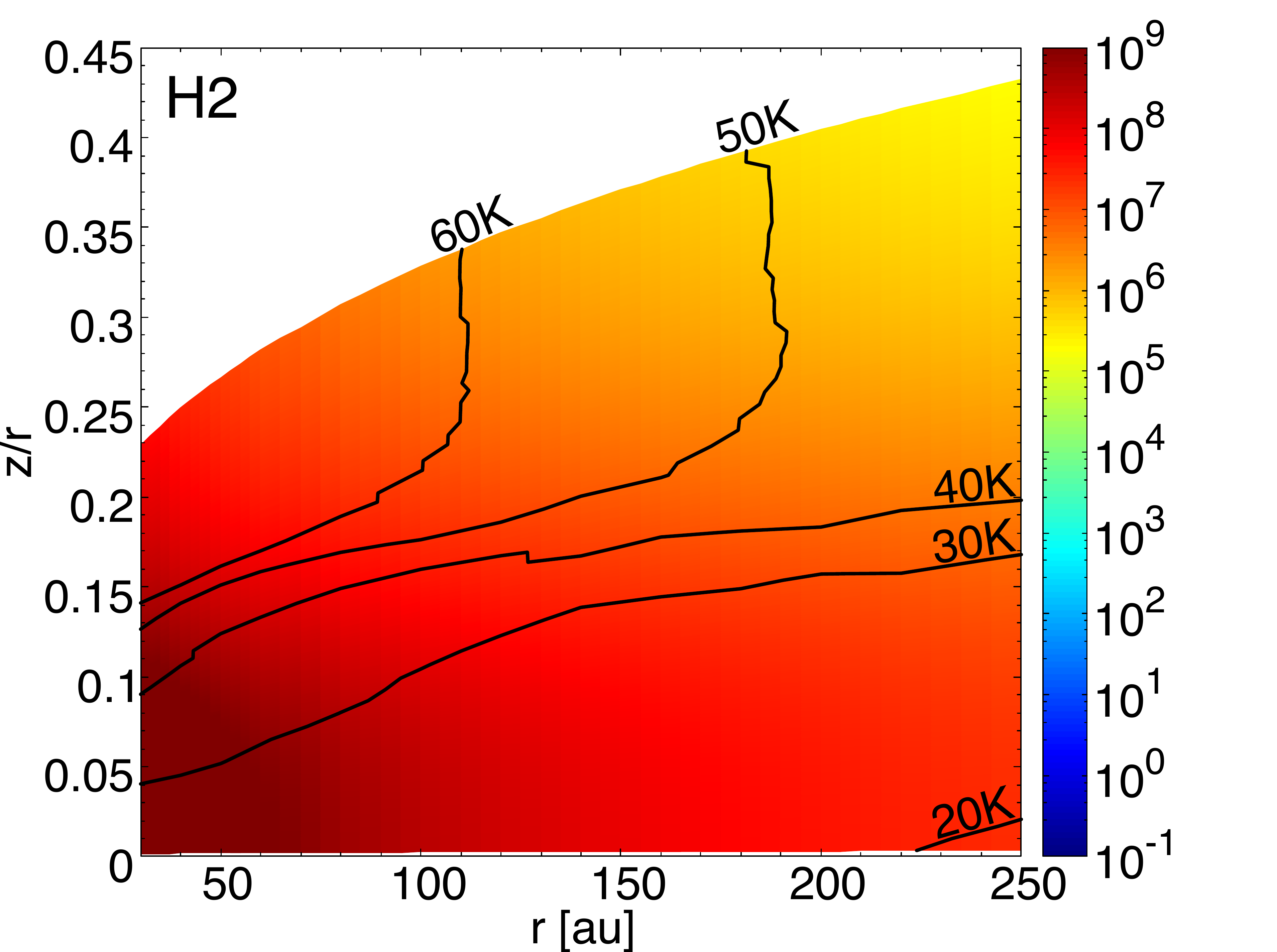} 
\end{subfigure}

\begin{subfigure}{.33\linewidth}
  \centering
  \includegraphics[width=1.05\linewidth]{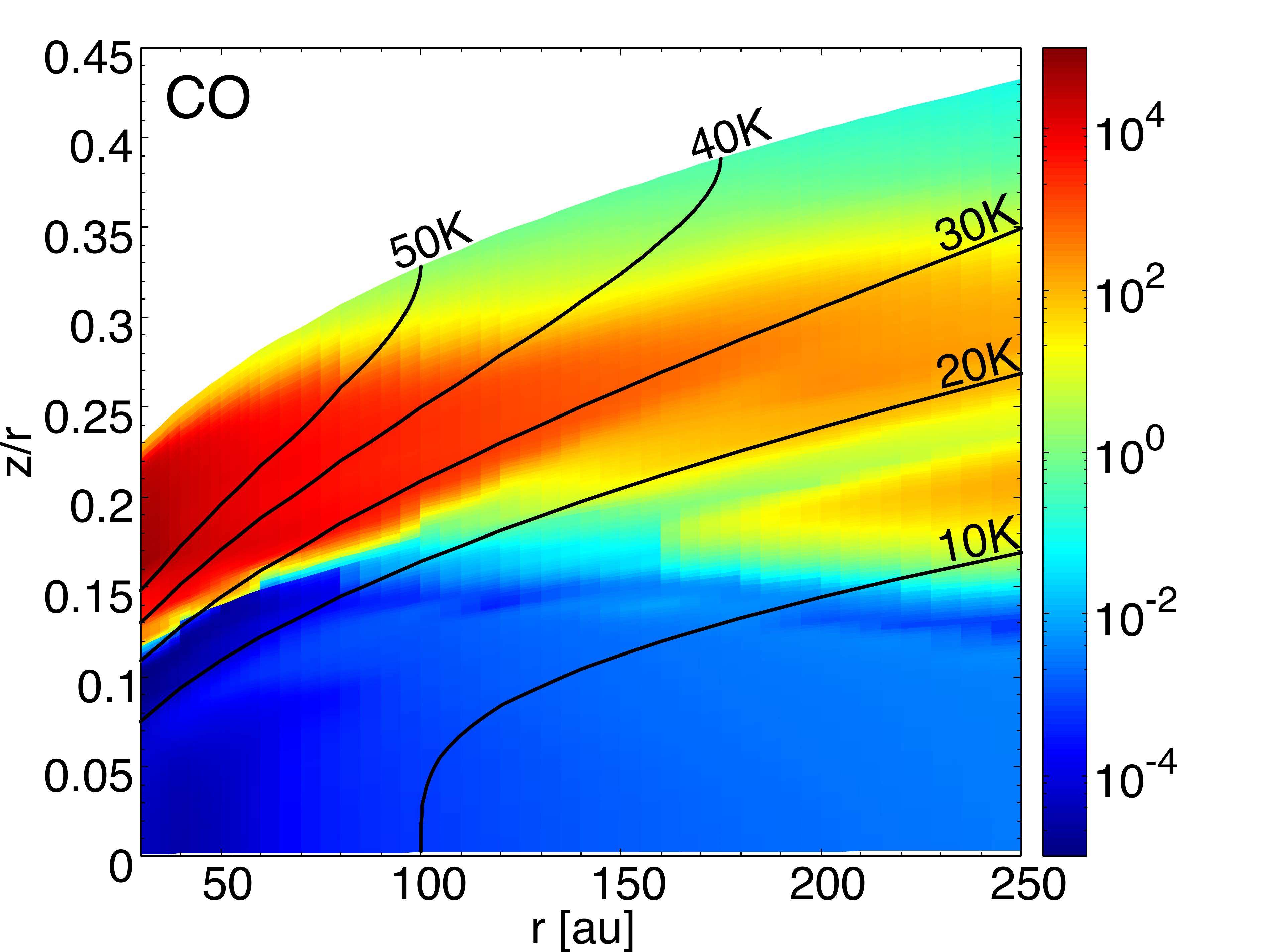} 
\end{subfigure}
\begin{subfigure}{.33\linewidth}
  \centering
  \includegraphics[width=1.05\linewidth]{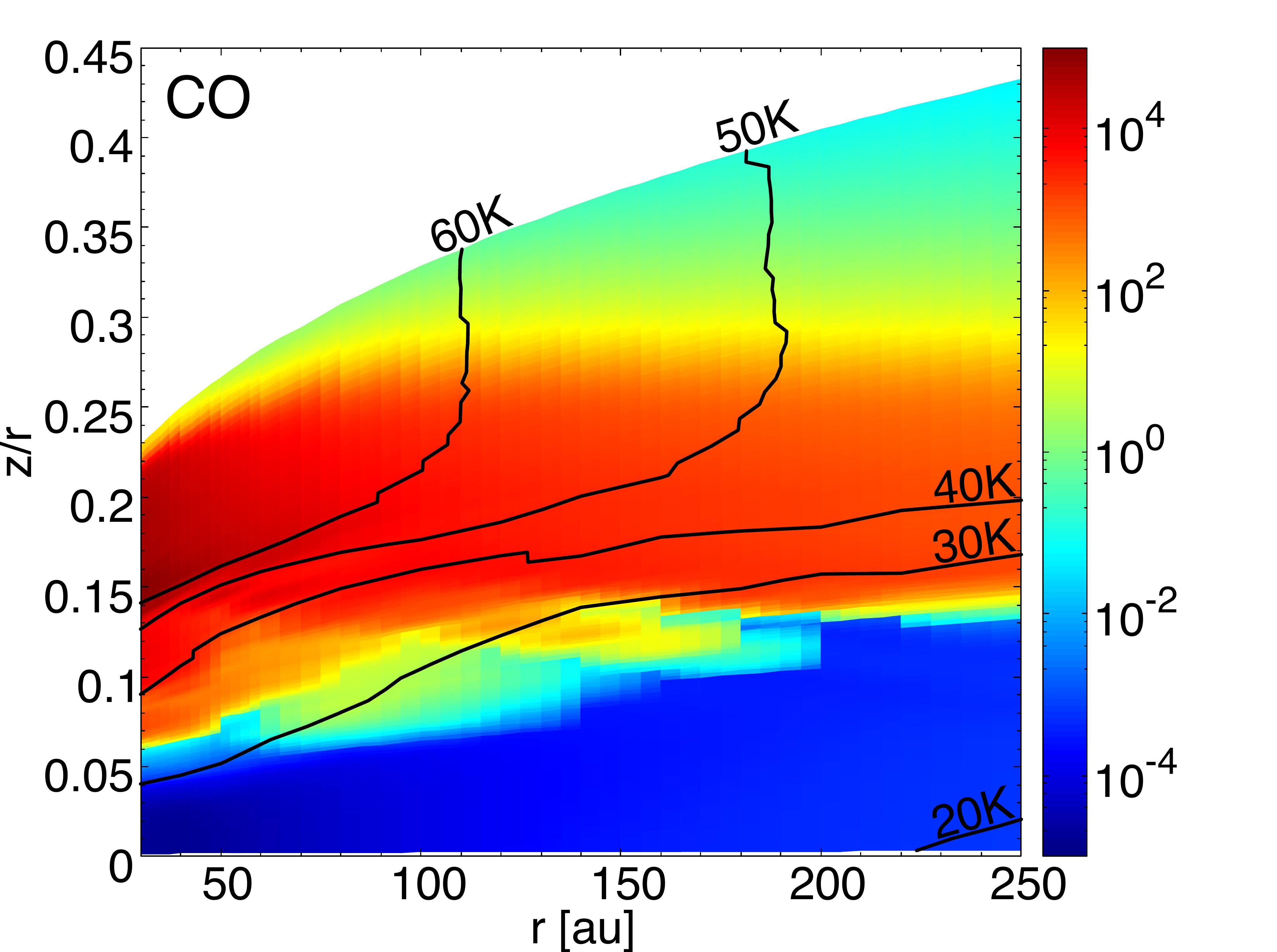}  
\end{subfigure}
\begin{subfigure}{.33\linewidth}
  \centering
  \includegraphics[width=1.05\linewidth]{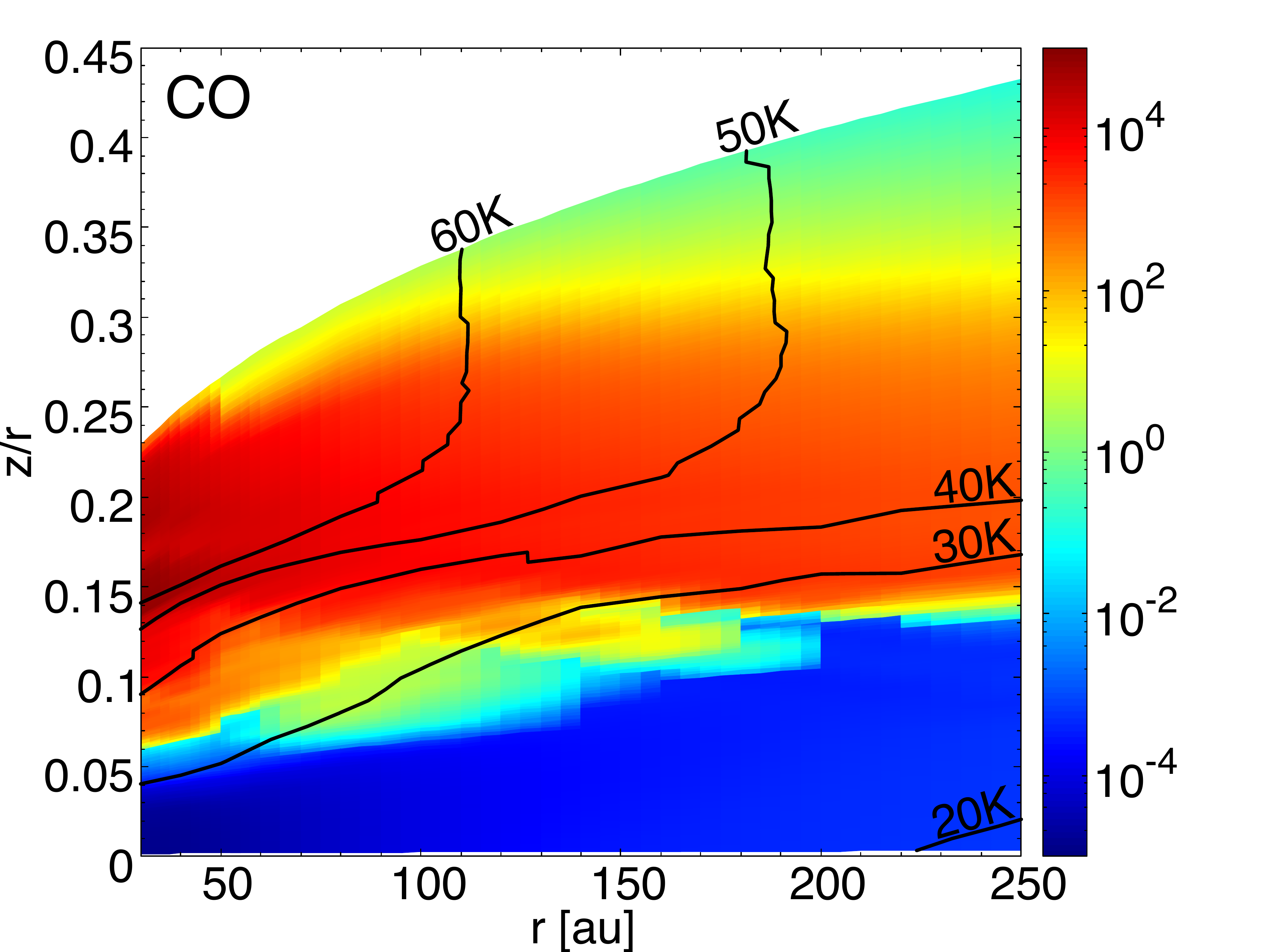} 
\end{subfigure}

\begin{subfigure}{.33\linewidth}
  \centering
  \includegraphics[width=1.05\linewidth]{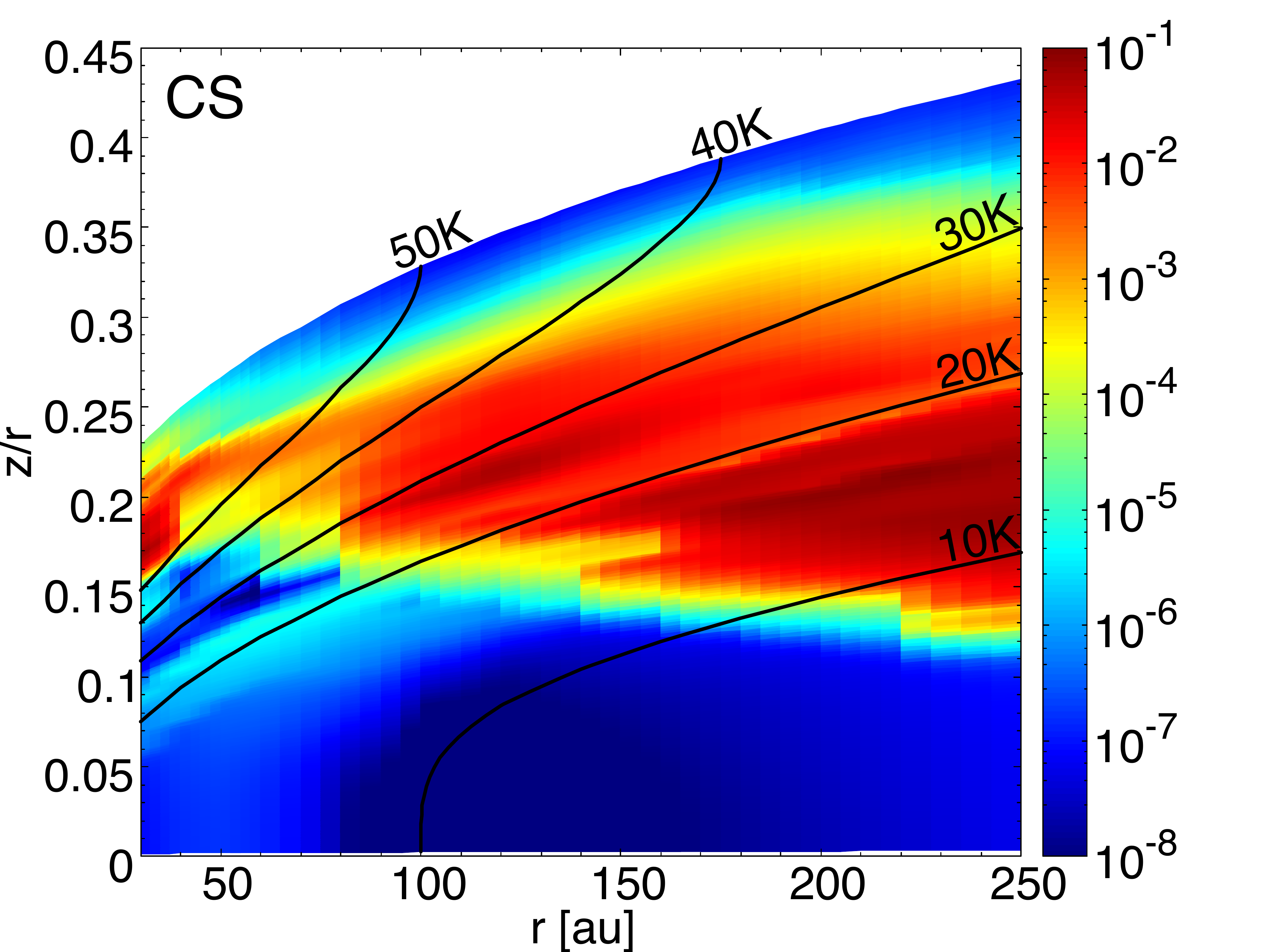} 
\end{subfigure}
\begin{subfigure}{.33\linewidth}
  \centering
  \includegraphics[width=1.05\linewidth]{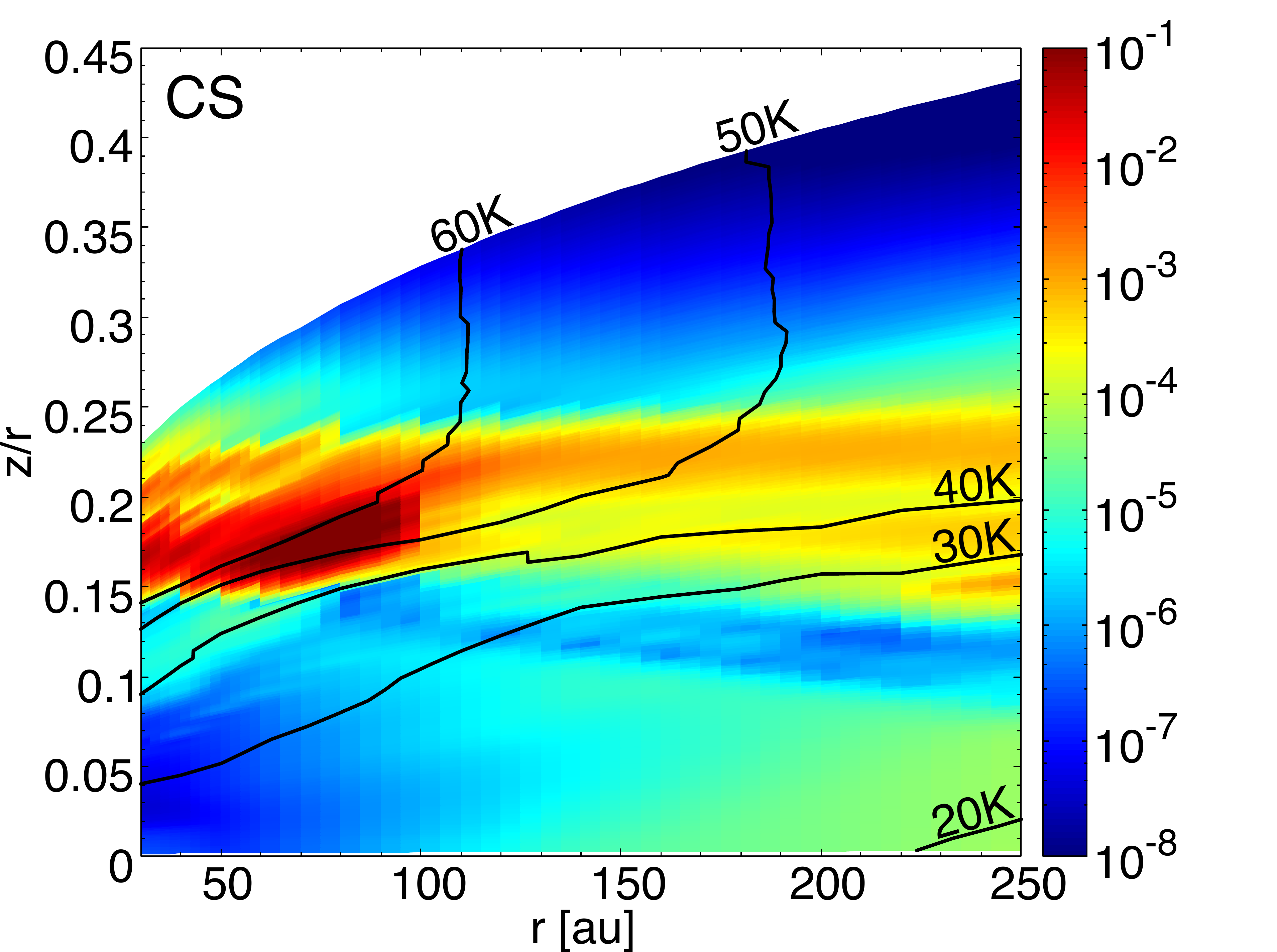} 
\end{subfigure}
\begin{subfigure}{.33\linewidth}
  \centering
  \includegraphics[width=1.05\linewidth]{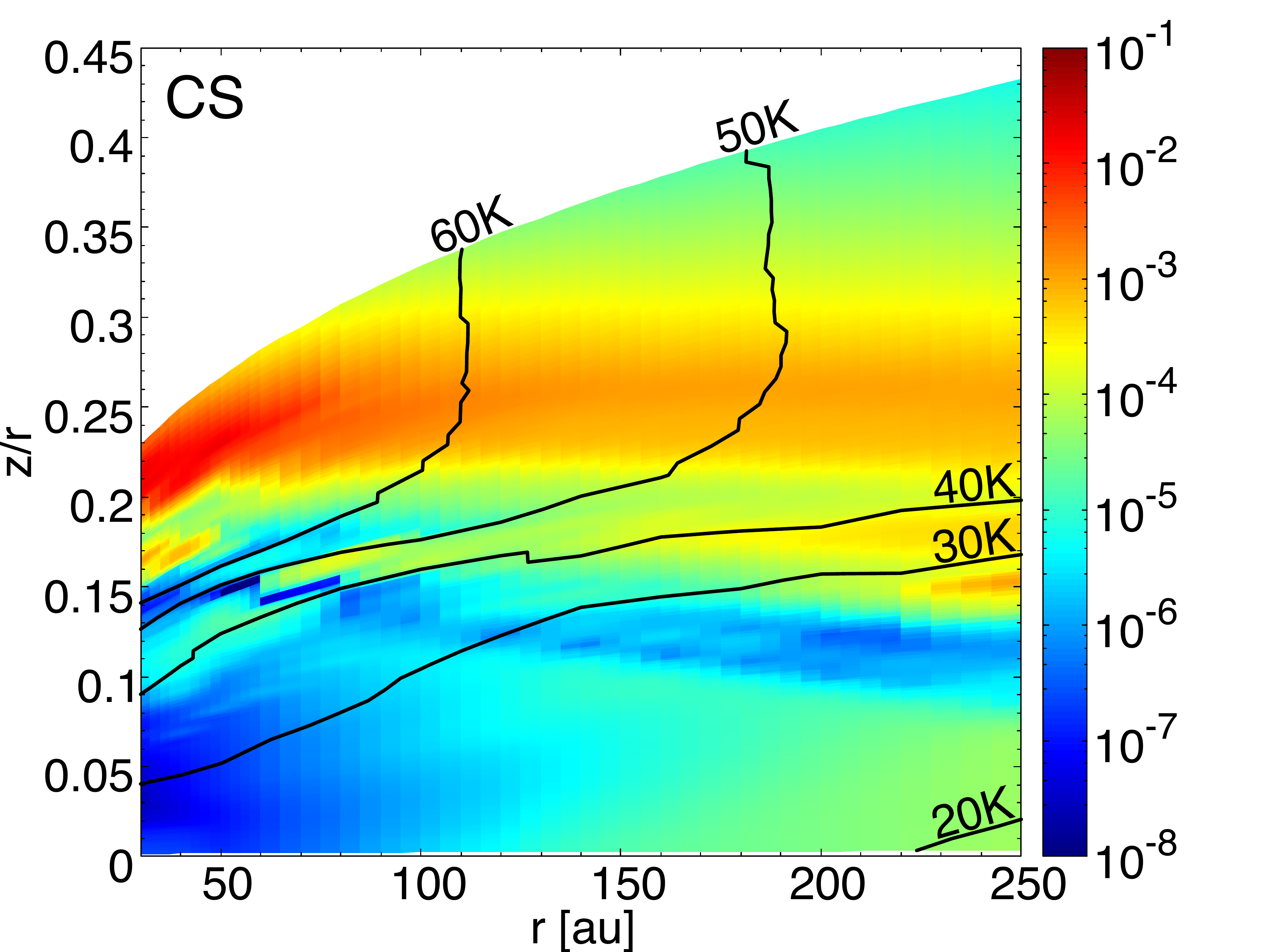} 
\end{subfigure}

\begin{subfigure}{.33\linewidth}
  \centering
  \includegraphics[width=1.05\linewidth]{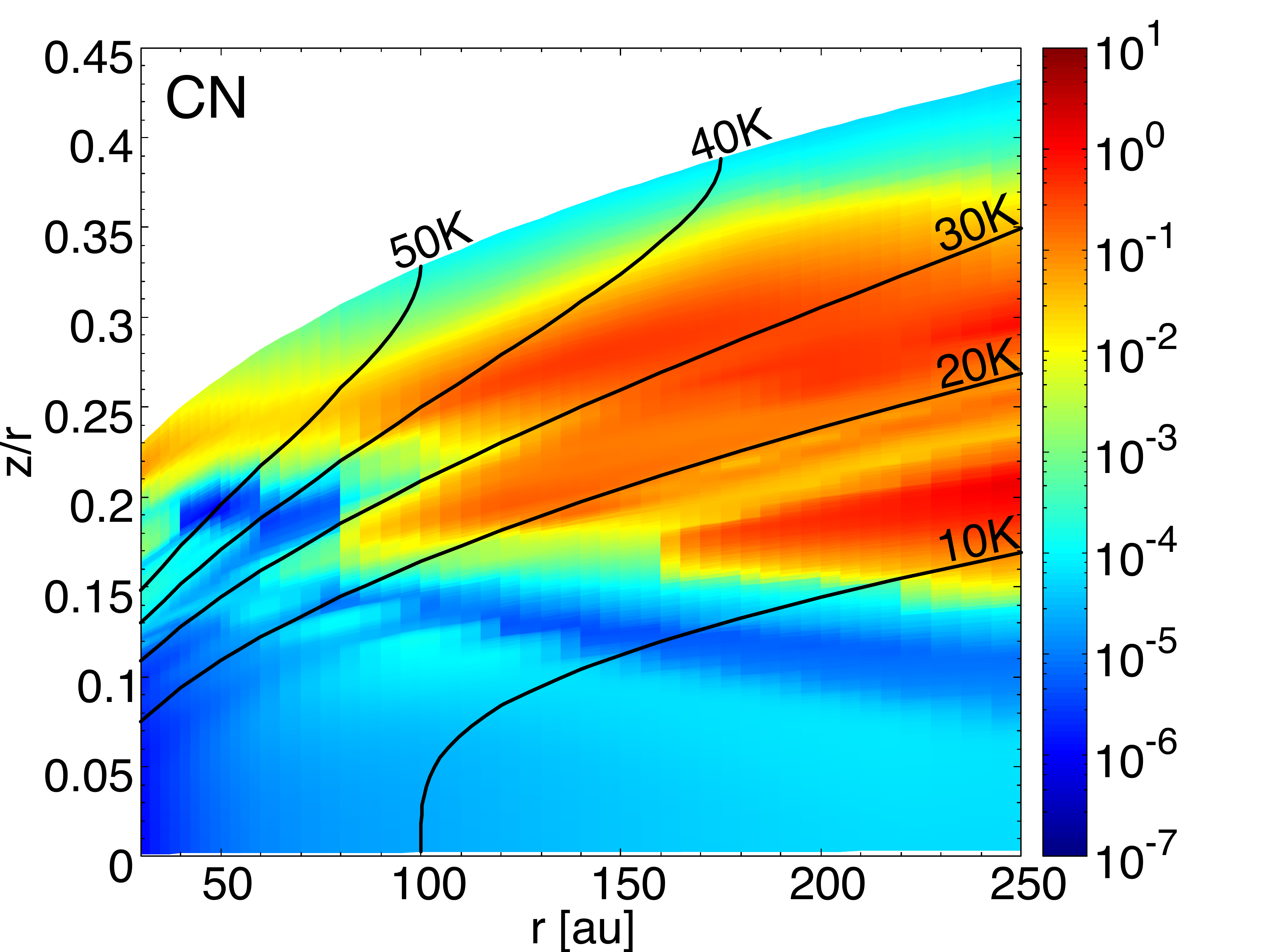} 
   \subcaption{LUV-LH-$\mathrm{T_{g}}$} 
\end{subfigure}
\begin{subfigure}{.33\linewidth}
  \centering
  \includegraphics[width=1.05\linewidth]{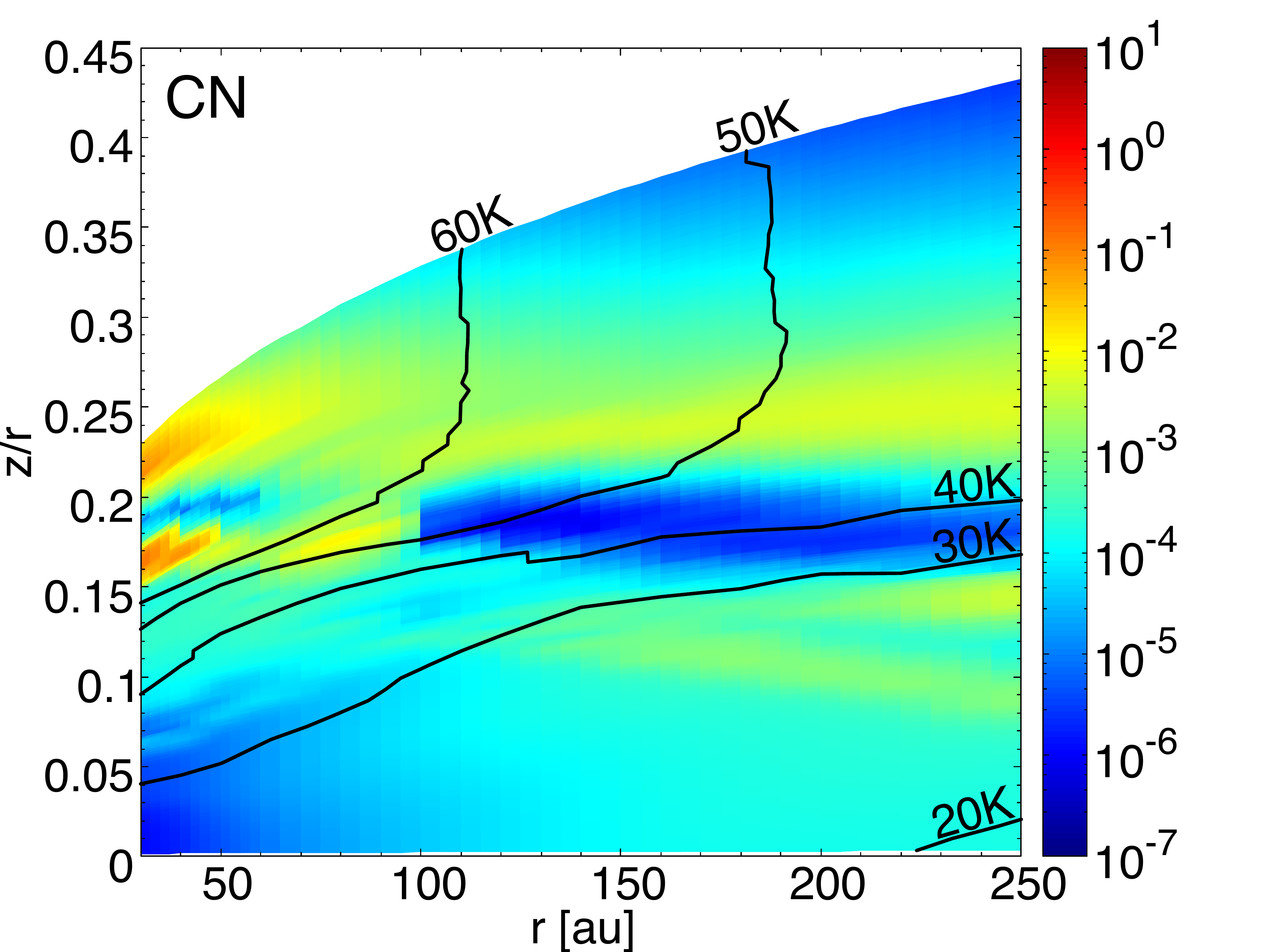} 
   \subcaption{LUV-LH-$\mathrm{T_{a}}$}  
\end{subfigure}
\begin{subfigure}{.33\linewidth}
  \centering
  \includegraphics[width=1.05\linewidth]{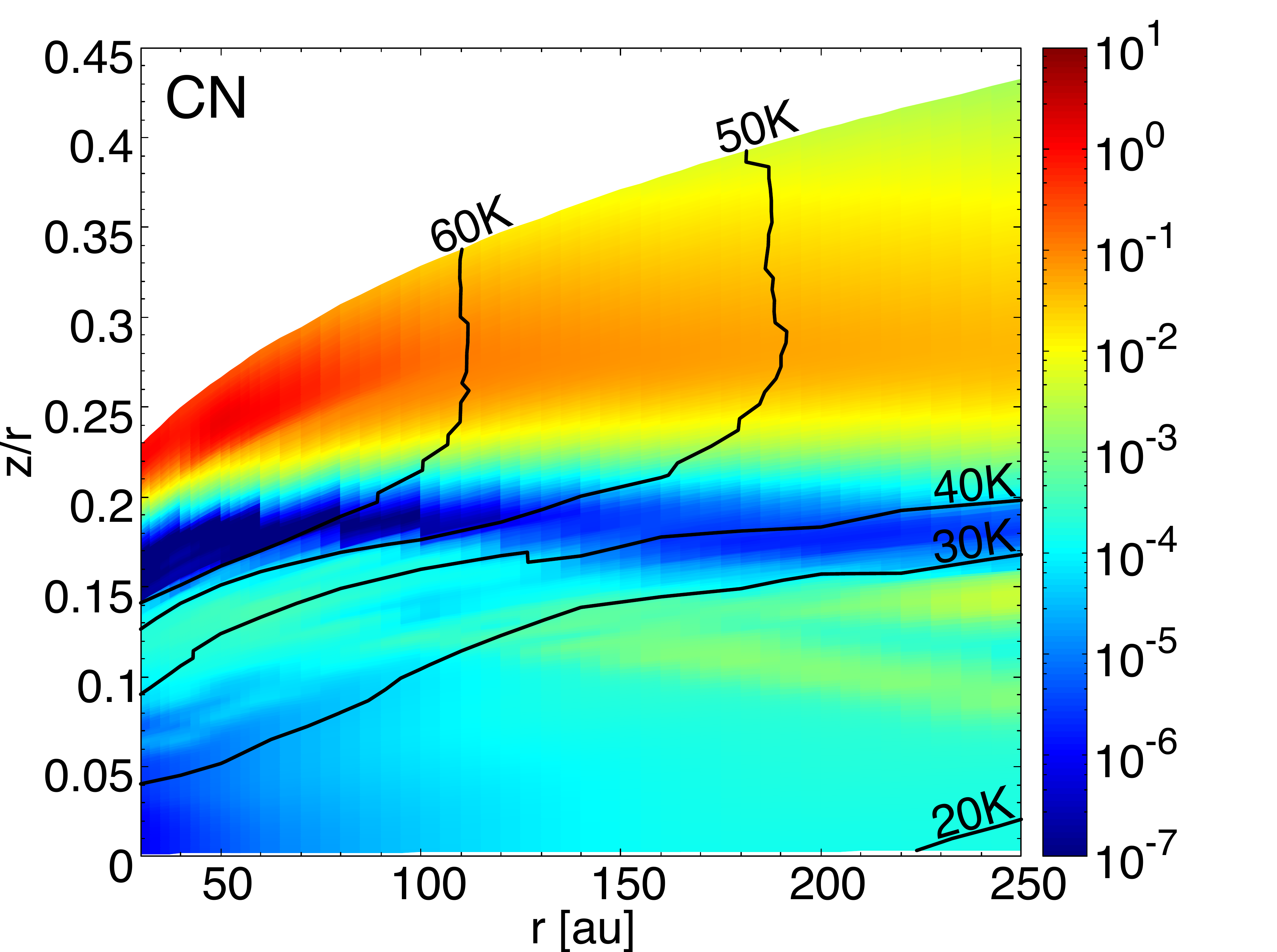} 
 \subcaption{LUV-B14-$\mathrm{T_{a}}$} 
\end{subfigure}
\caption{ Density [cm$^{-3}$] of H2, CO, CS and CN in the gas phase of the single-grain models in LUV regime. Left column shows the results of \sltg, middle column is the results of \slteff and right column is the results of \slb. Black contours represent the dust temperature (T$_\mathrm{dust}$ = T$_\mathrm{g}$ in the left column and T$_\mathrm{dust}$ = T$_\mathrm{a}$ in the middle and right columns).}
\label{fig:s-maps-luv}
\end{figure*}

In \autoref{fig:s-maps-luv} we present maps of number density for 
H$_2$, CO, CS and CN in the case of LUV single-grain models. Again, the 
similarity with HUV models is clear. The profiles around the midplane 
are identical between LUV and HUV. The impact of the different flux on 
the upper layers is visible. The peak of CO, CS and CN are located at 
higher altitudes and are globally wider in LUV models. We note one 
major difference for CN between LUV and HUV models. \shteff exhibits a 
clear spike of CN density around $2 H$ while, conversely, \slteff exhibits 
a drop at the same location.

\begin{figure*}
\begin{subfigure}{.32\linewidth}
  \centering
  \includegraphics[width=1.00\linewidth]{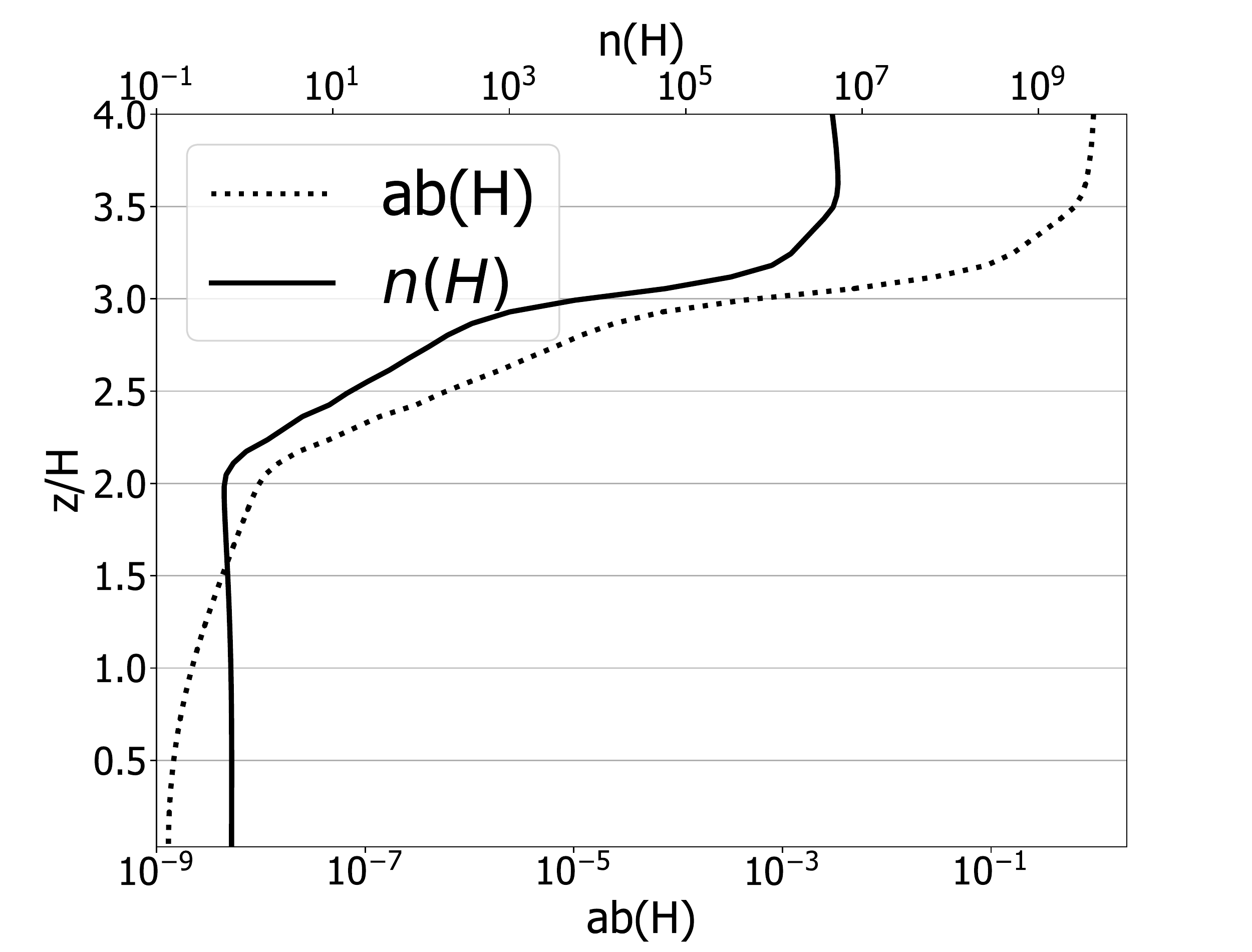}
\end{subfigure}
\begin{subfigure}{.32\linewidth}
  \centering
  \includegraphics[width=1.00\linewidth]{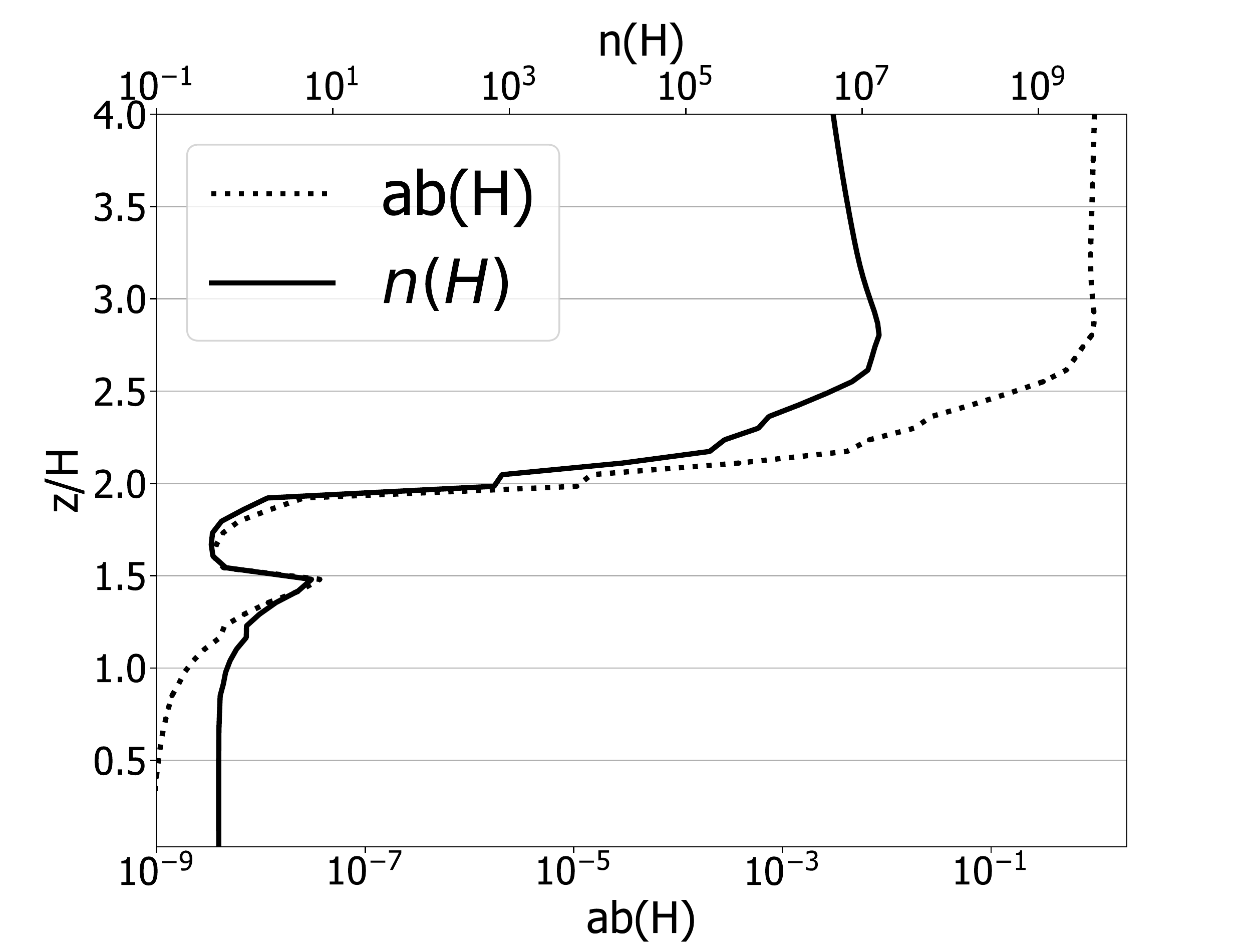}

\end{subfigure}
\begin{subfigure}{.32\linewidth}
  \centering
  \includegraphics[width=1.00\linewidth]{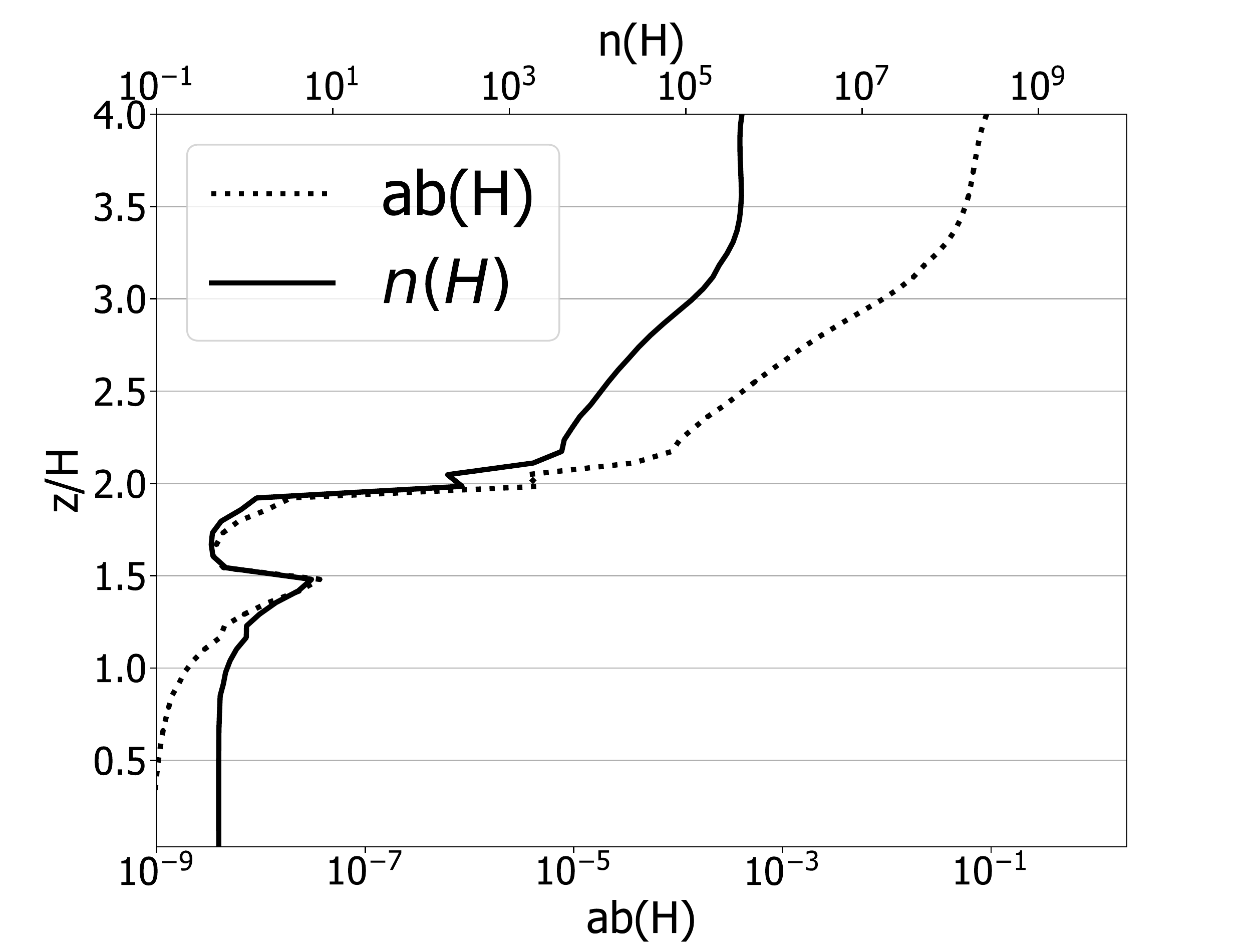}
\end{subfigure}

\begin{subfigure}{.32\linewidth}
  \centering
  \includegraphics[width=1.00\linewidth]{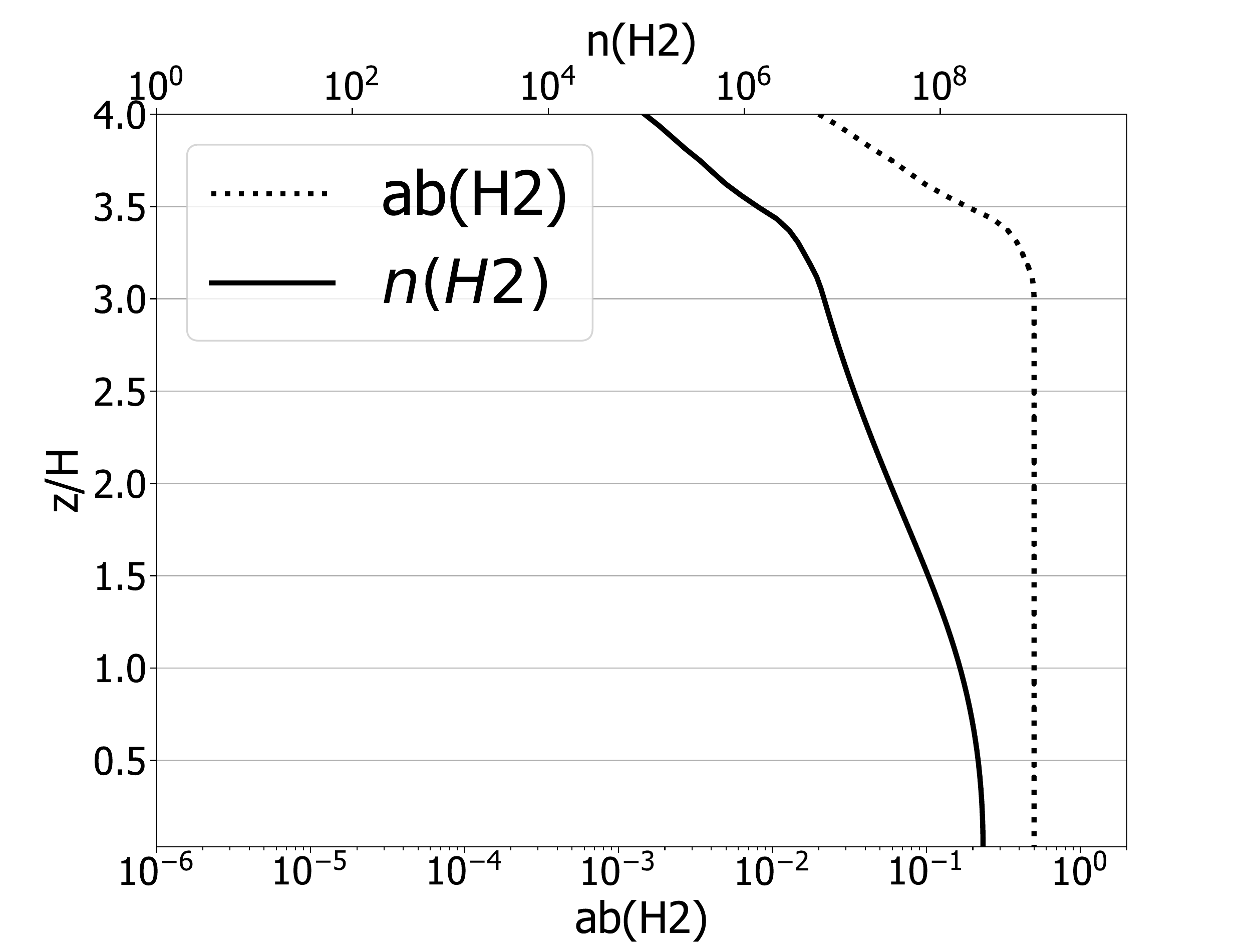}
\end{subfigure}
\begin{subfigure}{.32\linewidth}
  \centering
  \includegraphics[width=1.00\linewidth]{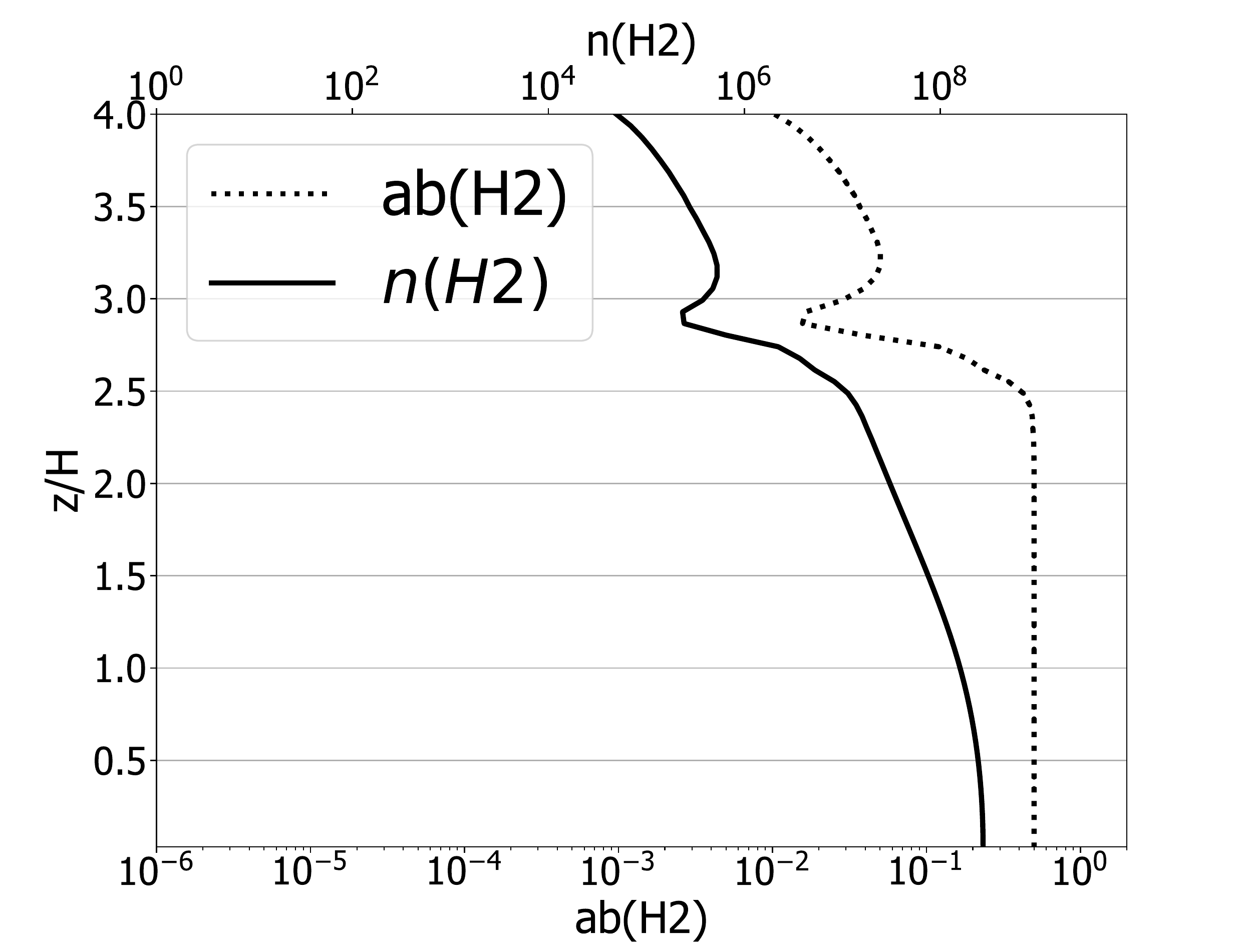}
 
\end{subfigure}
\begin{subfigure}{.32\linewidth}
  \centering
  \includegraphics[width=1.00\linewidth]{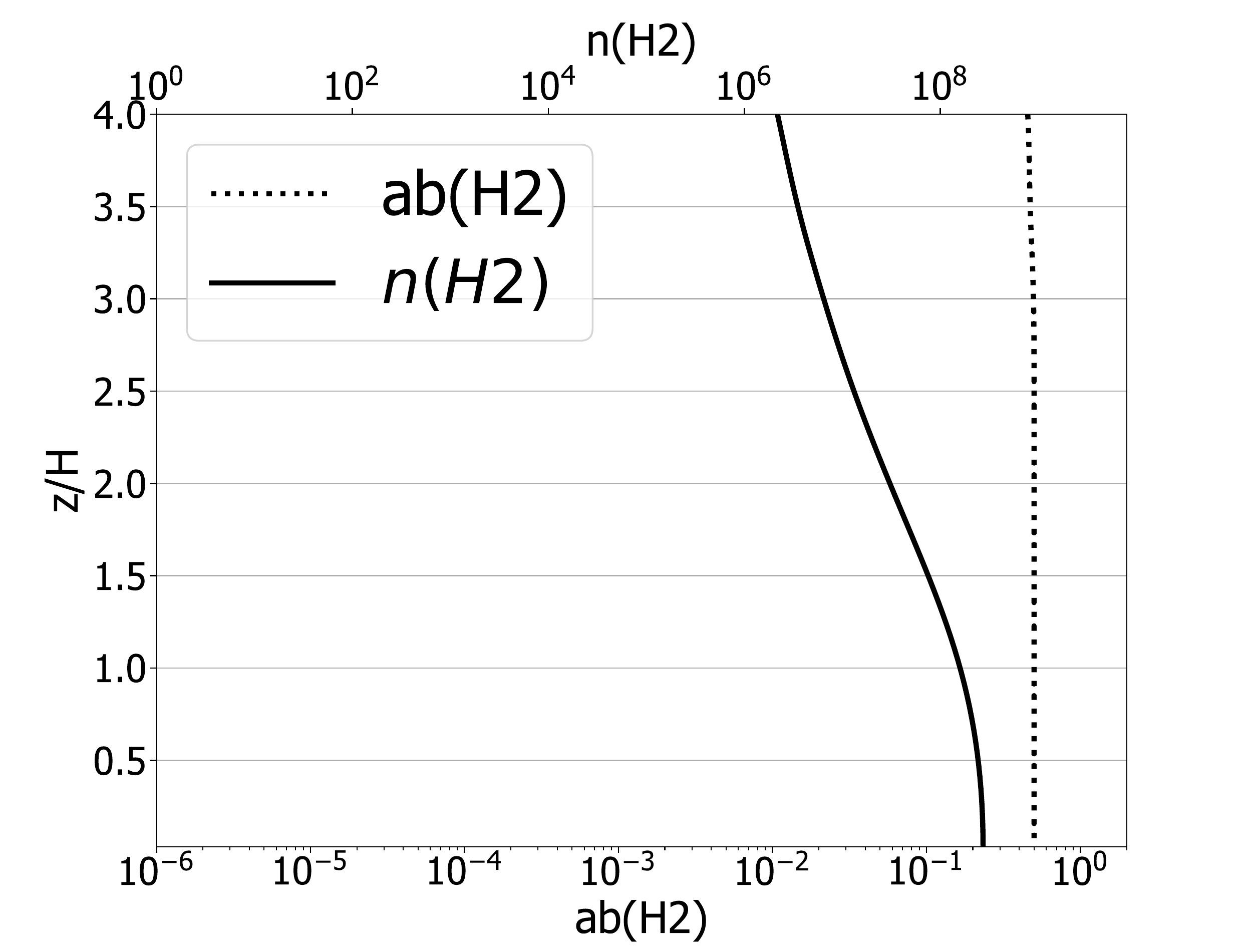}
\end{subfigure}

\begin{subfigure}{.32\linewidth}
  \centering
  \includegraphics[width=1.00\linewidth]{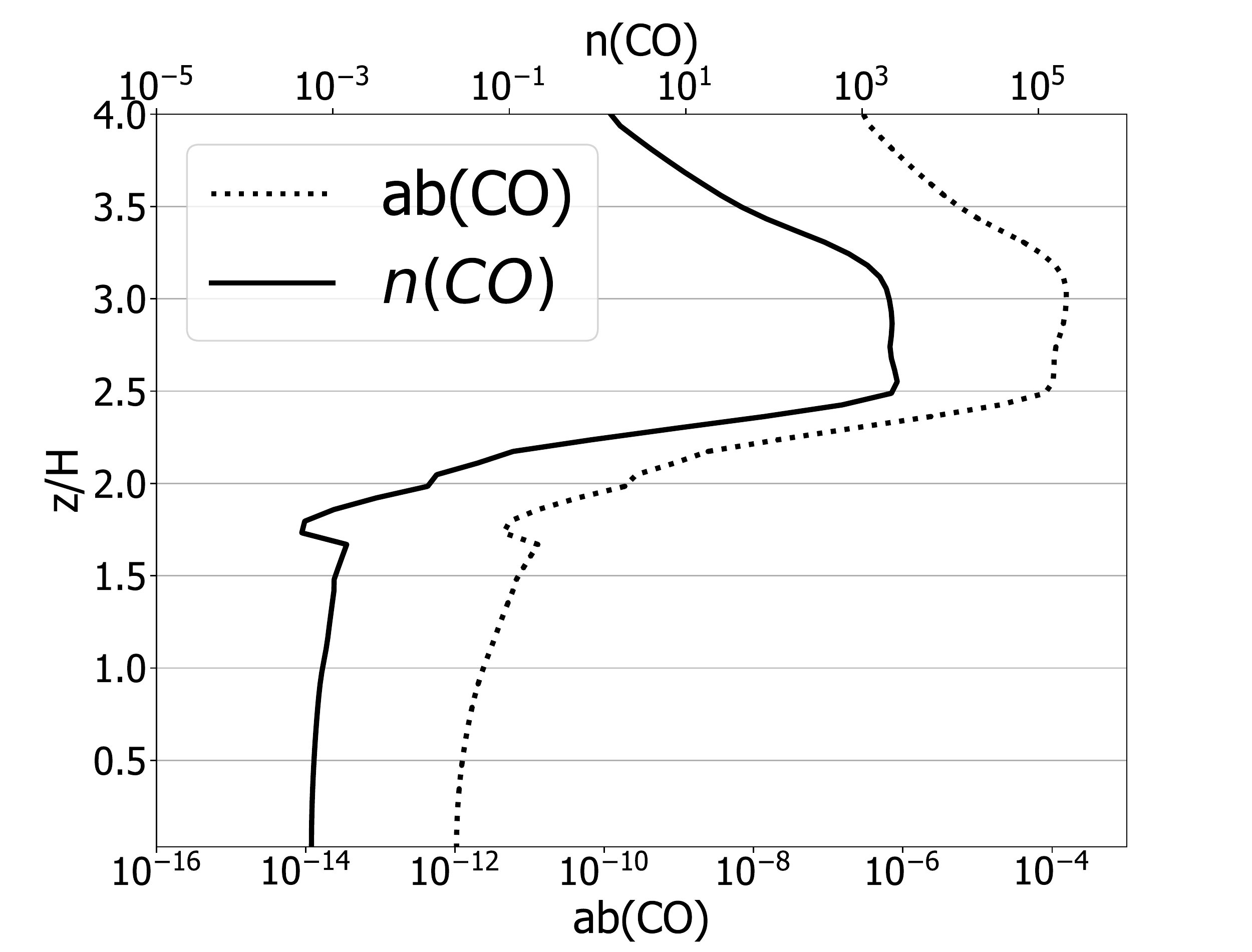}
\end{subfigure}
\begin{subfigure}{.32\linewidth}
  \centering
  \includegraphics[width=1.00\linewidth]{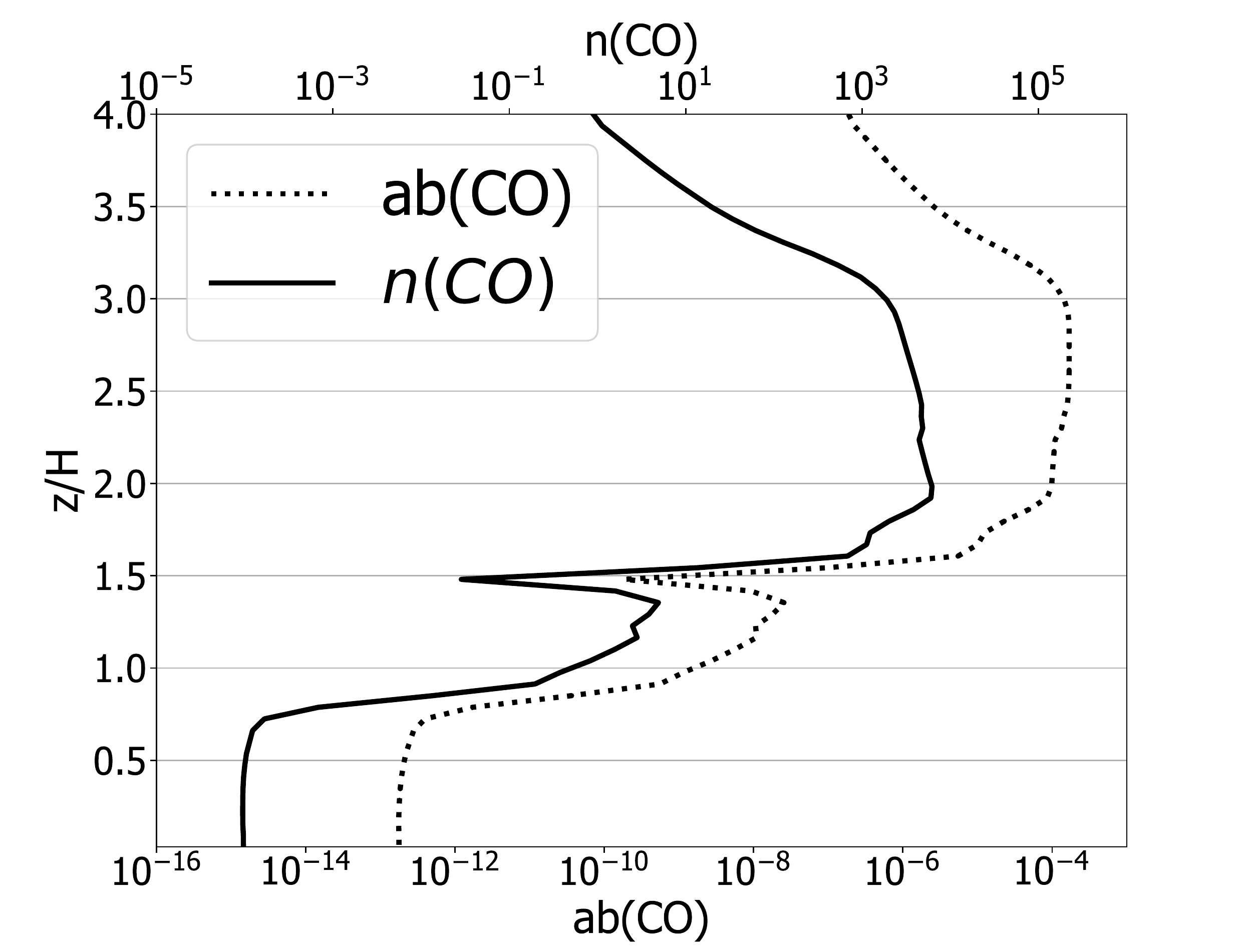}
  
\end{subfigure}
\begin{subfigure}{.32\linewidth}
  \centering
  \includegraphics[width=1.00\linewidth]{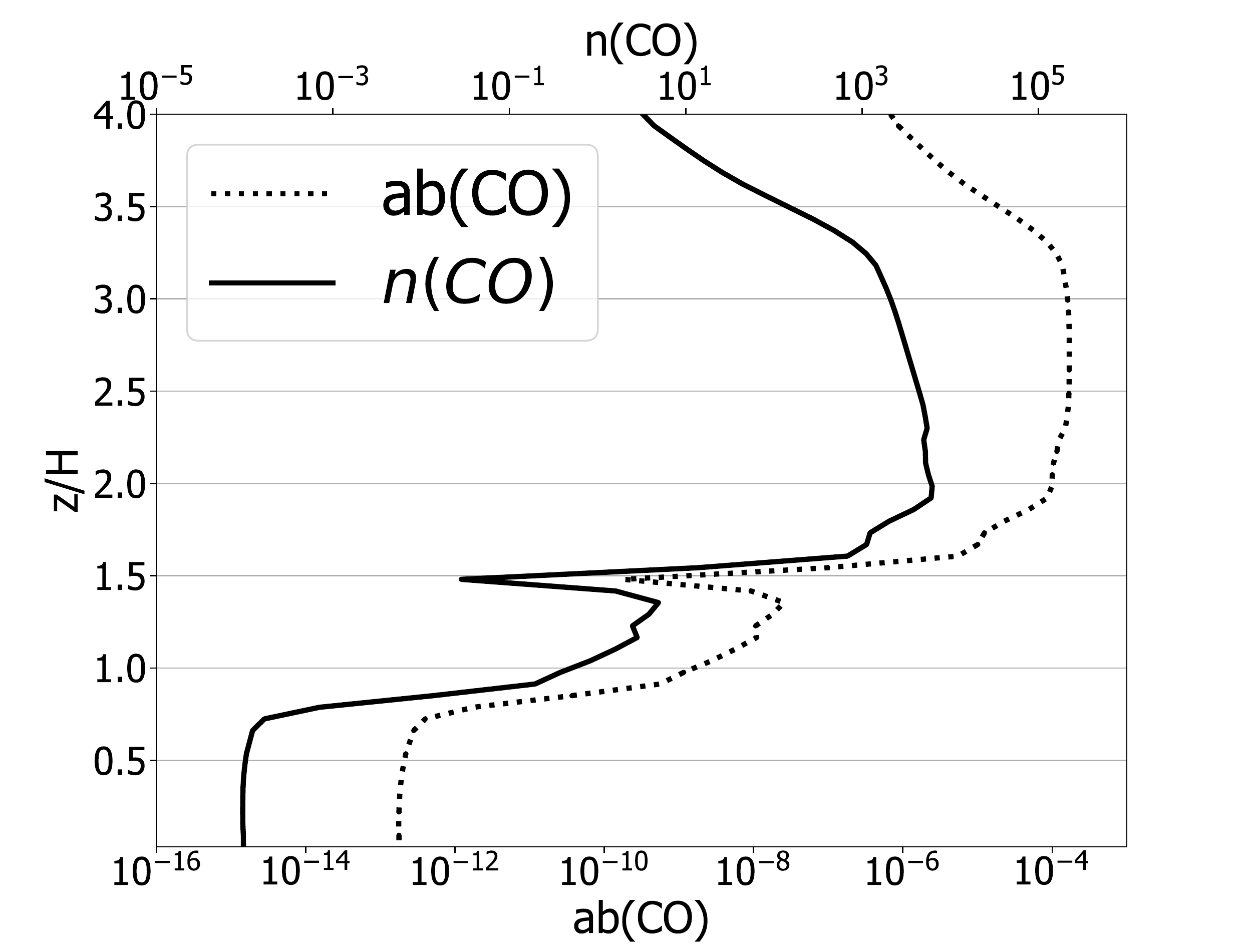}

\end{subfigure}

\begin{subfigure}{.32\linewidth}
  \centering
  \includegraphics[width=1.00\linewidth]{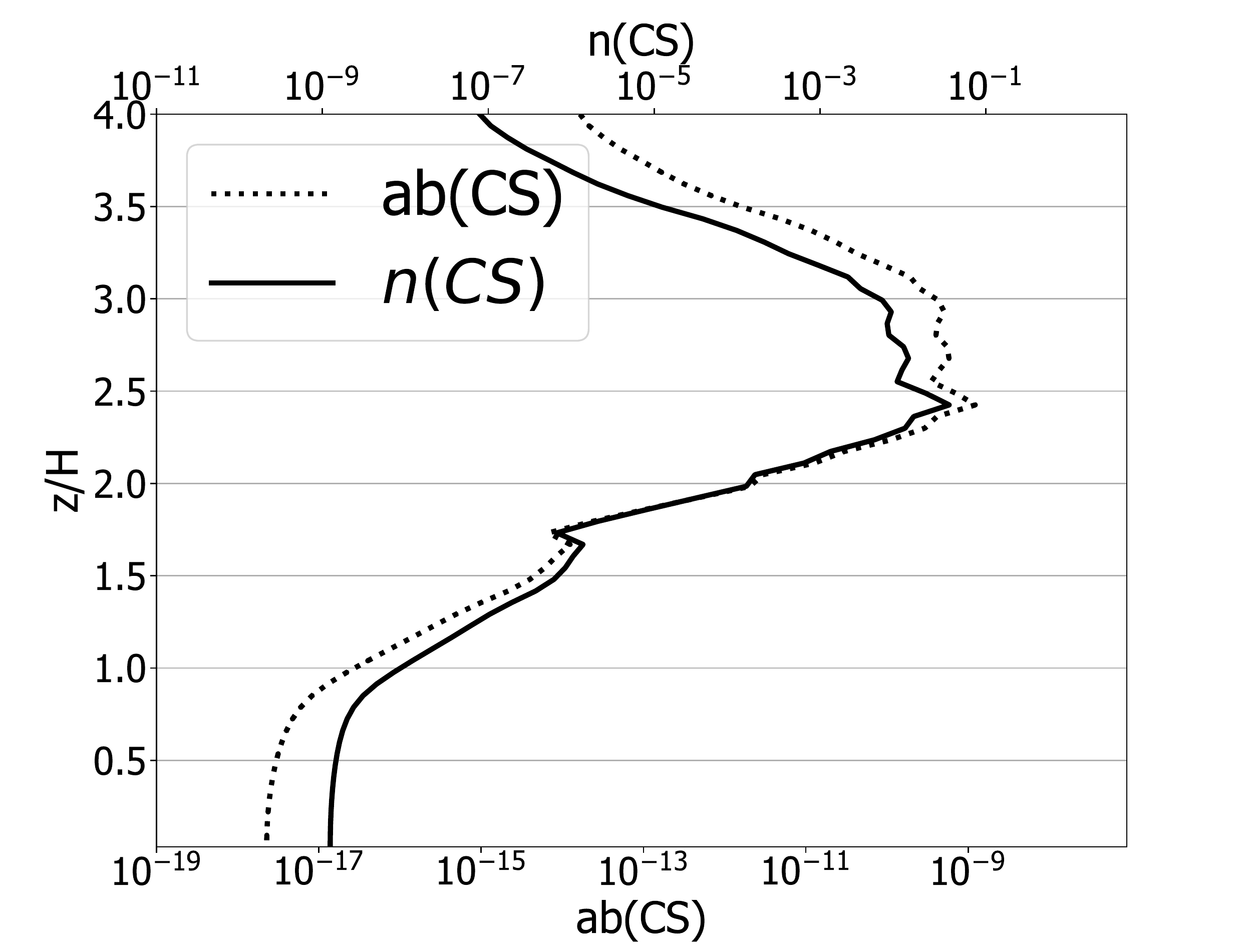}
\end{subfigure}
\begin{subfigure}{.32\linewidth}
  \centering
  \includegraphics[width=1.00\linewidth]{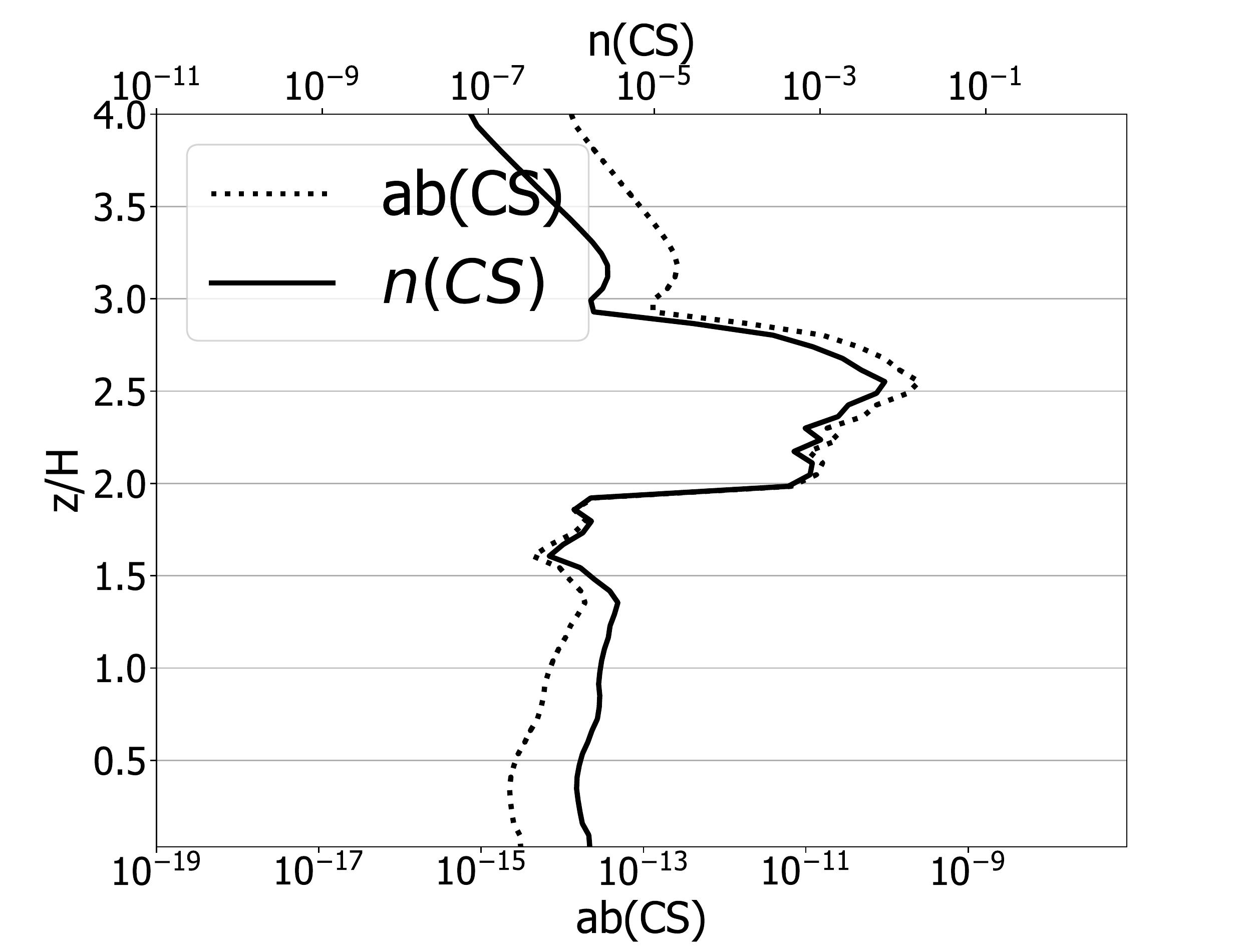}

\end{subfigure}
\begin{subfigure}{.32\linewidth}
  \centering
  \includegraphics[width=1.00\linewidth]{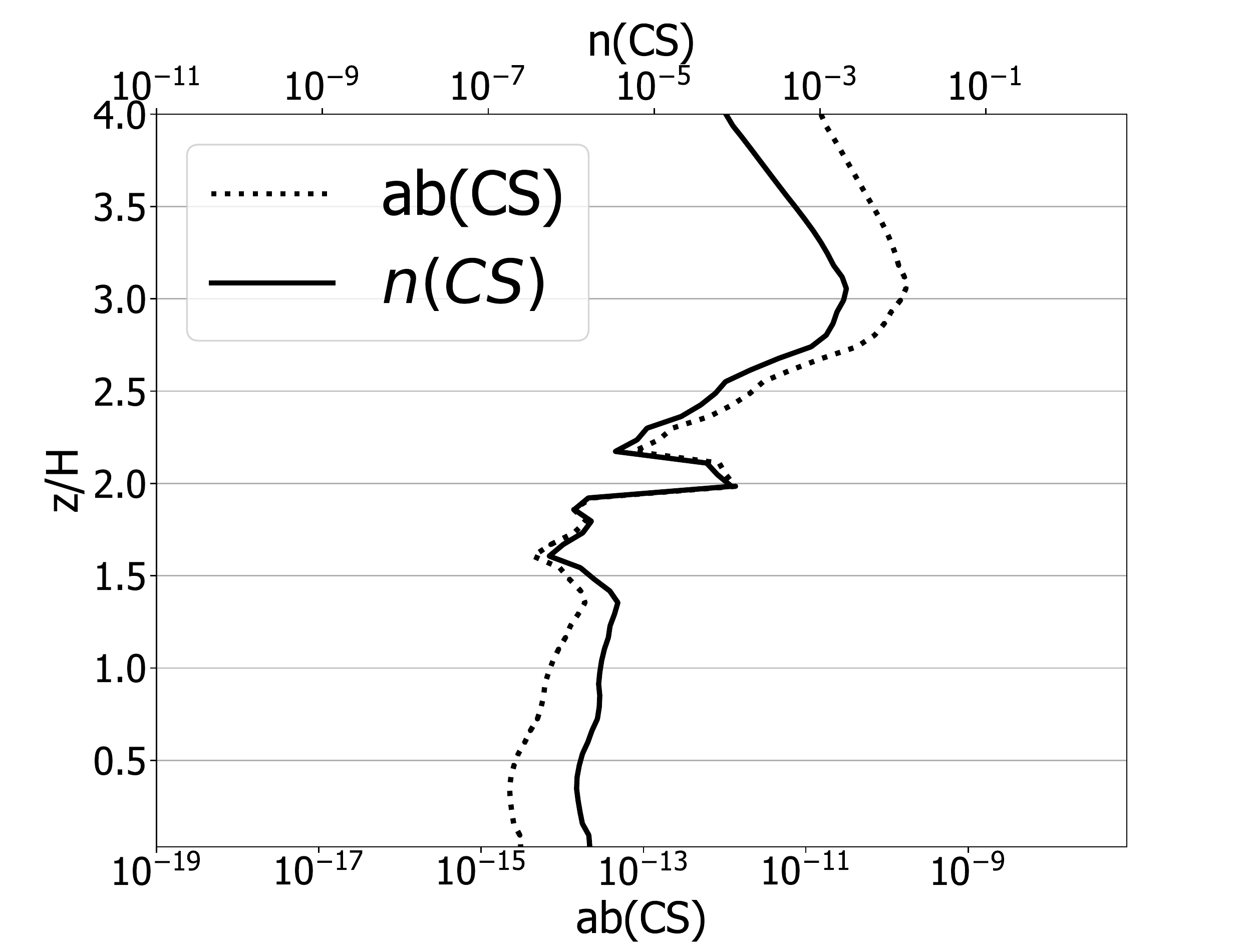}
\end{subfigure}

\begin{subfigure}{.32\linewidth}
  \centering
  \includegraphics[width=1.00\linewidth]{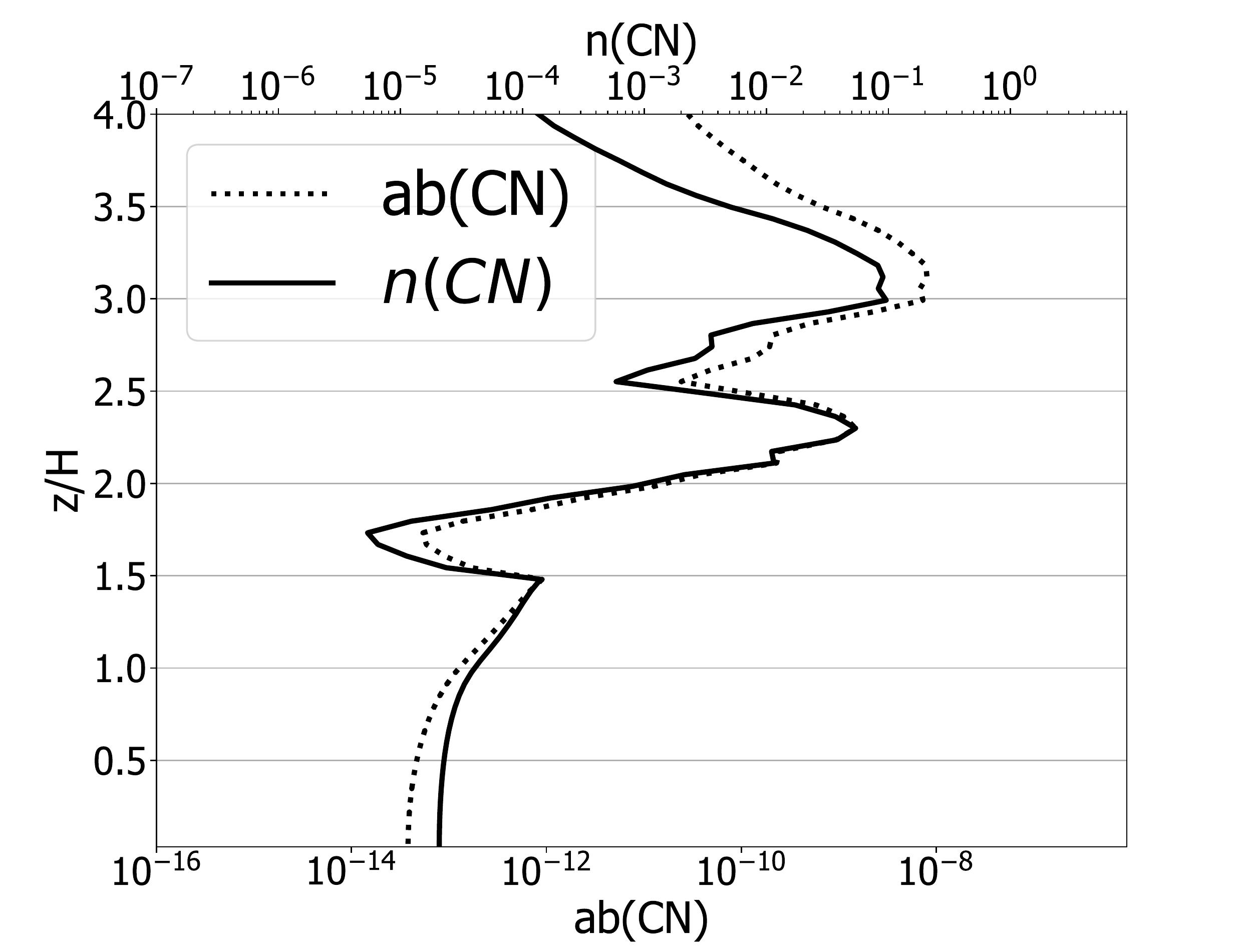}
   \subcaption{LUV-LH-$\mathrm{T_{g}}$}   
\end{subfigure}
\begin{subfigure}{.32\linewidth}
  \centering
  \includegraphics[width=1.00\linewidth]{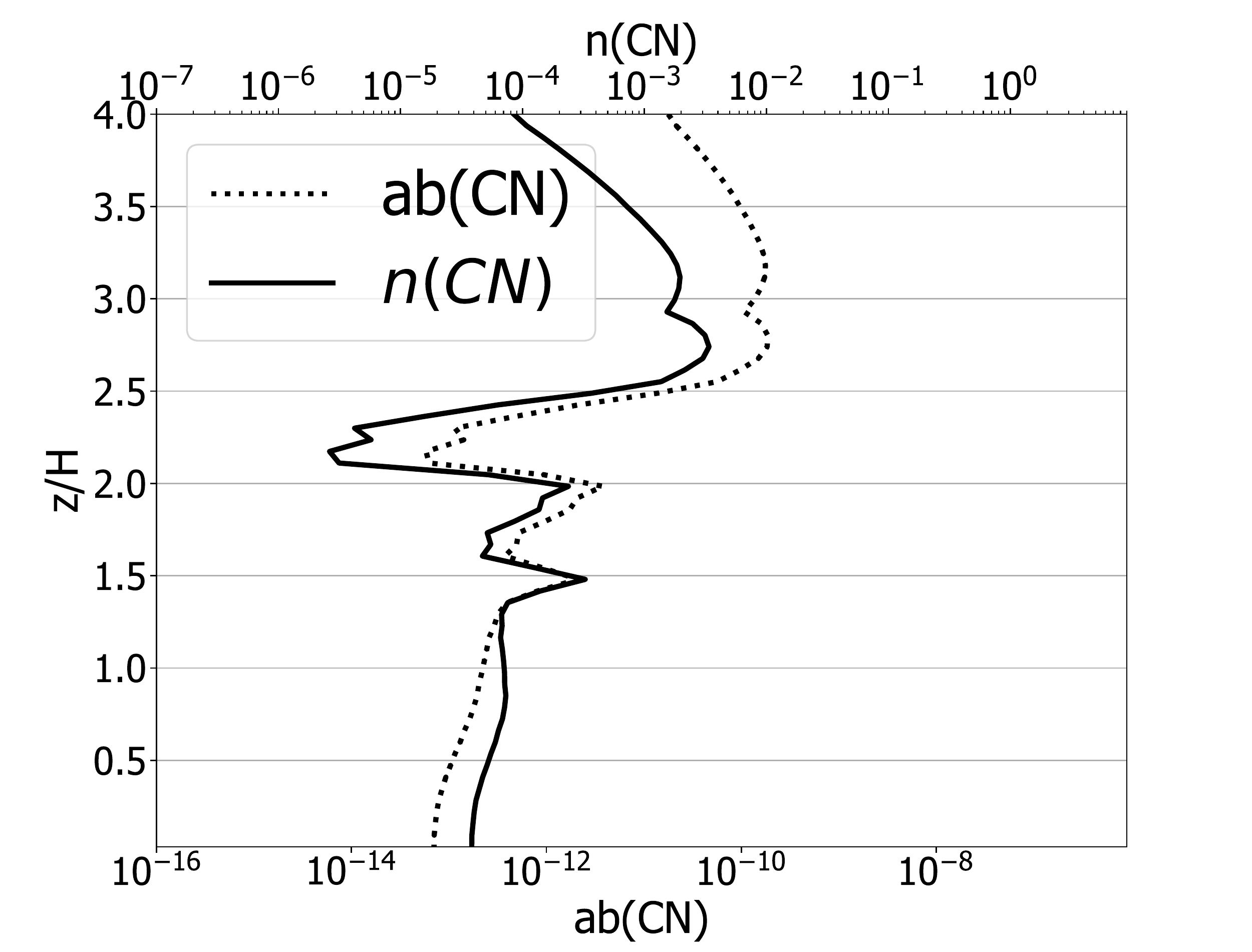}
   \subcaption{LUV-LH-$\mathrm{T_{a}}$}   
\end{subfigure}
\begin{subfigure}{.32\linewidth}
  \centering
  \includegraphics[width=1.00\linewidth]{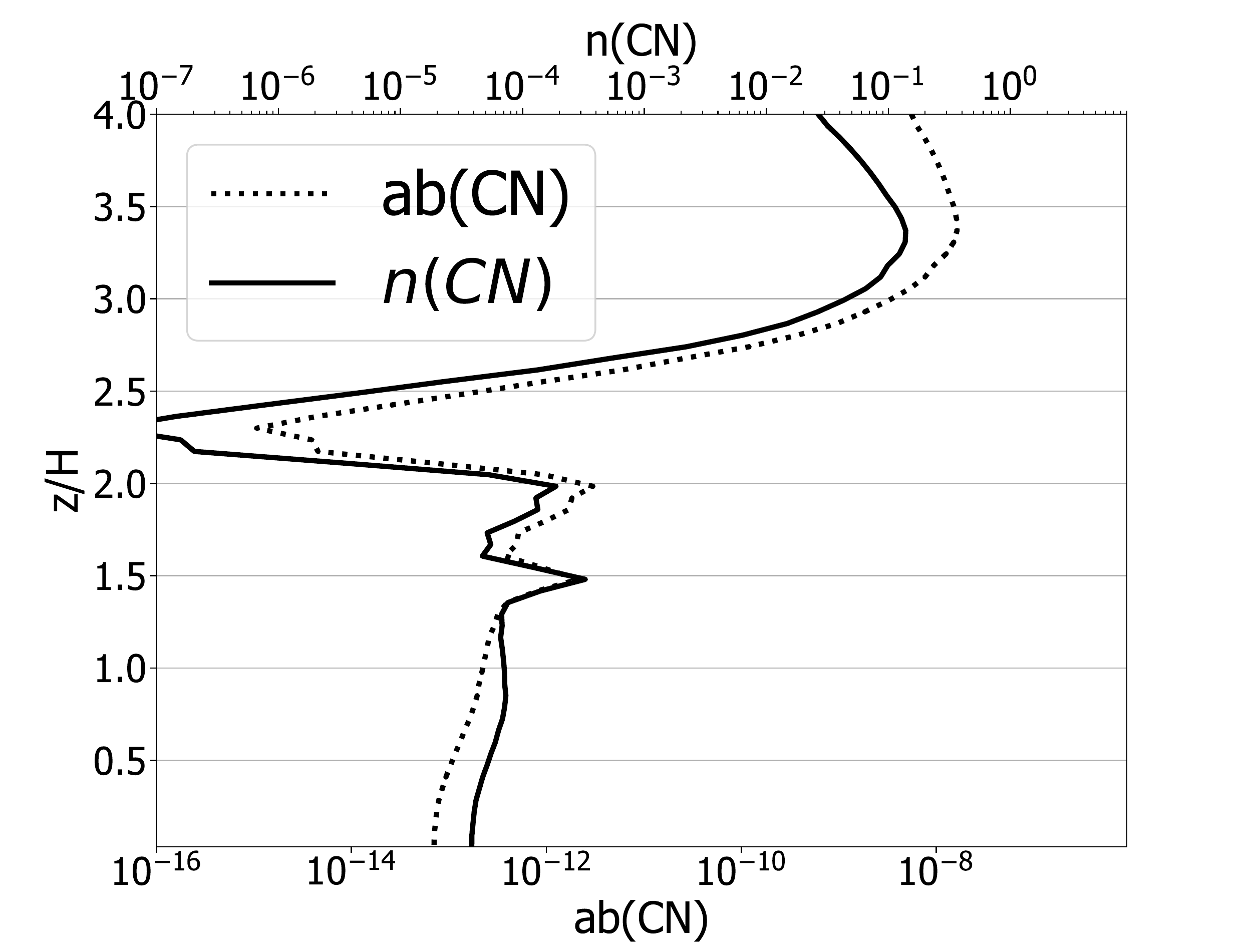}
   \subcaption{LUV-B14-$\mathrm{T_{a}}$} 
\end{subfigure}

\caption{Vertical profiles of H, $\mathrm{H_2}$, CO, CS and CN at 100 au from the star of the LUV single-grain models. The dotted line is the abundance relative to H and the solid line is the density [$cm^{-3}$].}
\label{fig:s-100profile_low}
\end{figure*}

\begin{figure*}
\begin{subfigure}{.32\linewidth}
  \centering
  \includegraphics[width=1.00\linewidth]{figures/SINGLE/LUV_HL_Tg/100AU/H_LUV_HL_Tg.pdf}
\end{subfigure}
\begin{subfigure}{.32\linewidth}
  \centering
  \includegraphics[width=1.00\linewidth]{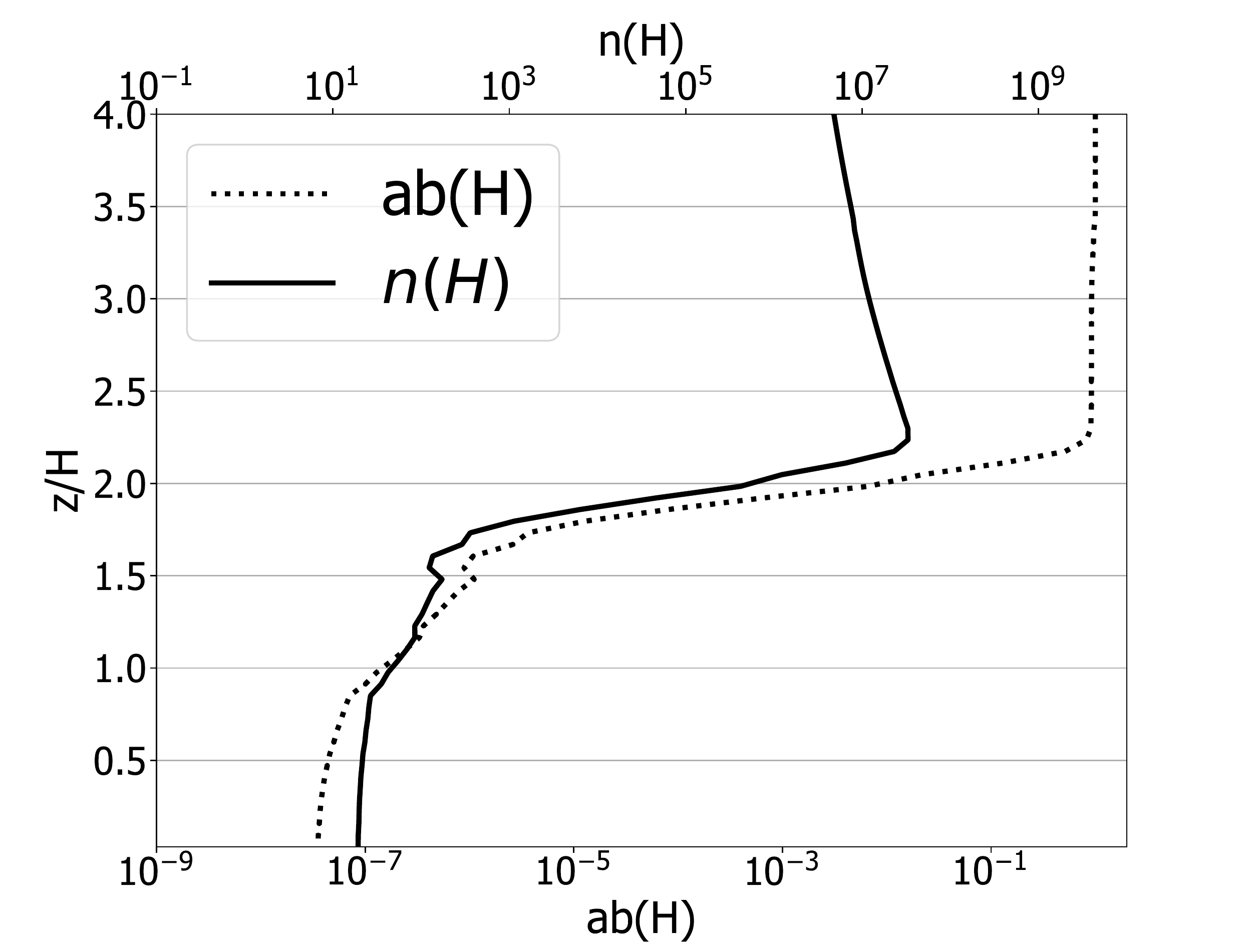}
\end{subfigure}
\begin{subfigure}{.32\linewidth}
  \centering
  \includegraphics[width=1.00\linewidth]{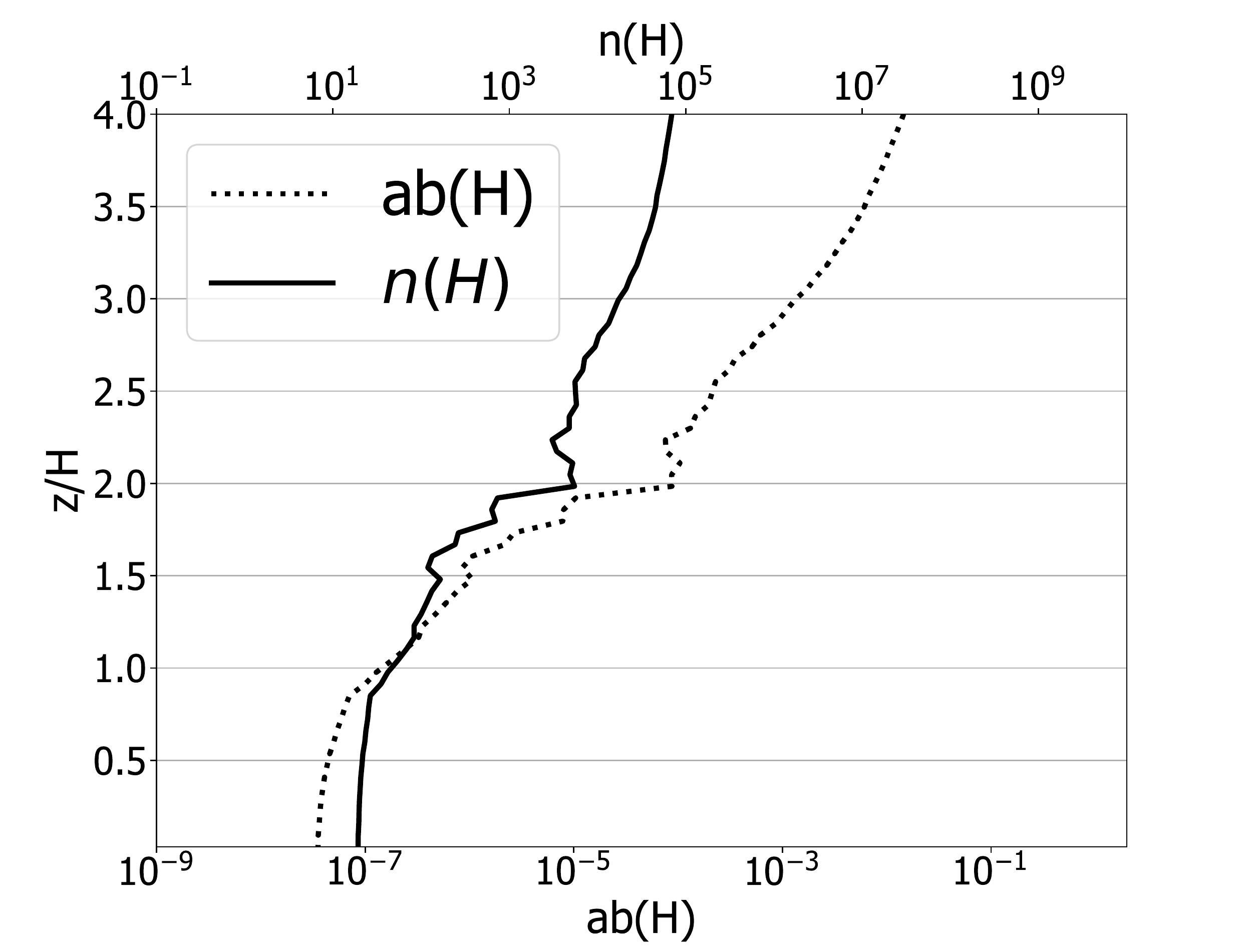}  
\end{subfigure}

\begin{subfigure}{.32\linewidth}
  \centering
  \includegraphics[width=1.00\linewidth]{figures/SINGLE/LUV_HL_Tg/100AU/H2_LUV_HL_Tg.pdf}
\end{subfigure}
\begin{subfigure}{.32\linewidth}
  \centering
  \includegraphics[width=1.00\linewidth]{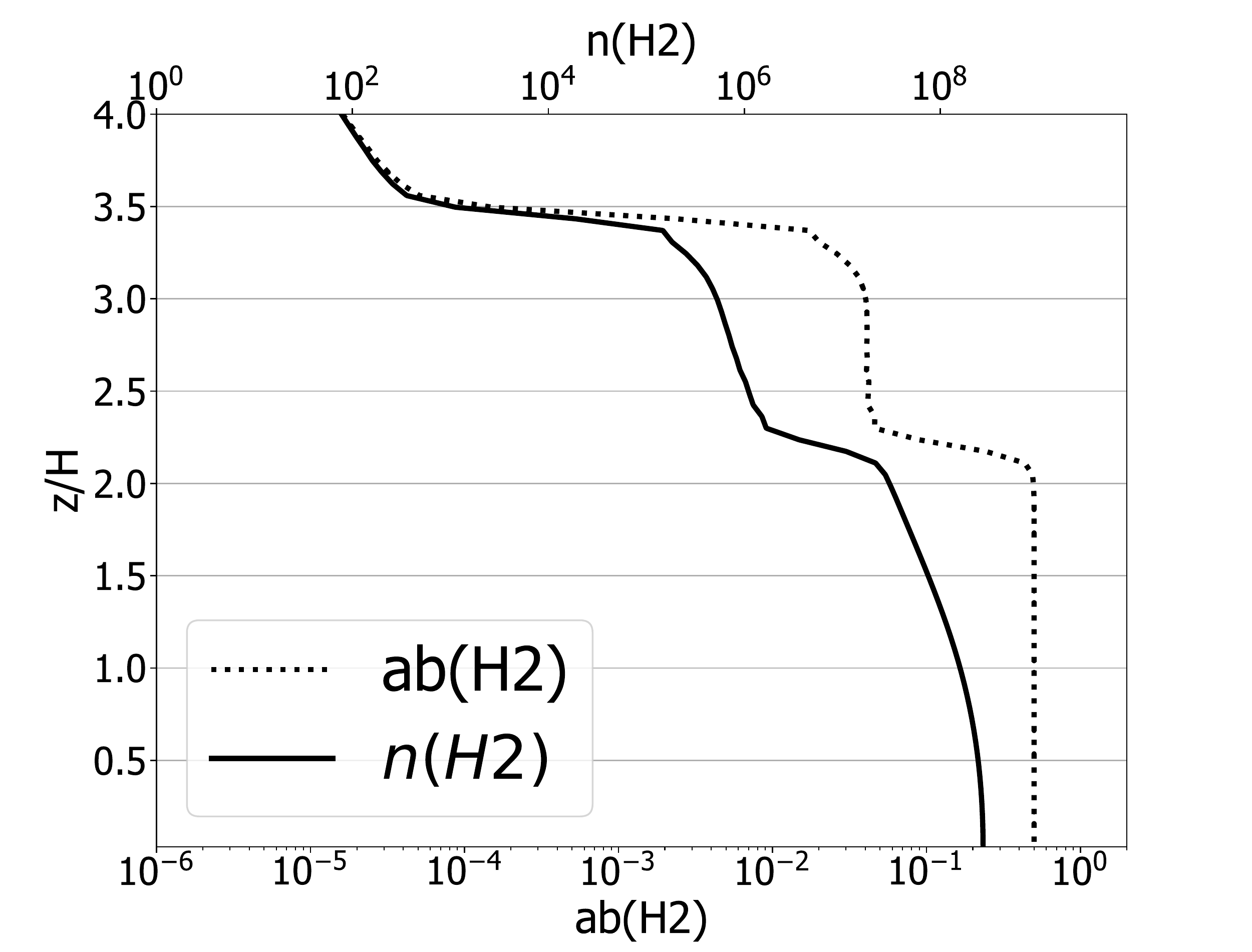}
\end{subfigure}
\begin{subfigure}{.32\linewidth}
  \centering
  \includegraphics[width=1.00\linewidth]{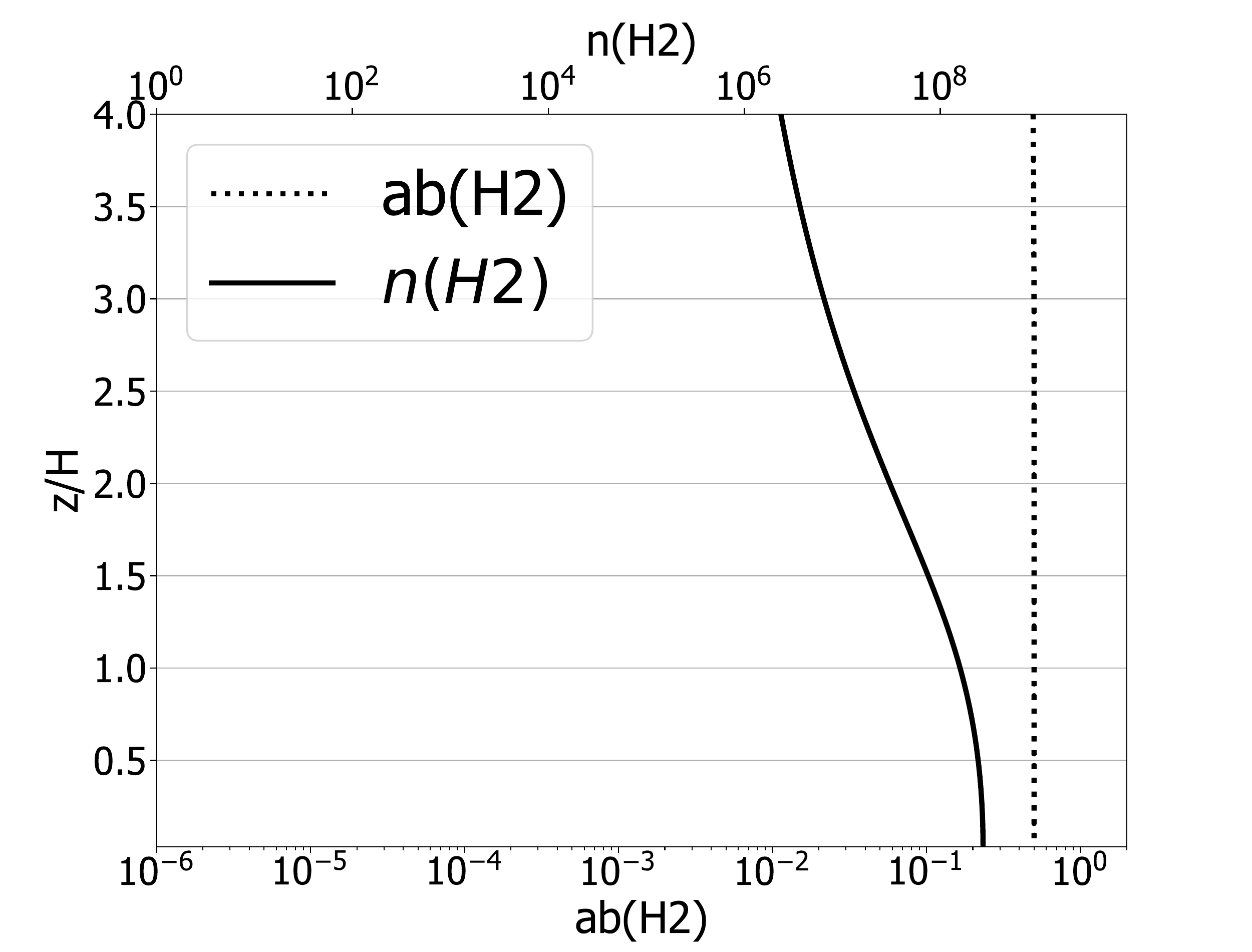} 
\end{subfigure}

\begin{subfigure}{.32\linewidth}
  \centering
  \includegraphics[width=1.00\linewidth]{figures/SINGLE/LUV_HL_Tg/100AU/CO_LUV_HL_Tg.pdf}
\end{subfigure}
\begin{subfigure}{.32\linewidth}
  \centering
  \includegraphics[width=1.00\linewidth]{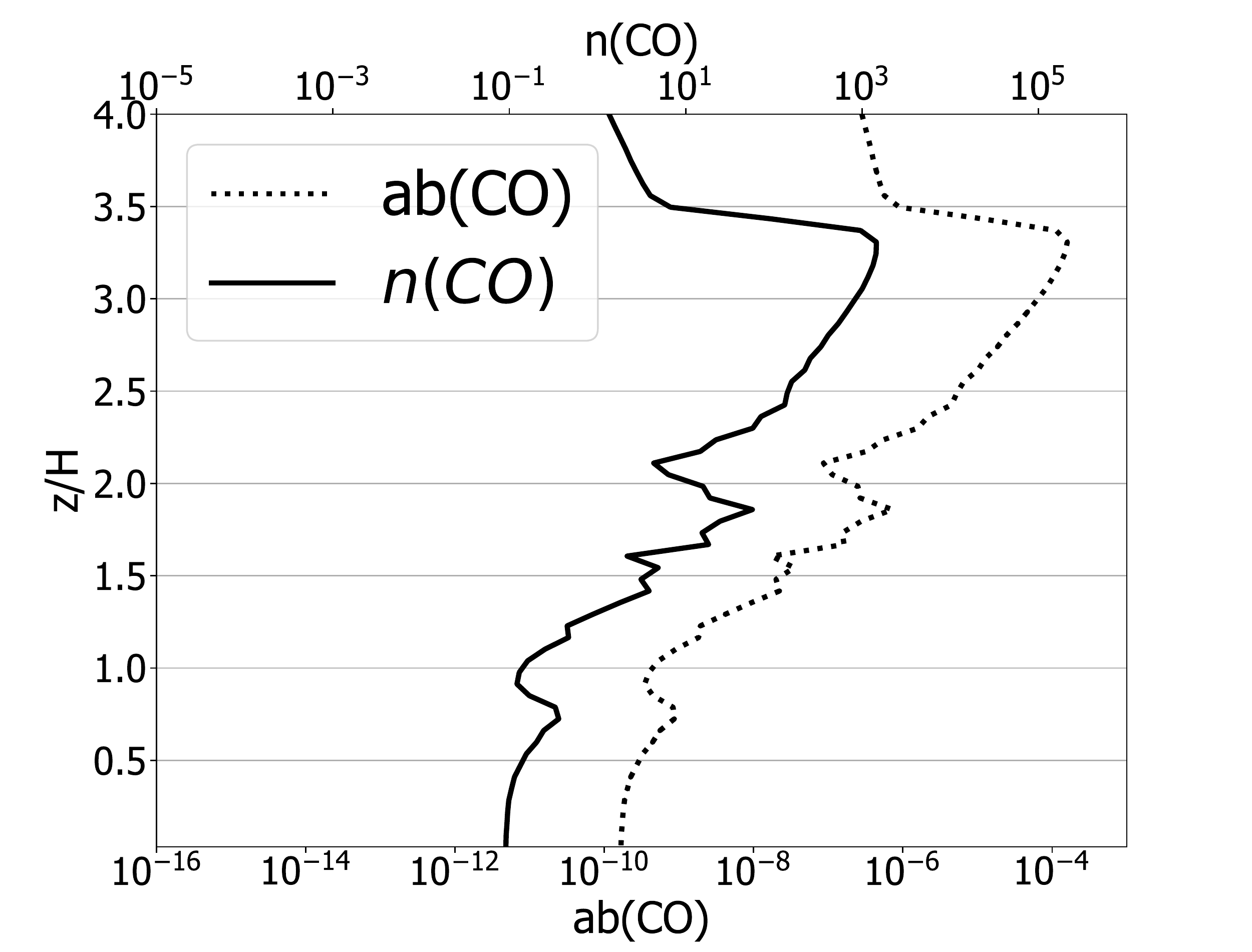}
\end{subfigure}
\begin{subfigure}{.32\linewidth}
  \centering
  \includegraphics[width=1.00\linewidth]{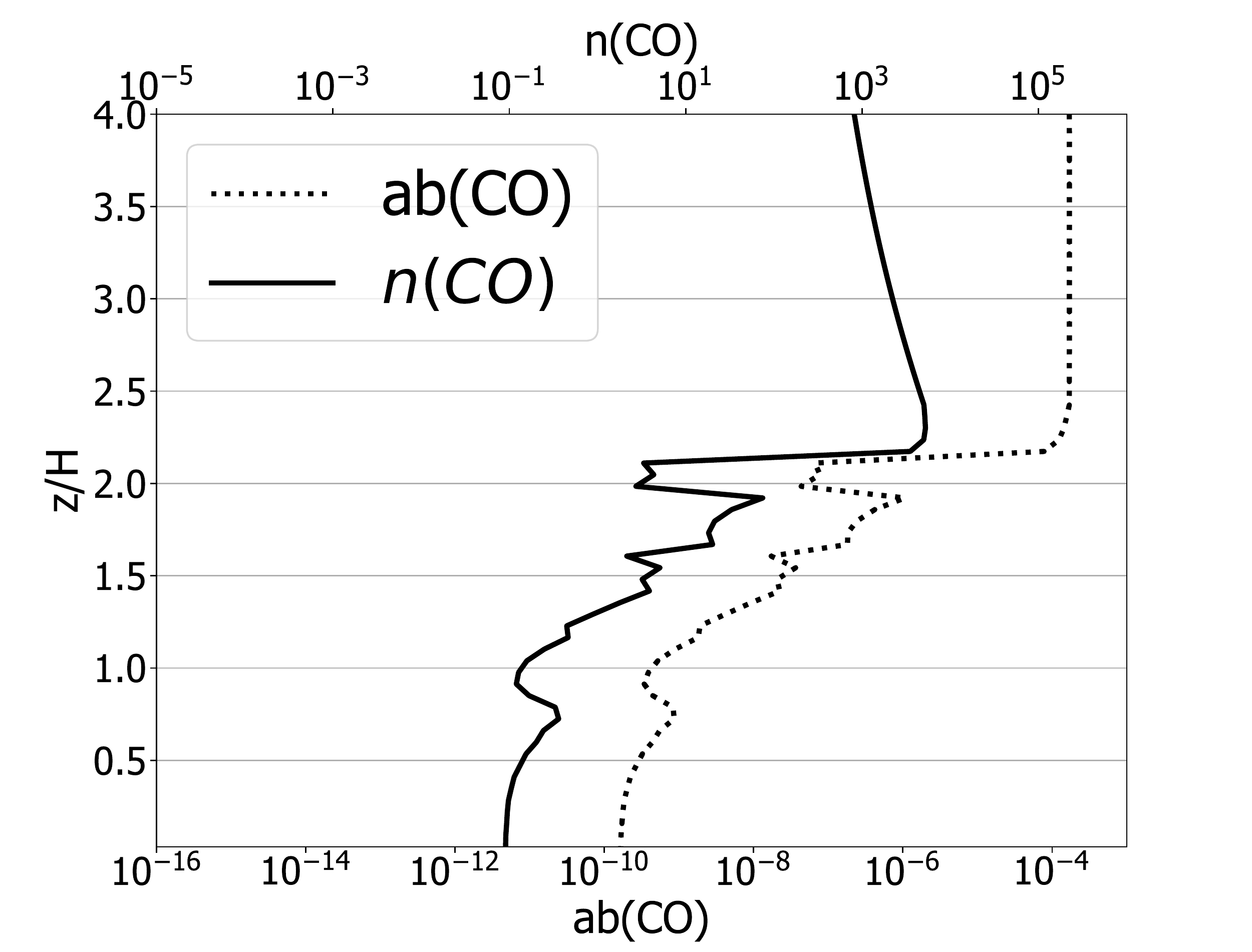} 
\end{subfigure}

\begin{subfigure}{.32\linewidth}
  \centering
  \includegraphics[width=1.00\linewidth]{figures/SINGLE/LUV_HL_Tg/100AU/CS_LUV_HL_Tg.pdf}
\end{subfigure}
\begin{subfigure}{.32\linewidth}
  \centering
  \includegraphics[width=1.00\linewidth]{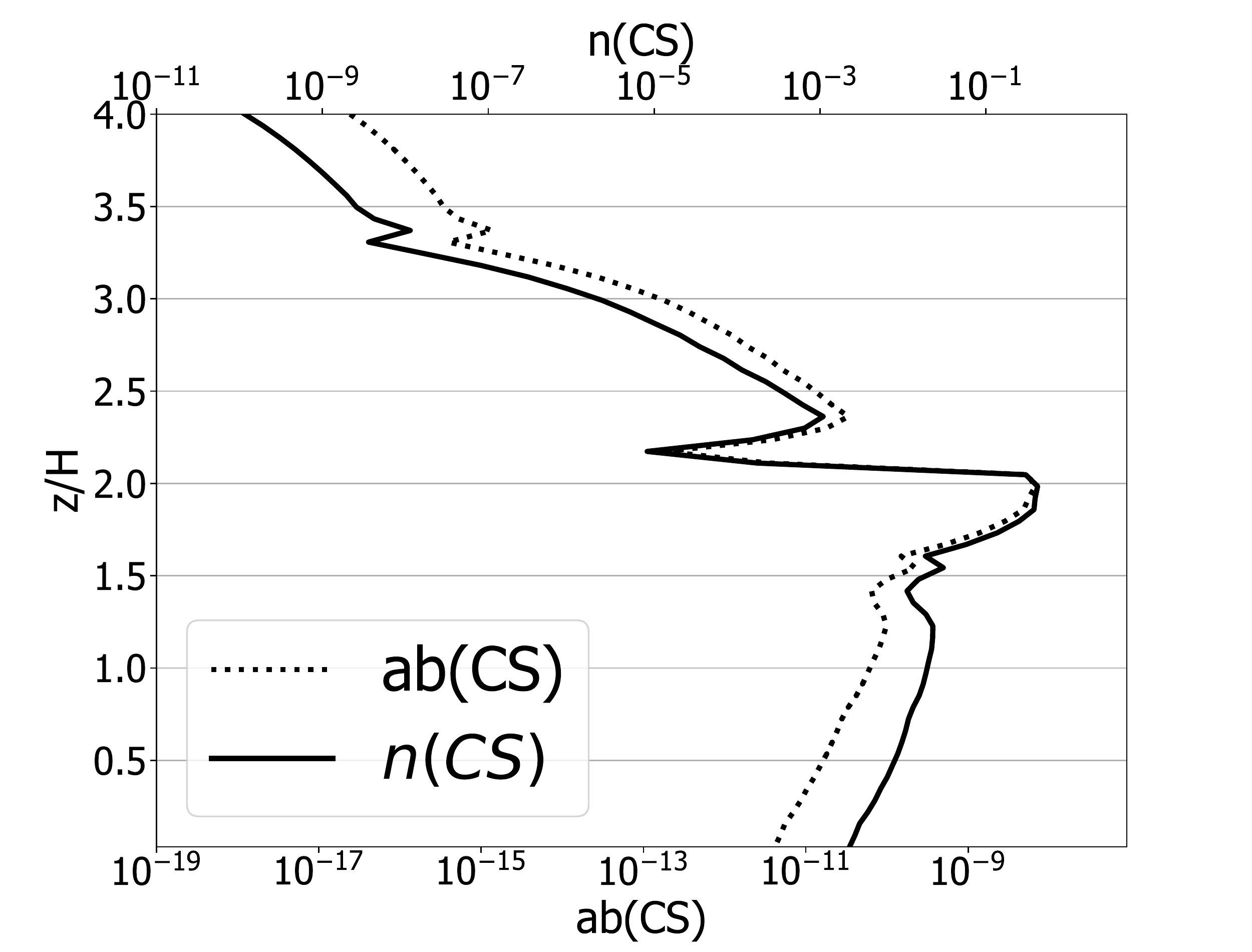}
\end{subfigure}
\begin{subfigure}{.32\linewidth}
  \centering
  \includegraphics[width=1.00\linewidth]{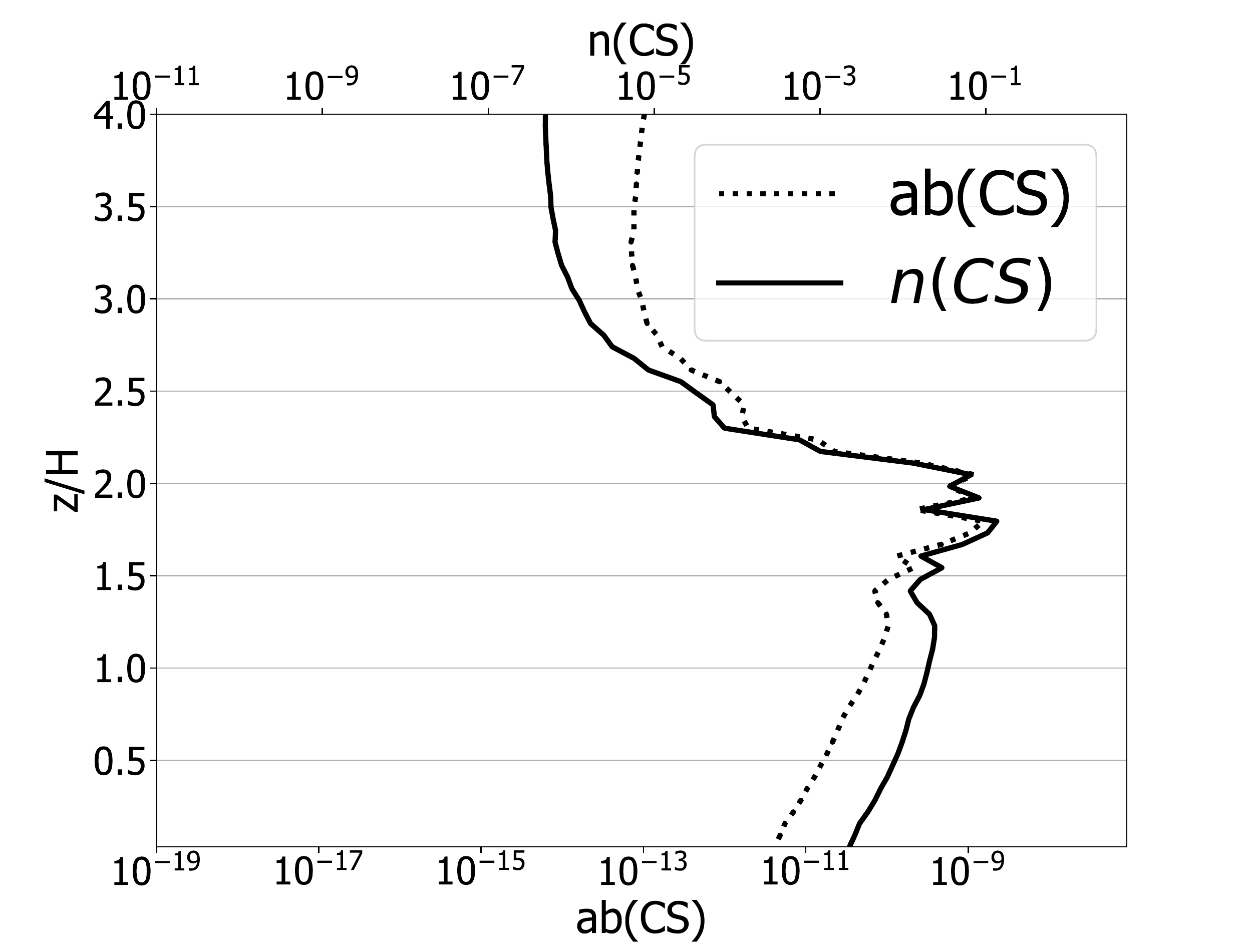} 
\end{subfigure}

\begin{subfigure}{.32\linewidth}
  \centering
  \includegraphics[width=1.00\linewidth]{figures/SINGLE/LUV_HL_Tg/100AU/CN_LUV_HL_Tg.pdf}
   \subcaption{LUV-LH-$\mathrm{T_{g}}$}   
\end{subfigure}
\begin{subfigure}{.32\linewidth}
  \centering
  \includegraphics[width=1.00\linewidth]{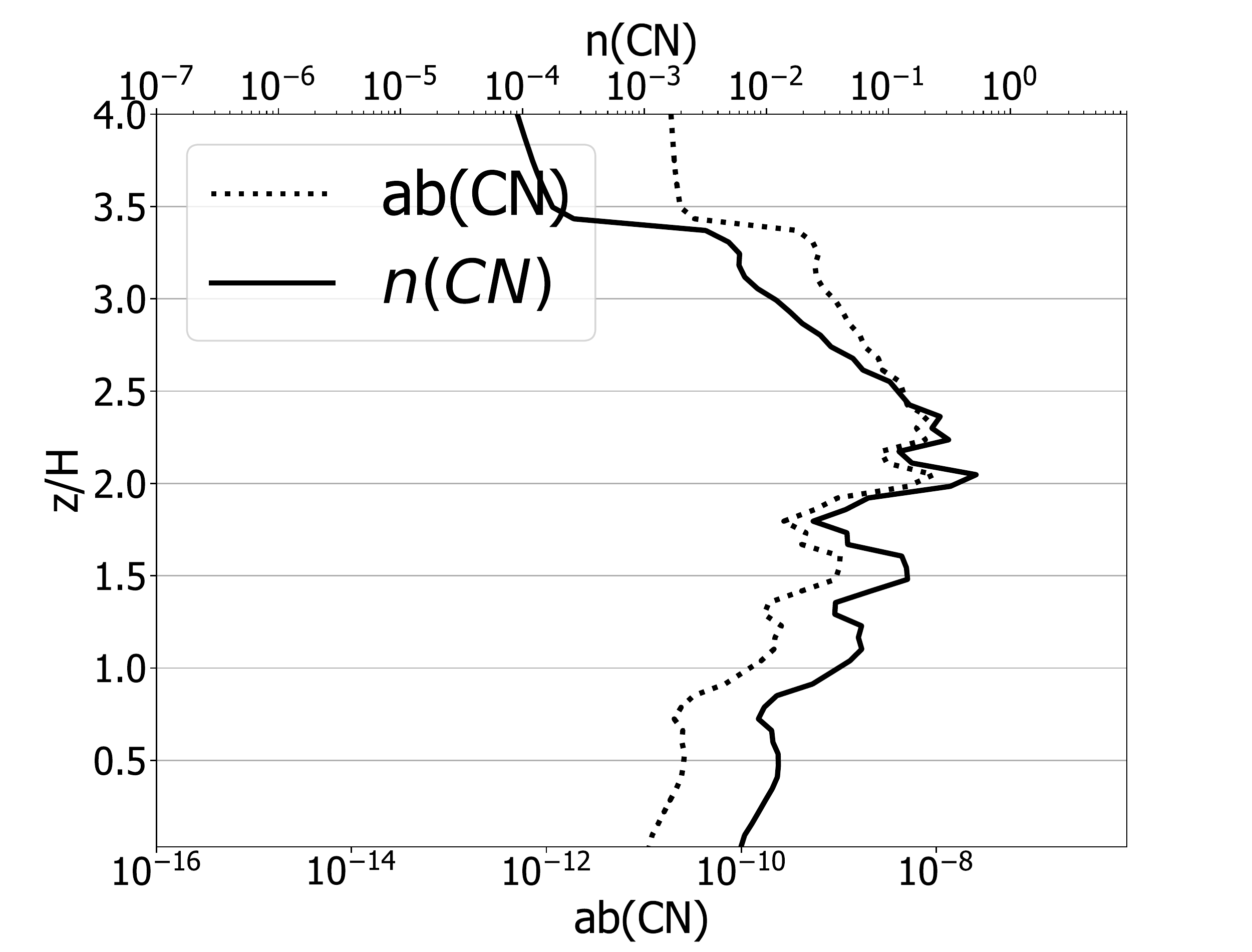}
   \subcaption{M-LUV-LH} 
\end{subfigure}
\begin{subfigure}{.32\linewidth}
  \centering
  \includegraphics[width=1.00\linewidth]{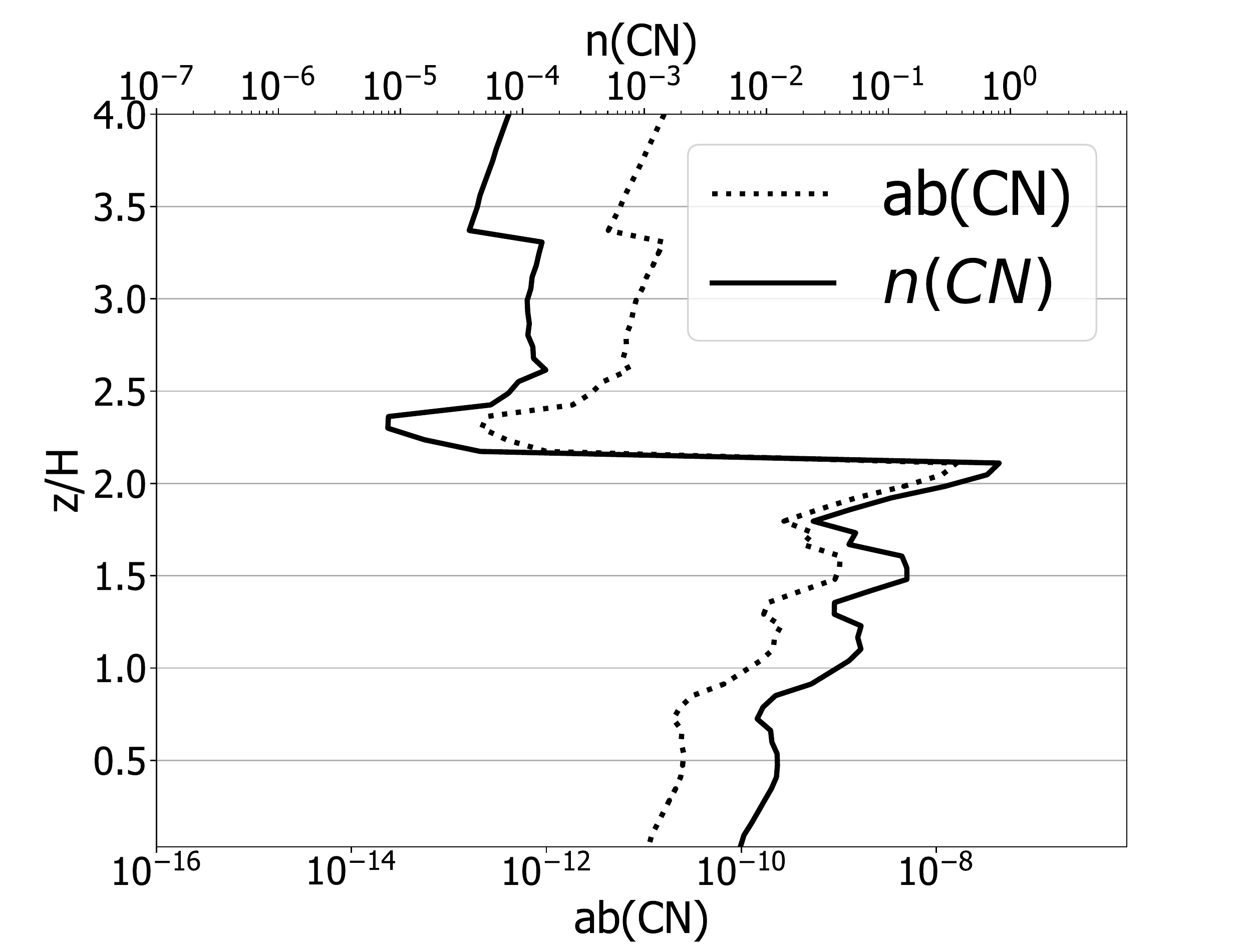} 
   \subcaption{M-LUV-B14} 
\end{subfigure}

\caption{Vertical profiles of H, $\mathrm{H_2}$, CO, CS and CN at 100 au from the star of the LUV single-grain model \shtg on the left column, and multi-grain models on middle and right columns. The dotted line is the abundance relative to H and the solid line is the density [$cm^{-3}$].}
\label{fig:m-100profile_low}
\end{figure*}

Figs.\,\ref{fig:s-100profile_low} and \ref{fig:m-100profile_low} show 
the vertical profiles at 100 au of the fractional abundances and number 
densities of gas-phase H, H$_2$, CO, CS and CN in LUV single-grain 
models and multi-grain models, respectively.

\begin{table*}
\centering
\caption{LUV cases: column densities of different species at 100 au for all models. \label{tab:luv_sigma}}
\begin{tabular}{l c c c c c}
\hline
\noalign{\smallskip}
& \revisionA{H} & \revisionA{H$_2$} & \revisionA{CO} & \revisionA{CS} & \revisionA{CN} \\
\noalign{\smallskip}
\hline	\hline
\noalign{\smallskip}
\revisionA{LUV-LH-Tg}  &  $9.70.10^{20}$ & $8.22.10^{22}$ & $4.11.10^{17}$ & $2.34.10^{12}$ & $1.01.10^{13}$	\\
\noalign{\smallskip}
\hline
\noalign{\smallskip}
\revisionA{LUV-HL-Ta} & $7.89.10^{21}$ & $7.87.10^{22}$ & $1.21.10^{18}$ & $3.75.10^{12}$ & $1.87.10^{12}$\\
\noalign{\smallskip}
\hline
\noalign{\smallskip}
\revisionA{LUV-B14-Ta} &      $9.68.10^{19}$ & $8.26.10^{22}$ & $1.47.10^{18}$ & $2.96.10^{11}$ & $2.25.10^{13}$	\\
\noalign{\smallskip}
\hline
\noalign{\smallskip}
\revisionA{M-LUV-LH}  &	     $6.70.10^{21}$ & $7.93.10^{22}$ & $1.79.10^{17}$ & $3.55.10^{13}$ & $5.87.10^{13}$ \\
\noalign{\smallskip}
\hline
\noalign{\smallskip}
\revisionA{M-LUV-B14} &     $1.34.10^{19}$ & $8.27.10^{22}$ & $1.15.10^{18}$ & $1.33.10^{13}$ & $4.67.10^{13}$ \\
\end{tabular}
\end{table*}

\end{appendix}

\end{document}